\documentstyle[preprint,aps]{revtex}
\begin{document}
\tightenlines
\draft

\title{Tetrad Gravity: III) Asymptotic Poincar\'e Charges, the Physical
Hamiltonian and Void Spacetimes.}

\author{Luca Lusanna}

\address{
Sezione INFN di Firenze\\
L.go E.Fermi 2 (Arcetri)\\
50125 Firenze, Italy\\
E-mail LUSANNA@FI.INFN.IT}

\author{and}

\author{Roberto De Pietri}

\address
{Centre de Physique Theorique\\
CNRS Luminy, Case 907\\
F-13288 Marseille, France\\
E-mail DEPIETRI@CPT.UNIV-MRS.FR}

\maketitle
\begin{abstract}

After a review on asymptotic flatness, a general discussion of asymptotic
weak and strong Poincar\'e charges in metric gravity is given with special
emphasis on the boundary conditions needed to define the proper Hamiltonian
gauge transformations and to get a differentiable Dirac Hamiltonian. Lapse
and shift functions are parametrized in a way which allows to identify their
asymptotic parts with the lapse and shift functions of Minkowski spacelike
hyperplanes. After having added the strong (surface integrals)
Poincar\'e charges to the Dirac Hamiltonian, it becomes the sum of a
differentiable Hamiltonian and of the weak (volume integrals) Poincar\'e
charges. By adding the ten Dirac extra variables at spatial
infinity, which identify special families of foliations with leaves asymptotic
(in a direction-independent way) to Minkowski spacelike hyperplanes, metric
gravity is extended to englobe Dirac's ten extra first
class constraints which identify the weak Poincar\'e charges with the
momenta conjugate to the extra variables. This opens the path to a
consistent deparametrization of general relativity to parametrized
Minkowski theories restricted to spacelike hyperplanes. The requirement of
absence of supertranslations restricts: i) the boundary conditions on the fields
and the gauge transformations to those identifying the family of
Christodoulou-Klainermann spacetimes; ii) the allowed 3+1 splittings of
spacetime to those whose spacelike leaves correspond to the Wigner hyperplanes
of Minkowski parametrized theories [on these leaves, named Wigner-Sen-Witten
hypersurfaces, there is a rule of parallel transport determined by the
Sen-Witten connection]. This approach is extended to tetrad gravity with its
interpretation of the superhamiltonian constraint as a generator of gauge
transformations: the last gauge variable is the momentum
$\rho (\tau ,\vec \sigma )$
conjugate to the conformal factor $q(\tau ,\vec \sigma )$ of the 3-metric
($q$ has to be determined as a solution of the constraint, namely of the
Lichnerowicz equation). In the 3-orthogonal gauges, the further addition of the
natural gauge fixing $\rho (\tau ,\vec \sigma )\approx 0$ leads to a
reduced phase space, which is parametrized by the canonical variables
$r_{\bar a}(\tau ,\vec \sigma )$, $\pi_{\bar a}(\tau ,\vec \sigma )$,
$\bar a=1,2$, defining the Hamiltonian
kinematical gravitational field. The evolution in
the parameter labelling the leaves of the foliation is generated by
the reduced ADM energy in the rest frame. Then, ``void
spacetimes", gauge equivalent to Minkowski spacetime in rectangular
coordinates, are defined and their realization in the 3-orthogonal gauges is
given by adding by hand the two pairs of second class
constraints $r_{\bar a}(\tau ,\vec \sigma )\approx 0$, $\pi_{\bar a}
(\tau ,\vec \sigma )\approx 0$, which are compatible with Einstein's
equations in absence of matter. These void spacetimes are the maximal
extension of the non-inertial Galilean reference frames of Newtonian gravity
to Einstein general relativity. Some comments on the quantization of the theory
are done.

\vskip 1truecm
\noindent \today
\vskip 1truecm

\end{abstract}

\newpage

\vfill\eject

\section
{Introduction}

In the first two papers\cite{russo1,russo2}, quoted as I and II respectively,
a new formulation of tetrad gravity
was given and its Dirac observables were found in 3-orthogonal coordinates on
the Cauchy hypersurfaces $\Sigma_{\tau}$ (assumed diffeomorphic to $R^3$) of
the 3+1 splitting of a globally hyperbolic spacetime $M^4$, asymptotically
flat at spatial infinity. This description is assumed valid, in a
variational sense, for an interval $\triangle \tau$ of the time parameter
$\tau$ labelling the leaves of the foliation associated with the 3+1
splitting, after which conjugate points of the 3-geometry on $\Sigma_{\tau}$
and/or 4-dimensional singularities develop due to Einstein's equations.

In II there was a preliminary discussion of the Hamiltonian
group of gauge transformations, whose generators are the 14 first class
constraints of the model. Also some preliminary statement about the
parametrization of the lapse and shift functions at spatial infinity was
made. However, only some indications on the boundary conditions for the fields
of tetrad gravity were given and there was no statement about the boundary
conditions on gauge transformations needed to define proper gauge
transformations; only some general remarks on the necessity of using weighted
Sobolev spaces to avoid isometries and Gribov ambiguities were done.

In the conclusions of II it was emphasized the difference between the
Hamiltonian group ${\bar {\cal G}}$ of gauge transformations and the group of
spacetime diffeomorphisms $Diff\, M^4$ in connection with the definition of
observables. There was the distinction between ``Hamiltonian kinematical
gravitational fields" (equivalence classes of spacetimes modulo the group of
Hamiltonian gauge transformations: in general every equivalence class contains
many gauge-equivalent 4-geometries, namely many standard ``kinematical
gravitational fields", elements of $Riem\, M^4/Diff\, M^4$) and ``Hamiltonian
Einstein or dynamical gravitational fields", which are those kinematical ones
which also satisfy the Hamilton-Dirac equations. It was shown that the
Hamiltonian dynamical gravitational fields coincide with the standard
``Einstein or dynamical gravitational fields", namely with a single 4-geometry
whose 4-metrics are solution of Einstein's equations; this 4-geometry is
parametrized in terms of one conformal 3-geometry. Therefore, on the space
of solutions of Einstein's equations the spacetime diffeomorphisms in $Diff\,
M^4$ become dynamical symmetries of Einstein's equations satisfying their
associated Jacobi equations (or linearized Einstein equations) and contain as a
subset the allowed Hamiltonian gauge transformations, namely those gauge
transformations which are dynamical symmetries of the Hamilton-Dirac equations.
There was also reported the criticism of Bergmann \cite{be} about general
covariance.

In this paper, after having identified the problems (like the existence of
supertranslations among the asymptotic symmetries) present in spacetimes
asymptotically flat at spatial infinity, we study the differentiability of the
Dirac Hamiltonian, how to define proper Hamiltonian gauge transformations and
the asymptotic conserved weak and strong Poincar\'e charges (the analogue of the
non Abelian charges of Yang-Mills theory) in metric gravity, and then in
tetrad gravity,  following the
ADM linearized theory \cite{adm}, Regge-Teitelboim \cite{reg}and
Beig-O'Murchadha\cite{reg1}.

We shall see that the final set of boundary conditions on the fields and the
gauge transformations, which allow the elimination of supertranslations, the
preservation of this property under the allowed gauge transformations
and a good definition of $\Sigma_{\tau}$-adapted tetrads at spatial
infinity, identify the family of spacetimes defined by Christodoulou and
Klainermann \cite{ckl} as
the one which admits a standard Hamiltonian formulation. All the quantities
must have a direction-independent limit at spatial infinity like in
Yang-Mills theory. In particular, following the approach of Dirac
\cite{p22,dirac} to add 10 degrees of freedom at spatial infinity
and 10 extra first class constraints to make them gauge variables, we get that
the allowed 3+1 splittings of $M^4$ must be foliations whose spacelike leaves
$\Sigma_{\tau}$ tend asymptotically to Minkowski spacelike hyperplanes in a
direction-independent way. The requirement of absence of supertranslations
further restricts the foliations to those (named Wigner-Sen-Witten) with the
leaves $\Sigma_{\tau}$ asymptotically normal to the ADM 4-momentum (definition
of rest frame). On these hypersurfaces there is a rule of parallel transport
determined by the Sen-Witten connection. The asymptotic Minkowskian
Wigner hyperplanes\cite{lus1} identify ``geometrical and dynamical" preferred
observers at spatial infinity and there is a realization of Bergmann's
expectation that it should be possible to restrict proper coordinate
transformations to contain an invariant Poincar\'e subgroup plus
asymptotically trivial diffeomorphisms.
The evolution in the parameter labelling the leaves of
the foliation and identified with the rest-frame time of a decoupled observer
(point clock) sitting in the ``external center of mass" of the universe, is
governed by the ADM energy.

There is a justification why the superhamiltonian constraint has to be solved
in the conformal factor of the 3-metric, becoming the reduced Lichnerowicz
equation, so that the last gauge variable of tetrad gravity is its
conjugate momentum. Since the ADM theory is independent from the choice of the
3+1 splitting of spacetime, it turns out that the transition from an allowed
3+1 splitting to another one is realized with a gauge transformation
generated by the superhamiltonian constraint.

Then, there is the definition of the equivalence class of ``void spacetimes",
namely those spacetimes in which the physical degrees of freedom of the
gravitational field have been explicitly eliminated. These spacetimes are
gauge equivalent to Minkowski spacetime in rectangular coordinates due to the
Hamiltonian group of gauge transformations. They turn out to have the spacelike
Cauchy surfaces 3-conformally flat, to have vanishing Poincar\'e charges and
to require a different Hamiltonian treatment without adding Dirac's extra
variables. In accord with the fact that in parametrized Minkowski theories
Wigner hyperplanes may be defined only in presence of matter, void spacetimes
do not allow the definition of Wigner-Sen-Witten hypersurfaces.

This concludes the study of the Hamiltonian formulation and canonical
reduction of
classical tetrad gravity in absence of matter. All its conceptual problems have
found an explanation and there is an indication on how to quantize the theory
in terms of Dirac's observables after a complete breaking of general covariance.
The study of the theory in presence of matter and its linearization in the
3-orthogonal gauge
will be done in future papers. The main obstacle in this standard
Hamiltonian treatment is the lack of solutions of the Lichnerowicz equation:
this explains why research has shifted towards either Ashtekar's programme or
superstring theory. We hope that after this clarification of the conceptual
aspects of tetrad gravity and after having found the bridge both to parametrized
Minkowski and Newtonian theories, the search of exact and/or approximate
solutions  of the Lichnerowicz equation will start again, also because our
results imply a unification of the techniques between Yang-Mills theory and
tetrad gravity. Moreover, the problems with the gauge fixings in numerical
gravity are deeply connected with a better understanding of these solutions.

In Section II there is a review of the properties of spacetimes asymptotically
flat at either null or spatial infinity in metric gravity.

In Section III, after a review of Poincar\'e charges and of the problem of
supertranslations in metric gravity, we give the boundary conditions on the
3-metric and the parametrization of the lapse and shift functions allowing to
get a differentiable Dirac Hamiltonian and to define proper gauge
transformations. Weak and strong Poincar\'e charges emerge naturally. Two
scenarios for metric gravity are delineated, one without and the other with
Dirac's ten extra degrees of freedom at spatial infinity. In the second case,
ten extra first class constraints are added so that the ten extra degrees of
freedom are eliminable gauge variables. Therefore, one has the same physical
degrees of freedom of the gravitational field and, moreover, one can make
contact with parametrized Minkowski theories on spacelike hyperplanes, which
are the natural limit for $G \rightarrow 0$ when matter is present.

In Section IV it is shown that the analogue of Minkowski spacelike hyperplanes
are hypersurfaces, tending to them at spatial infinity in a
direction-independent way, and having a rule of
parallel transport, which is determined by the Sen-Witten connection for those
hypersurfaces (named Wigner-Sen-Witten hypersurfaces) tending to Wigner
hyperplanes in Minkowski spacetime in absence of gravity.

In Section V the previous results are extended to the formulation of tetrad
gravity given in I and II, and, then, are specialized to the 3-orthogonal
gauges. After the introduction of a natural gauge fixing for the
superhamiltonian constraint, the canonical degrees
of freedom of the gravitational field in this special gauge are found and it is
shown that the ADM energy plays the role of the physical Hamiltonian for the
evolution in the parameter labelling the leaves of the foliation defining the
3+1 splitting of spacetime.

In Section VI we define ``void spacetimes" as those special spacetimes in
which there are no canonical degrees of freedom of the gravitational field.
They are gauge equivalent to Minkowski spacetime in rectangular coordinates by
using the Hamiltonian gauge transformation group according to constraint theory.
They contain the analogue of the inertial forces of Newtonian gravity appearing
when non-orthogonal (maybe time-dependent) coordinates are used for Euclidean
space (Galilean non-inertial reference frames).

In the Conclusions there is a summary of the main results, the description of
our way out from the ``problem of time", and some remarks on how to quantize
(at least formally) tetrad gravity in the
special 3-orthogonal gauge, after a complete
breaking of general covariance, without problems with the definition of the
physical scalar product and with the implication that volumes, areas and
lengths on the Cauchy surfaces of the mathematical spacetime (in this
completely fixed gauge) are going to be quantized (even if in a way different
from Ashtekar's one). There is also a comment on Komar-Bergmann's proposal of
identifying a posteriori the ``physical points" of spacetime with their
``individuating fields".

In Appendix A there is the application of the second Noether theorem to ADM
metric gravity and the explicit check that the ADM action is not invariant under
spacetime diffeomorphisms.

In Appendix B there is a review of proposals for the reduced phase space of
metric gravity.

In Appendix C there is a review of spinors on $M^4$ and $\Sigma_{\tau}$.

In Appendix D, for the sake of completeness, there is the application of the
second Noether theorem to the Einstein-Hilbert action, the definition of Komar
superpotential and of the energy-momentum pseudotensors of metric gravity.

In Appendix E there is the expression of 3- and 4-tensors in void spacetimes.

In Appendix F there is the definition of a set of null tetrads natural for the
Hamiltonian formulation of tetrad gravity.

In Appendix G there is the connection with the post-Newtonian approximation.

\vfill\eject

\section{Asymptotic Flatness at Spatial Infinity.}

The definition of an ``isolated system" in general relativity [see Ref.
\cite{p14} for a review] is a difficult problem, since there is neither a
background flat metric ${}^4\eta$ nor a natural global inertial coordinate
system allowing to define a preferred radial coordinate $r$ and a limit
${}^4g_{\mu\nu}\, \rightarrow \, {}^4\eta_{\mu\nu} +O(1/r)$ for $r\,
\rightarrow \, \infty$ along either spatial or null directions. Usually, one
considers an asymptotic Minkowski metric ${}^4\eta_{\mu\nu}$ in rectangular
coordinates [${}^4\eta_{\mu\nu}=\epsilon (+---)$, with $\epsilon =+1$ for the
particle physics convention and $\epsilon =-1$ for the general relativity
one] and tries to get asymptotic statements with various types of
definitions of $r$. However, it is difficult to correctly specify the limits
for $r\, \rightarrow \, \infty$ in a meaningful, coordinate independent way.

Therefore, Penrose \cite{p2} introduced the notions of asymptotic flatness
at null infinity (i.e. along null geodesics) and of asymptotic
simplicity with his conformal completion approach [a spacetime $(M^4,{}^4g)$
is asymptotically simple if: i) all its null geodesics are complete; ii) there
exists a smooth function $\Omega \geq 0$ going to zero in both directions along
null geodesics; iii) there exists a smooth extension ${\tilde M}^4=M^4\cup
{\cal I}$ of $M^4$ with boundary ${\cal I}={\cal I}^{+}\cup {\cal I}^{-}$,
${\cal I}^{+}\cap {\cal I}^{-}=0$; iv) $\Omega$ extends smoothly to ${\cal I}$,
where $\Omega =0$, $d\Omega \not= 0$; v) ${}^4{\tilde g}= \Omega^2\, {}^4g$
is a smooth 4-metric on ${\tilde M}^4$; vi) each null geodesic acquires an
endpoint in the past on ${\cal I}^{-}$ and in the future on ${\cal I}^{+}$].
See chapter 11 of Ref.\cite{wald} for the conformal completion of Minkowski
spacetime [through a conformal isometry it becomes an open region (with
boundary $\Omega =0$) of the Einstein static spacetime $S^3\times R$; the
boundary contains two points $i^{\pm}$ for the future and past timelike
infinity, one point $i^o$ for spacelike infinity, and two regions ${\cal I}
^{\pm}$ for future and past null infinity] and for a review of
asymptotic flatness. See also Ref.\cite{p0}
for definitions of ``asymptotically simple and weakly asymptotically simple"
spacetimes, intended to ensure that the asymptotic structure be globally the
same as that of Minkowski spacetime.

Instead, a coordinate-independent definition of asymptotic flatness at spatial
infinity in terms of the ``large distance" behaviour of initial data on a
Cauchy surface was introduced by Geroch \cite{p1}.

Then, the two definitions of asymptotic flatness at null and spatial infinity
were unified in the SPI formalism of Ashtekar and Hanson\cite{p3}
(see the bibliography of this reference for other
approaches like the projective one \cite{p4}, which,  however, has
problems with Schwarzschild spacetime).
 Essentially, in the SPI approach, the spatial infinity of the
spacetime $M^4$ is compactified to a point $i^o$ [instead in the projective
approach of Ref.\cite{p4}
a timelike hyperboloid is the boundary of $M^4$] and fields on $M^4$ have
direction-dependent limits at $i^o$ [this state of affairs implies a peculiar
differential structure on $\Sigma_{\tau}$ and awkward differentiability
conditions of the 4-metric]. Subsequently, in Ref.\cite{p5}, a new kind of
completion, neither conformal nor projective, is developed
by Ashtekar and Romano: now the boundary
of $M^4$ is a unit timelike hyperboloid like in the projective approach,
which, however, has a well defined contravariant normal in the completion
[different conformal rescalings of the 4-metric ${}^4g \mapsto {}^4\tilde g=
\Omega^2\, {}^4g$ ($\Omega \geq 0$, $\Omega =0$ is the boundary 3-surface of the
unphysical spacetime ${\tilde M}^4$) and
of the normal $n^{\mu} \mapsto {\tilde n}^{\mu}=
\Omega^{-4}\, n^{\mu}$ are needed]; now, there is
no need  of awkward differentiability conditions. While in the SPI framework
each hypersurface $\Sigma_{\tau}$ has the sphere at spatial infinity
compactified at the same point $i^o$, which is the vertex for both future
${\cal I}^{+}$ (scri-plus) and past ${\cal I}^{-}$ (scri-minus) null infinity,
these properties are lost in the new approach: each $\Sigma_{\tau}$ has as
boundary at spatial infinity the sphere cut by $\Sigma_{\tau}$ in the
timelike hyperboliod; there is no relation between the timelike hyperboloid at
spatial infinity and ${\cal I}^{\pm}$. This new approach simplifies
the  analysis of Ref.\cite{p6} of uniqueness (modulo the logarithmic
translations of Bergmann\cite{p7}) of the completion at spacelike infinity.

There is a coordinate-dependent formalism
of Beig and Schmidt\cite{p8} [developed to avoid the
awkward differentiability conditions of the SPI framework and using polar
coordinates like the standard hyperbolic ones for Minkowski spacetime and
agreeing with them at ``first order in 1/r"], whose relation to the new
completion is roughly the same as that between Penrose's coordinate-independent
approach to null infinity\cite{p2} and Bondi's approach\cite{p9} based on null
coordinates. The class of spacetimes studied in Ref.\cite{p8} [called
``radially smooth of order m" at spatial infinity] have 4-metrics of the type
\hfill\break
\hfill\break
$ds^2=d\rho^2(1+{{{}^1\sigma}\over {\rho}}+{{{}^2\sigma}\over {\rho^2}}+..)^2+
\rho^2({}^oh_{rs}+{1\over {\rho}}\, {}^1h_{rs}+..)d\phi^rd\phi^s$,\hfill\break
\hfill\break
where
${}^oh_{rs}$ is the 3-metric on the unit timelike hyperboloid and ${}^n\sigma$,
${}^nh_{rs}$, are functions on it. There are coordinate charts $x^{\sigma}$ in
$(M^4,{}^4g)$ where the 4-metric becomes \hfill\break
\hfill\break
${}^4g_{\mu\nu}={}^4\eta_{\mu\nu}+
\sum^m_{n=1} {1\over {\rho^n}}\, {}^nl_{\mu\nu}({{x^{\sigma}}\over {\rho}})+
O(\rho^{-(m+1)})$.\hfill\break
\hfill\break
See Ref.\cite{rich}  for the status of the conformal field equations, derived
from Einstein's equations, which arise in the study of the compatibility of
Penrose's conformal completion approach with Einstein's equations; the final
outcome is a description in which $i^o$ blows up to a 2-sphere at spatial
infinity (interpretable as the space of spacelike directions at $i^o$),
which intersects future ${\cal I}^{+}$ and past ${\cal I}^{-}$ null infinity
in two sets $I^{\pm}$. It is an open question whether the concepts of
asymptotic simplicity and conformal completion are too strong requirements.
For instance, the Christodoulou and Klainerman \cite{ckl} result on the
nonlinear gravitational stability of Minkowski spacetime implies a peeling
behaviour of the conformal Weyl tensor near null infinity which is weaker than
the peeling behaviour implied by asymptotic simplicity [see Ref.\cite{p9,p2}]
and this could mean that asymptotic simplicity can be established only, if at
all, with conditions stronger than those required by these authors.
In Ref.\cite{ckl} one studies the existence of global, smooth, nontrivial
solutions to Einstein's equations without matter, which look, in the large,
like the Minkowski spacetime (without singularities and black holes; since
the requirements needed for the existence of a conformal completion are not
satisfied, it is possible to evade the singularity theorems), are close
to Minkowski spacetime in all directions in a precise manner (for developments
of the initial data sets uniformly close to the trivial one) and admit
gravitational radiation in the Bondi sense. These author's
 reformulate Einstein's
equations with the ADM variables (there are four constraint equations plus the
equations for $\partial_{\tau}\, {}^3g_{rs}$ and $\partial_{\tau}\, {}^3K_{rs}$;
see Section V of I for their phase space counterpart), put the shift functions
equal to zero (the lapse function is assumed equal to 1 at spatial infinity,
but not everywhere because, otherwise, one should have a finite time breakdown)
and add the maximal slicing condition ${}^3K=0$. Then, they assume the
existence of a coordinate system $\vec \sigma$ near spatial infinity on the
Cauchy surfaces $\Sigma_{\tau}$ and of smoothness properties for ${}^3g_{rs}$,
${}^3K_{rs}$, such that for $r=\sqrt{ {\vec \sigma}^2}\, \rightarrow \, \infty$
the initial data set $(\Sigma_{\tau},{}^3g_{rs},{}^3K_{rs})$ is ``strongly
asymptotically flat", namely [$f(\vec \sigma )$ is $o_m(r^{-k})$ if $\partial
^l f =o(r^{-k-l})$ for $l=0,1,..,m$ and $r\, \rightarrow \, \infty$]

\begin{eqnarray}
{}^3g_{rs}&=&(1+{M\over r}) \delta_{rs} +o_4(r^{-3/2}),\nonumber \\
{}^3K_{rs}&=& o_3(r^{-5/2}),
\label{II1}
\end{eqnarray}

\noindent where the leading term in ${}^3g_{rs}$ is called the Schwarzschild
part of the 3-metric, also in absence of matter; this asymptotic behaviour
ensures the existence of the ADM energy and angular momentum and the
vanishing of the ADM momentum (``center-of-mass frame"). The addition of a
technical global smallness assumption on the strongly asymptotically flat
initial data leads to a unique, globally hyperbolic, smooth and geodesically
complete solution of Einstein's equations without matter, which is
globally asymptotically flat in the sense that its Riemann curvature tensor
approaches zero on any causal or spacelike geodesic. It is also shown that the
2-dimensional space of the dynamical degrees of freedom of the gravitational
field at a point (the reduced configuration space, see the Conclusions of II)
is the space of trace-free symmetric 2-covariant tensors on a 2-plane. A serious
technical difficulty (requiring the definition of an `optical function' and
reflecting the presence of gravitational radiation in any nontrivial
perturbation of Minkowski spacetime) derives from the `mass term' in the
asymptotic Schwartzschild part of the 3-metric: it has the long range effect
of changing the asymptotic position of the null geodesic cone relative to the
maximal (${}^3K=0$) foliation (these cones are expected to diverge
logarithmically from their positions in flat spacetime and to have their
asymptotic shear drastically different from that in Minkowski spacetime).

Other reviews of the ``problem of consistency", i.e. whether the geometric
assumptions inherent in the existing definitions of asymptotic flatness are
compatible with Einstein equations, are given in Refs.\cite{p14,p15}, while in
Ref.\cite{p16} a review is given about spacetimes with gravitational radiation
(nearly all the results on radiative spacetimes are at null infinity, where,
for instance, the SPI requirement of vanishing of the pseudomagnetic part of the
Weyl tensor to avoid supertranslations is too strong and destroys radiation).
See Ref.\cite{bicak} for a recent review of the state of the art.

Let us now consider the problem of asymptotic symmetries \cite{wald} and of
the associated conserved asymptotic charges containing the ADM Poincar\'e
charges (which will be discussed in the next Section).

Like null infinity admits an infinite-dimensional group (the BMS group
\cite{p9}) of ``asymptotic symmetries", the SPI formalism admits an even bigger
group, the SPI group \cite{p3}, of such symmetries. Both BMS and SPI algebras
have an invariant 4-dimensional subalgebra of translations, but they also have
invariant infinite-dimensional Abelian subalgebras (including the translation
subalgebra) of so called ``supertranslations" [angle (or direction)-dependent
translations]. Therefore, there is an infinite number of copies of Poincar\'e
subalgebras in both BMS and SPI algebras, whose Lorentz parts are conjugate
through supertranslations [the quotient of BMS and SPI groups with respect to
supertranslations is isomorphic to a Lorentz group]. All this implies that
there  is no unique definition of Lorentz generators and that in general
relativity one cannot define intrinsically angular momentum and the Poincar\'e
spin Casimir, so important for the classification of particles in Minkowski
spacetime.
In Ref.\cite{mcca} it is shown that the only known Casimirs of the BMS group
are $p^2$ and one its generalization involving supertranslations. While
Poincar\'e asymptotic symmetries correspond to the ten Killing fields of the
Minkowski spacetime (to which an asymptotically flat spacetime tends
asymptotically in some way depending on the chosen definition of asymptotic
flatness), supertranslations are ``angle-dependent translations", which come
just as close to satisfying Killing's equations asymptotically as any Poincar\'e
transformation \cite{wald}.
The problem seems to be that all known function spaces, used
for the 4-metric and for Klein-Gordon and electromagnetic fields, do not put
any restriction on the asymptotic angular behaviour of the fields, but only
restrict their radial decrease. Due to the relevance of the Poincar\'e group
for particle physics in Minkowski spacetime, and also to have a good definition
of angular momentum in general relativity [see Refs.\cite{p10,wald,p3} for
this topic], one usually restricts the class of spacetimes with boundary
conditions such that supertranslations are not allowed to exist. In the SPI
framework \cite{p3}, one asks that the pseudomagnetic part of the limit of the
conformally rescaled Weyl tensor vanishes at $i^o$.
In Ref.\cite{p11} a 3+1 decomposition is made of the SPI
framework; after having reexpressed the conserved quantities at $i^o$ in terms
of canonical initial data, it is shown that to remove ambiguities connected with
the supertranslations one must use stronger boundary conditions again implying
the vanishing of the pseudomagnetic part of the Weyl tensor.

A related approach to these problems is given by Anderson in Ref.\cite{reg2}.
He proved a slice theorem for the action of spacetime
diffeomorphisms asymptotic to Poincar\'e transformations on the set of
asymptotically flat solutions of Einstein's equations in the context of spatial
infinity, maximal slicing and asymptotic harmonic coordinates (as gauge
conditions). There is a heuristic extension of the method of reduction of
dynamical systems with symmetries to the diffeomorphism group. The standard
method considers a symplectic manifold $(P,\omega )$ with an action of a Lie
group G on P. One assumes that the action of G on P admits a momentum map
$J: P\rightarrow g^{*}$, $J(x)=\xi^{*}$ [$g^{*}$ is the dual of the Lie algebra
$g$ of G], such that, for each vector $\xi \in g$, the quantity
$J(\xi )(.)=< J,\xi >(.) :P\rightarrow R$ is the
Hamiltonian function (constant of the motion) generating the symplectic action
on P. The space (``level set" for the action of G on P) $J^{-1}(\xi )\subset P$ is
a manifold except at points where the action of G has a nontrivial isotropy
group (where at most quadratic singularities occur;
see II for this problematic). If $G_{\xi}$ is the
isotropy group of $\xi \in g$ with respect to the coadjoint action of G on
$g^{*}$, then the space $J^{-1}(\xi )/G_{\xi}$ is a stratified symplectic
manifold with at most quadratic singularities (the reduced phase space or the
space of dynamical degrees of freedom).

For metric general relativity in the spatially compact case this programme has
been carried out in Ref.\cite{ise}. If $M^4=\Sigma \times R$ with $\Sigma$ a
compact, orientable 3-manifold, if ${}^4g$ is a Lorentz metric on $M^4$
satisfying Einstein's equations and if $Riem\, ({}^3g)$ is the set of
Riemannian metrics on $\Sigma$, then the group G is $Diff\, M^4$ and the
momentum map  consists of the ADM
supermomentum and superhamiltonian constraints $\phi =\lbrace {\tilde {\cal
H}},{}^3{\tilde {\cal H}}^r\rbrace : T^{*}Riem\, ({}^3g) \rightarrow \Lambda^o
_d\times \Lambda^1_d$ [the dual of the space of infinitesimal diffeomorphisms
interpreted as the space of lapses and shifts since they are the coefficients of
the constraints in the Dirac Hamiltonian], even if the action of $Diff\,
\Sigma$ on the constraint set is not a true group action since there are
structure functions (see Section V of I).
The space of solutions of Einstein's equations is fibered
over $\phi^{-1}(0)\in T^{*}Riem\, ({}^3g)$, which is smooth at $({}^3g,
{}^3{\tilde \Pi})$ if and only if the initial data $({}^3g,{}^3\tilde \Pi )$
correspons to a solution ${}^3g$ with no Killing field. The reduced phase space
$\phi^{-1}(0)/Diff\, M^4$ has been constructed\cite{ise} and turns out to be a
stratified symplectic ILH manifold.

In the spatially asymptotically flat case, the group G becomes the group of
those diffeomorphisms which preserve the conditions for asymptotic flatness and
the nature of this group depends strongly on the precise asymptotic conditions.
Apart from the compactification schemes of Geroch\cite{p1} and of
Ashtekar-Hansen\cite{p3}, 3 main types of asymptotic conditions have been
studied: i) the ``finite energy" condition of O'Murchadha\cite{p12}; ii) the
York ``quasi isotropic" (QI) gauge conditions\cite{yoyo} (see Appendix B);
iii) the conditions of the type introduced by Regge-Teitelboim\cite{reg}
with the ``parity conditions" further refined by Beig-O'Murchadha\cite{reg1}
(see the next Section) plus the gauge conditions of maximal slices
and ``3-harmonic" asymptotic coordinates [their existence was shown
in Ref.\cite{p12}]. These 3 types of asymptotic conditions have quite different
properties. In the case of the finite energy conditions of i), one finds that
the group which leaves the asymptotic conditions invariant is a semidirect
product $S\, |\times L$, where L is the Lorentz group and S consists of
diffeomorphisms $\eta$ such that roughly $D^2\eta \in L^2$ [i.e. S contains
space- and time- translations]. Under these conditions, it does not appear to be
possible to talk about Hamiltonian dynamics. For a general element of the Lie
algebra of $S\, |\times L$, the corresponding momentum integral does not
converge, although for the special case of space- and time-translations the ADM
4-momentum is well defined.

York QI gauge conditions of Ref.\cite{yoyo} have the desirable feature that ``no
supertranslations" are allowed, but a more detailed analysis reveals that
without extra conditions, the transformations corresponding to boosts are not
well behaved; in any case, the QI asymptotic conditions do not give a well
defined angular and boost momentum and therefore are suitable only for the study
of diffeomorphisms asymptotic to space- and time- translations.

To get a well defined momentum for rotations and boosts, Anderson defines
asymptotic conditions which contain the parity conditions of Ref.\cite{reg1},
but he replaces the 3-harmonic coordinates used in this paper
with York's QI conditions. The space of transformations $Diff_P\, M^4$ which
leaves invariant the space of solutions of the
Einstein equations satisfying the parity conditions is a semidirect product
$Diff_P\, M^4=Diff_S\, M^4\, |\times P$, where P is the
Poincar\'e group and $Diff_S\, M^4$
denotes the space of diffeomorphisms which are asymptotic to supertranslations,
which in this case are O(1) with odd leading term. When the QI conditions are
added, the $Diff_S\, M^4$ part is restricted to $Diff_I\, M^4$,
the space of diffeomorphisms
which tend to the identity at spatial infinity [this result cannot be obtained
with the finite energy conditions\cite{p12} or from boost theorems
\cite{chris}]. In this way one obtains a realization of Bergmann's criticism
\cite{be} of general covariance: the group of coordinate transformations is
restricted to contain an invariant Poincar\'e subgroup plus asymptotically
trivial diffeomorphisms (similar to the gauge transformations of
electromagnetism). It can be shown that, when using the parity conditions, the
lapses and shifts corresponding to the group of supertranslations S have
``zero momentum".
Thus, assuming the QI conditions, the ADM momentum appears as the
momentum map with respect to the Poincar\'e group. Note that the classical form
of the ADM momentum is correct only using the restrictive assumption of parity
conditions, which are nontrivial restrictions ``not only" on the gauge freedom
``but also" on the asymptotic dynamical degrees of freedom of the gravitational
field [this happens also with Ashtekar-Hansen asymptotic condition on the Weyl
tensor]. Call the total
momentum map $\phi_E=\phi +E: T^{*}Riem\, ({}^3g) \rightarrow (\Lambda^o_d
\times \Lambda^1_d)\times p^{*}$, where $\Lambda^o_d\times \Lambda^1_d$ is the
3+1 version of the dual of the Lie algebra of
$Diff_I\, M^4$ and $p^{*}$ is the dual
of the Poincar\'e Lie algebra p; E will consist of certain integrals over
spheres at spatial infinity. One now expects (assuming the QI conditions) that:
i) $\phi^{-1}(0)/Diff_I\, M^4$ is a symplectic manifold (no isometry can be in
$Diff_I\, M^4$);
ii) for $\xi \in p^{*}$, the spaces $\phi^{-1}_E(0\times \xi )$ and $\phi_E
^{-1}(0\times \xi )/Diff_I\, M^4$ are manifolds except at points corresponding
to flat space or spaces with rotational symmetries
[$Diff_{P_{\xi}}\, M^4=Diff_I\, M^4
\times P_{\xi}$, where $P_{\xi}$ denotes the isotropy group of $\xi$ in P].

By assuming the validity of the conjecture on global existence of solutions of
Einstein's equations and of maximal slicing (in the framework of York
reduction; see Appendix C of II)
and working with Sobolev spaces with ``radial smoothness", Anderson
demonstrates a ``slice" theorem, according to which, assumed the parity and QI
conditions [which exclude the logarithmic translations of Bergmann\cite{p7}],
for every solution ${}^3g_o$ of Einstein's equations one has that: i) the gauge
orbit of ${}^3g_o$ is a closed $C^1$ embedded submanifold of the manifold of
solutions; ii) there exists a submanifold containing ${}^3g_o$ which is a
``slice for the action of $Diff_I\, M^4$" (see Appendix B of Anderson's paper
for the definition of slice).

York's QI conditions should be viewed as a slice condition which fixes part of
the gauge freedom at spatial infinity: i)  the $O(1/r^2)$ part of the trace of
${}^3{\tilde \Pi}^{rs}$ must vanish; ii) if ${}^3g={}^3f+{}^3h$ (${}^3f$ is a
flat metric) and if ${}^3h={}^3h_{TT}={}^3h_T+L_f(W)$ is the York decomposition
of ${}^3h$ with respect to ${}^3f$ (see Appendix C of II), then the O(1) part
of the longitudinal quantity W must vanish. In this way, one selects a QI
asymptotically flat metric ${}^3g_{QI}$ and a ``preferred frame at spatial
infinity" (remember Bergmann's criticism to general covariance\cite{be},
quoted in the Conclusions of II), i.e. preferred spacelike
hypersurfaces $\Sigma_{QI}$ [they can be mapped onto the space $\Sigma_{\cal K}$
of cross sections of the unit timelike hyperboloid ${\cal K}$ (with which
Beig\cite{p8} completes $M^4$ at spatial infinity), corresponding to the set
of intersections of ${\cal K}$ by spatial hyperplanes in $R^4$], as shown in
a lemma of Ref.\cite{reg2}.

Then, in the Anderson paper \cite{reg2}
there are comments on the open problem of the
validity of the conjecture of existence of maximal slicings [see
Appendix C of II; in this case the QI conditions are natural], due to the
discovery of topological obstructions which invalidate the demonstrations of
certain theorems.
Anderson\cite{reg2} showed that to put real control on boosts and rotations,
one should add the parity conditions to the weighted Sobolev spaces [this
reduces the class of solutions of Einstein's solutions]

Let us add other results connected with the previous problematic and with the
existence of the ADM Lorentz generators.

In Ref.\cite{chris} on the boost problem in general relativity,
Christodoulou and Murchadha show
(using weighted Sobolev spaces) that a very large class of asymptotically
flat initial data for Einstein's equations have a development which includes
complete spacelike surfaces boosted relative to the initial surface.
Furthermore, the asymptotic fall off [${}^3g-{}^3f\in W^{2,s,\delta+1/2}
(\Sigma )$, ${}^3K\in W^{2,s-1,\delta +1/2}(\Sigma )$, $s\geq 4$, $\delta > -2$]
is preserved along these boosted surfaces and there exist a global system of
harmonic coordinates on such a development.

As noted in Ref.\cite{p5}, the results of Ref.\cite{chris} suffice to establish
the existence of a large class off spacetimes which are asymptotically flat at
$i^o$ (in the sense of Ref.\cite{p3}) in all spacelike directions along a family
of Cauchy surfaces related to one another by ``finite" boosts (it is hoped that
new results will allow to put control also on ``infinite" boosts). The
situation is unsettled with regard the existence of spacetimes admitting both
$i^o$ (in the sense of Ref.\cite{p3}) as well as smooth ${\cal I}^{\pm}$.

In Ref.\cite{p12}, by making use of the previous paper
\cite{chris} and of weighted Sobolev
spaces, it is shown that the 1/r behaviour of the metric in the asymptotically
flat case (used in the boost and positive energy theorems) can be relaxed to
$r^{-1/2+\epsilon}$ without destroying the existence of a well defined,
conserved, Lorentz-covariant,timelike, future-pointing energy-momentum 4-vector.
The 1/r behaviour is connected to the fact that the ADM conserved quantities
are expressed as surface integrals at infinity which must be finite. However,
one can use the Gauss theorem to turn the surface integrals into volume
integrals: the leading term in the volume integral expression vanishes, due to
the constraints, and thus the volume integral expressions for the energy-
momentum converges (on the constraint manifold) using the weaker fall off
condition $r^{-1/2+\epsilon}$ [in electrodynamics one has $Q=e\int_{\partial V}
\vec E\cdot d\vec S\, {\buildrel \circ \over =}\, Q^{'}=
e\int_V div\, \vec E d^3x \approx \int_V \rho d^3x$ ($Q$ and $Q^{'}$ are the
strong and weak improper charges, see the paper c) in Ref. \cite{sha});
so that $\vec E =O(1/r^2)$ required by the finiteness of Q is replaced by the
$\vec E$-independent requirement $\rho =O(1/r^3)$ by using the Gauss law
constraint]. The falloff conditions of this paper are ${}^3g-{}^3f\in
W^{2,s,\delta}(R^3)$, ${}^3K\in W^{2,s-1,\delta +1}(R^3)$, $s\geq 4$, $\delta
> -1$.

In Ref.\cite{p13}, Chru\'sciel says that for asymptotically flat metrics
${}^3g={}^3f+O(r^{-\alpha})$, ${1\over 2} < \alpha \leq 1$, it is not proved
the ``asymptotic symmetry conjecture" that, given any two coordinate systems of
the previous type, all twice-differentiable coordinate transformations
preserving
these boundary conditions are of type $y^{\mu}=\Lambda^{\mu}{}_{\nu}x^{\nu}+
\zeta^{\mu}$ [a Lorentz transformation + a supertranslation $\zeta =O(r^{1-
\alpha})$]: this would be needed for the ADM 4-momentum to be Lorentz covariant.
By defining $P_{\mu}$ in terms of Cauchy data on a 3-end N [a spacelike
3-surface $\Sigma$ minus a ball] on which ${}^3g={}^3f+O(r^{-\alpha})$, one
can evaluate the invariant mass $m(N)=\sqrt{\epsilon P^{\mu}P_{\mu}}$. Then,
provided the hypersurfaces $x^o=const.$ ($N_1$), $y^o=const.$ ($N_2$), lie
within a ``finite" boost of each other or if the metric is a no-radiation
metric, one can show the validity of the ``invariant mass conjecture"
$m(N_1)=m(N_2)$ for
metrics satisfying vacuum Einstein equations. The main limitation is the lack
of knowledge of long-time behaviour of Einstein's equations. Ashtekar-Hansen
and Beig-O'Murchadha requirements are much stronger and restrictive than what
is compatible with Einstein's equations.

Since there is no agreement among the various viewpoints on the
coordinate-independent definition of asymptotic flatness at spatial infinity,
since we are interested in the coupling of general relativity to the standard
SU(3)xSU(2)XU(1) model and since we wish to recover the theory in Minkowski
spacetime if we put G=0 (the deparametrization problem of general relativity,
only partially solved in Ref.\cite{isha} by using coordinate gauge conditions
and studied in Ref.\cite{russo4}), in this paper we shall use a
coordinate-dependent approach and we shall work in the framework of Refs.
\cite{reg,reg1}, in which supertranslations may be eliminated
and there is a well defined Poincar\'e asymptotic symmetry group due to the
choice of the boundary conditions and of a certain class of gauge-fixings for
the supermomentum constraints of metric gravity. This will also be connected
with Bergmann's remarks\cite{be} on the existence of preferred coordinate
systems breaking general covariance (see the Conclusions of II and next
Sections).

In particular, the chosen boundary conditions and gauge-fixings
will imply an angle
(i.e. direction)-independent asymptotic limit of the canonical variables, just
as it is needed in Yang-Mills theory to have well defined covariant non-Abelian
charges\cite{ym,lusa} [as shown in Ref.\cite{lusa}, one needs a set of
Hamiltonian (not manifestly covariant except in the reformulation on spacelike
hypersurfaces) boundary conditions both for the fields and the gauge
transformations in the Hamiltonian gauge group ${\bar {\cal G}}$, implying
angle-independent limits at spatial infinity; it is also suggested that the
elimination of Gribov ambiguity requires the use of the following weighted
Sobolev spaces\cite{moncrief}
: ${\vec A}_a, {\vec E}_a \in W^{p,s-1,\delta -1}$, ${\vec B}_a
\in W^{p,s-2,\delta +2}$, ${\bar {\cal G}} \in W^{p,s,\delta}$, with
$p > 3$, $s \geq 3$, $0 \leq \delta \leq 1-{3\over p}$].
This is an important point for a future unified
description of general relativity and of the standard  model.

In particular, following Ref.\cite{hh} (see also Appendix A), we will assume
that at spatial infinity there is a 3-surface $S_{\infty}$ [not necessarily
a timelike hyperboloid at this stage of development], which intersects
orthogonally the Cauchy surfaces $\Sigma_{\tau}$ [their normals $l^{\mu}(\tau
,\vec \sigma )$ at spatial infinity, $l^{\mu}_{(\infty )\Sigma}$, are tangent
to $S_{\infty}$] along 2-surfaces $S^2_{\tau ,\infty}$. Since we will identify
special families of hypersurfaces $\Sigma_{\tau}$ asymptotic to Minkowski
hyperplanes at spatial infinity, these families can be mapped onto the space of
cross sections of the unit timelike hyperboloid by using the quoted
Anderson's lemma \cite{reg2}.

\vfill\eject

\section{Poincar\'e Charges in Metric Gravity.}

Before discussing the asymptotic Poincar\'e charges, let us summarize what is
known about the non Abelian charges and the Hamiltonian group of gauge
transformations for Yang-Mills theory.

As emphasized in Ref.\cite{lusa}, in the Hamiltonian formulation of a gauge
theory one has to make a choice of the boundary conditions of the canonical
variables and of the parameters of the gauge transformations [the infinitesimal
ones are generated by the first class constraints of the theory] in such a way
to give a meaning to integrations by parts, to the functional derivatives
(and therefore to Poisson brackets) and to the ``proper" gauge transformations
connected with the identity [the ``improper" ones, including the ``rigid or
global or first kind" gauge transformations related to the non-Abelian charges,
have to be treated separately; when there are topological numbers like
winding number, they label disjoint sectors of gauge transformations and one
speaks of ``large" gauge transformations]. In particular, the boundary
conditions must be such that the variation of the final Dirac Hamiltonian $H_D$
must be linear in the variations of the canonical variables [the coefficients
are the Dirac-Hamilton equations of motion] and this may require a redefinition
of $H_D$, namely $H_D$ has to be replaced by ${\tilde H}_D=H_D+H_{\infty}$,
where $H_{\infty}$ is a suitable integral on the surface at spatial infinity.
When this is accomplished, one has a good definition of functional derivatives
and Poisson brackets. Then, one must
consider the most general generator of gauge transformations of the theory (it
includes $H_D$ as a special case), in which there are arbitrary functions
(parametrizing infinitesimal gauge transformations) in front of all the first
class constraints and not only in front of the primary ones. Also the variations
 of this generator must be linear in the variations of the canonical variables:
this implies that all the surface terms coming from integration by parts must
vanish with the given boundary conditions on the canonical variables or must be
compensated by the variation of $H_{\infty}$. In this
way, one gets boundary conditions on the parameters of the infinitesimal gauge
transformations identifying the ``proper" ones, which transform the canonical
variables among themselves without altering their boundary conditions [the
symplectic vector fields associated with the proper gauge transformations map
the function space of the canonical variables into itself]. Let us remark that
in this way one is defining Hamiltonian boundary conditions which are not
manifestly covariant; however, in Minkowski spacetime a Wigner covariant
formulation is obtained by reformulating the theory on spacelike
hypersurfaces \cite{lus3,re} and then restricting it to spacelike hyperplanes.

In the Yang-Mills case\cite{lusa}, with the Hamiltonian gauge transformations
restricted to go to the identity in an angle-independent way at spatial infinity
, so to have well defined covariant non-Abelian charges, the ``proper" gauge
transformations are those which are connected to the identity and generated by
the Gauss law first class constraints at the infinitesimal level. The
``improper" ones are a priori of four types: \hfill\break
\hfill\break
i) ``global or rigid or first kind"
ones (the gauge parameters fields tend to constant at spatial infinity)
connected with the group G (isomorphic to the structure group of the
Yang-Mills principal bundle) generated by the ``non-Abelian charges";
\hfill\break
ii) the global or rigid ones in the ``center of the gauge group G" [triality
when G=SU(3)]; \hfill\break
iii) gauge transformations with non-vanishing winding number $n\in Z$
(``large" gauge transformations not connected with the identity; zeroth
homotopy group of the gauge group); \hfill\break
iv) other ``improper non rigid" gauge
transformations. Since this last type of gauge transformations does not play any
role in Yang-Mills dynamics, it was assumed \cite{lusa}
that the choice of the function
space for the gauge parameter fields $\alpha_a(\tau ,\vec \sigma )$ (describing
the component of the gauge group connected with the identity) be such that for
$r\, \rightarrow \infty$ one has \hfill\break
\hfill\break
$\alpha_a(\tau ,\vec \sigma )\, \rightarrow
\alpha_a^{(rigid)}+\alpha_a^{(proper)}(\tau ,\vec \sigma )$ \hfill\break
\hfill\break
with constant
$\alpha_a^{(rigid)}$ and with $\alpha_a^{(proper)}(\tau ,\vec \sigma )$ tending
to zero in a direction-independent way.

In metric gravity, the Hamiltonian gauge group is not connected with a principal
bundle (it contains 3-diffeomorphisms and its algebra has structure functions
and not structure constants). However, the asymptotic Poincar\'e charges (not
uniquely defined when supertranslations are allowed) and the eventuality of
supertranslations (whose generators, the ``supertranslation charges", should
vanish, i.e. have ``zero momentum" according to Anderson\cite{reg2}, when the
boundary conditions have well defined parity properties)  are the
counterpart of the Yang-Mills non-Abelian charges and also of the Abelian
electric charge. While the electric charge is a physical observable, the
hypothesis of quark confinement requires the existence only of color singlets,
namely i) physical observables must commute with the non-Abelian charges; ii)
the SU(3) color charges of isolated systems have to vanish themselves.
In both cases, inside local
quantum field theory, one speaks of superselection sectors determined by the
charges and does not consider them as generators of gauge transformations.

In
Ref.\cite{p18} the same possibility is opened for the asymptotic Poincar\'e
charges of asymptotically flat metric gravity: \hfill\break
\hfill\break
i) in the usual interpretation
\cite{p19} some observer is assumed to sit at or just outside the boundary at
spatial infinity but he is not explicitly included in the action functional;
this observer merely supplies a coordinate chart on the boundaries (perhaps,
through his `parametrization clock'), which we may use to fix the gauge of our
system at the boundary (the asymptotic lapse function; see later on); if one
wishes, this external observer may construct his clock to yield zero Poincar\'e
charges (so to recover a Machian interpretation\cite{p20} also in noncompact
universes with boundary; there is a strong
similarity with the results of Einstein-Wheeler cosmology \cite{ciuf}, based
on a closed compact universe without boundaries, for which Poincar\'e charges
are not defined), in which case every connection with particle physics is lost;
\hfill\break
ii) Marolf's proposal \cite{p18}
is to consider the system in isolation without the
utilization of any structure outside the boundary at spatial infinity and to
consider, at the quantum level, superselection rules for the asymptotic
Poincar\'e Casimirs, in particular for the ADM invariant mass [see Refs.
\cite{p21} for similar conclusions from different motivations].\hfill\break
\hfill\break
In Ref.\cite{giul}, also Giulini considers a matter of physical interpretation
whether all 3-diffeomorphisms of $\Sigma_{\tau}$ into itself must be considered
as gauge transformations. In the asymptotically flat open case, there is in
this paper a discussion of ``large" diffeomorphisms, but the gauge
transformations generated by the superhamiltonian constraint are not considered;
after a 1-point compactification ${\bar \Sigma}_{\tau}$ of $\Sigma_{\tau}$,
there is a study of the quotient space $Riem\, {\bar \Sigma}_{\tau}/ Diff_F\,
{\bar \Sigma}_{\tau}$, where $Diff_F\, {\bar \Sigma}_{\tau}$ are those
3-diffeomorphisms whose pullback goes to the identity at spatial infinity (the
point of compactification) where a ``privileged oriented frame" is chosen.
The Poincar\'e charges are not considered as generators of gauge
transformations; instead, there
is a study of the decomposition of ${\bar \Sigma}_{\tau}$ into its prime
factors as a 3-manifold, of the induced decomposition of $Diff_F\, {\bar
\Sigma}_{\tau}$ and of the evaluation of the homotopy groups of $Diff_f\,
{\bar \Sigma}_{\tau}$.

We shall take the point of view that the asymptotic Poincar\'e charges are not
generators of gauge transformations like in Yang-Mills theory (the ADM energy
will be the physical Hamiltonian for the evolution in $\tau$), that there are
superselection sectors labelled by the asymptotic Poincar\'e Casimirs and that
the parameters of the gauge transformations of ADM metric gravity have a clean
separation between a rigid part (differently from Yang-Mills theory it has both
a constant and a term linear in $\vec \sigma$) and a proper one restricted
not to contain supertranslations [namely we assume the absence of ``improper
non-rigid" gauge transformations like in Yang-Mills theory].

Let us now define the ``proper" gauge transformations of the ADM metric
gravity, whose canonical formalism was reviewed in Section V of I [in Appendix A
there are other properties of metric gravity deriving by the application of
the second Noether theorem \cite{sha} to the ADM action]. In Refs.
\cite{witt,dew,hh}, as shown in Appendix A, it is noted that, in asymptotically
flat spacetimes, the surface integrals arising in the transition from the
Hilbert action to the ADM action and, then, from this to the ADM phase space
action are connected with the ADM energy-momentum of the gravitational
field of the linearized theory of metric gravity \cite{adm}, if the lapse and
shift functions have certain asymptotic behaviours at spatial infinity.
Extra complications for the differentiability of the ADM canonical
Hamiltonian come from the presence of the second spatial derivatives of the
3-metric inside the ${}^3R$ term of the superhamiltonian constraint .

Regge and Teitelboim
\cite{reg} gave a set of boundary conditions for the ADM canonical variables
${}^3g_{rs}(\tau ,\vec \sigma )$, ${}^3{\tilde \Pi}^{rs}(\tau ,\vec \sigma )$,
so that it is possible to define 10 surface integrals associated with the
conserved Poincar\'e charges of the spacetime (the translation charges
are the ADM energy-momentum) and to show that the functional derivatives and
Poisson brackets are well defined in metric gravity; however, there is no
statement about gauge transformations and supertranslations in this paper.

A more complete analysis, including also a discussion of supertranslations in
the ADM canonical formalism, has been given by Beig and O'Murchadha\cite{reg1}
(extended to Ashtekar's formalism in Ref.\cite{reg3}). They consider
3-manifolds $\Sigma_{\tau}$ diffeomorphic to $R^3$ as in our case, so that there
exist ``global coordinate systems". If $\{ \sigma^{\check r}\}$ is one of
these global coordinate systems on $\Sigma_{\tau}$, the 3-metric ${}^3g_{\check
r\check s}(\tau ,\sigma^{\check t})$ [evaluated in this coordinate system] is
assumed asymptotically Euclidean in the following sense: if $r=\sqrt{\delta
_{\check r\check s}\sigma^{\check r}\sigma^{\check s}}$ [one could put $r=
\sqrt{{}^3g_{\check r\check s}\sigma^{\check r}\sigma^{\check s}}$ and get the
same kind of decomposition], then one assumes

\begin{eqnarray}
{}^3g_{\check r\check s}(\tau ,\vec \sigma )&=&\delta_{\check r\check s}+
{1\over r}\,\, {}^3s_{\check r\check s}(\tau ,{{\sigma^{\check n}}\over r})+
{}^3h_{\check r\check s}(\tau ,\vec \sigma ),\quad r\, \rightarrow \infty ,
\nonumber \\
&&{}\nonumber \\
&&{}^3s_{\check r\check s}(\tau ,{{\sigma^{\check n}}\over r})={}^3s_{\check
r\check s}(\tau ,-{{\sigma^{\check n}}\over r}),\quad EVEN\, PARITY,\nonumber \\
&&{}^3h_{\check r\check s}(\tau ,\vec \sigma )= O((r^{-(1+\epsilon )}),\quad
\epsilon > 0,\, for\, r\, \rightarrow \infty ,\nonumber \\
&&\partial_{\check u}\, {}^3h_{\check r\check s}(\tau ,\vec \sigma )=O(r^{-
(2+\epsilon )}).
\label{III1}
\end{eqnarray}

\noindent The functions ${}^3s_{\check r\check s}(\tau ,{{\sigma^{\check n}}
\over r})$ are $C^{\infty}$ on the sphere $S^2_{\tau ,\infty}$ at spatial
infinity; if they would be of odd parity, the ADM energy would vanish. The
difference ${}^3g_{\check r\check s}(\tau ,\vec \sigma )-\delta_{\check r\check
s}$ cannot fall off faster that 1/r, because otherwise the ADM energy would be
zero and the positivity energy theorem \cite{p17} would imply that the only
solution of the constraints is flat spacetime. For the ADM momentum
one assumes the following boundary conditions

\begin{eqnarray}
{}^3{\tilde \Pi}^{\check r\check s}(\tau ,\vec \sigma )&=&{1\over {r^2}}\,\,
{}^3t^{\check r\check s}(\tau ,{{\sigma^{\check n}}\over r})+{}^3k^{\check
r\check s}(\tau ,\vec \sigma ),\quad r\, \rightarrow \infty ,\nonumber \\
&&{}\nonumber \\
&&{}^3t^{\check r\check s}(\tau ,{{\sigma^{\check n}}\over r})=-{}^3t^{\check
r\check s}(\tau ,-{{\sigma^{\check n}}\over r}),\quad ODD\, PARITY,\nonumber \\
&&{}^3k^{\check r\check s}(\tau ,\vec \sigma )=O(r^{-(2+\epsilon )}),\quad
\epsilon > 0,\quad r\, \rightarrow \infty .
\label{III2}
\end{eqnarray}

\noindent If ${}^3{\tilde \Pi}^{\check r\check s}(\tau ,\vec \sigma )$ were to
fall off faster than $1/r^2$, the ADM linear momentum would vanish and we could
not consider Lorentz transformations. In this way, the integral $\int_{\Sigma
_{\tau}}d^3\sigma [{}^3{\tilde \Pi}^{\check r\check s} \delta \, {}^3g_{\check
r\check s}](\tau ,\vec \sigma )$ is well defined and finite [since the integrand
is of order $O(r^{-3})$, a possible logarithmic divergence is avoided due to
the odd parity of ${}^3t^{\check r\check s}$].

These boundary conditions imply that functional derivatives and Poisson
brackets  are well defined \cite{reg1} [in a more rigorous treatment one should
use appropriate weighted Sobolev spaces].

The supermomentum and superhamiltonian constraints [see Eqs.(79) of I and
Appendix A]  ${}^3{\tilde
{\cal H}}^{\check r}(\tau ,\vec \sigma )\approx 0$ and ${\tilde {\cal H}}(\tau ,
\vec \sigma )\approx 0$, are even functions of $\vec \sigma$ of order $O(r
^{-3})$. Their smeared version with the lapse and shift functions, appearing
in the canonical Hamiltonian  $H_{(c)ADM}=\int d^3\sigma [
N {\tilde {\cal H}}+N_{\check r}\, {}^3{\tilde {\cal H}}^{\check r}](\tau ,\vec
\sigma )$, will give a finite and differentiable $H_{(c)ADM}$ if we set
\cite{reg1}

\begin{eqnarray}
N(\tau ,\vec \sigma )&=&m(\tau ,\vec \sigma )=s(\tau ,\vec \sigma )+n(\tau
,\vec \sigma )=k(\tau ,{{\sigma^{\check n}}\over r})+O(r^{-\epsilon}),\quad
\epsilon >0,\quad r\, \rightarrow \infty ,\nonumber \\
N_{\check r}(\tau ,\vec \sigma )&=&m_{\check r}(\tau ,\vec \sigma )=s_{\check
r}(\tau ,\vec \sigma )+n_{\check r}(\tau ,\vec \sigma )=k_{\check r}(\tau ,
{{\sigma^{\check n}}\over r})+O(r^{-\epsilon}),\nonumber \\
&&{}\nonumber \\
&&s(\tau ,\vec \sigma )=k(\tau ,{{\sigma^{\check n}}\over r})=-k(\tau ,{{\sigma
^{\check n}}\over r}),\quad ODD\, PARITY,\nonumber \\
&&s_{\check r}(\tau ,\vec \sigma )=k_{\check r}(\tau ,{{\sigma^{\check n}}
\over r})=-k_{\check r}(\tau ,{{\sigma^{\check n}}\over r}),\quad\quad
ODD\, PARITY.
\label{III3}
\end{eqnarray}

With these boundary conditions one gets differentiability, i.e. $\delta
H_{(c)ADM}$ is linear in $\delta \, {}^3g_{\check r\check s}$ and $\delta \,
{}^3{\tilde \Pi}^{\check r\check s}$, with the coefficients being the
Dirac-Hamilton equations of metric gravity. Therefore, since N and $N_{\check
r}$ are a special case of the parameter fields of the most general
infinitesimal gauge transformations generated by the first class constraints
${\tilde {\cal H}}$, ${}^3{\tilde {\cal H}}^{\check r}$, with generator
$G=\int d^3\sigma [ \alpha {\tilde {\cal H}}+\alpha_{\check r}\,
{}^3{\tilde {\cal H}}^{\check r}](\tau ,\vec \sigma )$, the ``proper" gauge
transformations preserving Eqs.(\ref{III1}) and (\ref{III2}) have the
multiplier fields $\alpha (\tau ,\vec \sigma )$ and $\alpha_{\check r}(\tau
,\vec \sigma )$ with the same boundary conditions (\ref{III3}) of $m(\tau
,\vec \sigma)$ and $m_{\check r}(\tau ,\vec \sigma )$. Then, the Hamilton
equations imply that also
the Dirac multipliers $\lambda_N(\tau ,\vec \sigma )$ and $\lambda^{\vec N}
_{\check r}(\tau ,\vec \sigma )$ have these boundary conditions  [$\lambda_N\,
{\buildrel \circ \over =}\, \delta N$, $\lambda^{\vec N}_{\check r}\, {\buildrel
\circ \over =}\, \delta N_{\check r}$]. Instead, the momenta ${\tilde \Pi}^N
(\tau ,\vec \sigma )$ and ${\tilde \Pi}^{\check r}_{\vec N}(\tau ,\vec \sigma
)$, conjugate to N and $N_{\check r}$, must be of $O(r^{-(3+\epsilon )})$ to
have $H_{(D)ADM}$ finite.

The angle-dependent
functions $s(\tau ,\vec \sigma )=k(\tau ,{{\sigma^{\check n}}\over r})$
and $s_{\check r}(\tau ,\vec \sigma )=k_{\check r}(\tau ,{{\sigma^{\check n}}
\over r})$ on $S^2_{\tau ,\infty}$ (boundary of $\Sigma_{\tau}$ at spatial
infinity) are ``odd time and space supertranslations". The piece $\int
d^3\sigma [s\, {\tilde {\cal H}}+s_{\check r}\, {}^3{\tilde {\cal H}}^{\check r}
](\tau ,\vec \sigma ) \approx 0$ of the Dirac Hamiltonian is the Hamiltonian
generator of supertranslations (the ``zero momentum" of supertranslations
of Ref.\cite{reg2}). Their contribution to
gauge transformations is to alter the angle-dependent asymptotic terms
${}^3s_{\check r\check s}$ and ${}^3t^{\check r\check s}$ in ${}^3g_{\check
r\check s}$ and ${}^3{\tilde \Pi}^{\check r\check s}$.

With $N=m$, $N_{\check r}=m_{\check r}$ one can verify
the validity of the smeared form of Eqs.(81) of I:

\begin{eqnarray}
&&\lbrace H_{(c)ADM}[m_1,
m_1^{\check r}], H_{(c)ADM}[m_2,m_2^{\check r}]\rbrace =\nonumber \\
&=&H_{(c)ADM}
[m_2^{\check r}\, {}^3\nabla_{\check r} m_1 -m_1^{\check r}\, {}^3\nabla
_{\check r} m_2,\, {\cal L}_{{\vec m}_2}\, m_1^{\check r}+m_2\,
{}^3\nabla^{\check r} m_1 -m_1\, {}^3\nabla^{\check r} m_2],
\label{III3a}
\end{eqnarray}

\noindent with $m^{\check r}={}^3g^{\check r\check s} m_{\check s}$ and with
$H_{(c)ADM}[m,m^{\check r}]=\int d^3\sigma [m {\tilde {\cal H}}+m^{\check r}\,
{}^3{\tilde {\cal H}}_{\check r}](\tau ,\vec \sigma )=
\int d^3\sigma [m {\tilde {\cal H}}+m_{\check r}\,
{}^3{\tilde {\cal H}}^{\check r}](\tau ,\vec \sigma )$.

When the functions
$N(\tau ,\vec \sigma )$ and $N_{\check r}(\tau ,\vec \sigma )$ [and also
$\alpha (\tau ,\vec \sigma )$, $\alpha_{\check r}(\tau ,\vec \sigma )$] do not
have the asymptotic behaviour of $m(\tau ,\vec \sigma )$ and $m_{\check r}(\tau
,\vec \sigma )$ respectively, one speaks of ``improper" gauge transformations,
because $H_{(D)ADM}$ is not differentiable even at the constraint
hypersurface [in Ref.\cite{reg1} there is a non-rigorous proof that this
hypersurface is a manifold; this can be possible only if the spacetime has no
isometries].

At this point one has identified:

a) Certain coordinate systems on the spacelike 3-surface $\Sigma
_{\tau}$, which hopefully define a minimal atlas
${\cal C}_{\tau}$ for the spacelike
hypersurfaces $\Sigma_{\tau}$ foliating the asymptotically flat spacetime $M^4$.
With the ${\cal C}_{\tau}$'s and the parameter $\tau$ as $\Sigma_{\tau}$-adapted
coordinates of $M^4$ one should build an atlas ${\cal C}$ of allowed coordinate
systems of $M^4$.

b) A set of boundary conditions on the fields on $\Sigma_{\tau}$ (i.e. a
function space for them) ensuring that the 3-metric on $\Sigma_{\tau}$ is
asymptotically Euclidean in this minimal atlas (modulo 3-diffeomorphisms, see
the next point).

c) A set of ``proper" gauge transformations generated infinitesimally by the
first class constraints, which leave the fields on $\Sigma_{\tau}$ in the
chosen function space. Since the gauge transformations generated by the
supermomentum constraints ${}^3{\tilde {\cal H}}^{\check r}(\tau ,\vec \sigma
)\approx 0$ are the lift to the space of the tensor fields on $\Sigma_{\tau}$
(which contains the phase space of metric gravity) of the 3-diffeomorphisms
$Diff\, \Sigma_{\tau}$ of $\Sigma_{\tau}$ into itself, the restriction of
$N(\tau ,\vec \sigma )$, $N_{\check r}(\tau ,\vec \sigma )$ to $m(\tau ,\vec
\sigma )$, $m_{\check r}(\tau ,\vec \sigma )$, ensures that these
3-diffeomorphisms are restricted to be compatible with the chosen minimal
atlas for $\Sigma_{\tau}$ [this is the problem of the coordinate transformations
preserving Eq.(\ref{III1})].

More difficult is the interpretation\cite{wa,anton} of
the gauge transformations generated by the superhamiltonian constraint
${\tilde {\cal H}}(\tau ,\vec \sigma )\approx 0$, which
are the phase space adaptation of the time diffeomorphisms of $M^4$ (leaving
invariant the Hilbert action $S_H$) after the removal of the surface term
connecting the Hilbert action $S_H$ to the ADM action $S_{ADM}$ [which is
quasi-invariant under the gauge transformations generated by ${\tilde {\cal H}}
(\tau ,\vec \sigma )$ but not under $Diff\, M^4$, as shown in Appendix A].
Since the superhamiltonian constraint determines the
conformal factor of the 3-metric (see the Conclusions of II and Section VI),
the gauge transformations generated by ${\tilde {\cal H}}(\tau ,\vec \sigma )$
will induce a change of the momentum conjugate to this conformal factor, namely
of the extrinsic curvature of $\Sigma_{\tau}$, so that they generate the
transitions  from one allowed 3+1 splitting of $M^4$ to another allowed one.
This suggests that, in absence of
supertranslations, the functions $N$, $\alpha$, $\lambda_N$, should go like
$O(r^{-(2+\epsilon )})$ and not like $O(r^{-\epsilon })$ (in the case of
proper gauge transformations).

Let us remark at this point that the addition of gauge-fixing constraints to the
superhamiltonian and supermomentum constraints (see the end of Section V of I)
must happen in the chosen function space for the fields on $\Sigma_{\tau}$.
Therefore, the time constancy of these gauge-fixings will generate secondary
gauge-fixing constraints for the restricted lapse and shift functions
$m(\tau ,\vec \sigma )$, $m_{\check r}(\tau ,\vec \sigma )$.

By using the original ADM results\cite{adm}, Regge and Teitelboim\cite{reg}
wrote the expression of the ten conserved Poincar\'e charges, by allowing the
functions $N(\tau ,\vec \sigma )$, $N_{\check r}(\tau ,\vec \sigma )$, to have
a linear behaviour in $\vec \sigma$ for $r\, \rightarrow \infty$.
These charges are surface integrals at
spatial infinity, which have to added to the Dirac Hamiltonian so that it
becomes differentiable. Therefore, also the parameter fields
$\alpha (\tau ,\vec \sigma )$, $\alpha_{\check r}(\tau ,\vec \sigma )$, of
arbitrary (also improper)
gauge transformations should acquire this behaviour. In this way one
is enlarging the allowed 3-diffeomorphisms of $\Sigma_{\tau}$ into itself
(this would require an enlargement of the minimal atlas of $\Sigma_{\tau}$) and
one should change the function space of the tensor fields ${}^3g_{\check
r\check s}$, ${}^3{\tilde \Pi}^{\check r\check s}$. The added ``Poincar\'e
transformations at infinity" generated by the ten charges are often
considered as ``extra improper gauge transformations". However, if one takes
Marolf's viewpoint that the Poincar\'e charges are not generators of improper
gauge transformations but that they determine superselection sectors, one has
not to enlarge the allowed 3-diffeomorphisms and the minimal atlas of $\Sigma
_{\tau}$.

These results, the wish to have a clear separation between proper and
improper gauge transformations (like in Yang-Mills theory) and the hope to
solve the deparametrization problem of general relativity suggest that
it is reasonable to break the lapse and shift functions in the parts associated
with the proper and improper gauge transformations respectively (with no
``improper non rigid" gauge transformations like in Yang-Mills theory)

\begin{eqnarray}
N(\tau ,\vec \sigma )\, &=& N_{(as)}(\tau
,\vec \sigma )+m(\tau ,\vec \sigma ),\nonumber \\
N_{\check r}(\tau ,\vec \sigma )\, &=&
N_{(as) \check r}(\tau ,\vec \sigma )+m_{\check r}(\tau ,\vec \sigma ),
\label{III4}
\end{eqnarray}

\noindent and then to assume that the improper parts $N_{(as)}$, $N_{(as) r}$,
behave as the lapse and shift functions associated with spacelike hyperplanes
in Minkowski spacetime, as already  anticipated in II.

As shown in II, there are two independent definitions of
lapse and shift functions which can be associated with a 3+1 splitting of
Minkowski spacetime by means of a foliation with spacelike hypersurfaces.
However, for foliations with spacelike hyperplanes the two definitions coincide
and produce the following result [$(\mu )$ are flat Minkowski indices in
rectangular coordinates; ${\tilde \lambda}_{(\mu )}(\tau )$, ${\tilde \lambda}
_{(\mu )(\nu )}(\tau )$, are arbitrary Dirac multipliers]

\begin{eqnarray}
N_{(flat)}(\tau ,\vec \sigma )&=&N_{[z](flat)}(\tau ,\vec \sigma )=
-{\tilde \lambda}
_{(\mu )}(\tau )l^{(\mu )}-l^{(\mu )}{\tilde \lambda}_{(\mu )(\nu )}(\tau )b
^{(\nu )}_{\check s}(\tau ) \sigma^{\check s},\nonumber \\
N_{(flat )}(\tau ,\vec \sigma )&=&N_{[z](flat)\check r}
(\tau ,\vec \sigma )=-{\tilde \lambda}
_{(\mu )}(\tau )b^{(\mu )}_{\check r}(\tau )-b^{(\mu )}_{\check r}(\tau ){\tilde
\lambda}_{(\mu )(\nu )}(\tau ) b^{(\nu )}_{\check s}(\tau ) \sigma^{\check s}.
\label{III5}
\end{eqnarray}

In Ref.\cite{p22} and in the book in Ref.\cite{dirac} (see also Ref.\cite{reg}),
Dirac introduced asymptotic Minkowski rectangular coordinates

\begin{equation}
z^{(\mu )}
_{(\infty )}(\tau ,\vec \sigma )=x^{(\mu )}_{(\infty )}(\tau )+b^{(\mu )}
_{(\infty )\, \check r}(\tau ) \sigma^{\check r}
\label{III5b}
\end{equation}

\noindent in $M^4$ at spatial infinity $S_{\infty}=\cup_{\tau} S^2_{\tau
,\infty}$ [here $\{ \sigma^{\check r} \}$ are the previous global coordinate
systems of the atlas ${\cal C}_{\tau}$ of $\Sigma_{\tau}$, not matching
the spatial coordinates $z^{(i)}_{(\infty )}(\tau ,\vec \sigma )$].
For each value of $\tau$, the coordinates $x^{(\mu )}
_{(\infty )}(\tau )$ labels a point, near spatial infinity
chosen as origin. On it there is a flat tetrad $b^{(\mu )}_{(\infty )\, A}
(\tau )= (\, l^{(\mu )}_{(\infty )}=b^{(\mu )}_{(\infty )\, \tau}=\epsilon
^{(\mu )}{}_{(\alpha )(\beta )(\gamma )} b^{(\alpha )}_{(\infty )\, \check 1}
(\tau )b^{(\beta )}_{(\infty )\, \check 2}(\tau )b^{(\gamma )}_{(\infty )\,
\check 3}(\tau );\, b^{(\mu )}_{(\infty )\, \check r}(\tau )\, )$, with
$l^{(\mu )}_{(\infty )}$ $\tau$-independent, satisfying $b^{(\mu )}_{(\infty )\,
A}\, {}^4\eta_{(\mu )(\nu )}\, b^{(\nu )}_{(\infty )\, B}={}^4\eta_{AB}$ for
every $\tau$ [at this level there is no
reason to assume that $l^{(\mu )}_{(\infty )}$ is
tangent to $S_{\infty}$, as the normal $l^{\mu }$ to $\Sigma_{\tau}$, see
Appendix A]. There will be transformation coefficients $b^{\mu}_A(\tau ,\vec
\sigma )$ from the  adapted coordinates $\sigma^A=(\tau ,\sigma
^{\check r})$ to coordinates $x^{\mu}=z^{\mu}(\sigma^A)$ in an atlas of $M^4$,
such that in a chart at spatial infinity one has $z^{\mu}(\tau ,\vec \sigma )
\rightarrow \delta^{\mu}_{(\mu )} z^{(\mu )}(\tau ,\vec \sigma )$ and $b^{\mu}
_A(\tau ,\vec \sigma ) \rightarrow
\delta^{\mu}_{(\mu )} b^{(\mu )}_{(\infty )A}(\tau )$
[for $r\, \rightarrow \, \infty$ one has ${}^4g_{\mu\nu}\, \rightarrow \,
\delta^{(\mu )}_{\mu}\delta^{(\nu )}_{\nu}{}^4\eta_{(\mu )(\nu )}$ and
${}^4g_{AB}=b^{\mu}_A\, {}^4g_{\mu\nu} b^{\nu}_B\, \rightarrow \,
b^{(\mu )}_{(\infty )A}\, {}^4\eta_{(\mu )(\nu )} b^{(\nu )}_{(\infty )B}=
{}^4\eta_{AB}$ ]. In this way one defines the atlas ${\cal C}$ of the
allowed coordinate systems of $M^4$.

Dirac\cite{p22} and, then, Regge and
Teitelboim\cite{reg} proposed that the asymptotic Minkowski rectangular
coordinates $z^{(\mu )}_{(\infty )}(\tau ,\vec \sigma )=x^{(\mu )}_{(\infty )}
(\tau )+b^{(\mu )}_{(\infty ) \check r}(\tau )\sigma^{\check r}$ should define
10 new independent degrees of freedom at the spatial boundary $S_{\infty}$
[with ten associated conjugate momenta],
as it happens for Minkowski parametrized theories\cite{lus1} when the extra
configurational variables $z^{(\mu )}(\tau ,\vec \sigma )$ (describing the
embedded spacelike hypersurface and not existing in curved spacetimes) are
reduced to 10 degrees of freedom by the restriction
to spacelike hyperplanes [defined by $z^{(\mu )}(\tau ,\vec \sigma )\approx
x^{(\mu )}_s(\tau )+b^{(\mu )}_{\check r}(\tau )\sigma^{\check r}$], but with
these 10 degrees of freedom being gauge variables (independence from the choice
of the hyperplane) due to 10 surviving first class constraints.

In Minkowski parametrized theories for isolated systems restricted to spacelike
hyperplanes (see II) it can
be shown\cite{lus1} that these 20 variables  are:\hfill\break
\hfill\break
i) $x^{(\mu )}_s(\tau ), p^{(\mu )}_s$ [$\{ x^{(\mu )}_s,p^{(\nu )}_s\} =
-{}^4\eta^{(\mu )(\nu )}$], parametrizing the origin of the coordinates on the
family of spacelike hyperplanes. The four constraints ${\cal H}^{(\mu )}(\tau )
\approx p_s^{(\mu )} -p_{sys}^{(\mu )}\approx0$  say that
$p_s^{(\mu )}$ is determined by the 4-momentum of the isolated system.
\hfill\break
ii) $b^{(\mu )}_A(\tau )$ (with the $b^{(\mu )}_r(\tau )$'s being three
orthogonal spacelike unit vectors generating the fixed $\tau$-independent
timelike unit normal $b^{(\mu )}_{\tau}=l^{(\mu )}$ to the hyperplanes)
and $S^{(\mu )(\nu )}_s=-S^{(\nu )(\mu )}_s$ with the orthonormality
constraints $b^{(\mu )}_A\, {}^4\eta_{(\mu )(\nu )} b^{(\nu )}_B={}^4\eta_{AB}$.
The non-vanishing Dirac brackets enforcing the orthonormality constraints
\cite{hanson,lus1} for the $b^{(\mu )}_A$'s
are\hfill\break
\hfill\break
$\{ b^{(\rho )}_A, S^{(\mu )(\nu )}_s \}={}^4\eta^{(\rho )(\mu )} b^{(\nu )}_A
-{}^4\eta^{(\rho )(\nu )} b^{(\mu )}_A$,\hfill\break
\hfill\break
$\{ S^{(\mu )(\nu )}_s,S^{(\alpha )(\beta )}_s \} =C^{(\mu )(\nu )(\alpha
)(\beta )}_{(\gamma )(\delta )} S^{(\gamma )(\delta )}_s$ \hfill\break
\hfill\break
with $C^{(\mu )(\nu
)(\alpha )(\beta )}_{(\gamma )(\delta )}$ the structure constants of the
Lorentz algebra. Then one has that $p^{(\mu )}_s$, $J^{(\mu )(\nu )}_s=x
^{(\mu )}_sp^{(\nu )}_s-x^{(\nu )}_sp^{(\mu )}_s+S^{(\mu )(\nu )}_s$, satisfy
the algebra of the Poincar\'e group, with $S^{(\mu )(\nu )}_s$ playing the
role of the spin tensor. The other six constraints $
{\cal H}^{(\mu )(\nu )}(\tau )\approx S^{(\mu )(\nu )}
_s-S^{(\mu )(\nu )}_{sys}\approx 0$ say that
$S_s^{(\mu )(\nu )}$ coincides the spin tensor of the isolated system.

Let us remark that,
for each configuration of an isolated system there is a privileged
family of hyperplanes (the Wigner hyperplanes orthogonal to $p^{(\mu )}_s$,
existing when $\epsilon p^2_s > 0$) corresponding to the intrinsic rest-frame
of the isolated system. If we choose these hyperplanes with suitable
gauge fixings, we remain with only  the four constraints ${\cal H}^{(\mu )}(\tau
)\approx 0$, which can be rewritten as\hfill\break
\hfill\break
$\sqrt{\epsilon p^2_s} \approx [invariant\, mass\, of\, the\,
isolated\, system\, under\, investigation]= M_{sys}$; \hfill\break
${\vec p}_{sys}=[3-momentum\, of\, the\, isolated\, system\,
inside\, the\, Wigner\, hyperplane]\approx 0$.\hfill\break
\hfill\break
There is no more a restriction on $p_s^{(\mu )}$, because
$u^{(\mu )}_s(p_s)=p^{(\mu )}_s/p^2_s$ gives the orientation of the Wigner
hyperplanes containing the isolated system with respect to an arbitrary
given external observer.

In this special gauge we have $b^{(\mu )}_A\equiv L^{(\mu )}{}_A(p_s,{\buildrel
\circ \over p}_s)$ (the standard Wigner boost for timelike Poincar\'e orbits),
$S_s^{(\mu )(\nu )}\equiv S_{sys}^{(\mu )(\nu )}$,
${\tilde \lambda}_{(\mu )(\nu )}(\tau )\equiv 0$, and the only
remaining canonical variables are the non-covariant Newton-Wigner-like
canonical ``external" center-of-mass
coordinate ${\tilde x}^{(\mu )}_s(\tau )$ (living on the
Wigner hyperplanes) and $p^{(\mu )}_s$.
Now 3 degrees of freedom of the isolated system [an ``internal"
center-of-mass 3-variable ${\vec \sigma}_{sys}$ defined inside the Wigner
hyperplane and conjugate to ${\vec p}_{sys}$] become gauge variables [the
natural gauge fixing is ${\vec \sigma}_{sys}\approx 0$, so that it coincides
with the origin $x^{(\mu )}_s(\tau )=z^{(\mu )}(\tau ,\vec \sigma =0)$ of the
Wigner hyperplane], while the ${\tilde x}^{(\mu )}(\tau )$
is playing the role of a kinematical external
center of mass for the isolated system and may be interpreted as a decoupled
observer with his parametrized clock (point particle clock).
All the fields living on the Wigner hyperplane are now either Lorentz scalar
or with their 3-indices transforming under Wigner rotations (induced by Lorentz
transformations in Minkowski spacetime) as any Wigner spin 1 index. Let us
remark that the constant $x^{(\mu )}_s(0)$ [and, therefore, also ${\tilde x}
^{(\mu )}_s(0)$] is arbitrary, reflecting the arbitrariness in the absolute
location of the origin of the ``internal" coordinates on each hyperplane in
Minkowski spacetime.

One obtains
in this way a new kind of instant form of the dynamics,
the  ``Wigner-covariant 1-time rest-frame instant form"\cite{lus1} with a
universal breaking of Lorentz covariance.
It is the special relativistic generalization of
the non-relativistic separation of the center of mass from the relative motion
[$H={{ {\vec P}^2}\over {2M}}+H_{rel}$]. The role of the center of mass is
taken by the Wigner hyperplane, identified by the point ${\tilde x}^{(\mu )}
(\tau )$ and by its normal $p^{(\mu )}_s$. The
invariant mass $M_{sys}$ of the system replaces the non-relativistic  Hamiltonian
$H_{rel}$ for the relative degrees of freedom, after the addition of the
gauge-fixing $T_s-\tau \approx 0$ [identifying the time parameter $\tau$,
labelling the leaves of the foliation,  with
the Lorentz scalar time of the center of mass in the rest frame,
$T_s=p_s\cdot {\tilde x}_s/M_{sys}$; $M_{sys}$  generates the
evolution in this time].

The determination of ${\vec \sigma}_{sys}$ may be done with the group
theoretical methods of Ref.\cite{pauri}: given a realization on the phase space
of a given system of the ten Poincar\'e generators one can build three
3-position variables only in terms of them, which in our case of a system
on the Wigner hyperplane with ${\vec p}_{sys}\approx 0$ are: i) a canonical
center of mass (the ``internal" center of mass ${\vec \sigma}_{sys}$); ii)
a non-canonical M\o ller center of energy ${\vec \sigma}^{(E)}_{sys}$; iii)
a non-canonical Fokker-Price center of inertia ${\vec \sigma}^{(FP)}_{sys}$. Due
to ${\vec p}_{sys}\approx 0$, we have ${\vec \sigma}_{sys} \approx
{\vec \sigma}^{(FP)}_{sys} \approx {\vec \sigma}^{(E)}_{sys} =
\{ boost generator / energy \}$. By adding the
gauge fixings ${\vec \sigma}_{sys}\approx 0$ one can show that the origin
$x_s^{(\mu )}(\tau )$ becomes  simultaneously the Dixon center of mass of
an extended object and both the Pirani and Tulczyjew centroids (see Ref.
\cite{mate} for the application of these methods to find the center of mass
of a configuration of the Klein-Gordon field after the preliminary work of
Ref.\cite{lon}). With similar methods one can construct three ``external"
collective positions (all located on the Wigner hyperplane): i) the ``external"
canonical non-covariant center of mass ${\tilde x}_s^{(\mu )}$; ii) the
``external" non-canonical and non-covariant M\o ller center of energy
$R^{(\mu )}_s$; iii) the ``external" covariant non-canonical Fokker-Price center
of inertia $Y^{(\mu )}_s$ (when
there are the gauge fixings ${\vec \sigma}_{sys}\approx 0$ it also coincides
with the origin $x^{(\mu )}_s$). It turns out that the Wigner hyperplane is
the natural setting for the study of the Dixon multipoles of extended
relativistic systems\cite{dixon} and for defining the canonical relative
variables with respect to the center of mass. The Wigner hyperplane with its
natural Euclidean metric structure offers a natural solution to the problem of
boost for lattice gauge theories and realizes explicitly the Machian aspect of
dynamics that only relative motions are relevant.

Analogously, in Dirac's approach to metric gravity
one expects that the 20 extra variables of the Dirac proposal
should be replaced by a set of this kind: $x^{(\mu )}_{(\infty )}(\tau )$,
$p^{(\mu )}_{(\infty )}$, $b^{(\mu )}_{(\infty ) A}(\tau )$ [with $b^{(\mu )}
_{(\infty ) \tau }=l^{(\mu )}_{(\infty )}$ $\tau$-independent and coinciding
with the asymptotic normal to $\Sigma_{\tau}$, tangent to $S_{\infty}$],
$S^{(\mu )(\nu
)}_{(\infty )}$, with the previous Dirac brackets implying the orthonormality
constraints $b^{(\mu )}_{(\infty ) A}\, {}^4\eta_{(\mu )(\nu )} b^{(\nu )}
_{(\infty ) B}={}^4\eta_{AB}$. Moreover, $p^{(\mu )}_{(\infty )}$ and
$J^{(\mu )(\nu )}_{(\infty )}=x^{(\mu )}_{(\infty )}p^{(\nu )}_{(\infty )}-
x^{(\nu )}_{(\infty )}p^{(\mu )}_{(\infty )}+S^{(\mu )(\nu )}_{(\infty )}$
should satisfy a Poincar\'e algebra. In analogy with Minkowski parametrized
theories restricted to spacelike hyperplanes, one expects to have 10 extra
first class constraints of the type\hfill\break
\hfill\break
$p^{(\mu )}_{(\infty )}-P^{(\mu )}_{ADM}\approx 0$,\hfill\break
$S^{(\mu )(\nu )}_{(\infty )}-S^{(\mu )(\nu )}_{ADM}\approx 0$ \hfill\break
\hfill\break
with $P^{(\mu )}_{ADM}$, $S^{(\mu )(\nu )}_{ADM}$ related to the
ADM Poincar\'e charges and 10 extra Dirac multipliers ${\tilde \lambda}
_{(\mu )}(\tau )$, ${\tilde \lambda}_{(\mu )(\nu )}(\tau )$, in front of them
in the Dirac Hamiltonian. The origin $x^{(\mu )}_{(\infty )}$ is going to play
the role of an ``external" decoupled observer with his parametrized clock.
The main problem with respect to Minkowski
parametrized theory on spacelike hyperplanes is that it is not known which
could be the ADM spin part $S_{ADM}^{(\mu )(\nu )}$ of the ADM Lorentz charge
$J^{(\mu )(\nu )}_{ADM}$.

The way out from these problems is based on the following observation. If we
replace $p^{(\mu )}_{(\infty )}$ and $S^{(\mu )(\nu )}_{(\infty )}$, whose
Poisson algebra is the direct sum of an Abelian algebra of translations and of
a Lorentz algebra, with the new variables (with  indices adapted
to $\Sigma_{\tau}$) \hfill\break
\hfill\break
$p^A_{(\infty )}=b^A_{(\infty )(\mu )}p^{(\mu )}
_{(\infty )}$, \hfill\break
$J^{AB}_{(\infty )}\, {\buildrel {def} \over =}\,
b^A_{(\infty )(\mu )}b^B_{(\infty )(\nu )}
S^{(\mu )(\nu )}_{(\infty )}$ [$\not= b^A_{(\infty )(\mu )}b^B_{(\infty )(\nu )}
J^{(\mu )(\nu )}_{(\infty )}$], \hfill\break
\hfill\break
the Poisson brackets for $p^{(\mu )}
_{(\infty )}$, $b^{(\mu )}_{(\infty ) A}$, $S^{(\mu )(\nu )}_{(\infty )}$
[one has $\lbrace b^A_{(\infty )(\gamma )},S^{(\nu )(\rho )}_{(\infty )}
\rbrace =\eta^{(\nu )}_{(\gamma )}b_{(\infty )}^{A (\rho )}-\eta^{(\rho )}
_{(\gamma )}b_{(\infty )}^{A(\nu )}$], imply

\begin{eqnarray}
&&\lbrace p^A_{(\infty )},p^B_{(\infty )}\rbrace =0,\nonumber \\
&&\lbrace p^A_{(\infty )},J^{BC}_{(\infty )}\rbrace ={}^4g^{AC}_{(\infty )}p^B
_{(\infty )}-{}^4g^{AB}_{(\infty )} p^C_{(\infty )}, \nonumber \\
&&\lbrace J^{AB}
_{(\infty )},J^{CD}_{(\infty )}\rbrace =-(\delta^B_E\delta^C_F\, {}^4g^{AD}
_{(\infty )}+\delta^A_E\delta^D_F\, {}^4g^{BC}_{(\infty )}-\delta^B_E\delta^D
_F\, {}^4g^{AC}_{(\infty )}-\delta^A_E\delta^C_F\, {}^4g^{BD}_{(\infty )})J
^{EF}_{(\infty )}=\nonumber \\
&&=-C^{ABCD}_{EF} J^{EF}_{(\infty )},
\label{III6}
\end{eqnarray}

\noindent where ${}^4g^{AB}_{(\infty )}=b^A_{(\infty )(\mu )}\,
{}^4\eta^{(\mu )(\nu )} b^B_{(\infty )(\nu )}={}^4\eta^{AB}$ since the
$b^{(\mu )}_{(\infty )A}$ are flat tetrad in both kinds of indices.
Therefore, we get the algebra of a realization of the Poincar\'e group [this
explains the notation $J^{AB}_{(\infty )}$] with all the structure constants
inverted in the sign (transition from a left to a right action).

This  implies that the Poincar\'e generators ${P}^A_{ADM}$, ${J}^{AB}_{ADM}$
in $\Sigma_{\tau}$-adapted coordinates
should define in the asymptotic Dirac rectangular coordinates a momentum
${P}^{(\mu )}_{ADM}=b^{(\mu )}_A  P^A_{ADM}$
and only an  ADM spin tensor ${S}^{(\mu )(\nu )}_{ADM}$ [to define an angular
momentum tensor $J^{(\mu )(\nu )}_{ADM}$ one should find an ``external center
of mass of the gravitational field" $X^{(\mu )}_{ADM} [{}^3g,
{}^3{\tilde \Pi}]$ (see Ref.\cite{lon,mate}
for the Klein-Gordon case) conjugate to
$P^{(\mu )}_{ADM}$, so that $J^{(\mu )(\nu )}_{ADM}=X^{(\mu )}_{ADM}P^{(\nu )}
_{ADM}-X^{(\nu )}_{ADM}P^{(\mu )}_{ADM}+S^{(\mu )(\nu )}_{ADM}$].

Therefore we shall assume the existence of a global coordinate system $\{ \sigma
^{\check r} \}$ on $\Sigma_{\tau}$
, in which we have

\begin{eqnarray}
N(\tau ,\vec \sigma )\, &=& N_{(as)}(\tau
,\vec \sigma )+m(\tau ,\vec \sigma ),\nonumber \\
N_{\check r}(\tau ,\vec \sigma )\, &=&
N_{(as) \check r}(\tau ,\vec \sigma )+m_{\check r}(\tau ,\vec \sigma ),
\nonumber \\
&&{}\nonumber \\
N_{(as)}(\tau ,\vec \sigma )&=&-{\tilde \lambda}_{(\mu )}(\tau )l^{(\mu )}
_{(\infty )}-l^{(\mu )}_{(\infty )}{\tilde \lambda}_{(\mu )(\nu )}(\tau )
b^{(\nu )}_{(\infty) \check s}(\tau ) \sigma^{\check s}=\nonumber \\
&=&-{\tilde \lambda}_{\tau}(\tau )-{1\over 2}{\tilde \lambda}_{\tau \check
s}(\tau )\sigma^{\check s},\nonumber \\
N_{(as) \check r}(\tau ,\vec \sigma )&=&-b^{(\mu )}_{(\infty ) \check r}(\tau )
{\tilde \lambda}_{(\mu )}(\tau )-b^{(\mu )}_{(\infty ) \check r}(\tau ){\tilde
\lambda}_{(\mu )(\nu )}(\tau ) b^{(\nu )}_{(\infty ) \check s}(\tau ) \sigma
^{\check s}=\nonumber \\
&=&-{\tilde \lambda}_{\check r}(\tau )-{1\over 2}{\tilde \lambda}_{\check
r\check s}(\tau ) \sigma^{\check s},
\label{III7}
\end{eqnarray}

\noindent with $m(\tau ,\vec \sigma )$, $m_{\check r}(\tau ,\vec \sigma )$,
given by Eqs.(\ref{III3}) [they still contain odd supertranslations].

Let us remark that the quoted restriction to constant improper gauge
transformations [$\alpha^{(rigid)}=const.$] of Yang-Mills theory
is here replaced by the assumed forms of $N_{(as)}(\tau ,\vec \sigma )$,
$N_{(as)\check r}(\tau ,\vec \sigma )$.

This very strong assumption implies that one is selecting
asymptotically at spatial infinity only coordinate systems in which the lapse
and shift functions have behaviours similar to those of
Minkowski spacelike hyperplanes, so that the allowed foliations of the 3+1
splittings of the spacetime $M^4$ are restricted to have the leaves $\Sigma
_{\tau}$ approaching these Minkowski hyperplanes at spatial infinity in a way
independent from the direction if supertranslations are absent.
But this is coherent with Dirac's choice of
asymptotic rectangular coordinates [modulo 3-diffeomorphisms not changing the
nature of the coordinates] and with the assumptions used to define the
asymptotic Poincar\'e charges. It is  also needed to eliminate
coordinate transformations not becoming the identity at spatial infinity
[they are not associated with the gravitational fields of isolated systems
\cite{ll}].

By using ${\tilde \lambda}_A(\tau )=\{ {\tilde \lambda}_{\tau}(\tau );
{\tilde \lambda}_{\check r}(\tau ) \}$, ${\tilde \lambda}_{AB}(\tau )=
-{\tilde \lambda}_{BA}(\tau )$, $n(\tau ,\vec \sigma )$, $n_{\check r}(\tau
,\vec \sigma )$ as new configuration variables [replacing $N(\tau ,\vec \sigma
)$ and $N_{\check r}(\tau ,\vec \sigma )$] in the ADM Lagrangian (see Section
V of I and Appendix A) only produces the replacement of the first class
constraints \hfill\break
\hfill\break
${\tilde \pi}^N(\tau ,\vec \sigma )\approx 0$, ${\tilde \pi}
_{\vec N}^{\check r}(\tau ,\vec \sigma )\approx 0$, \hfill\break
\hfill\break
with the new first class constraints \hfill\break
\hfill\break
${\tilde \pi}^n(\tau ,\vec \sigma )\approx 0$, ${\tilde \pi}
_{\vec n}^{\check r}(\tau ,\vec \sigma )\approx 0$, ${\tilde \pi}^A(\tau )
\approx 0$, ${\tilde \pi}^{AB}(\tau )=-{\tilde \pi}^{BA}(\tau )\approx 0$,
\hfill\break
\hfill\break
corresponding to the vanishing of the canonical
momenta conjugate to the new configuration variables [we assume the Poisson
brackets $\{ {\tilde \lambda}_A(\tau ),{\tilde \pi}^B(\tau )\} =\delta^B_A$,
$\{ {\tilde \lambda}_{AB}(\tau ), {\tilde \pi}^{CD}(\tau ) \} =\delta^C_A
\delta^D_B-\delta^D_A \delta^C_B$]. The only change in the
Dirac Hamiltonian $H_{(D)ADM}$ of Eq.(\ref{a2}) is \hfill\break
\hfill\break
$\int d^3\sigma [\lambda_N {\tilde \pi}^N+\lambda
^{\vec N}_{\check r} {\tilde \pi}_{\vec N}^{\check r}](\tau ,\vec \sigma )\,
\mapsto \,
\zeta_A(\tau ) {\tilde \pi}^A(\tau )+\zeta_{AB}(\tau ) {\tilde \pi}^{AB}(\tau )
+\int d^3\sigma [\lambda_n {\tilde \pi}^n+\lambda^{\vec n}_{\check r} {\tilde
\pi}_{\vec n}^{\check r}](\tau ,\vec \sigma )$ \hfill\break
\hfill\break
with $\zeta_A$, $\zeta_{AB}$ Dirac's multipliers.

It seems impossible to have a reformulation of metric gravity
corresponding to the fully parametrized Minkowski theory on arbitrary
spacelike hypersurfaces, first because on them there is not a unique definition
of lapse and shift functions (see II)
and second because the coefficients $b^{\mu}_A$
are not tetrads in curved spacetimes like the $b^{(\mu )}_A=z^{(\mu )}_A$ in
Minkowski spacetime. The existence of this holonomic basis for vector fields
on it allows to use the coordinates $z^{(\mu )}(\tau ,\vec \sigma )$ as
configurational variables in parametrized Minkowski theories. Instead in the
ADM theory the configuration variables are only $N$, $N_{\check r}$, ${}^3g
_{\check r\check s}$, because the $b^{\mu}_A$ are now a non-holonomic basis.

Deferring to the next Section a discussion about the privileged observers
associated with these asymptotic Minkowskian hyperplanes,  let us come back to
the definition of the asymptotic Poincar\'e charges in metric gravity.
The splitting (\ref{III7}) of the lapse and shift functions and the
addition to the canonical Hamiltonian $H_{(c)ADM}$ of an appropriate surface
integral $H_{\infty}$,
defined in Refs.\cite{reg,reg1}, produce a modified canonical
Hamiltonian which is differentiable and finite: namely its variation
is linear in $\delta \, {}^3g_{\check r\check s}$, $\delta \, {}^3{\tilde \Pi}
^{\check r\check s}$. In our notation, by using the surface integral of
Eqs.(\ref{a3}) and $N_{(as)}(\tau ,\vec \sigma )$ and $N_{(as) \check r}
(\tau ,\vec \sigma )$ of Eqs.(\ref{III7}), the modified
canonical Hamiltonian is [we use the global coordinate system $\{ \sigma^{\check
r} \}$ of Eq.(\ref{III7}); we remember that $k=c^3/16\pi G$ with G the Newton
constant]

\begin{eqnarray}
{\hat H}_{(c)ADM}&=&\int d^3\sigma [N {\tilde {\cal H}}+N_{\check r}\,
{}^3{\tilde {\cal H}}^{\check r}](\tau ,\vec \sigma ) =\nonumber \\
&=&\int d^3\sigma [(N_{(as)}+m) {\tilde {\cal H}}+(N_{(as)\check r}+
m_{\check r})\, {}^3{\tilde {\cal H}}^{\check r}](\tau ,\vec \sigma )
\mapsto\nonumber \\
\mapsto {\hat H}^{'}_{(c)ADM}&&={\hat H}^{'}_{(c)ADM}[N,N^{\check r}]=
{\hat H}_{(c)ADM}+H_{\infty}=\nonumber \\
&=&\int d^3\sigma [N {\tilde {\cal H}}
+N_{\check r}\,
{}^3{\tilde {\cal H}}^{\check r}](\tau ,\vec \sigma )-\nonumber \\
&-&\int_{S^2_{\tau ,\infty}}d^2\Sigma_{\check u} \{
\epsilon k\sqrt{\gamma} \, {}^3g
^{\check u\check v}\, {}^3g^{\check r\check s} [N (\partial_{\check r}\,
{}^3g_{\check v\check s}-\partial_{\check v}\, {}^3g_{\check r\check s})+
\nonumber \\
&+&\partial_{\check u}N ({}^3g_{\check r\check s}-\delta_{\check
r\check s})-\partial_{\check r}N ({}^3g_{\check s\check v}-\delta
_{\check s\check v})]-2N_{\check r}\, {}^3{\tilde \Pi}^{\check r\check u}
\} (\tau ,\vec \sigma )=\nonumber \\
&=&\int d^3\sigma [(N_{(as)}+m) {\tilde {\cal H}}
+(N_{(as)\check r}+m_{\check r})\,
{}^3{\tilde {\cal H}}^{\check r}](\tau ,\vec \sigma )-\nonumber \\
&-&\int_{S^2_{\tau ,\infty}}d^2\Sigma_{\check u} \{
\epsilon k\sqrt{\gamma} \, {}^3g
^{\check u\check v}\, {}^3g^{\check r\check s} [N_{(as)} (\partial_{\check r}\,
{}^3g_{\check v\check s}-\partial_{\check v}\, {}^3g_{\check r\check s})+
\nonumber \\
&+&\partial_{\check u}N_{(as)} ({}^3g_{\check r\check s}-\delta_{\check
r\check s})-\partial_{\check r}N_{(as)} ({}^3g_{\check s\check v}-\delta
_{\check s\check v})]-2N_{(as) \check r}\, {}^3{\tilde \Pi}^{\check r\check u}
\} (\tau ,\vec \sigma )=\nonumber \\
&=&\int d^3\sigma [(N_{(as)}+m) {\tilde {\cal H}}+(N_{(as)\check r}+
m_{\check r})\, {}^3{\tilde {\cal H}}^{\check r}](\tau ,\vec \sigma )+
\nonumber \\
&+&{\tilde \lambda}_{(\mu )}(\tau ) P^{(\mu )}_{ADM}+
{\tilde \lambda}_{(\mu )(\nu )}(\tau ) S^{(\mu )(\nu )}_{ADM}=\nonumber \\
&=&\int d^3\sigma [(N_{(as)}+m) {\tilde {\cal H}}+(N_{(as)\check r}+
m_{\check r})\, {}^3{\tilde {\cal H}}^{\check r}](\tau ,\vec \sigma )+
\nonumber \\
&+&{\tilde \lambda}_A(\tau )P^A_{ADM}+{1\over 2}{\tilde \lambda}_{AB}(\tau )
J^{AB}_{ADM} \approx \nonumber \\
&\approx& {\tilde \lambda}_A(\tau )
P^A_{ADM}+{1\over 2}{\tilde \lambda}_{AB}(\tau ) J^{AB}_{ADM}.
\label{III10}
\end{eqnarray}

\noindent Indeed, by putting $N=N_{(as)}$, $N_{\check r}=N_{(as)\check r}$ in
the surface integrals,  the added term (\ref{a3}) becomes the given linear
combination of the strong ADM Poincar\'e charges $P^A_{ADM}$, $J^{AB}_{ADM}$
\cite{reg,reg1} first identified in the linearized theory \cite{adm}
[they are called ``strong conserved
improper charges" in analogy with the strong (surface integrals) Yang-Mills
non Abelian charges, whose conservation is not a consequence of the equations
of motion \cite{lusa} in that case]:

\begin{eqnarray}
P^{\tau}_{ADM}&=&\epsilon k
\int_{S^2_{\tau ,\infty}}d^2\Sigma_{\check u}
[\sqrt{\gamma}\,\, {}^3g^{\check u\check v}\, {}^3g^{\check r\check s}
(\partial_{\check r}\, {}^3g_{\check v\check s}-\partial_{\check v}\, {}^3g
_{\check r\check s})](\tau ,\vec \sigma ),\nonumber \\
P^{\check r}_{ADM}&=&-2 \int_{S^2_{\tau ,\infty}}d^2\Sigma
_{\check u} \, {}^3{\tilde \Pi}^{\check r\check u}(\tau ,\vec \sigma ),
\nonumber \\
J_{ADM}^{\tau \check r}&=&\epsilon k
\int_{S^2_{\tau ,\infty}}d^2\Sigma_{\check u} \sqrt{\gamma}\,\,
{}^3g^{\check u\check v}\, {}^3g^{\check n\check s}\cdot \nonumber \\
&\cdot& [\sigma^{\check r} (\partial_{\check n}\, {}^3g_{\check v\check s}-
\partial_{\check v}\, {}^3g_{\check n\check s})+\delta^{\check r}_{\check v}
({}^3g_{\check n\check s}-\delta_{\check n\check s})-\delta^{\check r}
_{\check n}({}^3g_{\check s\check v}-\delta_{\check s\check v})]
(\tau ,\vec \sigma ),\nonumber \\
J_{ADM}^{\check r\check s}&=&\int_{S^2_{\tau ,\infty}}d^2\Sigma_{\check u}
[\sigma^{\check r}\, {}^3{\tilde \Pi}^{\check s\check u}-
\sigma^{\check s}\, {}^3{\tilde \Pi}^{\check r\check u}]
(\tau ,\vec \sigma ),\nonumber \\
&&{}\nonumber \\
P^{(\mu )}_{ADM}&=&
l^{(\mu )} P^{\tau}_{ADM}+b^{(\mu )}_{(\infty ) \check r}(\tau ) P^{\check
r}_{ADM}=b^{(\mu )}_{(\infty )A}(\tau ) P^A_{ADM},\nonumber \\
S^{(\mu )(\nu )}_{ADM}&=&
[l^{(\mu )}_{(\infty )}b^{(\nu )}_{(\infty
) \check r}(\tau )-l^{(\nu )}_{(\infty )}b^{(\mu )}_{(\infty ) \check r}(\tau
)] J^{\tau \check r}_{ADM}+\nonumber \\
&+&[b^{(\mu )}_{(\infty ) \check r}(\tau )b^{(\nu )}_{(\infty ) \check s}(\tau
)-b^{(\nu )}_{(\infty ) \check r}(\tau )b^{(\mu )}_{(\infty ) \check s}(\tau )]
J^{\check r\check s}_{ADM}=\nonumber \\
&=&[b^{(\mu )}_{(\infty )A}(\tau )b^{(\nu )}_{(\infty )B}(\tau )-b^{(\nu )}
_{(\infty )A}(\tau )b^{(\mu )}_{(\infty )B}(\tau )] J^{AB}_{ADM},\nonumber \\
&&{}\nonumber \\
&&{\tilde \lambda}_A(\tau )={\tilde \lambda}_{(\mu )}(\tau )b^{(\mu )}
_{(\infty )A}(\tau ),\quad\quad {\tilde \lambda}_{(\mu )}(\tau )=b^A
_{(\infty )(\mu )}(\tau ){\tilde \lambda}_A(\tau ),\nonumber \\
&&{\tilde \lambda}_{AB}(\tau )={\tilde \lambda}_{(\mu )(\nu )}(\tau )
[b^{(\mu )}_{(\infty )A}b^{(\nu )}_{(\infty )B}-b^{(\nu )}
_{(\infty )A}b^{(\mu )}_{(\infty )B}](\tau )=2[{\tilde \lambda}_{(\mu )(\nu )}
b^{(\mu )}_{(\infty )A}b^{(\nu )}_{(\infty )B}](\tau ),\nonumber \\
&&{\tilde \lambda}_{(\mu )(\nu )}(\tau )={1\over 4}[b^A_{(\infty)(\mu )}b^B
_{(\infty )(\nu )}-b^B_{(\infty )(\mu )}b^A_{(\infty )(\nu )}](\tau ){\tilde
\lambda}_{AB}(\tau )=\nonumber \\
&&={1\over 2} [b^A_{(\infty )(\mu )}b^B_{(\infty )(\nu )}{\tilde \lambda}
_{AB}](\tau ).
\label{III11}
\end{eqnarray}

\noindent Here $J^{\tau \check r}_{ADM}=-J^{\check r\tau}_{ADM}$ by definition
and the inverse asymptotic tetrads are defined by $b^A_{(\infty )(\mu )}
b^{(\nu )}_{(\infty )B}=\delta^A_B$, $b^A_{(\infty )(\mu )}b^{(\nu )}
_{(\infty )A}=\delta^{(\nu )}_{(\mu )}$.

As shown in Ref.\cite{reg,reg1}, the parity conditions of Eqs.(\ref{III1})
are necessary to have a well defined and finite 3-angular-momentum $J_{ADM}
^{\check r\check s}$: in Appendix B of Ref.\cite{reg1} there is an explicit
example of initial data satisfying the constraints but not the parity conditions
, for which the 3-angular-momentum is infinite [moreover, it is shown that the
conditions of the SPI formalism to kill supertranslations and pick out a
unique asymptotic Poincar\'e group (the vanishing of the first-order asymptotic
part of the pseudomagnetic Weyl tensor) may give infinite 3-angular-momentum
if the parity conditions are not added].

The definition of the boosts $J_{ADM}^{\tau \check r}$ given in Ref.
\cite{reg1} is not only differentiable like the one in Ref.\cite{reg}, but also
finite. As seen in Section II, the problem of boosts is still open. However,
for any isolated system the boost part of the conserved Poincar\'e group cannot
be an independent variable [only the Poincar\'e Casimirs (giving the invariant
mass and spin of the system) are relevant and not the Casimirs of the
Lorentz subgroup]. At the end of this Section this point will be clarified by
giving the explicit realization of the Poincar\'e generators in the rest-frame
instant form (they are independent from the ADM boosts).

Let us  use Eqs.(\ref{a5}) with $N=N_{(as)}$,
$N_{\check r}=N_{(as)\check r}$, and  Eqs.(\ref{III7}) to
rewrite Eq.(\ref{III10}) in the following form

\begin{eqnarray}
{\hat H}^{'}_{(c)ADM}&=&\int d^3\sigma [ m {\tilde {\cal H}}+m_{\check r}\,
{}^3{\tilde {\cal H}}^{\check r}](\tau ,\vec \sigma )+\nonumber \\
&+&{\tilde \lambda}_{\tau}(\tau ) [-\int d^3\sigma {\tilde {\cal H}}(\tau
,\vec \sigma )+P^{\tau}_{ADM}]+{\tilde \lambda}_{\check r}(\tau )[-\int
d^3\sigma \,
{}^3{\tilde {\cal H}}^{\check r}(\tau ,\vec \sigma )+P^{\check r}_{ADM}]
+\nonumber \\
&+&{\tilde \lambda}_{\tau \check r}(\tau )[-{1\over 2}\int d^3\sigma
\sigma^{\check r}\, {\tilde {\cal H}}(\tau ,\vec \sigma )+J^{\tau \check r}
_{ADM}]+\nonumber \\
&+&{\tilde \lambda}_{\check r\check s}(\tau )[-{1\over 2}\int d^3\sigma
\sigma^{\check s}\, {}^3{\tilde {\cal H}}^{\check r}(\tau ,\vec \sigma )+
J^{\check r\check s}_{ADM}]=\nonumber \\
&=&\int d^3\sigma [ m {\tilde {\cal H}}+m_{\check r}\,
{}^3{\tilde {\cal H}}^{\check r}](\tau ,\vec \sigma )+\nonumber \\
&+&\int d^3\sigma \{ \epsilon N_{(as)} [k\sqrt{\gamma}\, {}^3g^{\check r\check
s} ({}^3\Gamma^{\check u}_{\check r\check v}\, {}^3\Gamma^{\check v}_{\check
s\check u}-{}^3\Gamma^{\check u}_{\check r\check s}\, {}^3\Gamma^{\check v}
_{\check v\check u})-\nonumber \\
&-&{1\over {2k\sqrt{\gamma} }} {}^3G_{\check r\check s\check u\check v}\,
{}^3{\tilde \Pi}^{\check r\check s}\,
{}^3{\tilde \Pi}^{\check u\check v}]+\nonumber \\
&+&\epsilon k ({}^3g_{\check v\check s}-\delta_{\check v\check s}) \partial_
{\check r}[\sqrt{\gamma} \partial_{\check u}N_{(as)} ({}^3g^{\check r\check s}
\, {}^3g^{\check u\check v}-{}^3g^{\check u\check r}\, {}^3g^{\check v\check
s})]-\nonumber \\
&-&2 N_{(as)\check r}\, {}^3\Gamma^{\check r}_{\check s\check u}\, {}^3{\tilde
\Pi}^{\check s\check u}+2 \partial_{\check u}N_{(as) \check r}\, {}^3{\tilde
\Pi}^{\check r\check u} \} (\tau ,\vec \sigma )=\nonumber \\
&=& \int d^3\sigma [ m {\tilde {\cal H}}+m_{\check
r} {}^3{\tilde {\cal H}}^{\check r}](\tau ,\vec \sigma )+
{\tilde \lambda}_{(\mu )}(\tau ) {\hat P}^{(\mu )}_{ADM}+{\tilde \lambda}
_{(\mu )(\nu )}(\tau ) {\hat S}^{(\mu )(\nu )}_{ADM} =\nonumber \\
&=&\int d^3\sigma [ m {\tilde {\cal H}}+m_{\check r}\,
{}^3{\tilde {\cal H}}^{\check r}](\tau ,\vec \sigma )+
{\tilde \lambda}_A(\tau ) {\hat P}^A_{ADM}+{1\over 2}{\tilde \lambda}
_{AB}(\tau ){\hat J}^{AB}_{ADM}\approx \nonumber \\
&\approx& {\tilde \lambda}_A(\tau ) {\hat P}^A_{ADM}+{1\over 2}{\tilde
\lambda}_{AB}(\tau ) {\hat J}^{AB}_{ADM},\nonumber \\
{\hat H}^{'}_{(D)ADM}&=&{\hat H}^{'}_{(c)ADM}[m,m^{\check r}]+\nonumber \\
&+&\int d^3\sigma [\lambda_n {\tilde
\pi}^n+\lambda^{\vec n}_r {\tilde \pi}^r_{\vec n}](\tau ,\vec \sigma )+
\zeta_A(\tau ) {\tilde \pi}^A(\tau )+\zeta_{AB}(\tau ) {\tilde \pi}
^{AB}(\tau ),
\label{III12}
\end{eqnarray}

\noindent with the following ``weak conserved improper charges"
${\hat P}^A_{ADM}$, ${\hat J}_{ADM}^{AB}$ [these volume expressions
(the analogue of the weak Yang-Mills non Abelian charges) for the ADM
4-momentum are used in Ref.\cite{positive}
in the study of the positiviteness of the energy;
the weak charges are Noether charges]

\begin{eqnarray}
{\hat P}^{\tau}_{ADM}&=&
\int d^3\sigma \epsilon
[k\sqrt{\gamma}\,\, {}^3g^{\check r\check s}({}^3\Gamma^{\check u}
_{\check r\check v}\, {}^3\Gamma^{\check v}_{\check
s\check u}-{}^3\Gamma^{\check u}_{\check r\check s}\, {}^3\Gamma^{\check v}
_{\check v\check u})-\nonumber \\
&-&{1\over {2k\sqrt{\gamma} }} {}^3G_{\check r\check s\check u\check v}\,
{}^3{\tilde \Pi}^{\check r\check s}\,
{}^3{\tilde \Pi}^{\check u\check v}](\tau ,\vec \sigma ),\nonumber \\
{\hat P}^{\check r}_{ADM}&=&-
2\int d^3\sigma \, {}^3\Gamma
^{\check r}_{\check s\check u}(\tau ,\vec \sigma )\, {}^3{\tilde \Pi}^{\check
s\check u}(\tau ,\vec \sigma ),\nonumber \\
{\hat J}^{\tau \check r}_{ADM}&=&-{\hat J}^{\check r\tau}_{ADM}=
\int d^3\sigma  \epsilon \{ \sigma^{\check r}\nonumber \\
&&[k\sqrt{\gamma}\,\,  {}^3g^{\check n\check s}({}^3\Gamma^{\check u}_{\check
n\check v}\, {}^3\Gamma^{\check v}_{\check s\check u}-{}^3\Gamma^{\check u}
_{\check n\check s}\, {}^3\Gamma^{\check v}_{\check v\check u})-{1\over
{2k\sqrt{\gamma}}} {}^3G_{\check n\check s\check u\check v}\, {}^3{\tilde
\Pi}^{\check n\check s}\, {}^3{\tilde \Pi}^{\check u\check v}]+\nonumber \\
&+& k \delta^{\check r}_{\check u}({}^3g_{\check v\check s}-\delta
_{\check v\check s}) \partial_{\check n}[\sqrt{\gamma}({}^3g^{\check n\check s}
\, {}^3g^{\check u\check v}-{}^3g^{\check n\check u}\, {}^3g^{\check s\check
v})] \} (\tau ,\vec \sigma ),\nonumber \\
{\hat J}^{\check r\check s}_{ADM}&=&
\int d^3\sigma  [(\sigma^{\check r}\, {}^3\Gamma^{\check s}
_{\check u\check v}-\sigma^{\check s}\, {}^3\Gamma^{\check r}_{\check u\check
v})\, {}^3{\tilde \Pi}^{\check u\check v}](\tau ,\vec \sigma ),\nonumber \\
&&{}\nonumber \\
{\hat P}^{(\mu )}_{ADM}&=&
l^{(\mu )}_{(\infty )} {\hat P}^{\tau}_{ADM} + b^{(\mu )}_{(\infty ) \check
r}(\tau ) {\hat P}^{\check r}_{ADM}=b^{(\mu )}_{(\infty )A}(\tau ) {\hat P}^A
_{ADM},\nonumber \\
{\hat S}^{(\mu )(\nu )}_{ADM}&=&
[l^{(\mu )}_{(\infty )}b^{(\nu )}_{(\infty ) \check r}(\tau )-l^{(\nu )}
_{(\infty )}b^{(\mu )}_{(\infty ) \check r}(\tau )] {\hat J}^{\tau \check r}
_{ADM} +\nonumber \\
&+&[b^{(\mu )}_{(\infty ) \check r}(\tau )b^{(\nu )}_{(\infty ) \check
s}(\tau )-b^{(\nu )}_{(\infty ) \check r}(\tau )b^{(\mu )}_{(\infty ) \check
s}(\tau )] {\hat J}_{ADM}^{\check r\check s}=\nonumber \\
&=&[b^{(\mu )}_{(\infty )A}b^{(\nu )}_{(\infty )B}-b^{(\nu )}_{(\infty )A}b
^{(\mu )}_{(\infty )B}](\tau ) {\hat J}^{AB}_{ADM}.
\label{III13}
\end{eqnarray}

In both Refs.\cite{reg,reg1} it is shown that the canonical Hamiltonian
${\hat H}^{'}_{(c)ADM}[N,N^{\check r}]$ of Eq.(\ref{III10}) with arbitrary
$N$, $N^{\check r}$, has the same Poisson brackets as in Eq.(\ref{III3a})
for $N=m$, $N^{\check r}=m^{\check r}$ [``proper" gauge transformations]

\begin{eqnarray}
&&\{ {\hat H}^{'}_{(c)ADM}[N_1,N_1^{\check r}],
{\hat H}^{'}_{(c)ADM}[N_2,N_2^{\check r}] \}=
{\hat H}^{'}_{(c)ADM}[N_3,N_3^{\check r}],\nonumber \\
&&{}\nonumber \\
N_3&=&N_2^{\check r}\partial_{\check r} N_1-N_1^{\check r}\partial_{\check r}
N_2,\nonumber \\
N_3^{\check r}&=&{\cal L}_{{\vec N}_2} N_1^{\check r}+N_2 \partial^{\check r}
N_1-N_1 \partial^{\check r} N_2=\nonumber \\
&=&-N_1^{\check s}\partial_{\check s} N_2^{\check r}+N_2^{\check s}\partial
_{\check s} N_1^{\check r}+N_2 \partial^{\check r}
N_1-N_1 \partial^{\check r} N_2,\nonumber \\
&&{}\nonumber \\
if&& N_i(\tau ,\vec \sigma )=m_i(\tau ,\vec \sigma )-{\tilde \lambda}
_{i\tau}(\tau )-{1\over 2}
{\tilde \lambda}_{i\tau \check u}(\tau ) \sigma^{\check u},\quad
i=1,2,\nonumber \\
and&& N_{i\check r}(\tau ,\vec \sigma )=m_{i\check r}(\tau ,\vec \sigma )-
{\tilde \lambda}_{i\check r}(\tau )-{1\over 2}
{\tilde \lambda}_{i\check r\check u}(\tau )
 \sigma^{\check u},\quad i=1,2,\nonumber \\
&&\Downarrow \nonumber \\
N_3&=&m_3-{\tilde \lambda}_{3\tau}-{1\over 2}{\tilde \lambda}_{3\tau \check u}
\sigma^{\check u},\nonumber \\
N_3^{\check r}&=&-\epsilon {}^3g^{\check r\check s}[m_{3\check s}-
{\tilde \lambda}_{3\check s}-{1\over 2}
{\tilde \lambda}_{3\check s\check u}\sigma^{\check u}],\nonumber \\
&&{}\nonumber \\
&&with\nonumber \\
&&{}\nonumber \\
{\tilde \lambda}_{3\tau}&=&-{{\epsilon}\over 2}
\delta^{\check r\check s}[{\tilde \lambda}
_{1\check r}{\tilde \lambda}_{2\tau \check s}-{\tilde \lambda}_{2\check r}
{\tilde \lambda}_{1\tau \check s}],\nonumber \\
{\tilde \lambda}_{3\tau \check u}&=&-{{\epsilon}\over 2}
\delta^{\check r\check s}[{\tilde \lambda}
_{1\check r\check u}{\tilde \lambda}_{2\tau \check s}-{\tilde \lambda}_{2\check
r\check u}{\tilde \lambda}_{1\tau \check s}],\nonumber \\
m_3&=&-\epsilon
{}^3g^{\check r\check s}\Big(m_{2\check s}[\partial_{\check r}m_1-
{\tilde \lambda}_{2\tau \check r}]-m_{1\check s}[\partial_{\check r}m_2-
{1\over 2}{\tilde \lambda}_{2\tau \check r}]+\nonumber \\
&&+\partial_{\check r}m_2[{\tilde \lambda}_{1\check s}+
{1\over 2}{\tilde \lambda}_{1\check
s\check u}\sigma^{\check u}]-\partial_{\check r}m_1[{\tilde \lambda}_{2\check s}
+{1\over 2}{\tilde \lambda}_{2\check s\check u}\sigma^{\check u}]\Big) -
\nonumber \\
&&-{{\epsilon}\over 2}
({}^3g^{\check r\check s}-\delta^{\check r\check s})\Big( {\tilde \lambda}
_{1\tau \check r}[{\tilde \lambda}_{2\check s}+
{1\over 2}{\tilde \lambda}_{2\check s\check
u}\sigma^{\check u}]-{\tilde \lambda}_{2\tau \check r}[{\tilde \lambda}_{1\check
s}+{1\over 2}{\tilde \lambda}_{1\check s\check u}\sigma^{\check u}]\Big) ,
\nonumber \\
{\tilde \lambda}_{3\check r}&=&{1\over 2}\Big(
{\tilde \lambda}_{1\tau}{\tilde \lambda}_{2\tau
\check r}-{\tilde \lambda}_{2\tau}{\tilde \lambda}_{1\tau \check r}-\epsilon
\delta^{\check m\check n}[{\tilde \lambda}_{1\check r\check m}{\tilde \lambda}
_{2\check n}-{\tilde \lambda}_{2\check r\check m}{\tilde \lambda}_{1\check n}]
\Big) ,\nonumber \\
{\tilde \lambda}_{3\check r\check u}&=&{1\over 2}\Big(
{\tilde \lambda}_{1\tau \check u}{\tilde
\lambda}_{2\tau \check r}-{\tilde \lambda}_{2\tau \check u}{\tilde \lambda}
_{1\tau \check r}-\epsilon
\delta^{\check m\check n}[{\tilde \lambda}_{1\check r\check m}
{\tilde \lambda}_{2\check n\check u}-{\tilde \lambda}_{2\check r\check m}
{\tilde \lambda}_{1\check n\check u}]\Big) ,\nonumber \\
m_{3\check r}&=&m_2[\partial_{\check r}m_1-
{1\over 2}{\tilde \lambda}_{1\tau \check r}]-
m_1[\partial_{\check r}m_2-
{1\over 2}{\tilde \lambda}_{2\tau \check r}]+\nonumber \\
&&+\partial_{\check r}m_2[{\tilde \lambda}_{1\tau}+
{1\over 2}{\tilde \lambda}_{1\tau
\check u}\sigma^{\check u}]-\partial_{\check r}m_1[{\tilde \lambda}_{2\tau}+
{1\over 2}{\tilde \lambda}_{2\tau \check u}\sigma^{\check u}]-\nonumber \\
&&-{{\epsilon}\over 2}
({}^3g^{\check m\check n}-\delta^{\check m\check n})\Big([{\tilde \lambda}_{1
\check m}+
{1\over 2}{\tilde \lambda}_{1\check m\check u}\sigma^{\check u}]{\tilde \lambda}
_{2\check r\check n}-[{\tilde \lambda}_{2\check m}+
{1\over 2}{\tilde \lambda}_{2\check m
\check u}\sigma^{\check u}]{\tilde \lambda}_{1\check r\check n}\Big) -
\nonumber \\
&&-\epsilon
{}^3g^{\check m\check n}\Big( m_{1\check m}[\partial_{\check n}m_{2\check r}
-{1\over 2}
{\tilde \lambda}_{2\check r\check n}]-m_{2\check m}[\partial_{\check n}
m_{1\check r}-{1\over 2}{\tilde \lambda}_{1\check r\check n}]+\nonumber \\
&&+{1\over 2}
[\partial_{\check m}m_{1\check r}{\tilde \lambda}_{2\check n\check u}-
\partial_{\check m}m_{2\check r}{\tilde \lambda}_{1\check n\check u}]\sigma
^{\check u} \Big) -\nonumber \\
&&-\epsilon
{}^3g_{\check r\check s}\, {}^3g^{\check t\check n}\, \partial_{\check t}\,
{}^3g^{\check s\check m} \Big( [m_{1\check n}-{\tilde \lambda}_{1\check n}-
{1\over 2}{\tilde \lambda}_{1\check n\check u}\sigma^{\check u}] [m_{2\check m}-
{\tilde \lambda}_{2\check m}-{1\over 2}{\tilde \lambda}_{2\check m\check u}
\sigma^{\check u}]-\nonumber \\
&&-[m_{2\check n}-{\tilde \lambda}_{2\check n}-
{1\over 2}{\tilde \lambda}_{2\check n\check u}\sigma^{\check u}] [m_{1\check m}-
{\tilde \lambda}_{1\check m}-{1\over 2}{\tilde \lambda}_{1\check m\check u}
\sigma^{\check u}]\Big) ,\nonumber \\
&&{}\nonumber \\
\int d^3\sigma&& [m_3{\tilde {\cal H}}+m_3^{\check r}\, {}^3{\tilde {\cal H}}
_{\check r}](\tau ,\vec \sigma )+{\tilde \lambda}_{3A}(\tau ){\hat P}^A_{ADM}+
{1\over 2}{\tilde \lambda}_{3AB}(\tau ){\hat J}^{AB}_{ADM}=\nonumber \\
&=&\int d^3\sigma_1 d^3\sigma_2 \Big[ m_1(\tau ,{\vec \sigma}_1)m_2(\tau ,{\vec
\sigma}_2) \{ {\tilde {\cal H}}(\tau ,{\vec \sigma}_1),{\tilde {\cal H}}(\tau
,{\vec \sigma}_2) \} +\nonumber \\
&+&[m_1(\tau ,{\vec \sigma}_1)m_2^{\check r}(\tau ,{\vec \sigma}_2)-m_2(\tau
,{\vec \sigma}_1)m_1^{\check r}(\tau ,{\vec \sigma}_2)] \{ {\tilde {\cal H}}
(\tau ,{\vec \sigma}_1),{}^3{\tilde {\cal H}}_{\check r}(\tau ,{\vec \sigma}_2)
\} +\nonumber \\
&+&m_1^{\check r}(\tau ,{\vec \sigma}_1)m_2^{\check s}(\tau ,{\vec \sigma}_2)
\{ {}^3{\tilde {\cal H}}_{\check r}(\tau ,{\vec \sigma}_1),{}^3{\tilde {\cal H}}
_{\check s}(\tau ,{\vec \sigma}_2) \} \Big] +\nonumber \\
&+&\int d^3\sigma \Big[ \Big( {\tilde \lambda}_{1A}(\tau )m_2(\tau ,\vec \sigma
)-{\tilde \lambda}_{2A}(\tau )m_1(\tau ,\vec \sigma )\Big) \{ {\hat P}^A_{ADM},
{\tilde {\cal H}}(\tau ,\vec \sigma ) \} +\nonumber \\
&+&\Big( {\tilde \lambda}_{1A}(\tau )m_2^{\check r}(\tau ,\vec \sigma )-{\tilde
\lambda}_{2A}(\tau )m_1^{\check r}(\tau ,\vec \sigma )\Big) \{ {\hat P}^A_{ADM},
{}^3{\tilde {\cal H}}_{\check r}(\tau ,\vec \sigma ) \} +\nonumber \\
&+&{1\over 2}
\Big( {\tilde \lambda}_{1AB}(\tau )m_2(\tau ,\vec \sigma )-{\tilde \lambda}
_{2AB}(\tau )m_1(\tau ,\vec \sigma )\Big) \{ {\hat J}^{AB}_{ADM}, {\tilde {\cal
H}}(\tau ,\vec \sigma ) \} +\nonumber \\
&+&{1\over 2}
\Big( {\tilde \lambda}_{1AB}(\tau )m_2^{\check r}(\tau ,\vec \sigma )-
{\tilde \lambda}_{2AB}(\tau )m_1^{\check r}(\tau ,\vec \sigma )\Big) \{ {\hat
J}^{AB}_{ADM}, {}^3{\tilde {\cal H}}_{\check r}(\tau ,\vec \sigma ) \} \Big] +
\nonumber \\
&+&{\tilde \lambda}_{1A}(\tau ){\tilde \lambda}_{2B}(\tau ) \{ {\hat P}^A_{ADM},
{\hat P}^B_{ADM} \} +
{1\over 4}{\tilde \lambda}_{1AB}(\tau ){\tilde \lambda}_{2CD}(\tau )
\{ {\hat J}^{AB}_{ADM},{\hat J}^{CD}_{ADM} \} +\nonumber \\
&+&{1\over 2}
\Big( {\tilde \lambda}_{1A}(\tau ){\tilde \lambda}_{2CD}(\tau )-
{\tilde \lambda}_{2A}(\tau ){\tilde \lambda}_{1CD}(\tau )\Big) \{ {\hat P}^A
_{ADM},{\hat J}^{CD}_{ADM} \} .
\label{III13a}
\end{eqnarray}

\noindent
This implies:
\hfill\break
i) the Poisson brackets of two proper gauge transformations [${\tilde \lambda}
_{iA}={\tilde \lambda}_{iAB}=0$, i=1,2] is a proper gauge transformation
[${\tilde \lambda}_{3A}={\tilde \lambda}_{3AB}=0$],
see Eq.(\ref{III3a});\hfill\break
ii) if $N_2=m_2$, $N_{2\check r}=m_{2\check r}$ [${\tilde \lambda}_{2A}=
{\tilde \lambda}_{2AB}=0$] correspond to a proper gauge
transformation and $N_1,N_{1\check r}$ [$m_1=m_{1\check r}=0$] to an improper
one, then we get a proper gauge transformation\hfill\break
\hfill\break
${\tilde \lambda}_{3A}={\tilde \lambda}_{3AB}=0$,\hfill\break
$m_3=-\epsilon {}^3g^{\check r\check s}(-{1\over 2}
m_{2\check s}{\tilde \lambda}_{1\tau \check r}+
\partial_{\check r}m_2[{\tilde \lambda}_{1\check s}+
{1\over 2}{\tilde \lambda}_{1\check s
\check u}\sigma^{\check u}])$,\hfill\break
$m_{3\check r}=-{1\over 2}
m_2{\tilde \lambda}_{1\tau \check r}+\partial_{\check r}m_2
[{\tilde \lambda}_{1\tau}+
{1\over 2}{\tilde \lambda}_{1\check s\check u}\sigma^{\check u}]
-{{\epsilon}\over 2}
{}^3g^{\check m\check n}(m_{2\check m}{\tilde \lambda}_{1\check r\check n}-
\partial_{\check m}m_{2\check r}{\tilde \lambda}_{1\check n\check u}\sigma
^{\check u})-$\hfill\break
$-\epsilon
{}^3g_{\check r\check s}\, {}^3g^{\check t\check n}\partial_{\check t}\,
{}^3g^{\check s\check m}(m_{2\check n}[{\tilde \lambda}_{1\check m}+
{1\over 2}{\tilde
\lambda}_{1\check m\check u}\sigma^{\check u}]-m_{2\check m}[{\tilde \lambda}
_{1\check n}+
{1\over 2}{\tilde \lambda}_{1\check n\check u}\sigma^{\check u}])$,
\hfill\break
\hfill\break
and Eqs.(\ref{III13a}) may be interpreted as saying that the 10
Poincar\'e charges are `gauge invariant' and Noether constants of motion

\begin{eqnarray}
&&\lbrace {\hat P}^{\tau}_{ADM},{\tilde {\cal H}}(\tau ,\vec \sigma )\rbrace
=-\partial_{\check r}\, {}^3{\tilde {\cal H}}^{\check r}(\tau ,\vec \sigma )
\approx 0,\nonumber \\
&&\lbrace {\hat P}^{\tau}_{ADM},{}^3{\tilde {\cal H}}_{\check r}(\tau ,\vec
\sigma )\rbrace =0,\nonumber \\
&&\lbrace {\hat P}^{\check r}_{ADM},{\tilde {\cal H}}(\tau ,\vec \sigma )
\rbrace =\epsilon
\partial_{\check s}[{}^3g^{\check r\check s}\, {\tilde {\cal H}}
(\tau ,\vec \sigma )]\approx 0,\nonumber \\
&&\lbrace {\hat P}^{\check r}_{ADM},{}^3{\tilde {\cal H}}_{\check s}(\tau ,\vec
\sigma )\rbrace =-\epsilon
\partial_{\check s}\, {}^3g^{\check r\check t}(\tau ,\vec
\sigma )\, {}^3{\tilde {\cal H}}_{\check t}(\tau ,\vec \sigma )+\nonumber \\
&&+\epsilon
{}^3g^{\check r\check t}(\tau ,\vec \sigma )\, {}^3g_{\check s\check w}(\tau
,\vec \sigma )\partial_{\check t}\, {}^3g^{\check w\check u}(\tau ,\vec \sigma )
\, {}^3{\tilde {\cal H}}_{\check u}(\tau ,\vec \sigma )\approx 0,\nonumber \\
&&\lbrace {\hat J}^{\tau \check r}_{ADM},{\tilde {\cal H}}(\tau ,\vec \sigma )
\rbrace =2\, {}^3{\tilde {\cal H}}^{\check r}(\tau ,\vec \sigma )-
2\partial_{\check
s}[\sigma^{\check r}\, {}^3{\tilde {\cal H}}^{\check s}(\tau ,\vec \sigma )]
\approx 0,\nonumber \\
&&\lbrace {\hat J}^{\tau \check r}_{ADM},{}^3{\tilde {\cal H}}_{\check s}(\tau
,\vec \sigma )\rbrace =- \delta^{\check r}_{\check s}\, {\tilde {\cal
H}}(\tau ,\vec \sigma )\approx 0,\nonumber \\
&&\lbrace {\hat J}^{\check r\check s}_{ADM},{\tilde {\cal H}}(\tau ,\vec
\sigma )\rbrace =\epsilon \partial_{\check u}\Big([{}^3g^{\check r\check u}
\sigma^{\check s}-{}^3g^{\check s\check u}\sigma^{\check r}]{\tilde {\cal H}}
\Big) (\tau ,\vec \sigma )\approx 0,\nonumber \\
&&\lbrace {\hat J}^{\check r\check s}_{ADM},{}^3{\tilde {\cal H}}_{\check
w}(\tau ,\vec \sigma )\rbrace =\Big( (\delta^{\check r}_{\check u}
\delta^{\check s}_{\check w}-\delta^{\check r}_{\check w}\delta^{\check s}
_{\check u}){}^3{\tilde {\cal H}}^{\check u}(\tau ,\vec \sigma )+\nonumber \\
&&+\sigma^{\check s}\Big[
-\epsilon \partial_{\check w}\, {}^3g^{\check r\check t}\,
{}^3{\tilde {\cal H}}_{\check t}+{}^3g_{\check w\check v}\, {}^3g^{\check
r\check m}\partial_{\check m}\, {}^3{\tilde {\cal H}}^{\check v}\Big]
(\tau ,\vec \sigma )-\nonumber \\
&&-\sigma^{\check r}\Big[
-\epsilon \partial_{\check w}\, {}^3g^{\check s\check t}\,
{}^3{\tilde {\cal H}}_{\check t}+{}^3g_{\check w\check v}\, {}^3g^{\check
s\check m}\partial_{\check m}\, {}^3{\tilde {\cal H}}^{\check v}\Big]
(\tau ,\vec \sigma ) \Big) \approx 0,\nonumber \\
&&\Downarrow \nonumber \\
&&\partial_{\tau}\, {\hat P}^A_{ADM}\, {\buildrel \circ \over =}\,
\lbrace {\hat P}^A_{ADM}, {\hat H}^{'}_{(D)ADM} \rbrace =
\lbrace {\hat P}^A_{ADM}, {\hat H}^{'}_{(c)ADM} \rbrace \approx 0,\nonumber \\
&&\partial_{\tau}\, {\hat J}^{AB}_{ADM}\, {\buildrel \circ \over
=}\, \lbrace {\hat J}^{AB}_{ADM}, {\hat H}^{'}_{(D)ADM} \rbrace =
\lbrace {\hat J}^{AB}_{ADM}, {\hat H}^{'}_{(c)ADM} \rbrace \approx 0.
\label{III13ba}
\end{eqnarray}

From Eqs.(\ref{III13ba}) we see that also the strong Poincar\'e charges are
constants of motion [we did not succeeded to show that they are conserved
independently from the first class constraints]

\begin{eqnarray}
P^{\tau}_{ADM}&=&{\hat P}^{\tau}_{ADM}+\int d^3\sigma {\tilde {\cal H}}(\tau
,\vec \sigma ),\nonumber \\
P^{\check r}_{ADM}&=&{\hat P}^{\check r}_{ADM}+\int d^3\sigma \, {}^3{\tilde
{\cal H}}^{\check r}(\tau ,\vec \sigma ),\nonumber \\
J^{\tau \check r}_{ADM}&=&{\hat J}^{\tau \check r}_{ADM}+{1\over 2} \int
d^3\sigma \sigma^{\check r}\, {\tilde {\cal H}}(\tau ,\vec \sigma ),\nonumber \\
J^{\check r\check s}_{ADM}&=&{\hat J}^{\check r\check s}_{ADM}+\int d^3\sigma
[\sigma^{\check s}\, {}^3{\tilde {\cal H}}^{\check r}(\tau ,\vec \sigma )-
\sigma^{\check r}\, {}^3{\tilde {\cal H}}^{\check s}(\tau ,\vec \sigma )],
\nonumber \\
&&\Downarrow \nonumber \\
&&\partial_{\tau}\, P^A_{ADM} \approx 0,\nonumber \\
&&\partial_{\tau}\, J^{AB}_{ADM} \approx 0;
\label{III13bb}
\end{eqnarray}

iii) the Poisson bracket of two improper gauge transformations
[$m_i=m_{i\check r}=0$, i=1,2] is an improper gauge transformation with the
previous ${\tilde \lambda}_{3A}$, ${\tilde \lambda}_{3AB}$ and with\hfill
\break
\hfill\break
$m_3=-{{\epsilon}\over 2}
({}^3g^{\check r\check s}-\delta^{\check r\check s})({\tilde \lambda}
_{1\tau \check r}[{\tilde \lambda}_{2\check s}+
{1\over 2}{\tilde \lambda}_{2\check s\check
u}\sigma^{\check u}]-{\tilde \lambda}_{2\tau \check r}[{\tilde \lambda}
_{1\check s}+
{1\over 2}{\tilde \lambda}_{1\check s\check u}\sigma^{\check u}])$,
\hfill\break
$m_{3\check r}=-{{\epsilon}\over 2}
({}^3g^{\check m\check n}-\delta^{\check m\check n})([{\tilde
\lambda}_{1\check m}+
{1\over 2}{\tilde \lambda}_{1\check m\check u}\sigma^{\check u}]
{\tilde \lambda}_{2\check r\check n}-[{\tilde \lambda}_{2\check m}+{1\over 2}
{\tilde \lambda}_{2\check m\check u}\sigma^{\check u}]{\tilde \lambda}
_{1\check r\check n})$.\hfill\break
\hfill\break
This implies that the 10 strong Poincar\'e charges [and, therefore, also the
weak ones] satisfy the Poincar\'e algebra
modulo the first class constraints, namely modulo the Hamiltonian group ${\bar
{\cal G}}$ of gauge transformations

\begin{eqnarray}
\{ {\hat P}^{\tau}_{ADM},{\hat J}^{\tau \check r}_{ADM} \} &=& -\epsilon
{\hat P}^{\check r}_{ADM},\nonumber\\
\{ {\hat P}^{\tau}_{ADM},{\hat J}^{\check r\check s}_{ADM} \} &=&0,\nonumber \\
\{ {\hat P}^{\check u}_{ADM}, {\hat J}^{\tau \check r}_{ADM} \} &=& -\epsilon
\delta^{\check u\check r}{\hat P}^{\tau}_{ADM}+\epsilon
\int d^3\sigma [({}^3g^{\check u\check
r}-\delta^{\check u\check r}) {\tilde {\cal H}}](\tau ,\vec \sigma ),
\nonumber \\
\{ {\hat P}^{\check u}_{ADM}, {\hat J}^{\check r\check s}_{ADM} \} &=& -\epsilon
\Big[ \delta^{\check u\check s}{\hat P}^{\check r}_{ADM}-\delta
^{\check u\check r}{\hat P}^{\check s}_{ADM}+\nonumber \\
&+&\int d^3\sigma [({}^3g^{\check u\check s}-\delta^{\check u\check s})
{}^3{\tilde {\cal H}}-({}^3g^{\check u\check r}-\delta^{\check u\check r})
{}^3{\tilde {\cal H}}^{\check s}](\tau ,\vec \sigma )\Big] ,\nonumber \\
\{ {\hat J}^{\tau \check r}_{ADM},{\hat J}^{\tau \check s}_{ADM} \} &=&\epsilon
{\hat J}^{\check r\check s}_{ADM},\nonumber \\
\{ {\hat J}^{\tau \check r}_{ADM}, {\hat J}^{\check u\check v}_{ADM} \} &=&
\epsilon \Big[ \delta^{\check r\check u}{\hat J}^{\tau \check v}_{ADM}
-\delta^{\check r\check v}{\hat J}^{\tau \check u}_{ADM}-\nonumber \\
&-&\int d^3\sigma [\Big( \sigma^{\check v}({}^3g^{\check r\check u}-\delta
^{\check r\check u})-\sigma^{\check u}({}^3g^{\check r\check v}-\delta^{\check
r\check v})\Big) {\tilde {\cal H}}](\tau ,\vec \sigma )\Big],\nonumber \\
\{ {\hat J}^{\check r\check s}_{ADM},{\hat J}^{\check u\check v}_{ADM} \} &=&
-\epsilon
[\delta^{\check r\check u}{\hat J}^{\check s\check v}_{ADM}+\delta^{\check
s\check v}{\hat J}_{ADM}^{\check r\check u}-\delta^{\check r\check v}{\hat J}
_{ADM}^{\check s\check u}-\delta^{\check s\check u}{\hat J}_{ADM}^{\check
r\check v}]+\nonumber \\
&+&\epsilon
\int d^3\sigma \Big[ \Big( \sigma^{\check s}({}^3g^{\check r\check v}-\delta
^{\check r\check v})-\sigma^{\check r}({}^3g^{\check s\check v}-\delta^{\check
s\check v})\Big) {}^3{\tilde {\cal H}}^{\check u}+\nonumber \\
&+&\Big( \sigma^{\check u}({}^3g^{\check v\check s}-\delta
^{\check v\check s})-\sigma^{\check v}({}^3g^{\check u\check s}-\delta^{\check
u\check s})\Big) {}^3{\tilde {\cal H}}^{\check r}-\nonumber \\
&-&\Big( \sigma^{\check s}({}^3g^{\check r\check u}-\delta
^{\check r\check u})-\sigma^{\check r}({}^3g^{\check s\check u}-\delta^{\check
s\check u})\Big) {}^3{\tilde {\cal H}}^{\check v}-\nonumber \\
&-&\Big( \sigma^{\check u}({}^3g^{\check r\check v}-\delta
^{\check r\check v})-\sigma^{\check v}({}^3g^{\check u\check r}-\delta^{\check
u\check r})\Big) {}^3{\tilde {\cal H}}^{\check s}\Big] (\tau ,\vec \sigma ),
\nonumber \\
&&\Downarrow\nonumber \\
\lbrace {\hat P}^A_{ADM},{\hat P}^B_{ADM} \rbrace &=& 0,
\nonumber \\
\lbrace {\hat P}^A_{ADM},{\hat J}^{BC}_{ADM} \rbrace &\approx& {}^4\eta^{AC}
{\hat P}^B_{ADM}-{}^4\eta^{AB} {\hat P}^C_{ADM},
\nonumber \\
\lbrace {\hat J}^{AB}_{ADM},{\hat J}^{CD}_{ADM}
\rbrace &\approx& - C^{ABCD}_{EF} {\hat J}^{EF}_{ADM},
\label{III13c}
\end{eqnarray}

\begin{eqnarray}
&&\Downarrow \nonumber \\
&&{}\nonumber \\
&&\lbrace P^A_{ADM},P^B_{ADM} \rbrace \approx 0,\nonumber \\
&&\lbrace P^A_{ADM}, J^{BC}_{ADM} \rbrace \approx {}^4\eta^{AC} P^B_{ADM}-
{}^4\eta^{AB} P^C_{ADM},\nonumber \\
&&\lbrace J^{AB}_{ADM}, J^{CD}_{ADM} \rbrace \approx - C^{ABCD}_{EF} J^{EF}
_{ADM},
\label{III13d}
\end{eqnarray}

\noindent in accord with Eqs. (\ref{III6}).

In Ref.\cite{reg1} it is noted that the terms depending on the constraints in
Eq.(\ref{III13c}) contain the Hamiltonian version of the supertranslation
ambiguity. Indeed, these terms depend on ${}^3g^{\check r\check s}(\tau ,\vec
\sigma )-\delta^{\check r\check s}$ and, by using Eq.(\ref{III1}), this
quantity may be rewritten as $-{1\over r}\, {}^3{\tilde s}^{\check r\check s}
(\tau ,{{\sigma^{\check n}}\over r})+{}^3{\tilde g}^{\check r\check s}(\tau
,\vec \sigma )$ with ${}^3{\tilde g}^{\check r\check s}(\tau ,\vec \sigma )$
going to zero at spatial infinity faster than $1/r$. Now the objects $\int
d^3\sigma \, {1\over r}\, {}^3{\tilde s}^{\check r\check s}
(\tau ,{{\sigma^{\check n}}\over r}) {\tilde {\cal H}}(\tau ,\vec \sigma )
\approx 0$,.... are generators of supertranslations with zero momentum
generalizing those [i.e. $\int d^3\sigma [s {\tilde {\cal H}}+s_{\check r}\,
{}^3{\tilde {\cal H}}](\tau ,\vec \sigma )$] appearing in the Dirac Hamiltonian.

To remove this gauge ambiguity in the Poincar\'e algebra and simultaneously to
kill the supertranslations, which forbid the existence of a unique
Poincar\'e group, the strategy of Ref.\cite{reg1} is to
add four gauge-fixings to the secondary
first class constraints ${\tilde {\cal H}}(\tau ,\vec \sigma )\approx 0$,
${}^3{\tilde {\cal H}}^{\check r}(\tau ,\vec \sigma )\approx 0$
to fix a coordinate system and therefore to build a realization of
the reduced phase space. In Appendix B there is a review of the main
realizations existing in the literature for metric gravity (some of them
have been already quoted in Section II). In Ref.\cite{reg1} one uses the
maximal slice condition and harmonic 3-coordinates.

Anderson's paper \cite{reg2}, quoted in Section II, shows that to have
``zero momentum" for the supertranslations [namely vanishing supertranslation
charges arising from the parts $s(\tau ,\vec \sigma )$, $s_{\check r}(\tau
,\vec \sigma )$ of $n(\tau ,\vec \sigma)$, $n_{\check r}(\tau ,\vec \sigma )$]
and also to have well defined Lorentz charges, one needs the parity
conditions in suitable function spaces, which do not imply a strong Poincar\'e
algebra, and a class ${\cal C}$ of coordinate systems of $M^4$ including
the gauges corresponding to York QI gauge conditions. In that paper it
is also shown that to preserve the boundary conditions containing the parity
conditions, one has to restrict $Diff\, M^4$ to the allowed transformations
$Diff_I\, M^4 \times P$ [namely to pseudo-diffeomorphisms tending to the identity
in a direction-independent way at spatial infinity plus the Poincar\'e group].

Instead of adding gauge fixings, we shall assume the existence of a restricted
class ${\cal C}$ of coordinate systems for $M^4$ associated with
Eqs.(\ref{III7}) [i.e. corresponding to $s(\tau ,\vec \sigma )=
s_{\check r}(\tau,\vec \sigma )={}^3t^{\check r\check s}(\tau ,{{\sigma^{\check
n}}\over r})=0$, ${}^3s_{\check r\check s}(\tau ,{{\sigma^{\check n}}\over r})
=M \delta_{\check r\check s}$] and that the gauge transformations are so
restricted that we cannot leave this class ${\cal C}$.
The four gauge-fixings then allow to choose a particular
coordinate system in the class ${\cal C}$ and to get a strong Poincar\'e
algebra.

Since supertranslations must be absent to have a unique Poincar\'e algebra, it
must be\hfill\break
\hfill\break
$s(\tau ,\vec \sigma )=s_{\check r}(\tau ,\vec \sigma )=0$, namely
$n(\tau ,\vec \sigma )=m(\tau ,\vec \sigma )$, $n_{\check r}(\tau ,\vec
\sigma )=m_{\check r}(\tau ,\vec \sigma )$,\hfill\break
\hfill\break
in every allowed coordinate system.
This suggests that, in a suitable class ${\cal C}$ of coordinate systems for
$M^4$ [then transformed to  coordinates adapted to the 3+1 splitting
of $M^4$ with a foliation with spacelike leaves $\Sigma_{\tau}$, whose allowed
coordinates systems are in
the previously defined atlas ${\cal C}_{\tau}$] asymptotic to
Minkowski coordinates and with the general coordinate
transformations suitably restricted at spatial infinity
so that it is not possible to go
out this class, one should have the following direction-independent boundary
conditions for the ADM variables for
$r\, \rightarrow \, \infty$ [$\epsilon > 0$]

\begin{eqnarray}
{}^3g_{\check r\check s}(\tau ,\vec \sigma )&=&(1+{M\over r})
\delta_{\check r\check s}+{}^3h_{\check r\check s}(\tau ,\vec \sigma ),
\quad\quad {}^3h_{\check r\check s}(\tau ,\vec \sigma )=O(r^{-(1+\epsilon )}),
\nonumber \\
{}^3{\tilde \Pi}^{\check r\check s}(\tau ,\vec \sigma )&=&{}^3k^{\check
r\check s}(\tau ,\vec \sigma )=O(r^{-(2+\epsilon )}),\nonumber \\
&&{}\nonumber \\
N(\tau ,\vec \sigma )&=& N_{(as)}(\tau ,\vec \sigma )
+n(\tau ,\vec \sigma ),\quad\quad n(\tau ,\vec
\sigma )\, = O(r^{-(2+\epsilon )}),\nonumber \\
N_{\check r}(\tau ,\vec \sigma )&=&N_{(as)\check r}(\tau ,\vec \sigma )+
n_{\check r}(\tau ,\vec \sigma ),\quad\quad
n_{\check r}(\tau ,\vec \sigma )\, = O(r^{-\epsilon}),\nonumber \\
&&{}\nonumber \\
N_{(as)}(\tau ,\vec \sigma )&=&
-{\tilde \lambda}_{\tau}(\tau )-{1\over 2}{\tilde \lambda}_{\tau \check
s}(\tau )\sigma^{\check s},\nonumber \\
N_{(as)\check r}(\tau ,\vec \sigma )&=&
-{\tilde \lambda}_{\check r}(\tau )-{1\over 2}{\tilde \lambda}_{\check r\check
s}(\tau ) \sigma^{\check s},\nonumber \\
\Rightarrow&& N_{(as)A}(\tau ,\vec \sigma )\, {\buildrel {def} \over =}\,
(N_{(as)}\, ;\, N_{(as) \check r}\, )(\tau ,\vec \sigma )
=-{\tilde \lambda}_A(\tau )-{1\over 2}{\tilde \lambda}_{A\check s}(\tau )
\sigma^{\check s},
\label{III8}
\end{eqnarray}

\noindent in accord with Regge-Teitelboim\cite{reg} and Beig-O'Murchadha
\cite{reg1}. We have assumed the angle-independent behaviour ${}^3s_{\check
r\check s}(\tau ,{{\sigma^{\check n}}\over r})=M\delta_{\check r\check s}$,
${}^3t^{\check r\check s}(\tau ,{{\sigma^{\check n}}\over r})= 0$. Since
this implies the vanishing of the ADM momentum, $P^{\check r}_{ADM}=0$,
we see that the elimination of
supertranslations seems to be connected with a definition of ``rest frame"
in the asymptotic Dirac coordinates
$z^{(\mu )}_{(\infty )}(\tau ,\vec \sigma )$.
Therefore, the previous boundary conditions on ${}^3g$, ${}^3{\tilde \Pi}$,
are compatible and can be replaced with the Christodoulou-Klainermann ones
of Eq.(\ref{II1}). To have a non-vanishing ADM
momentum one should have ${}^3t^{\check r\check s}(\tau ,\vec \sigma )= const.\,
\delta^{\check r\check s}$ in Eqs.(\ref{III2})
violating the parity conditions and creating
problems with supertranslations.

However, now Eq.(\ref{III13bb}) and $P^{\check r}_{ADM}=0$ imply

\begin{eqnarray}
{\hat P}^{\check r}_{ADM}&\approx& 0,\nonumber \\
P^{(\mu )}_{ADM}&=&b^{(\mu )}_{(\infty )\tau} P^{\tau}_{ADM}=l^{(\mu )}
_{(\infty )} P^{\tau}_{ADM},\nonumber \\
{\hat P}^{(\mu )}_{ADM}&\approx& l^{(\mu )}_{(\infty )} {\hat P}^{\tau}_{ADM}.
\label{III8a}
\end{eqnarray}

\noindent
But, as we have said, these results may be obtained in parametrized Minkowski
theories only after having done the restriction to
Wigner-like hypersurfaces by adding 6 suitable gauge fixing constraints, whose
time constancy implies ${\tilde \lambda}_{(\mu )(\nu )}(\tau )=0$
so that ${\tilde \lambda}_{AB}(\tau )=0$.

With these assumptions one has from Eqs.(6) of I
the following form of the line element

\begin{eqnarray}
ds^2&=& \epsilon \Big( [N_{(as)}+n]^2 - [N_{(as)\check r}+n_{\check r}]
{}^3g^{\check r\check s}[N_{(as)\check s}+n_{\check s}] \Big) (d\tau )^2-
\nonumber \\
&-&2\epsilon [N_{(as)\check r}+n_{\check r}] d\tau d\sigma^{\check r} -
\epsilon \, {}^3g_{\check r\check s} d\sigma^{\check r} d\sigma^{\check s}=
\nonumber \\
&=&\epsilon \Big( [N_{(as)}+n]^2 (d\tau )^2- \nonumber \\
&&{}^3g^{\check r\check s}
[{}^3g_{\check r\check u}d\sigma^{\check u}+(N_{(as)\check r}+n_{\check r})
d\tau ]
[{}^3g_{\check s\check v}d\sigma^{\check v}+(N_{(as)\check s}+n_{\check s})
d\tau ]\Big) .
\label{III9}
\end{eqnarray}

The Dirac Hamiltonian without supertranslations is

\begin{eqnarray}
{\hat H}^{"}_{(c)ADM}
&=&\int d^3\sigma [(N_{(as)}+ n) {\tilde {\cal H}}+(N_{(as)\check r}+
n_{\check r})\, {}^3{\tilde {\cal H}}^{\check r}](\tau ,\vec \sigma )+
\nonumber \\
&+&{\tilde \lambda}_A(\tau ) P^A_{ADM}+{1\over 2}{\tilde \lambda}_{AB}(\tau )
J^{AB}_{ADM}= \nonumber \\
&=&\int d^3\sigma [ n {\tilde {\cal H}}+n_{\check r}\,
{}^3{\tilde {\cal H}}^{\check r}](\tau ,\vec \sigma )+
{\tilde \lambda}_A(\tau ) {\hat P}^A_{ADM}+{1\over 2}{\tilde \lambda}
_{AB}(\tau ){\hat J}^{AB}_{ADM}, \nonumber \\
{\hat H}^{"}_{(D)ADM}&=&{\hat H}^{"}_{(c)ADM}+\int d^3\sigma [\lambda_n {\tilde
\pi}^n+\lambda^{\vec n}_r {\tilde \pi}^r_{\vec n}](\tau ,\vec \sigma )+
\zeta_A(\tau ) {\tilde \pi}^A(\tau )+\zeta_{AB}(\tau ) {\tilde \pi}
^{AB}(\tau ),
\label{III13b}
\end{eqnarray}

The conclusion of this discussion is a qualitative indication on the choice of
of the special class ${\cal C}$ of coordinate systems on $M^4$ and of
the function space ${\cal W}$ [an appropriate weighted
Sobolev space as for Yang-Mills theory\cite{lusa}]
for the field variables ${}^3g_{\check r\check
s}(\tau ,\vec \sigma )$, ${}^3{\tilde \Pi}^{\check r\check s}(\tau ,\vec
\sigma )$, $n(\tau ,\vec \sigma )$, $n_{\check r}(\tau ,\vec \sigma )$ and for
the parameters $\alpha (\tau ,\vec \sigma )$, $\alpha_{\check r}(\tau ,\vec
\sigma )$ [of which $n(\tau ,\vec \sigma )$, $n_{\check r}(\tau ,\vec \sigma
)$ are special cases] of allowed proper gauge transformations connected to
the identity [the rigid improper ones have been eliminated and replaced by
the new canonical variables ${\tilde \lambda}_A(\tau )$, ${\tilde
\lambda}_{AB}(\tau )$], generated by the secondary first class
constraints.

We must have:\hfill\break
i) The atlas ${\cal C}$ for $M^4$ equipped with the 3+1 splittings should
contain only coordinate systems approaching the Dirac asymptotic Minkowski
rectangular coordinates of Eq.(\ref{III5b}) at spatial infinity in a
direction-independent way.\hfill\break
ii) The allowed 3+1 splittings must have the leaves, i.e. the Cauchy spacelike
hypersurfaces $\Sigma_{\tau}$, approaching Minkowski hyperplanes at spatial
infinity in a direction-independent way. The leaves $\Sigma_{\tau}\approx R^3$
have an atlas ${\cal C}_{\tau}$ containing the global coordinate
systems $\{ \sigma^{\check r} \}$ in which Eq.(\ref{III8}) holds.\hfill\break
iii)As a consequence of what has been said and
of Eqs.(\ref{III8}), the space
${\cal W}$ should be defined by angle (or direction)-independent
boundary conditions for the field variables for $r\, \rightarrow \infty$
of the following  form:

\begin{eqnarray}
&&{}^3g_{\check r\check s}(\tau ,\vec \sigma )\, {\rightarrow}_{r\,
\rightarrow \infty}\, (1+{M\over r})
\delta_{\check r\check s}+{}^3h_{\check r\check s}(\tau
,\vec \sigma )=(1+{M\over r})
\delta_{\check r\check s}+o_4(r^{-3/2}),\nonumber \\
&&{}^3{\tilde \Pi}^{\check r\check s}(\tau ,\vec \sigma )\, {\rightarrow}
_{r\, \rightarrow \infty}\, {}^3k^{\check r\check s}(\tau ,\vec \sigma )=
o_3(r^{-5/2}),\nonumber \\
&&n(\tau ,\vec \sigma )\, {\rightarrow}_{r\,
\rightarrow \infty}\, O(r^{-(2+\epsilon )}),\quad \epsilon > 0,\nonumber \\
&&n_{\check r}(\tau ,\vec \sigma )\,
{\rightarrow}_{r\, \rightarrow \infty}\, O(r^{-\epsilon}),\quad
\epsilon > 0,\nonumber \\
&&{\tilde \pi}_n(\tau ,\vec \sigma )\, {\rightarrow}_{r\,
\rightarrow \infty}\, O(r^{-3}),\nonumber \\
&&{\tilde \pi}^{\check r}_{\vec n}(\tau
,\vec \sigma )\, {\rightarrow}_{r\, \rightarrow \infty}\, O(r^{-3}),\nonumber \\
&&\lambda_n(\tau ,\vec \sigma )\, {\rightarrow}_{r\, \rightarrow \infty}\,
O(r^{-(3+\epsilon )}),\nonumber \\
&&\lambda^{\vec n}_{\check r}(\tau ,\vec \sigma )\, {\rightarrow}_{r\,
\rightarrow \infty}\, O(r^{-\epsilon}),\nonumber \\
&&\alpha (\tau ,\vec \sigma )\, {\rightarrow}_{r\, \rightarrow \infty}\,
O(r^{-(3+\epsilon )}),\nonumber \\
&&\alpha_{\check r}(\tau ,\vec \sigma )\, {\rightarrow}_{r\, \rightarrow
\infty}\, O(r^{-\epsilon}),\nonumber \\
&&\Downarrow \nonumber \\
&&{\tilde {\cal H}}(\tau ,\vec \sigma )\, {\rightarrow}_{r\, \rightarrow
\infty}\, O(r^{-3}),\nonumber \\
&&{}^3{\tilde {\cal H}}^{\check r}(\tau ,\vec \sigma )\, {\rightarrow}_{r\,
\rightarrow \infty}\, O(r^{-3}).
\label{III14}
\end{eqnarray}

With these boundary conditions we have $\partial_{\check u}\, {}^3g_{\check
r\check s}=O(r^{-2})$ and not $O(r^{-(1+\epsilon )})$ [note that with this last
condition and $\epsilon < 1/2$ it is shown in Ref.\cite{fadde} that the ADM
action (but in the first order formulation)
becomes meaningless since the spatial integral diverges (in this
reference it is also noted that with these boundary conditions adapted to
asymptotic flatness at spatial infinity the Hilbert action may not produce a
consistent and finite variational principle)]; this is compatible with
the definition of gravitational radiation given by Christodoulou and
Klainermann, but not with the one of Ref.\cite{trautm}.

In this function space ${\cal W}$ supertranslations are not allowed by
definition and proper gauge transformations generated by the secondary
constraints map ${\cal W}$ into itself. A coordinate-independent
characterization of ${\cal W}$ should be given through an intrinsic definition
of a minimal atlas of coordinate charts ${\cal C}_{\tau}$ of $\Sigma_{\tau}$
such that the lifts to 3-tensors on $\Sigma_{\tau}$ in ${\cal W}$ of the
3-diffeomorphisms in $Diff\, \Sigma_{\tau}$  maps them into them. Therefore,
a unique asymptotic Poincar\'e group, modulo gauge transformations, is selected.
Moreover, in accord with Anderson\cite{reg2}
also $Diff\, M^4$ is restricted to $Diff_I\, M^4 \times P$, so to map the
class ${\cal C}$ of coordinate systems into itself. Now in $Diff_I\, M^4\times
P$ the allowed  proper pseudo-diffeomorphisms $Diff_I\, M^4$ are a normal
subgroup (they go to the identity in an angle-independent way at
spatial infinity), while the Poincar\'e group $P$ describes the rigid
improper gauge transformations (the non-rigid improper ones are assumed to be
absent) as in the quoted Bergmann proposal. Finally, following
Marolf, the Poincar\'e group is not interpreted as a group of improper gauge
transformations but only as a source of superselection rules, which however
seem to be consistent only in the rest frame $P^{\check r}_{ADM}=0$, if we
insist on the absence of supertranslations so to have the possibility to define
the ADM spin Casimir.

To summarize this discussion, after the modification of
metric gravity at the canonical level  with the
addition of the surface integrals and with the primary constraints resulting
from the assumed splitting of the lapse and shift functions, two possible
scenarios can be imagined (for the second one  the Lagrangian is unknown):

a) Consider as configurational variables \hfill\break
\hfill\break
$n_A(\tau ,\vec \sigma )=(n\, ;\,
n_{\check r}\, )(\tau ,\vec \sigma )$, ${\tilde \lambda}_A(\tau )$, ${\tilde
\lambda}_{AB}(\tau )$, ${}^3g_{\check r\check s}(\tau ,\vec \sigma )$,
\hfill\break
\hfill\break
with conjugate momenta \hfill\break
\hfill\break
${\tilde \pi}^A_n(\tau ,\vec \sigma )=({\tilde \pi}^n\, ;\,
{\tilde \pi}^{\check r}_{\vec n}\, )(\tau ,\vec \sigma )\approx 0$,
${\tilde \pi}^A(\tau )\approx 0$, ${\tilde \pi}^{AB}
(\tau )\approx 0$, ${}^3{\tilde \Pi}^{\check r\check s}
(\tau ,\vec \sigma )$ \hfill\break
\hfill\break
[the vanishing momenta are assumed to be the primary
constraints], and take the following Dirac Hamiltonian [it is finite and
differentiable] as the defining Hamiltonian:

\begin{eqnarray}
{\hat H}^{(1)}_{(D)ADM}&=&\int d^3\sigma [n_A\, {\tilde {\cal H}}^A+\lambda
_{n\, A} {\tilde \pi}^A_n](\tau ,\vec \sigma )+{\tilde \lambda}_A(\tau )
{\hat P}^A_{ADM}+
{1\over 2}{\tilde \lambda}_{AB}(\tau ) {\hat J}^{AB}_{ADM}+\nonumber \\
&+&\zeta_A(\tau ) {\tilde \pi}^A(\tau )+
\zeta_{AB}(\tau ) {\tilde \pi}^{AB}(\tau ),
\label{III15}
\end{eqnarray}

\noindent where $n_A=(n; n_{\check r})$,
${\tilde {\cal H}}^A=({\tilde {\cal H}}\, ;\, {}^3{\tilde
{\cal H}}^{\check r}\, )$ and where $\lambda_{n\, A}(\tau ,\vec \sigma )=
(\lambda_n\, ;\, \lambda^{\vec n}_{\check r}\, )(\tau ,\vec \sigma )$,
$\zeta_A(\tau )$, $\zeta_{AB}(\tau )$, are
Dirac multipliers associated with the primary constraints.\hfill\break
\hfill\break
For ${\tilde \lambda}
_{AB}(\tau ) =0$, ${\tilde \lambda}_A(\tau )=\epsilon \delta_{A\tau}$, one has
${\hat H}^{(1)}_{(D)ADM}\approx
\epsilon {\hat P}^{\tau}_{ADM}$ \cite{witt}.\hfill\break
\hfill\break
The time constancy of the primary constraints implies the following secondary
ones \hfill\break
\hfill\break
${\tilde {\cal H}}^A(\tau ,\vec \sigma )\approx 0$ [generators of proper
gauge transformations], \hfill\break
${\hat P}^A_{ADM}\approx 0$, ${\hat J}^{AB}_{ADM}\approx 0$ \hfill\break
\hfill\break
[either generators of improper gauge transformations
(in this case 10 conjugate degrees of freedom in the 3-metric
are extra gauge variables) or, following Marolf's proposal \cite{p18},
defining a superselection sector (like it happens for the vanishing of the
color charges for the confinement of quarks)], all of
which are constants of the motion. All the constraints are first class, so that:
\hfill\break
\hfill\break
i) ${\tilde \lambda}_A(\tau )$, ${\tilde \lambda}_{AB}(\tau )$ are arbitrary
gauge variables conjugate to ${\tilde \pi}^A(\tau )\approx 0$, ${\tilde
\pi}^{AB}(\tau )\approx 0$; \hfill\break
ii) the physical reduced phase space of canonical
metric gravity is restricted to have ``zero asymptotic Poincar\'e charges"
so that there is no natural Hamiltonian
for the evolution in $\tau$. \hfill\break
\hfill\break
This is the natural interpretation of ADM metric gravity which leads to the
Wheeler-DeWitt equation after quantization (see the Conclusions for the problem
of time in this scenario) and, in a sense, it is a Machian formulation of an
asymptotically flat noncompact (with boundary) spacetime $M^4$ in the same
spirit of Barbour's approach\cite{p20} and of the closed (without boundary)
Einstein-Wheeler universes.
However, in this case there is no solution to the problem of deparametrization
of metric gravity and no connection with parametrized Minkowski theories
restricted to spacelike hyperplanes.\hfill\break
\hfill\break
Let us remark that the scenario a)
corresponds to the exceptional orbit ${\hat P}^A_{ADM}=0$
of the asymptotic Poincar\'e group.\hfill\break
\hfill\break

b) According to the suggestion of Dirac, modify ADM metric gravity by
adding the 10 new canonical pairs
$x^{(\mu )}_{(\infty )}(\tau )$, $p^{(\mu )}_{(\infty )}$,
$b^{(\mu )}_{(\infty ) A}(\tau )$, $S^{(\mu )(\nu )}_{\infty}$
[with the previously given Dirac brackets implying the orthonormality
constraints for the b's]
to the metric gravity phase space with canonical basis $n_A(\tau ,\vec \sigma )
=(n\, ;\, n_{\check r}\, )(\tau ,\vec \sigma )$, ${\tilde \pi}^A_n(\tau ,\vec
\sigma )=({\tilde \pi}^n; {\tilde \pi}^{\check r}_{\vec n})
\approx 0$ (the primary constraints), ${}^3g_{\check r\check s}(\tau
,\vec \sigma )$, ${}^3{\tilde \Pi}^{\check r\check s}(\tau ,\vec \sigma )$, and
then: \hfill\break
i) add the 10 new primary constraints \hfill\break
\hfill\break
 $\chi^A=p^A_{(\infty )}-{\hat P}^A_{ADM}=
b^A_{(\infty )(\mu )}(\tau ) [p^{(\mu )}_{(\infty )}-b^{(\mu )}
_{(\infty )B}(\tau ) {\hat P}^B_{ADM}] \approx 0$,\hfill\break
\hfill\break
$\chi^{AB}=J^{AB}_{(\infty )}-{\hat J}^{AB}_{ADM}=b^A_{(\infty )
(\mu )}(\tau ) b^B_{(\infty )(\nu )}(\tau ) [S^{(\mu )(\nu )}_{(\infty )}-
b^{(\mu )}_{(\infty )C}(\tau ) b^{(\nu )}_{(\infty )D}(\tau ) {\hat J}^{CD}
_{ADM}] \approx 0$,\hfill\break
\hfill\break
$\{ \chi^A, \chi^{BC} \} \approx {}^4\eta^{AC} \chi^B - {}^4\eta^{AB} \chi^C
\approx 0\quad\quad$,
$\{ \chi^A,\chi^B \} \approx 0$,\hfill\break
$\{ \chi^{AB}, \chi^{CD} \} \approx - C^{ABCD}_{EF} \chi^{EF}\approx 0$.
\hfill\break
\hfill\break
$\{ \chi^A(\tau ), {\tilde \pi}^D_n(\tau ,\vec \sigma )\}=\{ \chi^{AB}(\tau ),
{\tilde \pi}^D_n(\tau ,\vec \sigma )\}=0$,\hfill\break
$\{ \chi^A(\tau ), {\tilde {\cal H}}^D(\tau ,\vec \sigma )\} \approx 0,\quad
\quad \{ \chi^{AB}(\tau ), {\tilde {\cal H}}^D(\tau ,\vec \sigma )\} \approx
0$,\hfill\break
\hfill\break
where
$p^A_{(\infty )}=b^A_{(\infty )(\mu )}p^{(\mu )}_{(\infty )}$, $J^{AB}_{(\infty
)}=b^A_{(\infty )(\mu )}b^B_{(\infty )(\nu )}S^{(\mu )(\nu )}_{(\infty )}$
[remember that $p^A_{(\infty )}$ and $J^{AB}_{(\infty )}$ satisfy a Poincar\'e
algebra];\hfill\break
ii) consider ${\tilde \lambda}_A(\tau )$, ${\tilde \lambda}_{AB}(\tau )$, as
Dirac multipliers [like $\lambda_{n A}(\tau ,\vec \sigma )$] for these 10
new primary constraints, and not as configurational
(arbitrary gauge) variables coming from the
lapse and shift functions [so that there are no conjugate momenta
${\tilde \pi}^A(\tau )$, ${\tilde \pi}^{AB}(\tau )$ and no associated Dirac
multipliers $\zeta_A(\tau )$, $\zeta_{AB}(\tau )$], in the
assumed Dirac Hamiltonian [it is finite and differentiable]

\begin{eqnarray}
H_{(D)ADM}&=& \int d^3\sigma [ n_A {\tilde {\cal H}}^A+\lambda_{n A} {\tilde
\pi}^A_n](\tau ,\vec \sigma )-\nonumber \\
&-&{\tilde \lambda}_A(\tau ) [p^A_{(\infty )}-{\hat P}^A
_{ADM}]-{1\over 2}{\tilde \lambda}_{AB}(\tau )[J^{AB}_{(\infty )}-
{\hat J}^{AB}_{ADM}]\approx 0,
\label{III16}
\end{eqnarray}

Now the reduced phase space is the ADM one and there is consistency with
Marolf's proposal regarding superselection sectors:
on the ADM variables there are only
the secondary first class constraints ${\tilde {\cal H}}^A(\tau ,\vec \sigma )
\approx 0$ [generators of proper gauge transformations], because the other
first class constraints $p^A_{(\infty )}-{\hat P}^A_{ADM}\approx 0$, $J^{AB}
_{(\infty )}-{\hat J}^{AB}_{ADM}\approx 0$ do not generate improper gauge
transformations but eliminate 10 of the extra 20 variables.  One has an
asymptotically flat at spatial infinity
noncompact (with boundary $S_{\infty}$) spacetime $M^4$
with non-vanishing asymptotic Poincar\'e
charges  and the possibility to deparametrize metric gravity so to obtain the
connection with parametrized Minkowski theories restricted to spacelike
hyperplanes [more exactly to Wigner hyperplanes due to the rest-frame condition
$P^{\check r}_{ADM}=0$ forced by the elimination of supertranslations].

While the gauge-fixings for the primary constraints ${\tilde \Pi}^A_n(\tau ,\vec
\sigma )\approx 0$ and the resulting ones for the secondary ones ${\tilde
{\cal H}}^A(\tau ,\vec \sigma )\approx 0$, implying the determination of the
$\lambda_{n A}(\tau ,\vec \sigma )$, follow the scheme outlined at the end of
Section V of I and in the Conclusions of II,
one has to clarify the meaning of the gauge-fixings for the extra
10 first class constraints.

Let us remark that the line element $ds^2$ of Eq.(\ref{III9}) becomes
asymptotically at spatial infinity

\begin{eqnarray}
ds^2_{(as)}&=&\epsilon \Big( [N^2_{(as)}-{\vec N}^2_{(as)}] (d\tau )^2 -2{\vec
N}_{(as)}\cdot d\tau d\vec \sigma -d{\vec \sigma}^2 \Big) +O(r^{-1})=
\nonumber \\
&=&\epsilon \Big( \Big[ {\tilde \lambda}^2_{\tau}-{\vec {\tilde
\lambda}}^2+({\tilde
\lambda}_{\tau}{\tilde \lambda}_{\tau s}-{\tilde \lambda}_r{\tilde \lambda}
_{rs})\sigma^s+{1\over 4}({\tilde \lambda}_{\tau u}{\tilde \lambda}_{\tau v}-
{\tilde \lambda}_{ru}{\tilde \lambda}_{rv})\sigma^u\sigma^v\Big] (d\tau )^2+
\nonumber \\
&+&2({\tilde \lambda}_r+{1\over 2}{\tilde \lambda}_{rs}\sigma^s) d\tau d\sigma^r
-d{\vec \sigma}^2 \Big) +O(r^{-1})=\nonumber \\
&=&\epsilon \Big( [{\tilde \lambda}^2_{\tau}-{\vec {\tilde \lambda}}+2({\tilde
\lambda}_{\tau}{{a_s}\over {c^2}}+\epsilon_{sru}{\tilde \lambda}_r{{\omega^u}
\over c})\sigma^s+{1\over {c^2}}({{a^ua^v}\over {c^2}}+\omega^u\omega^v-\delta
^{uv}{\vec \omega}^2)\sigma^u\sigma^v ] (d\tau )^2+\nonumber \\
&+&2[{\tilde \lambda}_r-\epsilon_{rsu}\sigma^s{{\omega^u}\over c}] d\tau
d\sigma^r - d{\vec \sigma}^2 \Big) +O(r^{-1}),\nonumber \\
&&{}\nonumber \\
&&{\tilde \lambda}_{\tau r}(\tau )=2{{a_r(\tau )}\over {c^2}},\quad acceleration
,\nonumber \\
&&{\tilde \lambda}_{rs}(\tau )=-2\epsilon_{rsu}{{\omega^u(\tau )}\over c},\quad
angular\, velocity\, of\, rotation.
\label{III16a}
\end{eqnarray}

Since we have ${\dot x}^{(\mu )}_s(\tau )\, {\buildrel \circ \over =}\,
b^{(\mu )}_{(\infty )A}{\tilde \lambda}^A(\tau )$, it follows that for ${\tilde
\lambda}_{\tau}(\tau )=\epsilon$, ${\tilde \lambda}_r(\tau )=0$, the origin
moves with 4-velocity $(\epsilon ;\vec 0)$ and has attached an accelerated
rotating coordinate system\cite{stephani}: \hfill\break
\hfill\break
$ds^2_{(as)}={}^4\eta_{AB}d\sigma^Ad\sigma^B +{1\over
{c^2}} [2\vec a\cdot \vec \sigma +({{a^ua^v}\over {c^2}}+\omega^u\omega^v-
\delta^{uv}{\vec \omega}^2)\sigma^u\sigma^v] (d\tau )^2-2\epsilon^{rsu}\sigma^s
{{\omega^u}\over c} d\tau d\sigma^u +O(r^{-1})$,\hfill\break
\hfill\break
which becomes inertial when ${\tilde \lambda}_{AB}(\tau )=0$.

To go to the Wigner-like hypersurfaces [the
analogue of the Minkowski Wigner hyperplanes with the asymptotic normal
$l^{(\mu )}_{(\infty )}=l^{(\mu )}_{(\infty )\Sigma}$ parallel to ${\hat P}
^{(\mu )}_{ADM}$ (i.e. $l^{(\mu )}_{(\infty )}=b^{(\mu )}_{(\infty ) \tau}=
{\hat P}^{(\mu )}_{ADM}/\sqrt{\epsilon {\hat P}^2_{ADM}}$);
see Eqs.(\ref{III8a})] one follows the procedure defined for
Minkowski spacetime: \hfill\break
i) one restricts oneself to spacetimes with $\epsilon p^2_{(\infty )}
={}^4\eta_{(\mu )(\nu )} p^{(\mu )}_{(\infty )}p^{(\nu )}
_{(\infty )} > 0$ [this is possible, because the positivity theorems for the
ADM energy imply that one has only timelike or light-like
orbits of the asymptotic Poincar\'e group]; \hfill\break
ii) one boosts at rest $b^{(\mu )}_{(\infty )A}(\tau )$ and
$S^{(\mu )(\nu )}_{(\infty )}$ with the Wigner boost $L^{(\mu )}{}_{(\nu )}
(p_{(\infty )}, {\buildrel \circ \over p}_{(\infty )})$; \hfill\break
iii) one adds the
gauge-fixings $b^{(\mu )}_{(\infty )A}(\tau ) \approx L^{(\mu )}{}_{(\nu )=A}
(p_{(\infty )}, {\buildrel \circ \over p}_{(\infty )})=\epsilon ^{(\mu )}_A
(u(p_{(\infty )}))$ [with $u^{(\mu )}(p_{(\infty )})=p^{(\mu )}_{(\infty )}/
\pm \sqrt{\epsilon p^2_{(\infty )} }$] and goes to Dirac brackets.\hfill\break
 In this way one gets \hfill\break
\hfill\break
$S^{(\mu )(\nu )}_{(\infty )} \equiv \epsilon^{(\mu )}_C(u(p_{(\infty )}))
\epsilon_D
^{(\nu )}(u(p_{(\infty )})) {\hat J}^{CD}_{ADM}=S^{(\mu )(\nu )}_{ADM}$ and
\hfill\break
$z^{(\mu )}_{(\infty )}(\tau ,\vec \sigma )=x^{(\mu )}_{(\infty )}(\tau )+
\epsilon^{(\mu )}_r(u(p_{(\infty )})) \sigma^r$.\hfill\break
\hfill\break
The origin $x^{(\mu )}_{(\infty )}$ is now replaced by the not covariant
``external" center-of-mass canonical variable\hfill\break
\hfill\break
${\tilde x}^{(\mu )}_{(\infty )}=x^{(\mu )}_{(\infty )}+{1\over 2} \epsilon^A
_{(\nu )}(u(p_{(\infty )})) \eta_{AB} {{\partial \epsilon^B_{(\rho )}(u(p
_{(\infty )}))}\over {\partial p_{(\infty )(\mu )}}} S^{(\nu )(\rho )}
_{(\infty )}$ \hfill\break
\hfill\break
and one has \hfill\break
\hfill\break
$J^{(\mu )(\nu )}_{(\infty )}=
{\tilde x}^{(\mu )}_{(\infty )}p^{(\nu )}_{(\infty )}-{\tilde x}^{(\nu )}
_{(\infty )}p^{(\mu )}_{(\infty )}+{\tilde S}^{(\mu )(\nu )}_{(\infty )}$
\hfill\break
\hfill\break
with ${\tilde S}^{(\mu )(\nu )}_{(\infty )}=S^{(\mu )(\nu )}_{(\infty )}-
{1\over 2} \epsilon^A_{(\rho )}(u(p_{(\infty )})) \eta_{AB} ({{\partial
\epsilon^B_{(\sigma )}(u(p_{(\infty )}))}\over {\partial p_{(\infty )(\mu )}}}
p^{(\nu )}_{(\infty )}-{{\partial \epsilon^B_{(\sigma )}(u(p_{(\infty )}))}
\over {\partial p_{(\infty )(\nu )}}} p^{(\mu )}_{(\infty )} ) S^{(\rho
)(\sigma )}_{(\infty )}$. \hfill\break
\hfill\break
As in the Minkowski case one defines \hfill\break
\hfill\break
${\bar S}^{AB}_{(\infty )}=\epsilon^A_{(\mu )}
(u(p_{(\infty )}))\epsilon^B_{(\nu )}(u(p_{(\infty )})) {\tilde S}^{(\mu )
(\nu )}_{(\infty )}$ \hfill\break
\hfill\break
and one obtains at the level of Dirac brackets

\begin{eqnarray}
{\bar S}^{\check r\check s}_{(\infty )}&\equiv& {\hat J}^{\check r\check s}
_{ADM},\nonumber \\
&&{}\nonumber \\
{\tilde \lambda}_{AB}(\tau )&=&0,\nonumber \\
&&{}\nonumber \\
-{\tilde \lambda}_A(\tau ) \chi^A &=&-{\tilde
\lambda}_A(\tau )\epsilon^A_{(\mu )}(u(p_{(\infty )})) [p^{(\mu )}_{(\infty )}
-\epsilon^{(\mu )}_B(u(p_{(\infty )})) {\hat P}^B_{AM}]=\nonumber \\
&=&-{\tilde \lambda}_A(\tau )\epsilon^A_{(\mu )}(u(p_{(\infty )})) [u^{(\mu )}
(p_{(\infty )}) (\epsilon_{(\infty )}-{\hat P}^{\tau}_{ADM})-\epsilon^{(\mu )}
_{\check r}(p_{(\infty )}){\hat P}^{\check r}_{ADM}]=\nonumber \\
&=&-{\tilde \lambda}_{\tau}(\tau ) [\epsilon_{(\infty )}-{\hat P}^{\tau}_{ADM}]
+{\tilde \lambda}_{\check r}(\tau ) {\hat P}^{\check r}_{ADM},\nonumber \\
&&{}\nonumber \\
\Rightarrow&& \epsilon_{(\infty )}-{\hat P}^{\tau}_{ADM} \approx 0,\quad\quad
{\hat P}^{\check r}_{ADM}\approx 0.
\label{I23}
\end{eqnarray}

\noindent in accord with Eqs.(\ref{III8a}).

Therefore, on the Wigner-like hypersurfaces [they will be named
Wigner-Sen-Witten hypersurfaces in the next Section and define the intrinsic
asymptotic rest frame of the gravitational field; strictly speaking, the
absence of supertranslations makes the scenario b) fully consistent only on
these hypersurfaces], the remaining four extra
constraints are:\hfill\break
\hfill\break
${\hat P}^{\check r}_{ADM}\approx 0$,\hfill\break
\hfill\break
$\epsilon_{(\infty )}=\sqrt{\epsilon p^2
_{(\infty )}}\approx {\hat P}^{\tau }_{ADM} \approx M_{ADM}=\sqrt{\epsilon
{\hat P}^2_{AM}}$.\hfill\break
\hfill\break
 Now the spatial indices have become spin-1 Wigner indices [they
transform with Wigner rotations under asymptotic Lorentz transformations].
As said for parametrized theories in Minkowski spacetime, in this special
gauge 3 degrees of freedom of the gravitational field
become gauge variables, while ${\tilde x}^{(\mu )}_{(\infty )}$
becomes a decoupled observer with his clock near spatial infinity.
These 3 degrees of freedom represent an ``internal" center-of-mass
3-variable ${\vec \sigma}_{ADM}[{}^3g,{}^3{\tilde \Pi}]$ inside the
Wigner-Sen-Witten hypersurface;
$\sigma^{\check r}=\sigma^{\check r}_{ADM}$ is a
variable representing the ``center of mass" of the 3-metric of the slice
$\Sigma_{\tau}$ of the asymptotically flat spacetime $M^4$ and is obtainable
from the weak Poincar\'e charges with the group-theoretical methods of
Ref.\cite{pauri} as it is done in Ref.\cite{mate} for the Klein-Gordon field on
the Wigner hyperplane. Due to ${\hat P}^r_{ADM}\approx 0$ we have\hfill\break
\hfill\break
$\sigma^r_{ADM} \approx -{\hat J}^{\tau r}_{ADM}/ {\hat P}^{\tau}_{ADM}$,
\hfill\break
\hfill\break
so that ${\vec \sigma}_{ADM}\approx 0$ is equivalent to the requirement that
the ADM boosts vanish.

When $\epsilon {\hat P}^2_{ADM} > 0$, with the asymptotic Poincar\'e Casimirs
${\hat P}^2_{ADM}$, ${\hat W}^2_{ADM}$ one can build the M\"oller radius
$\rho_{AMD}=\sqrt{-\epsilon {\hat W}^2_{ADM}}/{\hat P}^2_{ADM}c$,
which is an intrinsic
classical unit of length like in parametrized Minkowski theories, to be used
as an ultraviolet cutoff in a future attempt of quantization\cite{bari}.

By going from ${\tilde x}^{(\mu )}
_{(\infty )}$ [the non-covariant variable replacing $x^{(\mu )}_{(\infty )}$
after going to Dirac brackets with respect to the previous six pairs of second
class constraints] and $p^{(\mu )}_{(\infty )}$ to the canonical basis
\cite{lus1}\hfill\break
\hfill\break
$T_{(\infty )}=p_{(\infty )(\mu )}{\tilde x}^{(\mu )}_{(\infty )}/\epsilon
_{(\infty )}=p_{(\infty )(\mu )}x^{(\mu )}_{(\infty )}/\epsilon_{(\infty )}
{}{}{}$, \hfill\break
$\epsilon_{(\infty )}$, \hfill\break
$z^{(i)}_{(\infty )}=\epsilon_{(\infty )}
({\tilde x}^{(i)}_{(\infty )}-p^{(i)}_{(\infty )}{\tilde x}^{(o)}_{(\infty )}
/p^{(o)}_{(\infty )})$, \hfill\break
$k^{(i)}_{(\infty )}=p^{(i)}_{(\infty )}/\epsilon
_{(\infty )}=u^{(i)}(p^{(\rho )}_{(\infty )})$, \hfill\break
\hfill\break
one finds that the final reduction requires the gauge-fixings \hfill\break
\hfill\break
$T_{(\infty )}-\tau \approx 0$ and $\sigma^{\check r}_{ADM}\approx 0$.
\hfill\break
\hfill\break
Since $\{ T_{(\infty )},\epsilon_{(\infty )} \}=-\epsilon$, with the
gauge fixing $T_{(\infty )}-\tau \approx 0$ one gets ${\tilde \lambda}_{\tau}
(\tau )\approx \epsilon$,  and the final Dirac
Hamiltonian is

\begin{equation}
H_D=M_{ADM}+{\tilde \lambda}_{\check r}(\tau ) {\hat P}^{\check r}_{ADM},\quad
\quad M_{ADM}\approx {\hat P}^{\tau}_{ADM},
\label{I23a}
\end{equation}

\noindent with $M_{ADM}$ [the ADM mass of the universe] the
natural physical Hamiltonian to reintroduce an evolution in $T_{(\infty )}\equiv
\tau$: namely in the rest-frame time identified with the parameter $\tau$
labelling the leaves $\Sigma_{\tau}$ of the foliation of $M^4$. See the
Conclusions for comments on the problem of time in general relativity.

The final gauge fixings $\sigma^{\check r}_{ADM}\approx 0$
imply ${\tilde \lambda}_{\check r}(\tau )\approx 0$, $H_D=M_{ADM}$ and a
reduced theory with the ``external" center-of-mass
variables $z^{(i)}_{(\infty )}$, $k^{(i)}_{(\infty )}$ decoupled [therefore
the choice of the origin $x^{(\mu )}_{(\infty )}$ becomes irrelevant] and
playing the role of a ``point particle clock" for the time $T_{(\infty )}
\equiv \tau$ [see the Conclusions]. There would be a weak form of
Mach's principle, because only relative degrees of freedom would be present.
That $M_{ADM}$ is the correct Hamiltonian for getting a $\tau$-evolution
equivalent to Einstein's equations in spacetimes asymptotically flat at
spatial infinity is also shown in Ref.\cite{fermi}.

The condition ${\tilde \lambda}_{AB}(\tau )=0$ with ${\tilde \lambda}_{\tau}
(\tau )=\epsilon$, ${\tilde \lambda}_r(\tau )=0$
means that at spatial infinity there are no local (direction
dependent) accelerations and/or rotations [$\vec a=\vec \omega =0$]. The
asymptotic line element is \hfill\break
\hfill\break
$ds^2=\epsilon \Big( [1-{\vec {\tilde \lambda}}^2
(\tau )] (d\tau )^2 +2{\tilde \lambda}_r(\tau ) d\tau d\sigma^r -d(\vec \sigma
)^2\Big) +O(r^{-1})$\hfill\break
$=\epsilon \Big( {}^4\eta_{AB}d\sigma^Ad\sigma^B -{\vec
{\tilde \lambda}}^2(\tau ) (d\tau )^2+2{\vec {\tilde \lambda}}(\tau )\cdot
d\tau d\vec \sigma \Big) +O(r^{-1})$,\hfill\break
\hfill\break
 which, for ${\vec {\tilde \lambda}}(\tau
)=0$ reduces to the line element of an inertial system near spatial infinity
[``preferred asymptotic inertial observers"].

The asymptotic rest-frame instant form
realization of the Poincar\'e generators becomes (no more reference to the
boosts ${\hat J}^{\tau r}_{ADM}$)

\begin{eqnarray}
&&\epsilon_{(\infty )}=M_{ADM},\nonumber \\
&&p^{(i)}_{(\infty )},\nonumber \\
&&J^{(i)(j)}_{(\infty )}={\tilde x}^{(i)}_{(\infty )}p^{(j)}_{(\infty )}-
{\tilde x}^{(j)}_{(\infty )} p^{(i)}_{(\infty )} +\delta^{(i)\check r}\delta
^{(j)\check s}{\hat J}^{\check r\check s}_{ADM},\nonumber \\
&&J^{(o)(i)}_{(\infty )}=p^{(i)}_{(\infty )} {\tilde x}^{(o)}_{(\infty )}-
\sqrt{M^2_{ADM}+{\vec p}^2_{(\infty )}} {\tilde x}^{(i)}_{(\infty )}-
{ {\delta^{(i)\check r}{\hat J}^{\check r\check s}_{ADM} \delta^{(\check s(j)}
p^{(j)}_{(\infty )} }\over
{M_{ADM}+\sqrt{M^2_{ADM}+{\vec p}^2_{(\infty )}} } } .
\label{I24}
\end{eqnarray}

\vfill\eject

\section{Wigner-Sen-Witten 3-surfaces.}

In the previous Section the splitting from the lapse and shift functions of
their asymptotic parts [corresponding to improper gauge transformations like in
Yang-Mills theory] and the assumed form of
these asymptotic parts can be interpreted as a restriction on the foliations
realizing the 3+1 splittings of the spacetime $M^4$ [their leaves $\Sigma
_{\tau}$ must tend to Minkowski spacelike hyperplanes at spatial infinity
in a way independent from the direction]. This splitting of the lapse and
shift functions is equivalent, inside parametrized Minkowski theories, to the
gauge-fixings $z^{(\mu )}(\tau ,\vec \sigma )-x^{(\mu )}(\tau )-b^{(\mu )}
_r(\tau ) \sigma^r \approx 0$, which restrict arbitrary spacelike hypersurfaces
to spacelike hyperplanes. The  non asymptotic part of lapse and shift
functions $n(\tau ,\vec \sigma )=m(\tau ,\vec \sigma )$,
$n_r(\tau ,\vec \sigma )=m_r(\tau ,\vec \sigma )$, [after having put $s(\tau
,\vec \sigma )=s_{\check r}(\tau ,\vec \sigma )=0$ to kill supertranslations]
are fields (vanishing at spatial infinity) on the hypersurfaces $\Sigma_{\tau}$
associated to these restricted foliations of $M^4$ [which can be called
`Minkowski-compatible'; they become foliations with spacelike
hyperplanes of the Minkowski spacetime in rectangular coordinates when $G=0$],
which describe local deformations of these hypersurfaces.

Therefore, the boundary conditions defined in the previous Section
and the connection with parametrized Minkowski theories
restricted to spacelike hyperplanes, show that there
exist special families of spacelike hypersurfaces $\Sigma_{\tau}$, diffeomorphic
to $R^3$, in our class of asymptotically flat spacetimes $M^4$, which enjoy
the same formal properties of spacelike hyperplanes in Minkowski spacetime
[i.e. given an origin on each one of them and an adapted tetrad at this origin,
there is a natural parallel transport so that one can uniquely define
the adapted tetrads in all points of $\Sigma_{\tau}$ starting from the given
adapted one at the origin] and which asymptotically agree with Minkowski
spacelike hyperplanes. They  correspond to the ``non-flat
preferred observers of Bergmann\cite{be} (see also the Conclusions of II),
namely there would be a set of ``privileged observers" (privileged tetrads
adapted to $\Sigma_{\tau}$) of ``geometrical nature"
(since they depend on the intrinsic and extrinsic geometry of $\Sigma_{\tau}$;
on the solutions of Einstein's equations they also acquire a ``dynamical
nature" depending on the configuration of the gravitational field itself) and
not of ``static nature" like in the approaches of M\"oller\cite{p23},
Pirani \cite{p24}
and Goldberg\cite{gold}. These privileged observers are  associated with the
existence of the asymptotic Poincar\'e charges. A posteriori, namely after
having solved Einstein's equations, one could try to use these ``geometrical
and dynamical" privileged observers (privileged
non-holonomic coordinate systems replacing
the rectangular Minkowski coordinates of the flat case) in the same
way as, in metric gravity, are used the ``bimetric theories", like the one
of Rosen\cite{p25}, with a set of privileged static non-flat background
metrics. Since a congruence of timelike preferred observers means a tetrad
field adapted to $\Sigma_{\tau}$, tetrad
gravity is again preferred to metric gravity (the other reason being the
fermion fields associated with matter). This congruence of timelike preferred
observers[with asymptotic inertial observers when ${\tilde \lambda}_A(\tau )=
(\epsilon ;\vec 0)$ and ${\tilde \lambda}_{AB}(\tau )=0$,
see the end of previous Section] is a non-Machian element
of these noncompact spacetimes. The asymptotic worldlines of the congruence
may replace the static concept of ``fixed stars" in the study of the
precessional effects of gravitomagnetism on gyroscopes (dragging of inertial
frames) and seem to be naturally connected with the definition of
post-Newtonian coordinates (they require some concept of center of mass in
their definition) \cite{mtw}.

Even if we do not have a characterization [like $z^{(\mu )}(\tau ,\vec
\sigma )=x^{(\mu )}(\tau )+b^{(\mu )}_r(\tau )
\sigma^r$ for Minkowski hyperplanes]
of the hypersurfaces $\Sigma_{\tau}$ corresponding to arbitrary spacelike
hyperplanes in Minkowski spacetime and asymptotic to them in a
direction-independent way at spatial infinity,
we will see in this Section that the special families
of hypersurfaces $\Sigma_{\tau}$ (denoted WSW, for Wigner-Sen-Witten, in the
following), corresponding to the Wigner hyperplanes orthogonal to the
4-momentum of an isolated system and automatically selected by the requirement
of absence of supertranslations, can be defined as those hypersurfaces having
a certain rule of parallel transport and certain preferred adapted tetrad fields
whose direction-independent value near spatial infinity is a tetrad whose
timelike component is an angle-independent normal vector to $\Sigma_{\tau}$ at
spatial infinity $l^{(\mu )}_{(\infty )}=l^{(\mu )}_{(\infty )\Sigma}$, [it is
tangent to $S_{\infty}$], and is parallel to ${\hat P}^{(\mu )}_{ADM}$
(namely ${\hat P}^{(\mu )}_{ADM}$ is normal to $\Sigma_{\tau}$ at
spatial infinity) as required by Eqs.(\ref{III8a})
. Moreover, there are well defined equations determining these
preferred adapted tetrads.

This picture can be obtained by putting together partial results of various
authors. Ashtekar and Horowitz\cite{p26} pointed out the
existence in metric gravity of a privileged family of lapse and shift functions,
which can be extracted by the spinorial demonstration of Witten\cite{p27} of
the positivity of the ADM energy (see Appendix C for a review of spinors on
$M^4$ and on $\Sigma_{\tau}$). In our approach this family can be replaced by
a different one determined by gauge-fixings implying
${\tilde \lambda}_{AB}(\tau )=0$: this is the final condition of absence of
supertranslations, see Eqs.(\ref{III8a}).
Then, Frauendiener\cite{p28} translated this
fact in terms of privileged geometric adapted tetrads on each $\Sigma_{\tau}$ of
this set of spacelike hypersurfaces, enjoying the same properties of
tetrads adapted to Minkowski spacelike hyperplanes (he starts from the
Sen-Witten equation\cite{p27,sen,spinor,spinor1,rindler}
and uses ideas based on the Sparling 3-form \cite{p29,p30}).\hfill\break
\hfill\break
Let us review these arguments in more detail:\hfill\break

i) In his demonstration of the positivity energy theorem [i.e. $P_{ADM,(\mu )}
n^{(\mu )}\geq 0$ for all future pointing null vectors ($n^2=0$), which implies
$P_{ADM,(\mu )} n^{(\mu )}\geq 0$ for all future pointing asymptotic either
timelike or null translations, with $n^{(\mu )}=lim_{r\, \rightarrow \infty}\,
\sigma^{(\mu )}{}_{\tilde A{\tilde A}^{'}}
\xi^{\tilde A}\xi^{{\tilde A}^{'}}\equiv
\xi^{\tilde A}\xi^{{\tilde A}^{'}}$ (by the usual identification of spinor and
tensor indices $(\mu )\equiv \tilde A{\tilde A}^{'}$) for some SU(2) spinor
field on $\Sigma_{\tau}$], Witten \cite{p27} introduced SU(2) spinor fields on
$\Sigma_{\tau}$ [see also Refs.\cite{p35a,fadde}].
In the reformulation using the so called Nester-Witten 2-form
$F(\xi )$ \cite{p35a}, defined on the total space of the spin bundle over
$M^4$ as $F(\xi )=i \sigma_{(\mu )}^{\tilde A{\tilde A}^{'}} {\bar \xi}
_{{\tilde A}^{'}}\, {}^4\nabla_{(\nu )} \xi_{\tilde A} dx^{(\mu )}\wedge dx
^{(\nu )}$, one can show that $P_{ADM,(\mu )}n^{(\mu )}=lim_{r\, \rightarrow
\infty} 2k \int_{\Sigma
_{\tau}} F(\xi ) =2k \int_{S^2_{\tau ,\infty}} dF(\xi )$. As first noted by
Sparling\cite{p29} (see also the last chapter of Vol.2 of Ref.\cite{rindler})
there is a 3-form $\Gamma$ on the spin bundle, the so called Sparling 3-form,
such that \hfill\break
\hfill\break
$\Gamma = dF-{1\over 2} n^{\mu}\, {}^4G_{\mu\nu} X^{\nu}$ \hfill\break
\hfill\break
[$\, X
^{\mu}={1\over 6} \epsilon^{\mu}{}_{\alpha\beta\gamma} dx^{\alpha}\wedge
dx^{\beta}\wedge dx^{\gamma}$]; therefore, the vacuum Einstein equations can
be characterized by $d\Gamma =0$. In presence of matter Einstein equations
give $\Gamma \, {\buildrel \circ \over =}\, dF-{k\over 2} n^{\mu}\, {}^4T
_{\mu\nu} X^{\nu}$, so that
$P_{ADM,(\mu )}n^{(\mu )}\, {\buildrel \circ \over =}\,
2k \int_{S^2_{\tau ,\infty}} (\Gamma +{k\over 2} n^{\mu}\, {}^4T_{\mu\nu} X
^{\nu})$. Using the dominant energy condition\cite{he} for the positivity of
the second term, one can arrive at the result
$P_{ADM,(\mu )} n^{(\mu )} \geq 0$ if
the SU(2) spinor $\xi^{\tilde A}$ [$n^{(\mu )}\equiv \xi^{\tilde A}{\bar \xi}
^{{\tilde A}^{'}}$] satisfies the elliptic Sen-Witten equation for the
noncompact hypersurface $\Sigma_{\tau}$ [see Eq.(\ref{c2})]

\begin{equation}
{}^3{\cal D}_{\tilde A\tilde B}\psi^{\tilde B}={}^3{\tilde \nabla}_{\tilde
A\tilde B}\psi^{\tilde B}+{1\over {2\sqrt{2}}}\, {}^3K \psi_{\tilde A} =0.
\label{IV1}
\end{equation}

As stressed by Frauendiener and Mason\cite{p30}, the Sparling 3-form is a
Hamiltonian density for canonical general relativity (see also Ref.
\cite{p35b} on this point), while, when used quasi-locally, the 2-form F gives
rise to Penrose's formula\cite{p31}  for the angular momentum twistor of the
quasi-local mass construction.

As further evidence that these ideas are required for treatment of conserved
quantities in general relativity, it can be shown that the Sparling 3-form
can be extended to be one of a collection of 3-forms ``on the bundle of general
linear frames" which, when pulled back to spacetime, give rise to classical
formulae for the ``pseudo-energy-momentum tensor" of the gravitational field
\cite{p32} [see Ref.\cite{emp} for the Einstein complex, Ref.\cite{ll} for the
Landau-Lifschitz one and Ref.\cite{empt} for a review].
See also Ref.\cite{p33}, where the Sparling 3-form
is studied in arbitrary dimension and where it is contrasted with Yang-Mills
theory. In Ref.\cite{p34} there is the relationship of the Sparling 3-form to
the spin coefficient formalism. These papers show the connection of the
Poincar\'e charges with the standard theory of the Komar superpotentials and
of the energy-momentum pseudotensors, which is reviewed in Appendix D.

See Refs.\cite{soluz,p26} for the existence of solutions of the Sen-Witten
equation on noncompact spacelike hypersurfaces [for non-spacelike ones see the
last chapter of Vol.2 in Ref.\cite{rindler}, its references and Ref.
\cite{horotod}]. See Refs.\cite{bergq} for the non-unicity of Witten's
positivity proof as first noted in Ref.\cite{horope}: other equations different
from the Sen-Witten one can be used in variants of the proof.

In particular, in the paper of Reula in Ref.\cite{soluz}, used in
Ref.\cite{p28}, the problem of the existence of solutions of the Sen-Witten
equation (\ref{IV1}) has been formalized in the following way. An ``initial
data set" $(\Sigma_{\tau},{}^3g_{rs},{}^3K_{rs})$ for Einstein's
equations consists of a 3-dimensional manifold $\Sigma_{\tau}$ without
boundary equipped with a positive definite 3-metric ${}^3g_{rs}$ and a
second rank, symmetric tensor field ${}^3K_{rs}$. For simplicity it is
assumed that $\Sigma_{\tau}$ is diffeomorphic to $R^3$ and that ${}^3g
_{rs}$ and ${}^3K_{rs}$ are $C^{\infty}$ (i.e. smooth) tensor fields on
$\Sigma_{\tau}$. An initial data set is said to satisfy the ``local energy
condition" if the quantities \hfill\break
\hfill\break
$\mu ={1\over 2}[{}^3R+{}^3K_{rs}\, {}^3K
^{rs}-({}^3K)^2]$ and $J_{\mu}=\partial^{\nu}[{}^3K_{\mu\nu}-{}^3g
_{\mu\nu}\, {}^3K]$ \hfill\break
\hfill\break
[i.e. $[{}^3R-{{\epsilon}\over {2k\sqrt{\gamma}}}{\tilde {\cal H}}]
(\tau ,\vec \sigma )$ and $[ -\{ {1\over 2}{}^3{\tilde {\cal H}}^r+{}^3\Gamma
^r_{su}\, {}^3{\tilde \Pi}^{su} +{}^3{\tilde \Pi}^{rs}\partial_s ln\sqrt{\gamma}
\} /\sqrt{\gamma}](\tau ,\vec \sigma )$ in the ADM canonical metric gravity
formalism (Section V of I)] satisfy \hfill\break
\hfill\break
$\mu \geq {|\, J^{\mu}J_{\mu}\, |}^{1/2}$.\hfill\break
\hfill\break
An initial data set is ``asymptotically flat" if one can introduce an
asymptotically Euclidean coordinate system such that ${}^3g_{rs}-\delta_{rs}=
O(r^{-1})$ and $\partial_u\, {}^3g_{rs}=O(r^{-2})$ for $r\, \rightarrow \infty$
and, moreover, ${}^3K_{rs}=O(r^{-2})$ and ${}^3R_{rs}=O(r^{-3})$ for $r\,
\rightarrow \infty$ [they are compatible with Christodoulou-Klainermann
Eqs.(\ref{II1})]. Then one has the
following existence theorem(see also Ref.\cite{p26}): \hfill\break
\hfill\break
If $(\Sigma_{\tau},
{}^3g_{rs},{}^3K_{rs})$ is an initial data set that satisfies the local energy
condition and is asymptotically flat, then for any spinor field $\psi
^{\tilde A}_o$ that is ``asymptotically constant" (i.e. $\partial_r \psi_o
^{\tilde A}=0$ outside a compact subset of $\Sigma_{\tau}$; see also Ref.
\cite{horotod}) there exists a spinor field $\psi^{\tilde A}$ satisfying the
Sen-Witten equation (\ref{IV1}) and such that $\psi^{\tilde A}=\psi_o
^{\tilde A}+O(r^{-1})$ at spatial infinity.\hfill\break

ii)
In Ref.\cite{p26}, Ashtekar and Horowitz note that the Sen-Witten equation
enables one to transport rigidly constant spinors at infinity to the interior of
the 3-manifold on which the initial data are defined. By ``taking squares" of
the Sen-Witten spinors one can construct a ``preferred" family of lapse and
shifts and interpret them as the projections of 4-dimensional null evolution
vector fields $z^{\mu}_{\tau}(\tau ,\vec \sigma )=[Nl^{\mu}+N^{\mu}](\tau
,\vec \sigma )$, $N^{\mu}(\tau ,\vec \sigma )=[z^{\mu}_rN^r](\tau ,\vec
\sigma )$, $[l^{\mu}N_{\mu}](\tau ,\vec \sigma )=0$, $z^2_{\tau}(\tau ,\vec
\sigma )=0$, obtained by transporting rigidly the spacetime asymptotic
translations at spatial infinity. The preferred family
correspond to a ``gauge fixing prescription" for lapse and shift functions.
Next it is shown that,
on the phase space of general relativity, one can compute Hamiltonians
corresponding to these lapse and shifts. Although these Hamiltonians have a
complicated form  in terms of the usual canonical variables (involving volume
and surface integrals), they are simply the volume integrals of squares of
derivatives of the Witten spinors. In particular, the Hamiltonians generating
Witten-time translations are manifestly positive and differentiable.

To define the preferred 4-parameter family of lapses and shifts, they
proceed as follow. Since Sen-Witten's equation enables one to ``transport"
constant spinors $\psi_o^{\tilde A}$ at infinity to the ``interior" of
$\Sigma_{\tau}$ [the rotation of the spinor at infinity causes a rigid rotation
on the entire spinor field], consequently, there is available on
$\Sigma_{\tau}$ a ``distinguished" complex 2-dimensional vector space of
asymptotically constant spinor fields $\psi^{\tilde A}$. With each of these
spinor fields $\psi^{\tilde A}$ we shall associate a lapse-shift pair
$(N,N^{\mu})$ given by $N\equiv \psi^{+\, \tilde A}\psi_{\tilde A}$ and $N^{\mu}
\equiv -i\sqrt{2}\,  \psi^{+\, (\tilde A}\psi^{\tilde B)}$. Let
$\alpha^{\tilde A}$ and $\beta^{\tilde A}$ be two linearly independent Witten
transported spinor fields. Then, by consecutively substituting $\alpha^{\tilde
A}$, $\beta^{\tilde A}$, $(\alpha^{\tilde A}+\beta^{\tilde A})$ and $(\alpha
^{\tilde A}+i\beta^{\tilde A})$ for $\psi^{\tilde A}$ in the above prescription,
one obtains 4 pairs $(N_{(k)},N^{\mu}_{(k)})$ with $(k)=1,2,3,4$, of
lapse-shifts pairs each of which defines an asymptotic null translation.
Consider the real 4-dimensional vector space generated by these pairs: this
space is independent of the initial choice of $\alpha^{\tilde A}$ and $\beta
^{\tilde A}$. This is the preferred family
of lapses and shifts. Each element of this family defines an asymptotic
translation and is, in turn, determined by this translation.

These expressions are essentially spinorial, i.e. they depend on the
phases of the individual spinors whereas the original lapse-shift vector did
not . It is essential for a coherent point of view, therefore, to regard
the spinors as fundamental, and the lapse-shift vector as derived (this requires
supergravity, which motivated Witten, but is not justified in ordinary gravity).
The Witten argument required that the phases of the spinors making up the null
lapse-shift vector be assumed to be asymptotically constant along with the
lapse-shift vector: without this, the argument fails.

In terms of vectors, given a ``tetrad" at infinity, it is noted in Ref.
\cite{p26}that the SL(2,C) Sen-Witten equation then provides us with a tetrad
field everywhere on $\Sigma_{\tau}$. If we rotate the
tetrad at infinity, the entire field rotates rigidly by the same amount; the
freedom is that of global rather than local Lorentz transformations. It is in
this sense that we have a ``gauge fixation procedure". Note, however, that the
preferred tetrad fields depend on the choice of the variables $({}^3g_{rs},
{}^3K_{rs})$ on $\Sigma_{\tau}$; if we change the metric
${}^3g_{rs}$ near $\Sigma_{\tau}$, the tetrad fields change.

It can also be shown\cite{p26} that if ${}^3T^{\mu}$ is a vector field tangent
to $\Sigma_{\tau}$ (not necessarily spacelike)
with asymptotic value ${}^3T_{(\infty )}^{\mu}$, then
${}^3T^{\mu}$ is timelike (respectively, null, spacelike) everywhere,
if ${}^3T_{(\infty )}^{\mu}$ is
timelike (respectively, null, spacelike) at infinity.

Then, in Ref.\cite{p26} it is noted that, if $(M^4,{}^4\eta_{(\mu )(\nu )})$ is
the Minkowski spacetime, then, since the constant spinor fields in it
automatically satisfy Sen-Witten equation,
for any choice of $\Sigma_{\tau}$, the transport of
translations at infinity yields the translational Killing fields everywhere on
$M^4$. In a generic spacetime, however, the transport is tied to the choice of
$\Sigma_{\tau}$. Thus, it is only when we are given a foliation of a generic
spacetime that we can obtain 4 vector fields everywhere on the spacetime, and
they ``depend" on the choice of the foliation. The transport is well suited to
the canonical framework, however, because in this framework one deals only with
3-surfaces.\hfill\break

iii)
All these results can be rephrased in our language, by noting that the family
of preferred lapse and shift functions of Ref.\cite{p26} could be replaced
with the family which can be obtained from
Eq.(\ref{III8}) when ${\tilde \lambda}_{AB}(\tau )=0$ so that $N_{(as) A}(\tau ,
\vec \sigma )=N_{(as) A}(\tau )=- {\tilde \lambda}_A(\tau )$. Our 4 arbitrary
functions ${\tilde \lambda}_A(\tau )$ give the same multiplicity as in the
previous spinorial construction without relying on the special null evolution
vectors needed in it. Therefore, in our approach,
the ``gauge-fixing prescription" for
selecting the preferred family of lapse and shifts becomes the requirement of
absence of supertranslations according to Eqs.(\ref{III8a}), i.e.
${\tilde \lambda}_{AB}(\tau )=0$. But this implies ${\hat P}^{(\mu )}_{ADM}
\approx l^{(\mu )}_{(\infty )} {\hat P}^{\tau}_{ADM}$ and,
as a consequence, the allowed foliations and their leaves, i.e. the
spacelike hypersurfaces $\Sigma_{\tau}$,
could be called ``Wigner-Sen-Witten" (WSW) foliations and spacelike
hypersurfaces, being the analogues of the Wigner foliations and spacelike
hyperplanes of the parametrized Minkowski theories.\hfill\break

iv)
The final step, also to justify our change of the family of lapse and shift
functions, is to eliminate completely any reference to spinors and to
reformulate the properties of these WSW spacelike hypersurfaces $\Sigma_{\tau}$
only in terms of triads on $\Sigma_{\tau}$ and adapted tetrads (see Section III
of I for the transition to general tetrads). This has been done in Ref.
\cite{p28}, where Frauendiener , exploiting the fact that there is a unique
(up to a global sign) correspondence between a spinor and a triad on a spacelike
hypersurface, derives the necessary and sufficient conditions that have to be
satisfied by a triad in order to correspond to a spinor that satisfies the
Sen-Witten equation.

Given a SU(2) spinor $\psi^{\tilde A}$, one constructs the symmetric object
$\psi^{\tilde A}\psi^{\tilde B}$, which corresponds to a spatial 3-vector
${}^3m^{\mu}$ (instead, a SL(2,C) spinor corresponds to a selfdual bivector
\cite{rindler}); this vector is complex and null: ${}^3m_{\mu}\, {}^3m^{\mu}
=0$. Conversely, every complex spatial null vector defines a SU(2) spinor
$\psi^{\tilde A}$ up to a sign. The spinor $\sqrt{2} \psi^{\dagger \, (\tilde A}
\, \psi^{\tilde B)}$ is real and corresponds to a 3-vector ${}^3u^{\mu}$
orthogonal to ${}^3m_{\mu}$, ${}^3m_{\mu}\, {}^3u^{\mu}=0$. Writing ${}^3m
^{\mu}={1\over {\sqrt{2}}}({}^3x^{\mu}+i\, {}^3y^{\mu})$ and defining $a=\psi
^{\dagger}_{\tilde A}\, \psi^{\tilde A}$, one gets ${}^3x^{\mu}\, {}^3y_{\mu}=
{}^3x^{\mu}\, {}^3u_{\mu}={}^3y^{\mu}\, {}^3u_{\mu}=0$, ${}^3x^{\mu}\, {}^3x
_{\mu}={}^3y^{\mu}\, {}^3y_{\mu}={}^3u^{\mu}\, {}^3u_{\mu}=-
\epsilon a^2$. Therefore,
going to the holonomic basis of $\Sigma_{\tau}$, one has ${}^3u^r=-{1\over a}
\epsilon^{ruv}\, {}^3x^u\, {}^3y^v$ and one can define an oriented triad
${}^3e^{(W) r}_{(a)}$ on $\Sigma_{\tau}$: ${}^3e^{(W) r}_{(1)}={}^3x^r$,
${}^3e^{(W) r}_{(2)}={}^3y^r$, ${}^3e^{(W) r}_{(3)}={}^3u^r$. Since the SU(2)
spinor $i \sqrt{2} \partial_{(\tilde A}{}^{\tilde B}\, \psi_{\tilde C)\,
\tilde B}$ corresponds to the curl $\epsilon^{ruv}\partial^u\, {}^3v^v$ of a
3-vector ${}^3v^r$, one can use the Sen-Witten equation for $\psi^{\tilde A}$
to get an equation for $v_{\tilde A\tilde B}=\psi_{\tilde A}\psi_{\tilde B}
\equiv {}^3m_{\mu}$. Since the Sen-Witten equation corresponds to 4 real
equations for 4 real unknown, two of the 6 equations for ${}^3m_{\mu}$ cannot
be independent and generate constraints on the triad. The final result is that
to each solution $\psi^{\tilde A}$ of the Sen-Witten equation it corresponds a
triad on the WSW hypersurface $\Sigma_{\tau}$, satisfying a certain cyclic
condition, with two divergence free 3-vectors and with the third one having a
non-vanishing divergence proportional to the trace of the extrinsic curvature of
$\Sigma_{\tau}$ [on  a maximal WSW hypersurface (${}^3K=0$)
all three vectors are divergence free ]

\begin{eqnarray}
&&{}^3\nabla_r\, {}^3e^{(W) r}_{(1)}={}^3\nabla_r\, {}^3e^{(W) r}_{(2)}=0,
\nonumber \\
&&{}^3\nabla_r\, {}^3e^{(W) r}_{(3)}=-\alpha {}^3K,\nonumber \\
&&{}^3e^{(W) r}_{(1)}\, {}^3e^{(W) s}_{(3)}\, {}^3\nabla_r\, {}^3e^{(W)}_{(2)s}
+{}^3e^{(W) r}_{(3)}\, {}^3e^{(W) s}_{(2)}\, {}^3\nabla_r\, {}^3e^{(W)}_{(1)s}
+{}^3e^{(W) r}_{(2)}\, {}^3e^{(W) s}_{(1)}\,
{}^3\nabla_r\, {}^3e^{(W)}_{(3)s}=0.
\label{IV2}
\end{eqnarray}

\noindent There is a 2-1 correspondence between solutions to the Sen-Witten
equation and such triad fields on the WSW $\Sigma_{\tau}$. It may be checked
that the 4-dimensional freedom in the choice of a spinor at one point (at
spatial infinity) implies that a triad satisfying these conditions is unique up
to global frame ``rotations" and ``homotheties". In this sense, the geometry of
an initial data set ``uniquely" determines a ``triad" and hence together with
the associated surface normal [$l^{(\mu )}_{(\infty )}=l^{(\mu )}_{(\infty )
\Sigma}$ parallel to ${\hat P}^{(\mu )}_{ADM}$ at spatial infinity]
an ``adapted tetrad" ${}^4_{(\Sigma )}{\check E}^{(W)
(\mu )}_A$ in spacetime according to Eq.(40) of I [with $N, N^r$ given by
Eq.(\ref{III8}) with ${\tilde \lambda}_{AB}(\tau )=0$]. We may call these
triads  ``geometrical triads". In Ref.\cite{p28} it is shown: 1) these triads
do not exist for compact $\Sigma_{\tau}$; 2) with nontrivial topology for
$\Sigma_{\tau}$ there can be less than 4 real solutions and the triads cannot
be build; 3) the triads exist for asymptotically null surfaces, but the
corresponding tetrad will be degenerate in the limit of null infinity.

Moreover, in Ref,\cite{p28}, using the results of Ref.\cite{p32}, it is noted
that the Einstein energy-momentum pseudo-tensor\cite{emp} is a canonical object
only in the frame bundle over $M^4$, where it coincides with the Sparling
3-form. In order to bring this 3-form back to a 3-form (and then to an
energy-momentum tensor) over the spacetime $M^4$, one needs a section (i.e.
a tetrad) in the frame bundle. Only with the 3+1 decomposition of $M^4$ with
WSW foliations one gets that (after imposition of Einstein's equations together
with the local energy condition) one has a preferred (geometrical and
dynamical) adapted tetrad on the initial surface $\Sigma_{\tau}$.
\hfill\break

v)
Then one has the geometric problem of determining the Wigner-Sen-Witten
3-hypersurfaces in $M^4$ given a solution of the Hamilton equations. With a
set of cotriads ${}^3{\hat e}_{(a)r}$ solution of the equation of motion we can
construct the associated extrinsic curvature ${}^3K_{rs}$. Then Eqs.(\ref{IV2})
allow to find the associated triads ${}^3e^{(W)r}_{(a)}$ and we can define the
$\Sigma_{\tau}$-adapted tetrads (see Section III of I)
such that the asymptotic normal $l^{(\mu )}
_{(\infty )}$ is parallel to the weak ADM 4-momentum associated with the
solution ${}^3{\hat e}_{(a)r}$.

The Wigner-Sen-Witten hypersurfaces are
those surfaces admitting the given adapted tetrad
fields. By using Eq.(40) of I, we can obtain the $\Sigma_{\tau}$-adapted
cotetrads ${}^4_{(\Sigma )}{\check {\tilde E}}^{(\alpha )}_A(\tau ,\vec \sigma
)$ associated with a solution of the Hamilton equations ${}^3e_{(a)r}$, $N$,
$N_{(a)}$; then, with the transition coefficients $b^A_{\mu}(\tau ,\vec \sigma
)=\partial \sigma^A(z) / \partial z^{\mu}$, we obtain the $\Sigma_{\tau}$-
adapted cotetrads ${}^4_{(\Sigma )}{\check E}^{(\alpha )}_{\mu}(z(\tau ,\vec
\sigma ))$ of Eqs.(39) of I. Since $M^4$ is a globally hyperbolic spacetime,
we have ${}^4_{(\Sigma )}{\check E}^{(o)}_{\mu}(z(\tau ,\vec \sigma ))=\epsilon
l_{\mu}(z(\tau ,\vec \sigma ))=\epsilon N(\tau ,\vec \sigma ) \partial_{\mu}
\tau (z)$ [$l^{\mu}(\tau ,\vec \sigma )$ is the normal to $\Sigma_{\tau}$ in
$z^{\mu}(\tau ,\vec \sigma )$]. Therefore, from the equation $\partial_{\mu}
\tau (z)=l_{\mu}(z)/N(z)$ we can determine the function $\tau (z)$ associated
with the given solution. The WSW hypersurface $\Sigma_{\tau}$ associated with
the given solution is the set of points $z^{\mu}(\tau ,\vec \sigma )$ such
that $\tau (z) = \tau$. This allows to find the functions
$z_{wsw}^{\mu}(\tau ,\vec
\sigma )$ defining the embedding of the Wigner-Sen-Witten hypersurfaces in
$M^4$ and giving its Wigner-Sen-Witten foliations: by construction they satisfy
[on WSW hypersurfaces we have $b^{(\mu )}_{(\infty ) r}(\tau )=\epsilon^{(\mu )}
_r(u(p_{(\infty )})) \sigma^r$]\hfill\break
\hfill\break
$z^{\mu}_{wsw}(\tau ,\vec \sigma )\, {\rightarrow}_{|\vec \sigma |\rightarrow
\infty}\, \delta^{\mu}_{(\mu )} z^{(\mu )}_{(\infty )}(\tau ,\vec \sigma ) =
\delta^{\mu}_{(\mu )} [ x^{(\mu )}_{(\infty )}(\tau )+\epsilon^{(\mu )}_r(u(p
_{(\infty )})) \sigma^r]$\hfill\break
\hfill\break
with $x^{(\mu )}_{(\infty )}(0)$ arbitrary [it reflects the arbitrariness of the
absolute location of the origin of asymptotic coordinates (and, therefore, also
of the ``external" center of mass ${\tilde x}^{(\mu )}_{(\infty )}(0)$) near
spatial infinity]. See Ref.\cite{bai} and its interpretation of the center
of mass in general relativity (this paper contains the main references on the
problem starting from Dixon's definition\cite{dixon}): $x^{(\mu )}_{(\infty
)}(\tau )$ may be interpreted as the arbitrary ``reference" (or ``central")
timelike worldline of this paper.

This also allows the
determination of the coefficients $b^{\mu}_A(\tau ,\vec \sigma )=\partial
z^{\mu}(\tau ,\vec \sigma )/\partial \sigma^A$ allowing the transition from
general 4-coordinates to adapted 4-coordinates. Since ${}^4_{(\Sigma )}{\check
E}^{(\alpha )}_{\mu} dz^{\mu}= {}^4_{(\Sigma )}\theta^{(\alpha )}$ are
non-holonomic coframes (see Appendix A of II), there are not coordinate
hypersurfaces and lines for the associated non-holonomic coordinates
$z^{(\alpha )}$ \cite{petrov} on $M^4$; as shown in Ref.\cite{robinson} for
them we have ${}^4_{(\Sigma )}\theta^{(\alpha )}=dz^{(\alpha )} +z^{(\beta )}
\Big[ {}^4_{(\Sigma )}{\check E}^{(\alpha )}_{\mu}\, {{\partial \,
{}^4_{(\Sigma )}{\check E}^{\mu}_{(\beta )}}\over {\partial z^{(\gamma )}}}
\Big] dz^{(\gamma )}$.

Let us remark that in void spacetimes (see Section VI)
one has ${\hat P}_{ADM}^{(\mu )}=0$ and WSW hypersurfaces do not exist in them:
therefore it is only in presence of matter
that we can recover the Wigner hyperplane underlying a given Wigner-Sen-Witten
hypersurface. Since in parametrized Minkowski theories Wigner hyperplanes are
orthogonal to the total 4-momentum of the isolated matter-field system,
such hyperplanes are not defined in absence of fields and matter.

In conclusion it turns out that with WSW
Minkowski-compatible
foliations with spacelike hypersurfaces $\Sigma_{\tau}$ , preferred adapted
tetrads and cotetrads are associated. Therefore, there are ``preferred
geometrical observers" associated with the leaves $\Sigma_{\tau}$ of a WSW
foliation, which are determined by both the intrinsic and extrinsic (${}^3K$)
geometry of these $\Sigma_{\tau}$'s.

It is not clear whether there exists a characterization of the more general
foliations with ${\tilde \lambda}_{AB}(\tau )\not= 0$, which have  associated
unavoidable supertranslations and have ill-defined $\Sigma_{\tau}$-adapted
tetrads at spatial infinity due to the terms linear in $\vec \sigma$ in the
lapse and shift functions.

Let us finish this Section by quoting the formulation of general relativity as
a ``teleparallel" theory done by Nester in Refs.\cite{p35c}
in order to prove the positivity of gravitational energy with purely tensorial
methods. It could be connected either with a different notion of parallel
transport on the WSW hypersurfaces or
with the characterization of the Minkowski-compatible
hypersurfaces $\Sigma_{\tau}$ corresponding to arbitrary Minkowski
hyperplanes [$l^{\mu}_{(\infty )}=l^{(\mu )}_{(\infty )\Sigma}$
not parallel to $P^{(\mu )}_{ADM}$] and
not to Wigner hyperplanes.

Nester shows that by imposing certain
gauge conditions on tetrads one can obtain positivity of the ADM energy. His
conditions  are closely related to Eqs.(\ref{IV2}).
Specifically, he also imposes the cyclic condition but on global cotriads
rather than on global triads. Clearly, a global triad
defines a connection on an initial surface, by requiring that a parallel vector
field has constant coefficients with respect to the triad. This connection will
be metric compatible and integrable since it preserves the triad. Therefore,
its curvature will be zero, but the torsion will be nonzero. We see from the
present result, that on an initial data set satisfying the local energy
conditions (needed to prove the existence of Sen-Witten spinors)
there exists a ``preferred absolute parallelism".

While the orthonormal coframe ${}^3\theta^{(a)}={}^3e^{(a)}_rd\sigma^r={}^3e
_{(a)r}d\sigma^r$ determines the metric and the Riemannian geometry, a
given Riemannian geometry determines only an equivalence class of orthonormal
coframes: coframes are defined only modulo position-dependent rotations and,
under these gauge transformations, the spin connection transforms as a
SO(3) gauge potential (see Section III of I).
A gauge-fixing for the rotation freedom usually means a choice of a
representative of the spin connection inside its gauge orbit [like the
Coulomb gauge for the electromagnetic vector gauge potential $\vec A$]: this
would induce a choice of an associated coframe with respect to some standard
origin.
However, since coframes ${}^3\theta^{(a)}$ are more elementary of the
Levi-Civita spin connection ${}^3\omega^{(a)}{}_{(b)}$ [which is built in
terms of them], it is possible to define gauge-fixings directly at the level of
coframes [see Ref.\cite{p35c},papers b)].
The idea of these papers is that the choice of a preferred
coframe ${}^3\theta^{(a)}_{(P)}$ on the Riemannian parallelizable 3-manifold
$(\Sigma_{\tau},{}^3g)$
[with its associated metric compatible Levi-Civita connection
and parallel transport and vanishing torsion] may be associated with the
definition of a new kind of parallel transport on $\Sigma_{\tau}$, i.e. of a
``teleparallel" (or ``weitzenb\"ock or distant parallelism")
geometry on $\Sigma_{\tau}$, according to which a covariant vector is parallely
transported along a curve in $\Sigma_{\tau}$ if in each
point q of the curve it has the same components with respect to the local
coframe ${}^3\theta^{(a)}_{(P)}{|}_q$. The special coframe ${}^3\theta^{(a)}
_{(P)}$ is said ``orthoteleparallel" (OT) coframe.
With this structure $(\Sigma_{\tau},
\delta_{(a)(b)})$ is a 3-manifold with flat metric [the curvature vanish
because this parallel transport is path-independent (absolute parallelism) like
in Euclidean geometry] and the OT coframe ${}^3\theta^{(a)}_{(P)}$ is that
special coframe in which by construction also all the spin connection
coefficients vanish], but with a nonvanishing ``torsion" [it completely
characterizes this kind of geometry] \hfill\break
\hfill\break
${}^3T^{(a)}_{(P)}=d{}^3\theta^{(a)}
_{(P)}=-{1\over 2}\, {}^3C^{(a)}{}_{(b)(c)}\, {}^3\theta^{(b)}_{(P)} \wedge
{}^3\theta^{(c)}_{(P)}$, \hfill\break
\hfill\break
${}^3C^{(a)}
{}_{(b)(c)}=-d{}^3\theta^{(a)}({}^3e_{(b)},{}^3e_{(c)})$, ${}^3e_{(a)}={}^3e^r
_{(a)}\partial /\partial \sigma^r$.\hfill\break
\hfill\break
The Riemannian geometry $(\Sigma_{\tau},{}^3g)$
corresponds to a whole equivalence class of teleparallel geometries $(\Sigma
_{\tau},{}^3\theta
^{(a)}_{(P)})$, according to which coframe is chosen as the preferred OT one.

In Ref.\cite{p35c}b) it is pointed out that there exists a natural
(of elliptic type) gauge-fixing for the choice of a special OT coframe
${}^3\theta^{(a)}_{(P)}$ [${}^3e_{(P)(a)}$ is the dual OT frame]:

\begin{eqnarray}
&&\delta {\hat q}_{(P)}=0\quad \Rightarrow \quad d{*}{\hat q}_{(P)}=0
\nonumber \\
&&d{\tilde q}_{(P)}=0,\nonumber \\
&&{}\nonumber \\
{\tilde q}_{(P)}&=& i_{{}^3e_{(P)(a)}}\, {}^3T^{(a)}_{(P)} =i_{{}^3e_{(P)(a)}}\,
d{}^3\theta^{(a)}_{(P)}=-{}^3C^{(a)}{}_{(a)(b)} {}^3\theta^{(b)},\nonumber \\
&&(this\, 1-form\, is\, the\, algebraically\, irreducible\, trace\, of\, the\,
teleparralel\, torsion) ,\nonumber \\
&&{}\nonumber \\
{\hat q}_{(P)}&=&{1\over 2} {}^3\theta^{(a)}_{(P)} \wedge \delta_{(a)(b)}
{}^3T^{(b)}_{(P)} ={1\over 4} {}^3C_{(a)(b)(c)} {}^2\theta^{(a)}_{(P)} \wedge
{}^3\theta^{(b)}_{(P)} \wedge {}^3\theta^{(c)}_{(P)},\nonumber \\
&&(this\, 3-form\, is\, the\, totally\, antisymmetric\, part\, of\, the\,
teleparallel\, torsion),\nonumber \\
&&{}\nonumber \\
&&\Downarrow \nonumber \\
&&-{}^3C^{(a)}{}_{(b)(c)}={2\over 3}
(q^{(a)}_{(P)}{}_{(b)(c)}-q^{(a)}_{(P)}{}_{(c)(b)})+
{1\over 2}(\delta^{(a)}_{(b)} {\tilde q}_{(P)(c)}-\delta^{(a)}_{(c)}{\tilde q}
_{(P)(b)})+{2\over 3} {\hat q}^{(a)}_{(P)}{}_{(b)(c)},\nonumber \\
&&{}^3\Gamma^{(a)(b)}_{(P)}{}_{(c)}={2\over 3}(q_{(P)(c)}{}^{(b)(a)}-
q_{(P)(c)}{}^{(a)(b)})-{1\over 2}(\delta^{(a)}_{(c)}{\tilde q}^{(b)}_{(P)}-
\delta^{(b)}_{(c)}{\tilde q}^{(a)}_{(P)})+{1\over 3}{\hat q}^{(a)(b)}_{(P)}
{}_{(c)},\nonumber \\
&&where\, q_{(P)(a)(b)(c)}\quad is\quad
-{1\over 2}[{}^3C_{(a)(b)(c)}+{}^3C_{(b)(a)(c)}],\quad\quad
with\, all\, traces\, removed.
\label{IV3}
\end{eqnarray}

\noindent Here $\delta$ is the
codifferential [$\delta ={*}d{}*$ with ${*}$ the Hodge dual; $\epsilon_{123}=1$;
for noncompact $\Sigma_{\tau}$  suitable boundary conditions are
needed]; ${*}{\hat q}_{(P)}$ is a function and ${*}\delta {\hat q}_{(P)}$ a
1-form.

These are three conditions (one is the cyclic condition),
 which determine a special orthonormal
coframe on a 3-manifold [i.e. they determine the 3 Euler angles of the dual
frame with respect to a standard frame chosen as an identity cross section in
the orthonormal frame bundle $F(\Sigma_{\tau})$]
once appropriate boundary conditions
are fixed. For asymptotically flat 3-manifolds [${}^3g\rightarrow \delta +
O(1/r)$] the boundary condition is ${\tilde q}_{(P)}\rightarrow 0$ for $r=
\sqrt{{\vec x}^2}\rightarrow \infty$ and ${*}{\hat q}_{(P)}=0$.
When the first de Rahm cohomology group $H_1(\Sigma_{\tau})
=0$ vanishes, the closed 1-form ${\tilde q}_{(P)}$ is globally exact in this
gauge, ${\tilde q}_{(P)}=dF_{(P)}$, and determines a function $F_{(P)}$ up to a
constant, which may be suitably normalized at infinity, and it can be shown
\cite{p35c}c), that it is the best definition of the generalization of the
Newton potential: it is the scale factor which satisfies the superhamiltonian
constraint equation. With this gauge\cite{p35c}c), one gets a locally positive
representation for the Hamiltonian density allowing a new, ``strictly
tensorial" (in contrast to Witten's spinor method\cite{p27}) proof of positive
energy for Einstein's theory of gravity.

Given an orthonormal coframe ${}^3\theta^{(a)}$, the gauge conditions
(\ref{IV3}) become a nonlinear second-order elliptic system for the rotation
matrix defining an OT coframe ${}^3\theta^{(a)}_{(P)}=R^{(a)}{}_{(b)}{}^3
\theta^{(b)}$. In Ref.\cite{p35c}b) it is shown that the associated
linearized problem has a unique solution if $d{}^3\theta^{(a)}$ is not too
large and the second deRahm cohomology group $H_2(\Sigma_{\tau})=0$ vanishes
[for asymptotically flat spaces one should use the first paper in
Ref.\cite{soluz}]. In Ref.\cite{dimakis}
it is shown that for 3-manifolds the gauge conditions (\ref{IV3}) are
essentially equivalent to the ``linear" Dirac equation, for which unique
solutions exist. Hence for 3-manifolds special OT coframes exist except
possibly at those (isolated) points where the Dirac spinor vanishes.

\vfill\eject

\section{Poincar\'e Charges and the Physical Hamiltonian in Tetrad Gravity.}

In the formulation of tetrad gravity given in I and II we used the ADM action.
Therefore, all the discussion of Section III about the differentiability of
the Hamiltonian, the definition of Poisson brackets, the definition of proper
and improrper gauge transformations can be directly reformulated in tetrad
gravity. The only difference inside tetrad gravity
in the Hamiltonian treatment of quantities
depending upon ${}^3g_{rs}(\tau ,\vec \sigma )$, ${}^3{\tilde \Pi}^{rs}(\tau
,\vec \sigma )$,  is that now we have $\{ {}^3{\tilde
\Pi}^{rs}(\tau ,\vec \sigma ), {}^3{\tilde \Pi}^{uv}(\tau ,{\vec \sigma}^{'})
\} = \delta^3(\vec \sigma ,{\vec \sigma}^{'}) F^{rsuv}_{(a)(b)}(\tau ,\vec
\sigma ) \, {}^3{\tilde M}_{(a)(b)}(\tau ,\vec \sigma ) \approx 0$ [see
Eqs.(84) of I] and not $=0$. Therefore, constants of motion (functional
$F[{}^3g_{rs},{}^3{\tilde \Pi}^{rs}]$) of metric gravity remain such in tetrad
gravity, since they have weakly zero Poisson brackets with ${\tilde {\cal
H}}(\tau ,\vec \sigma )$, ${}^3{\tilde {\cal H}}^r(\tau ,\vec \sigma )$ [and,
therefore, with ${}^3{\tilde \Theta}_r(\tau ,\vec \sigma )$ and ${\hat {\tilde
{\cal H}}}_{(a)}(\tau ,\vec \sigma )$, see Eqs.(61), (62), (79) and (85) of I]
and also
with the other first class constraints ${\tilde \pi}^{\vec \varphi}_{(a)}(\tau
,\vec \sigma )\approx 0$, ${}^3{\tilde M}_{(a)}(\tau ,\vec \sigma )\approx 0$
[see Eqs. (54) and (55) of I].

As a consequence the weak and strong
Poincar\'e charges are still constants of motion in tetrad gravity and their
weak Poincar\'e algebra under Poisson brackets may only be modified by extra
terms containing ${}^3{\tilde M}_{(a)}(\tau ,\vec \sigma )\approx 0$. A more
complete study of these properties would require the study of the
quasi-invariances of the Lagrangian density of tetrad gravity [Eq.(51) of I]
under the gauge transformations generated by the 14 first class constraints of
the theory [using the second Noether theorem as it was done in Appendix A
for metric gravity].

The only lacking ingredients are the definition of
proper gauge transformations generated by the primary (without associated
secondary) first class constraints ${\tilde \pi}^{\vec \varphi}_{(a)}(\tau
,\vec \sigma )\approx 0$, ${}^3{\tilde M}_{(a)}(\tau ,\vec \sigma )\approx 0$,
and the boundary conditions for cotriads ${}^3e_{(a)r}(\tau ,\vec \sigma )$,
because the lapse and shift functions $N(\tau ,\vec \sigma )$, $N_{(a)}
(\tau ,\vec \sigma )={}^3e_{(a)}^r(\tau ,\vec \sigma ) N_r(\tau ,\vec \sigma )$
are treated in the same way as in metric gravity.

Therefore, we shall assume that there exist the same coordinate systems of
$M^4$ and $\Sigma_{\tau}$ as in metric gravity [for the sake of simplicity
the indices $\check r$ are replaced with $r$] and that the $\Sigma_{\tau}$
-adapted tetrads of Eqs.(39) of I, whose expression is
${}^4_{(\Sigma )}{\check E}^{\mu}_{(\mu )}$ with \hfill\break
\hfill\break
${}^4_{(\Sigma )}{\check E}^{\mu}_{(o)}=l^{\mu}\quad\quad$,
${}^4_{(\Sigma )}{\check E}^{\mu}_{(a )}={}^3e^s_{(a)}
b^{\mu}_s$ \hfill\break
\hfill\break
[$b^{\mu}_A$ are the transformation coefficients to $\Sigma_{\tau}$
-adapted coordinates], have a well defined angle-independent limit ${}^4
_{(\Sigma )}{\check E}^{\mu}_{(\infty )(\mu )}$ at spatial infinity,
such that \hfill\break
\hfill\break
${}^4_{(\Sigma )}{\check
E}^{\mu}_{(\infty )(o )}=\delta^{\mu}_{(\mu )} l^{(\mu )}_{(\infty )}=
\delta^{\mu}_{(\mu )} b^{(\mu )}_{(\infty )\tau}\quad\quad$,
${}^4_{(\Sigma )}{\check
 E}^{\mu}_{(\infty )(a)}=\delta^s_{(a)}\delta^{\mu}_{(\mu )} b^{(\mu
)}_{(\infty )s}(\tau )$ \hfill\break
\hfill\break
with the same asymptotic $b^{(\mu )}_{(\infty )A}(\tau )$ of Section III.

Let us remark that the $\Sigma_{\tau}$-adapted tetrads in adapted
coordinates of Eqs.(40) of I, are ${}^4_{(\Sigma )}{\check
{\tilde E}}^A_{(\mu )}$ with \hfill\break
\hfill\break
${}^4_{(\Sigma )}{\check
{\tilde E}}^{\tau}_{(\mu )}=({1\over N}; -{}^3e^r_{(a)}\, {}^3e^s_{(a)} {{N_s}
\over N})\quad\quad$,
${}^4_{(\Sigma )}{\check {\tilde E}}^r_{(\mu )}=(0;{}^3e^r_{(a)})$.
\hfill\break
\hfill\break
Due to the presence of the lapse function in the denominator which is linearly
increasing in $\vec \sigma$ [to have the possibility of defining
$J^{AB}_{ADM}$], these adapted tetrads exist without singularities at
spatial infinity only if ${\tilde \lambda}_{AB}(\tau )=0$, i.e. on WSW
hypersurfaces. The same happens for the adapted cotetrads
${}^4_{(\Sigma )}{\check {\tilde E}}^{(\mu )}_A$ with
${}^4_{(\Sigma )}{\check {\tilde E}}^{(o )}_A=(N; 0)$,
${}^4_{(\Sigma )}{\check {\tilde E}}^{(a )}_A=(N^{(a)}=N_{(a)}; {}^3e^{(a)}_r
={}^3e_{(a)r})$. Therefore, it seems again that tetrad gravity, without
supertranslations and with Poincar\'e charges, admits well defined adapted
tetrads and cotetrads (with components in adapted holonomic coordinates) only
after having been restricted to WSW hypersurfaces (rest frame), whose
asymptotic normals $l^{(\mu )}_{(\infty )}=l^{(\mu )}_{(\infty )\Sigma}$ ,
tangent to $S_{\infty}$, are
parallel to ${\hat P}^{(\mu )}_{ADM}=b^{(\mu )}_{(\infty )A} {\hat P}^A_{ADM}$
with ${\hat P}^r_{ADM}\approx 0$ [namely when one is inside the
Christodoulou-Klainermann class of solutions]. Let us remember from the end of
Section III that this implies the existence of an inertial system at spatial
infinity when ${\tilde \lambda}_A(\tau )=(\epsilon ;\vec 0)$ and ${\tilde
\lambda}_{AB}(\tau )=0$, namely the
absence of accelerations and rotations there [when ${\tilde \lambda}_A(\tau )
\not= 0$ there is a direction independent global acceleration of the origin
$x^{(\mu )}_{(\infty )}(\tau )$, since ${\dot x}^{(\mu )}_{(\infty )}(\tau )=
b^{(\mu )}_{(\infty )A}{\tilde \lambda}_A(\tau )$].

In tetrad gravity we shall assume the following boundary conditions consistent
with Eqs.(\ref{III8}) and (\ref{III14}) of metric gravity [$\epsilon > 0$]

\begin{eqnarray}
{}^3e_{(a)r}(\tau ,\vec \sigma )\,&& {\rightarrow}_{r\, \rightarrow
\infty }\, (1+{M\over {2r}})
\delta_{(a)r}+{}^3w_{(a)r}(\tau ,\vec \sigma ),\quad\quad
{}^3w_{(a)r}(\tau ,\vec \sigma )=o_4(r^{-3/2}),\nonumber \\
{}^3e^{r}_{(a)}(\tau ,\vec \sigma )\,&& {\rightarrow}_{r\, \rightarrow
\infty}\, (1-{M\over {2r}})
\delta_{(a)}^{r}+{}^3w^{r}_{(a)}(\tau ,\vec \sigma ),\quad\quad
{}^3w^{r}_{(a)}(\tau ,\vec \sigma )=o_4(r^{-3/2}),\nonumber \\
{}^3g_{rs}(\tau ,\vec \sigma )&&=[{}^3e_{(a)r}\, {}^3e_{(a)s}]
(\tau ,\vec \sigma )\, {\rightarrow}_{r\, \rightarrow \infty}\, (1+{M\over r})
\delta_{rs}+{}^3h_{rs}(\tau ,\vec \sigma ),\nonumber \\
{}^3h_{rs}(\tau ,\vec \sigma )&&={1\over r}  [\delta_{(a) r}\, {}^3w
_{(as)(a)s}(\tau ,\vec \sigma )+{}^3w_{(as)(a)r}(\tau ,\vec \sigma )\delta
_{(a)s}]+O(r^{-2})=o(r^{-3/2}),\nonumber \\
&&{}\nonumber \\
{}^3{\tilde \pi}^{r}_{(a)}(\tau ,\vec \sigma )\,&& {\rightarrow}_{r\,
\rightarrow \infty}\, o_3(r^{-5/2}),\nonumber \\
{}^3{\tilde \Pi}^{rs}(\tau ,\vec \sigma )&&={1\over 4}[{}^3e^{r}_{(a)}\,
{}^3{\tilde \pi}^{s}_{(a)}+{}^3e^{s}_{(a)}\, {}^3{\tilde \pi}
^{r}_{(a)}](\tau ,\vec \sigma )\, {\rightarrow}_{r\, \rightarrow
\infty}\, {}^3{\tilde k}^{rs}(\tau ,\vec \sigma )=o_3(r^{-5/2}),\nonumber \\
&&{}\nonumber \\
N(\tau ,\vec \sigma )&&=N_{(as)}(\tau ,\vec \sigma )+n(\tau ,\vec \sigma ),
\nonumber \\
n(\tau ,\vec \sigma )\,&& {\rightarrow}_{r\, \rightarrow \infty}\,
O(r^{-(2+\epsilon )}),\nonumber \\
N_{(as)}(\tau ,\vec \sigma )&&=-{\tilde \lambda}_{\tau}(\tau )-{1\over 2}
{\tilde \lambda}_{\tau s}(\tau ) \sigma^{s},\nonumber \\
&&{}\nonumber \\
N_{r}(\tau ,\vec \sigma )&&=N_{(as)r}(\tau ,\vec \sigma )+
n_{r}(\tau ,\vec \sigma ),\nonumber \\
n_{r}(\tau ,\vec \sigma )\,&& {\rightarrow}_{r\, \rightarrow \infty}\,
O(r^{-\epsilon}),\nonumber \\
N_{(as) r}(\tau ,\vec \sigma )&&=-{\tilde \lambda}_{r}(\tau )-
{1\over 2}{\tilde \lambda}_{rs}(\tau ) \sigma^{s},
\nonumber \\
N_{(a)}(\tau ,\vec \sigma )&&={}^3e^{r}_{(a)}(\tau ,\vec \sigma )
N_{r}(\tau ,\vec \sigma )=
\sum_r \delta^r_{(a)} e^{-q(\tau ,\vec \sigma )-{1\over {\sqrt{3}}}\sum
_{\bar a}\gamma_{\bar ar}r_{\bar a}(\tau ,\vec \sigma )}N_r(\tau ,\vec \sigma )
=\nonumber \\
&&=N_{(as)(a)}(\tau ,\vec \sigma )
+n_{(a)}(\tau ,\vec \sigma ),\nonumber \\
n_{(a)}(\tau ,\vec \sigma )&&=[{}^3e^{r}_{(a)}n_{r}](\tau ,\vec
\sigma )\, {\rightarrow}_{r\, \rightarrow \infty}\, O(r^{-\epsilon}),
\nonumber \\
&&{}\nonumber \\
{\tilde \pi}^n(\tau ,\vec \sigma )\,&& {\rightarrow}_{r\, \rightarrow \infty}\,
O(r^{-3}),\nonumber \\
{\tilde \pi}_{\vec n,(a)}(\tau ,\vec \sigma )\,&& {\rightarrow}_{r\,
\rightarrow \infty}\, O(r^{-3}),\nonumber \\
\lambda_n(\tau ,\vec \sigma )\,&& {\rightarrow}_{r\, \rightarrow \infty}\,
O(r^{-(3+\epsilon )}),\nonumber \\
\lambda_{\vec n,(a)}(\tau ,\vec \sigma )\,&& {\rightarrow}_{r\, \rightarrow
\infty}\, O(r^{-\epsilon}),\nonumber \\
\beta (\tau ,\vec \sigma )\,&& {\rightarrow}_{r\, \rightarrow \infty}\,
O(r^{-(3+\epsilon )}),\nonumber \\
\beta^r(\tau ,\vec \sigma )\,&& {\rightarrow}_{r\, \rightarrow \infty}\,
O(r^{-\epsilon}),\nonumber \\
&&{}\nonumber \\
{\hat {\cal H}}(\tau ,\vec \sigma )\,&& {\rightarrow}_{r\, \rightarrow \infty}\,
O(r^{-3}),\nonumber \\
{}^3{\tilde \Theta}_{\check r}(\tau ,\vec \sigma )\,&& {\rightarrow}_{r\,
\rightarrow \infty}\, O(r^{-3}),\nonumber \\
&&{}\nonumber \\
{}^3{\tilde M}_{(a)}(\tau ,\vec \sigma )\,&& {\rightarrow}_{r\, \rightarrow
\infty}\, O(r^{-6}),\nonumber \\
\alpha_{(a)}(\tau ,\vec \sigma )\,&& {\rightarrow}_{r\, \rightarrow \infty}\,
O(r^{-(1+\epsilon )}),\nonumber \\
{\hat \mu}_{(a)}(\tau ,\vec \sigma )\,&& {\rightarrow}_{r\, \rightarrow
\infty}\, O(r^{-(1+\epsilon )}),\nonumber \\
&&{}\nonumber \\
\varphi_{(a)}(\tau ,\vec \sigma )\,&& {\rightarrow}_{r\, \rightarrow \infty}\,
O(r^{-(1+\epsilon )}),\nonumber \\
{\tilde \pi}^{\vec \varphi}_{(a)}(\tau ,\vec \sigma )\,&& {\rightarrow}_{r\,
\rightarrow \infty}\, O(r^{-2}),\nonumber \\
\lambda^{\vec \varphi}_{(a)}(\tau ,\vec \sigma )\,&& {\rightarrow}_{r\,
\rightarrow \infty}\, O(r^{-(1+\epsilon )}).
\label{V1}
\end{eqnarray}

With these boundary conditions all proper gauge transformations [generated by
${\tilde {\cal H}}(\tau ,\vec \sigma )$ with parameter $\beta (\tau ,\vec
\sigma )\rightarrow O(r^{-(3+\epsilon )})$, ${\tilde \Theta}_r(\tau ,\vec
\sigma )$  with $\beta^{\check r}(\tau ,\vec \sigma ) \rightarrow
O(r^{-\epsilon}) $, ${}^3{\tilde M}_{(a)}(\tau ,\vec \sigma )$ with $\alpha
_{(a)}(\tau ,\vec \sigma )\rightarrow O(r^{-(1+\epsilon )})$,
${\tilde \pi}_{(a)}
^{\vec \varphi}(\tau ,\vec \sigma )$ with $\varphi_{(a)}(\tau ,\vec \sigma )
 \rightarrow O(r^{-(1+\epsilon )})$
for $r\, \rightarrow \infty$] go asymptotically to the identity.

Near spatial infinity there is a dynamical preferred observer [either the
canonical non-covariant Newton-Wigner-like position ${\tilde x}^{(\mu )}
_{(\infty )}(\tau )$ or the covariant non-canonical origin of asymptotic
Cartesian coordinates $x^{(\mu )}_{(\infty )}(\tau )$] with an associated
asymptotic inertial (or Lorentz) reference frame given by the asymptotic limit
of the $\Sigma_{\tau}$-adapted tetrads of Eqs.(40) of I: however, as said,
these asymptotic tetrads are well defined only in absence of supertranslations
on the rest-frame WSW hypersurfaces, where (modulo a rigid 3-rotation) we
get\hfill\break
\hfill\break
${}^4_{(\infty \Sigma )}{\check {\tilde E}}^{\tau}_{(\alpha )}=({1\over
{N_{(as)}(\tau )}}; \vec 0)$,
${}^4_{(\infty \Sigma )}{\check {\tilde E}}^r_{(\alpha )}=(-{{N^r_{(as)}(\tau )}
\over {N_{(as)}(\tau )}}; \delta^r_{(a)}),$\hfill\break
${}^4_{(\infty \Sigma )}{\check {\tilde E}}_{\tau}^{(\alpha )}=(N_{(as)}(\tau );
\delta^{(a)}_rN^r_{(as)}(\tau ) ),$
${}^4_{(\infty \Sigma )}{\check {\tilde E}}_r^{(\alpha )}=(0; \delta^{(a)}_r).$
\hfill\break
\hfill\break
The associated asymptotic triads are the possible asymptotic limits of the
Frauendiener triads.

Then, following the scenario b) of Section III, the differentiable and finite
Dirac Hamiltonian is assumed to be [from paper I and Eqs.(51) and (70) of II
we have $N_{\check r}\, {}^3{\tilde {\cal H}}^{\check r}\approx
-N_{(a)}\, {}^3e^{\check s}_{(a)}\, {}^3{\tilde \Theta}_{\check s}\approx -N
^{\check s}\, {}^3{\tilde \Theta}_{\check s}=-N_{(a)}\, {}^3e^s_{(a)}
\, {}^3{\tilde \Theta}_s \approx -N_{(a)}\, {}^3e^s_{(a)} {{\partial \xi^r}
\over {\partial \sigma^s}} {\tilde \pi}^{\vec \xi}_r =-N_u\, {}^3g^{us}
{{\partial \xi^r}\over {\partial \sigma^s}} {\tilde \pi}^{\vec \xi}_r=-{\tilde
N}^r {\tilde \pi}^{\vec \xi}_r$]

\begin{eqnarray}
{\hat H}_{(D)ADM}&=&\int d^3\sigma [n{\hat {\cal H}}-{\tilde n}^r {\tilde \pi}
^{\vec \xi}_r+\lambda^{\vec \varphi}_{(a)}{\tilde \pi}^{\vec \varphi}
_{(a)}+{\hat \mu}_{(a)}\, {}^3{\tilde M}_{(a)}+
\nonumber \\
&+&\lambda_n {\tilde \pi}^n +\lambda^{\vec n}_{(a)} {\tilde \pi}^{\vec n}
_{(a)}](\tau ,\vec \sigma )+\nonumber \\
&+&{\tilde \lambda}_A(\tau )[p^A_{(\infty )}-{\hat P}^A_{ADM}]+{1\over 2}{\tilde
\lambda}_{AB}(\tau ) [J^{AB}_{(\infty )}-{\hat J}^{AB}_{ADM}],
\label{V2}
\end{eqnarray}

\noindent with the same weak Poincar\'e charges of metric gravity, Eqs.
(\ref{III13}), expressed in terms of cotriads ${}^3e_{(a)r}$ and their
conjugate momenta ${}^3{\tilde \pi}^r_{(a)}$,
by using ${}^3g_{rs}={}^3e_{(a)r}\, {}^3e_{(a)s}$, ${}^3{\tilde \Pi}^{rs}=
{1\over 4} [{}^3e^r_{(a)}\, {}^3{\tilde \pi}^s_{(a)}+{}^3e^s_{(a)}\,
{}^3{\tilde \pi}^r_{(a)}]$ (see Eq.(84) of I).

Let us remark that, since we
are using the ADM expression for the energy ${\hat P}^{\tau}_{ADM}$,
we have not to show that it is
definite positive, because the ADM canonical approach to metric gravity is
contained in the one to tetrad gravity.

In the 3-orthogonal gauges of Section V of II and in the final canonical basis
[$q$, $\rho$, $r_{\bar a}$, $\pi_{\bar a}$], one has (before the
restriction to WSW hypersurfaces):\hfill\break
\hfill\break
i) $\alpha_{(a)}(\tau ,\vec \sigma )=\varphi_{(a)}(\tau ,\vec \sigma )=0$ [so
that $\lambda^{\vec \varphi}_{(a)}(\tau ,\vec \sigma )={\hat \mu}_{(a)}(\tau
,\vec \sigma )=0$ in the Dirac Hamiltonian], i.e. for the sake of simplicity
we choose the timelike congruence of observers' worldlines with the normal
$l^A(\tau ,\vec \sigma )$ to $\Sigma_{\tau}$ as 4-velocity field;
\hfill\break
ii) $N(\tau ,\vec \sigma )=[N_{(as)}+n](\tau ,\vec \sigma )=-{\tilde \lambda}
_{\tau}(\tau )-{1\over 2}
{\tilde \lambda}_{\tau s}(\tau )\sigma^s+n(\tau ,\vec \sigma )$ as the total
lapse function;\hfill\break
iii) The gauge fixings $\xi^r(\tau ,\vec \sigma )-\sigma^r \approx 0$ of
Section V of II (choice of the 3-orthogonal coordinates)  now do not imply
$n_r(\tau ,\vec \sigma )\approx 0$ as in II, due to the modification introduced
by the addition of the surface terms to the Dirac Hamiltonian. Instead they
imply the following
results for the total shift function [the Poincar\'e charges of
Eq.(\ref{III13}) have to be used in this equations]

\begin{eqnarray}
\partial_{\tau} [\xi^r(\tau ,\vec \sigma )&-&\sigma^r]\, {\buildrel \circ \over
=}\, \{ \xi^r(\tau ,\vec \sigma ), {\hat H}_{(D)ADM} \} = \Big[ n^{s}\,
{{\partial \xi^r}\over {\partial \sigma^s}}\Big] (\tau ,\vec
\sigma )-\nonumber \\
&-&{\tilde \lambda}_A(\tau ) \{ \xi^r(\tau ,\vec \sigma ),{\hat P}^A_{ADM} \} -
{1\over 2} {\tilde \lambda}_{AB}(\tau ) \{ \xi^r(\tau ,\vec \sigma ), {\hat J}
^{AB}_{ADM} \} \approx 0,\nonumber \\
\Rightarrow&& \quad n_r(\tau ,\vec \sigma )-{\hat n}_r(\tau ,\vec \sigma |r
_{\bar a}, \pi_{\bar a}, {\tilde \lambda}_A, {\tilde \lambda}_{AB}] \approx 0,
\nonumber \\
&&{}\nonumber \\
&&{\hat n}_r(\tau ,\vec \sigma |r
_{\bar a}, \pi_{\bar a}, {\tilde \lambda}_A, {\tilde \lambda}_{AB}] ={}^3g_{rs}
(\tau ,\vec \sigma )\Big[ {\tilde \lambda}_A(\tau ) \{ \xi^s(\tau ,\vec \sigma )
,{\hat P}^A_{ADM}\}+\nonumber \\
&+&{1\over 2}{\tilde \lambda}_{AB}(\tau ) \{ \xi^s(\tau ,\vec \sigma ),{\hat
J}^{AB}_{ADM}\} \Big],\nonumber \\
&&{}\nonumber \\
&&\partial_{\tau} \Big[ n_r(\tau ,\vec \sigma )-{\hat n}_r(\tau ,\vec \sigma |r
_{\bar a}, \pi_{\bar a}, {\tilde \lambda}_A, {\tilde \lambda}_{AB}]
\Big] \approx \nonumber \\
&&\approx [\lambda^{\vec n}_{(a)}\, {}^3{\hat e}_{(a)r}](\tau ,\vec \sigma )-
\nonumber \\
&&- \{ {\hat n}_r(\tau ,\vec \sigma |r
_{\bar a}, \pi_{\bar a}, {\tilde \lambda}_A, {\tilde \lambda}_{AB}] ,
{\hat H}_{(D)ADM} \} \approx 0,\nonumber \\
&&{}\nonumber \\
\Rightarrow&& \quad \lambda^{\vec n}_{(a)}(\tau ,\vec \sigma )\quad
determined.
\label{V2a}
\end{eqnarray}
\hfill\break

Therefore, the shift functions do not vanish in the 3-orthogonal gauges avoiding
the ``synchronous" coordinates with their tendency to develop coordinate
singularities in short times.\hfill\break
iv) After going to Dirac brackets with respect to all the second class
constraints implied by the previous gauge fixings (for the sake of simplicity
they will always be denoted $\{ .,. \}$]
we remain with the Dirac Hamiltonian\hfill\break
\hfill\break
${\hat H}_{(D)ADM,R}=\int d^3\sigma [n {\hat {\cal H}}_R+\lambda_n{\tilde \pi}
^n](\tau ,\vec \sigma )+
{\tilde \lambda}_A(\tau )[p^A_{(\infty )}-{\hat P}^A_{ADM}]+{1\over 2}{\tilde
\lambda}_{AB}(\tau ) [J^{AB}_{(\infty )}-{\hat J}^{AB}_{ADM}]$.
\hfill\break
\hfill\break
As shown in II, the surviving canonical variables in 3-orthogonal gauges are
$n$, ${\tilde \pi}^n$, $q$, $\rho$, $r_{\bar a}$, $\pi_{\bar a}$.

From Eqs.(99), (102), (84), (90), (95), (96) of II [$\gamma_{PP_1}$ is the
geodesic between P and $P_1$ for the 3-metric; $\phi =e^{q/2}$] we get

\begin{eqnarray}
{}^3{\hat e}_{(a)r}(\tau ,\vec \sigma ) &=& \delta_{(a)r} (e^{q+{1\over
{\sqrt{3}}} \sum_{\bar a}\gamma_{\bar ar}r_{\bar a}})(\tau ,\vec \sigma ),
\nonumber \\
{}^3{\hat g}_{rs}(\tau ,\vec \sigma )&=&[e^{2q+{2\over {\sqrt{3}}}
\sum_{\bar a}\gamma_{\bar ar}r_{\bar a}}](\tau ,\vec \sigma )
\delta_{rs},\nonumber \\
{}^3{\hat {\tilde \pi}}^r_{(a)}(\tau ,\vec \sigma )&=&\sum_s \int d^3\sigma_1
{\cal K}^r_{(a)s}(\vec \sigma ,{\vec \sigma}_1;\tau |q,r_{\bar a}]
(e^{-q-{1\over {\sqrt{3}}}\sum_{\bar a}\gamma_{\bar as}r_{\bar a}})
(\tau ,{\vec \sigma}_1) \nonumber \\
&&\Big[ {1\over 3}\rho +\sqrt{3} \sum_{\bar b}
\gamma_{\bar bs} \pi_{\bar b}\Big] (\tau ,{\vec \sigma}_1)\nonumber \\
{\rightarrow}_{\rho \rightarrow 0}&& \sqrt{3} \sum_s\sum_{\bar b} \gamma_{\bar
bs}\int d^3\sigma_1
{\cal K}^r_{(a)s}(\vec \sigma ,{\vec \sigma}_1;\tau |\phi ,r_{\bar a}]
(\phi^{-2}e^{-{1\over {\sqrt{3}}}\sum_{\bar a}\gamma_{\bar as}r_{\bar a}}
\pi_{\bar b})(\tau ,{\vec \sigma}_1),\nonumber \\
&&{}\nonumber \\
{\cal K}^r_{(a)s}(\vec \sigma ,{\vec \sigma}_1,\tau |q,r_{\bar a}]&=&
\delta^r_{(a)}\delta^r_s\delta^3(\vec \sigma ,{\vec \sigma}_1)+{\cal T}^r
_{(a)s}(\vec \sigma ,{\vec \sigma}_1,\tau |q,r_{\bar a}],\nonumber \\
&&{}\nonumber \\
{\cal T}^r_{(a)s}(\vec \sigma ,{\vec \sigma}_1;\tau |q,r_{\bar a}]
&=&{1\over 2}e^{-{1\over {\sqrt{3}}}\sum_{\bar c}\gamma_{\bar cr}r_{\bar c}
(\tau ,\vec \sigma )}\Big[ \sum_{w\not= s} \delta_{(k)w} e^{{1\over {\sqrt{3}}}
\sum_{\bar c}(\gamma_{\bar cw}-\gamma_{\bar cs})r_{\bar c}(\tau ,{\vec
\sigma}_1)}\cdot \nonumber \\
&&\cdot \Big( {{\partial q(\tau ,{\vec \sigma}_1)}\over {\partial \sigma_1^w}}+
{1\over {\sqrt{3}}} \sum_{\bar c}\gamma_{\bar cs}{{\partial r_{\bar c}(\tau ,
{\vec \sigma}_1)}\over {\partial \sigma_1^w}}\Big) e^{-q(\tau ,\vec \sigma )}
\delta^r_{(b)} T_{(b)(a)(k)}(\vec \sigma ,{\vec \sigma}_1;\tau )+\nonumber \\
&&+\delta_{(k)s} {{\partial}\over {\partial \sigma_1^s}} e^{-q(\tau, \vec
\sigma )} \delta^r_{(b)} T_{(b)(a)(k)}(\vec \sigma ,{\vec \sigma}_1;\tau )
\Big] ,\nonumber \\
&&{}\nonumber \\
e^{-q(\tau ,\vec \sigma )}&& \delta^r_{(b)} T_{(b)(a)(k)}(\vec \sigma ,{\vec
\sigma}_1;\tau ) =\nonumber \\
&=&e^{ {1\over {\sqrt{3}}}\sum_{\bar c}\gamma_{\bar cr}r_{\bar
c}(\tau ,\vec \sigma )} d^r_{\gamma_{PP_1}} \Big( P_{\gamma_{PP_1}}\, e^{\int
^{\vec \sigma}_{{\vec \sigma}_1}d\sigma_2^w\, {}^3{\hat \omega}_{w(c)}(\tau ,
{\vec \sigma}_2){\hat R}^{(c)} }\, \Big)_{(a)(k)}+\nonumber \\
&+&\sum_u\delta_{(a)u} e^{ {1\over {\sqrt{3}}} \sum_{\bar c}\gamma_{\bar cu}
r_{\bar c}(\tau ,\vec \sigma )} d^u_{\gamma_{PP_1}}(\vec \sigma ,{\vec
\sigma}_1) \nonumber \\
&&\delta^r_{(b)}\Big( P_{\gamma_{PP_1}}\, e^{\int
^{\vec \sigma}_{{\vec \sigma}_1}d\sigma_2^w\, {}^3{\hat \omega}_{w(c)}(\tau ,
{\vec \sigma}_2){\hat R}^{(c)} }\, \Big)_{(b)(k)},\nonumber \\
&&{}\nonumber \\
{}^3{\hat \omega}_{t(d)}(\tau ,\vec \sigma )
&=&\epsilon_{(d)(m)(n)} \delta_{(m)t}\delta_{(n)u}
(e^{ {1\over {\sqrt{3}}}\sum_{\bar a}(\gamma_{\bar at}-\gamma_{\bar au})
r_{\bar a} }
[\partial_uq+{1\over {\sqrt{3}}} \sum_{\bar b} \gamma_{\bar bt}
\partial_ur_{\bar b}] )(\tau ,\vec \sigma ),\nonumber \\
&&\Downarrow \nonumber \\
q(\tau ,\vec \sigma )\, &{\rightarrow}_{r\, \rightarrow \infty}&\, {M\over {2r}}
+o_4(r^{-3/2}),\nonumber \\
\phi (\tau ,\vec \sigma )&=& e^{q(\tau ,\vec \sigma )/2}\, {\rightarrow}_{r\,
\rightarrow \infty}\, 1+{M\over {4r}}+ o_4(r^{-3/2}),\nonumber \\
r_{\bar a}(\tau ,\vec \sigma )\, &{\rightarrow}_{r\, \rightarrow \infty}&\,
o_4(r^{-3/2}),\nonumber \\
\rho (\tau ,\vec \sigma )&=&{1\over 2}\phi (\tau ,\vec \sigma ) \pi_{\phi}(\tau
,\vec \sigma )\, {\rightarrow}_{r\, \rightarrow \infty}\, o_3(r^{-5/2}),
\nonumber \\
\pi_{\bar a}(\tau ,\vec \sigma )\, &{\rightarrow}_{r\, \rightarrow \infty}&\,
o_3(r^{-3}),\nonumber \\
{}^3{\hat \omega}_{r(a)}(\tau ,\vec \sigma )\, &{\rightarrow}_{r\, \rightarrow
\infty}&\, O(r^{-2}),
\label{V3}
\end{eqnarray}

\noindent and

\begin{eqnarray}
{}^3{\hat K}_{rs}(\tau ,\vec \sigma )
&=&{{\epsilon}\over {4k}} [e^{ {1\over {\sqrt{3}}}\sum_{\bar c}
(\gamma_{\bar cr}+\gamma_{\bar cs})r_{\bar c} } \sum_u(\delta_{ru}\delta_{(a)s}
+\delta_{su}\delta_{(a)r}-\delta_{rs}\delta_{(a)u})\nonumber \\
&&e^{ {1\over {\sqrt{3}}} \sum_{\bar c}\gamma_{\bar cu}r_{\bar c}}\, {}^3{\hat
{\tilde \pi}}^u_{(a)}](\tau ,\vec \sigma ),\nonumber \\
{}^3{\hat K}(\tau ,\vec \sigma )
&=&-{{\epsilon}\over {4k}} [e^{-2q} \sum_u \delta_{(a)u} e^{ {1\over
{\sqrt{3}}}\sum_{\bar c}\gamma_{\bar cu}r_{\bar c}}\, {}^3{\hat {\tilde \pi}}
^u_{(a)}](\tau ,\vec \sigma )=\nonumber \\
&=&-{{\epsilon}\over {4k}}e^{-3q(\tau ,\vec \sigma )} \{
\rho(\tau ,\vec \sigma )+\sum_u (e^{q+{1\over {\sqrt{3}}}\sum_{\bar a}
\gamma_{\bar au}r_{\bar a}})(\tau ,\vec \sigma ) \int d^3\sigma_1 \delta_{(a)u}
\nonumber \\
&&{\cal T}^u_{(a)s}(\vec \sigma ,{\vec \sigma}_1;\tau |q,r_{\bar a}]
(e^{-q-{1\over {\sqrt{3}}}\sum_{\bar b}\gamma_{\bar bs}r_{\bar a}})(\tau,
{\vec \sigma}_1)[{1\over 3}\rho+\sqrt{3}\sum_{\bar c}\gamma_{\bar cs}\pi_{\bar
c}](\tau ,{\vec \sigma}_1) \} ,\nonumber \\
{}^3{\hat {\tilde \Pi}}^{rs}(\tau ,\vec \sigma )&=&{1\over 4}[{}^3{\hat e}^r
_{(a)}\, {}^3{\hat {\tilde \pi}}^s_{(a)} +{}^3{\hat e}^s_{(a)}\, {}^3{\hat
{\tilde \pi}}^r_{(a)}](\tau ,\vec \sigma )=\nonumber \\
&=&{1\over 4}e^{-q(\tau ,\vec \sigma )}[e^{-{1\over {\sqrt{3}}}\sum_{\bar a}
\gamma_{\bar ar}r_{\bar a}} \delta^r_{(a)}\, {}^3{\hat {\tilde \pi}}^s_{(a)}+
e^{-{1\over {\sqrt{3}}}\sum_{\bar a}\gamma_{\bar as}r_{\bar a}} \delta^s_{(a)}\,
{}^3{\hat {\tilde \pi}}^r_{(a)}](\tau ,\vec \sigma ),\nonumber \\
&&{}\nonumber \\
{}^3{\hat \Gamma}^r_{uv}(\tau ,\vec \sigma )&=&\Big( -\delta_{uv}\sum_s
\delta^r_s e^{{2\over {\sqrt{3}}}\sum_{\bar a}(\gamma_{\bar au}-\gamma_{\bar
as})r_{\bar a}}[\partial_sq+{1\over {\sqrt{3}}}\sum_{\bar b}\gamma_{\bar bu}
\partial_sr_{\bar b}]+\nonumber \\
&+&\delta^r_u[\partial_vq+{1\over {\sqrt{3}}}\sum_{\bar b}\gamma_{\bar bu}
\partial_vr_{\bar b}]+\delta^r_v[\partial_uq+{1\over {\sqrt{3}}}\sum_{\bar b}
\gamma_{\bar bv}\partial_ur_{\bar b}] \Big) (\tau ,\vec \sigma ),\nonumber \\
{}^3{\hat \Gamma}^u_{uv}(\tau ,\vec \sigma )&=& 3\partial_vq(\tau ,\vec
\sigma ),\nonumber \\
{}^3{\hat G}_{rsuv}(\tau ,\vec \sigma ) &=& [{}^3{\hat g}_{ru}\, {}^3{\hat g}
_{sv}+{}^3{\hat g}_{rv}\, {}^3{\hat g}_{su} -{}^3{\hat g}_{rs}\, {}^3{\hat g}
_{uv}](\tau ,\vec \sigma )=\nonumber \\
&=&e^{4q(\tau ,\vec \sigma )}[e^{{2\over {\sqrt{3}}}\sum_{\bar a}(\gamma_{\bar
ar}+\gamma_{\bar as})r_{\bar a}} (\delta_{ru}\delta_{sv}+\delta_{rv}\delta
_{su})-e^{{2\over {\sqrt{3}}}\sum_{\bar a}(\gamma_{\bar ar}+\gamma_{\bar au})
r_{\bar a}} \delta_{rs}\delta_{uv}](\tau ,\vec \sigma ),\nonumber \\
&&{}
\label{V4}
\end{eqnarray}

\noindent where the second equation can be read as an integral equation to get
$\rho (\tau ,\vec \sigma )$, the momentum conjugate to the conformal factor,
in terms of ${}^3{\hat K}(\tau ,\vec \sigma )$, $q(\tau ,\vec \sigma )=
{1\over 6} ln\, {}^3{\hat g}(\tau ,\vec \sigma )$ and $r_{\bar a}(\tau ,\vec
\sigma )={{\sqrt{3}}\over 2} \sum_r \gamma_{\bar ar} ln\, [{}^3{\hat g}
_{rr}/{}^3\hat g](\tau ,\vec \sigma )$ [see Eqs.(101) of II],
in the 3-orthogonal gauges.

The expression of ${}^3{\hat K}_{rs}$ in these gauges replaces the
knowledge of the gravitomagnetic potential $\vec W$
in the York TT decomposition of the
extrinsic curvature in the conformal approach (see Appendix C of II; the three
degrees of freedom of $\vec W$ correspond to the three eliminated parameters
$\vec \xi$ of pseudo-diffeomorphisms).

We can now write explicitly the equations determining $n_r(\tau ,\vec \sigma )$
for ${\tilde \lambda}_{AB}(\tau )=0$, i.e. on the WSW hypersurfaces. From
Eqs.(79) or (90) of II and from ${}^3{\tilde \Pi}^{uv}={1\over 4}[{}^3e^u_{(a)}
\, {}^3{\tilde \pi}^v_{(a)}+{}^3e^v_{(a)}\, {}^3{\tilde \pi}^u_{(a)}]$
(see I) we get\hfill\break
\hfill\break
$\{ \xi^r(\tau ,\vec \sigma ), {}^3{\tilde \pi}^s_{(a)}(\tau ,{\vec \sigma}_1)
\} {|}_{\vec \xi =\vec \sigma}$\hfill\break
$=-{1\over 2}\, {}^3{\hat e}^s_{(b)}(\tau ,{\vec
\sigma}_1) \Big[ {}^3{\hat e}_{(b)w}(\tau ,{\vec \sigma}_1) \zeta^{(\hat
\omega )w}_{(a)(c)}({\vec \sigma}_1,\vec \sigma ,\tau )+{}^3{\hat e}
_{(a)w}(\tau ,{\vec \sigma}_1) \zeta^{(\hat \omega )w}_{(b)(c)}({\vec \sigma}
_1,\vec \sigma ,\tau )\Big] {}^3{\hat e}^r_{(c)}(\tau ,\vec \sigma )$
\hfill\break
\hfill\break
$\{ \xi^r(\tau ,\vec \sigma ),{}^3{\tilde \Pi}^{uv}(\tau ,{\vec \sigma}_1) \}
=-{1\over 4}[{}^3{\hat e}^u_{(a)} \delta^v_w+ {}^3{\hat e}^v_{(a)} \delta^u_w]
(\tau ,{\vec \sigma}_1) \zeta^{(\hat \omega )w}_{(a)(b)}({\vec \sigma}_1,\vec
\sigma ,\tau )\, {}^3{\hat e}^r_{(b)}(\tau ,\vec \sigma )$.\hfill\break
\hfill\break
By using Eqs.(\ref{III13}) we get\hfill\break
\hfill\break
$\{ \xi^r(\tau ,\vec \sigma ), {\hat P}^{\tau}_{ADM} \} =-{1\over {4k}}\,
{}^3G_{o(a)(b)(c)(d)} \int d^3\sigma_1 {{{}^3{\hat e}_{(d)s}\, {}^3{\hat e}
_{(b)v}\, {}^3{\hat {\tilde \pi}}^s_{(c)} }\over {\sqrt{\gamma} }}(\tau ,{\vec
\sigma}_1) \zeta^{(\hat \omega )v}_{(a)(e)}({\vec \sigma}_1,\vec \sigma ,\tau )
\, {}^3{\hat e}^r_{(e)}(\tau ,\vec \sigma )$\hfill\break
\hfill\break
$\{ \xi^r(\tau ,\vec \sigma ), {\hat P}^s_{ADM} \} = \int d^3\sigma_1\,
{}^3{\hat \Gamma}^s_{uv}(\tau ,{\vec \sigma}_1)\, {}^3{\hat e}^u_{(a)}(\tau
,{\vec \sigma}_1)\, \zeta^{(\hat \omega )v}_{(a)(b)}({\vec \sigma}_1,\vec
\sigma ,\tau )\, {}^3{\hat e}^r_{(b)}(\tau ,\vec \sigma )$.\hfill\break
\hfill\break
Therefore, by using Eqs.(\ref{V4}) we get [$\phi =e^{q/2}$]

\begin{eqnarray}
n_r(\tau ,\vec \sigma )&\approx& {\hat n}_r(\tau ,\vec \sigma |r_{\bar a},
\pi_{\bar a},{\tilde \lambda}_A,0]=\nonumber \\
&=&\Big[ \phi^2 e^{{1\over {\sqrt{3}}}\sum_{\bar a}\gamma_{\bar ar}r_{\bar a}}
\Big] (\tau ,\vec \sigma )
\Big( -{{ {\tilde \lambda}_{\tau}(\tau )}\over {4k}}\, {}^3G_{o(a)(b)(c)(d)}
\delta_{(b)v}\delta_{(d)r} \nonumber \\
&&\int d^3\sigma_1 \Big[ \phi^{-6} e^{{1\over
{\sqrt{3}}}\sum_{\bar a}(\gamma_{\bar ar}-\gamma_{\bar av})r_{\bar a}}\,
{}^3{\hat {\tilde \pi}}^r_{(c)}\Big] (\tau ,{\vec \sigma}_1)
\zeta^{(\hat \omega )v}_{(a)(e)}({\vec \sigma}_1.\vec \sigma ,\tau )\delta^r
_{(e)}+\nonumber \\
&+&{\tilde \lambda}_r(\tau )\int d^3\sigma_1 \delta^u_{(a)}\, {}^3{\hat \Gamma}
^r_{uv}(\tau ,{\vec \sigma}_1) \zeta^{(\hat \omega )v}_{(a)(b)}({\vec \sigma}
_1,\vec \sigma ,\tau ) \delta^r_{(b)}\Big),\nonumber \\
&&{}\nonumber \\
\Rightarrow&& N_{(a)}(\tau ,\vec \sigma ){|}_{{\tilde \lambda}_{AB}=0}
=[{}^3{\hat e}^r_{(a)}N_r](\tau ,\vec \sigma )
{|}_{{\tilde \lambda}_{AB}=0} \approx \nonumber \\
&\approx& \Big[{}^3{\hat e}^r_{(a)} \Big( -{\tilde \lambda}_r(\tau )
+{\hat n}_r[r
_{\bar a}, \pi_{\bar a}, {\tilde \lambda}_A, {\tilde \lambda}_{AB}]
\Big) \Big] (\tau ,\vec \sigma ).
\label{V4a}
\end{eqnarray}

The reduced Dirac Hamiltonian and the weak  and strong Poincar\'e charges of
Eqs.(\ref{III12}), (\ref{III13}) and (\ref{III11})
become [Eqs.(102) and (104) of II are used]

\begin{eqnarray}
{\hat H}_{(D)ADM,R}&=&\int d^3\sigma [n{\hat {\cal H}}_R+\lambda_n
{\tilde \pi}^n](\tau ,\vec \sigma )+\nonumber \\
&+&{\tilde \lambda}_A(\tau )
[p^A_{(\infty )}-{\hat P}^A_{ADM,R}]+{1\over 2}{\tilde \lambda}_{AB}(\tau )
[J^{AB}_{(\infty )}-{\hat J}^{AB}_{ADM,R}],\nonumber \\
&&{}\nonumber \\
{\hat P}^{\tau}_{ADM,R}&=&\epsilon \int d^3\sigma \Big( k \Big[ e^q \sum_r
e^{-{1\over {\sqrt{3}}}\sum_{\bar a}\gamma_{\bar ar}r_{\bar a}} \times
\nonumber \\
&&\Big( {1\over 3} \sum_{\bar b\bar c}(2\gamma_{\bar br}\gamma_{\bar cr}+\delta
_{\bar br\bar cr})\partial_rr_{\bar b}\partial_rr_{\bar c}-{2\over {\sqrt{3}}}
(\sum_{\bar b}\gamma_{\bar br}\partial_rr_{\bar b})\partial_rq -(\partial_rq)^2
-\nonumber \\
&-&\sum_ue^{{2\over {\sqrt{3}}}\sum_{\bar a}(\gamma_{\bar ar}-\gamma_{\bar au})
r_{\bar a}} [{2\over 3}\sum_{\bar b\bar c}\gamma_{\bar br}\gamma_{\bar cr}
\partial_ur_{\bar b}\partial_ur_{\bar c}+\nonumber \\
&+&\sqrt{3}(\sum_{\bar b}\gamma_{\bar br}
\partial_ur_{\bar b})\partial_uq+(\partial_uq)^2] \Big) \Big] (\tau ,\vec
\sigma )-\nonumber \\
&-&{{e^{-q(\tau ,\vec \sigma )}}\over {8k}}\Big[ (e^{-2q}[6 \sum_{\bar a}\pi^2
_{\bar a}-{1\over 3}\rho^2])(\tau ,\vec \sigma )+\nonumber \\
&+&2 (e^{-q}\sum_ue^{{1\over {\sqrt{3}}}\sum_{\bar a}\gamma_{\bar au}r_{\bar a}}
[2\sqrt{3}\sum_{\bar b}\gamma_{\bar bu}\pi_{\bar b}-{1\over 3}\rho ])(\tau
,\vec \sigma )\times \nonumber \\
&&\int d^3\sigma_1 \sum_r \delta^u_{(a)} {\cal T}^u_{(a)r}(\vec \sigma ,{\vec
\sigma}_1,\tau |q,r_{\bar a}] \Big( e^{-q-{1\over {\sqrt{3}}}\sum_{\bar a}
\gamma_{\bar ar}r_{\bar a}}[{{\rho}\over 3}+\sqrt{3}\sum_{\bar b}\gamma_{\bar
br} \pi_{\bar b}]\Big) (\tau ,{\vec \sigma}_1)+\nonumber \\
&+&\int d^3\sigma_1d^3\sigma_2 \Big( \sum_u e^{{2\over {\sqrt{3}}}\sum_{\bar a}
\gamma_{\bar au}+r_{\bar a}(\tau ,\vec \sigma )} \times \nonumber \\
&&\sum_r{\cal T}^u_{(a)r}(\vec \sigma ,{\vec
\sigma}_1,\tau |q,r_{\bar a}] \Big( e^{-q-{1\over {\sqrt{3}}}\sum_{\bar a}
\gamma_{\bar ar}r_{\bar a}}[{{\rho}\over 3}+\sqrt{3}\sum_{\bar b}\gamma_{\bar
br} \pi_{\bar b}]\Big) (\tau ,{\vec \sigma}_1)\times \nonumber \\
&&\sum_s {\cal T}^u_{(a)s}(\vec \sigma ,{\vec
\sigma}_2,\tau |q,r_{\bar a}] \Big( e^{-q-{1\over {\sqrt{3}}}\sum_{\bar a}
\gamma_{\bar as}r_{\bar a}}[{{\rho}\over 3}+\sqrt{3}\sum_{\bar c}\gamma_{\bar
cs} \pi_{\bar c}]\Big) (\tau ,{\vec \sigma}_2)+\nonumber \\
&+&\sum_{uv} e^{{1\over {\sqrt{3}}}\sum_{\bar a}(\gamma_{\bar au}+\gamma_{\bar
av})r_{\bar a}(\tau ,\vec \sigma )} (\delta^u_{(b)}\delta^v_{(a)}-\delta^u_{(a)}
\delta^v_{(b)})\times \nonumber \\
&&\sum_r {\cal T}^u_{(a)r}(\vec \sigma ,{\vec
\sigma}_1,\tau |q,r_{\bar a}] \Big( e^{-q-{1\over {\sqrt{3}}}\sum_{\bar a}
\gamma_{\bar ar}r_{\bar a}}[{{\rho}\over 3}+\sqrt{3}\sum_{\bar b}\gamma_{\bar
br} \pi_{\bar b}]\Big) (\tau ,{\vec \sigma}_1)\nonumber \\
&&\sum_s {\cal T}^v_{(b)s}(\vec \sigma ,{\vec
\sigma}_2,\tau |q,r_{\bar a}] \Big( e^{-q-{1\over {\sqrt{3}}}\sum_{\bar a}
\gamma_{\bar as}r_{\bar a}}[{{\rho}\over 3}+\sqrt{3}\sum_{\bar c}\gamma_{\bar
cs} \pi_{\bar c}]\Big) (\tau ,{\vec \sigma}_2)\, \Big)\, \Big]\,
\Big) ,\nonumber \\
{\hat P}^r_{ADM,R}&=&\int d^3\sigma e^{q(\tau ,\vec \sigma )}
e^{-{1\over {\sqrt{3}}}\sum_{\bar a}\gamma_{\bar ar}r_{\bar a}(\tau ,\vec
\sigma )} \Big( e^{-q(\tau ,\vec \sigma )} \nonumber \\
&&\Big[ \sum_u e^{-{1\over {\sqrt{3}}}\sum_{\bar a}\gamma_{\bar au}r_{\bar a}}
(\partial_rq +{1\over {\sqrt{3}}}\sum_{\bar b}\gamma_{\bar bu}\partial_rr_{\bar
b} ) ({{\rho}\over 3}+\sqrt{3}\sum_{\bar c}\gamma_{\bar cu}\pi_{\bar c})-
\nonumber \\
&-&2e^{-{1\over {\sqrt{3}}}\sum_{\bar a}\gamma_{\bar ar}r_{\bar a}}
(\partial_rq +{1\over {\sqrt{3}}}\sum_{\bar b}\gamma_{\bar br}\partial_rr
_{\bar b}) ({{\rho}\over 3}+\sqrt{3}\sum_{\bar c}\gamma_{\bar cr}\pi_{\bar c})
\Big] (\tau ,\vec \sigma )+\nonumber \\
&+&\sum_{uv} \int d^3\sigma_1 \Big[ (\partial_rq +{1\over {\sqrt{3}}}\sum_{\bar
b}\gamma_{\bar bu}\partial_rr_{\bar b}])(\tau ,\vec \sigma )\delta_{u(a)}
{\cal T}^u_{(a)v}(\vec \sigma ,{\vec \sigma}_1,\tau |q,r_{\bar a}] -
\nonumber \\
&-&(\partial_uq+{1\over {\sqrt{3}}}\sum_{\bar b}
\gamma_{\bar br} \partial_ur_{\bar b})(\tau ,\vec \sigma )
\Big( \delta_{r(a)}{\cal T}^u_{(a)v}+\delta_{u(a)}{\cal
T}^r_{(a)v}\Big) (\vec \sigma ,{\vec \sigma}_1,\tau |q,r_{\bar a}] \Big]
\nonumber \\
&&\Big( e^{-q-{1\over {\sqrt{3}}}\sum_{\bar a}\gamma_{\bar av}r_{\bar a}}
[{{\rho}\over 3}+\sqrt{3}\sum_{\bar c}\gamma_{\bar cv}\pi_{\bar c}]\Big) (\tau
,{\vec \sigma}_1) \Big),\nonumber \\
{\hat J}^{rs}_{ADM,R}&=&{1\over 2}\int d^3\sigma  e^{-2q(\tau ,\vec \sigma )}
\Big[ \sigma^s e^{-{1\over {\sqrt{3}}}\sum_{\bar a}\gamma_{\bar ar}r_{\bar a}
(\tau ,\vec \sigma )}\nonumber \\
&& \Big[ \sum_u
e^{-{2\over {\sqrt{3}}}\sum_{\bar a}\gamma_{\bar au}r_{\bar a}}
(\partial_rq +{1\over {\sqrt{3}}}\sum_{\bar b}\gamma_{\bar bu}\partial_rr
_{\bar b}) ({{\rho}\over 3}+\sqrt{3}\sum_{\bar c}\gamma_{\bar cu} \pi_{\bar c})
-\nonumber \\
&-&2e^{-{1\over {\sqrt{3}}}\sum_{\bar a}\gamma_{\bar ar}r_{\bar a}}
(\partial_rq+{1\over {\sqrt{3}}}\sum_{\bar b}\gamma_{\bar br}\partial_rr_{\bar
b}) ({{\rho}\over 3}+\sqrt{3}\sum_{\bar c}\gamma_{\bar cr}\pi_{\bar c})\Big]
(\tau ,\vec \sigma )-\nonumber \\
&-&\sigma^r e^{-{1\over {\sqrt{3}}}\sum_{\bar a}\gamma_{\bar as}r_{\bar
a}(\tau ,\vec \sigma )}\nonumber \\
&&\Big[ \sum_u
e^{-{2\over {\sqrt{3}}}\sum_{\bar a}\gamma_{\bar au}r_{\bar a}}
(\partial_sq +{1\over {\sqrt{3}}}\sum_{\bar b}\gamma_{\bar bu}\partial_sr
_{\bar b}) ({{\rho}\over 3}+\sqrt{3}\sum_{\bar c}\gamma_{\bar cu} \pi_{\bar c})
-\nonumber \\
&-&2e^{-{1\over {\sqrt{3}}}\sum_{\bar a}\gamma_{\bar as}r_{\bar a}}
(\partial_sq+{1\over {\sqrt{3}}}\sum_{\bar b}\gamma_{\bar bs}\partial_sr_{\bar
b}) ({{\rho}\over 3}+\sqrt{3}\sum_{\bar c}\gamma_{\bar cs}\pi_{\bar c})\Big]
(\tau ,\vec \sigma ) \Big] \Big]+\nonumber \\
&+&{1\over 2}\sum_{uv} \int d^3\sigma d^3\sigma_1 e^{-q(\tau ,\vec \sigma )}
\Big[ \sigma^s e^{-{1\over {\sqrt{3}}}\sum_{\bar a}\gamma_{\bar ar}r_{\bar a}
(\tau ,\vec \sigma )}\nonumber \\
&&\Big[ (\partial_rq+{1\over {\sqrt{3}}}\sum_{\bar b}\gamma_{\bar bu}\partial_rr
_{\bar b}](\tau ,\vec \sigma )
\delta_{u(a)}{\cal T}^u_{(a)v}(\vec \sigma ,{\vec \sigma}_1,\tau
|q,r_{\bar a}] ]-\nonumber \\
&-&(\partial_uq+{1\over {\sqrt{3}}}\sum_{\bar b}\gamma_{\bar br}\partial_ur
_{\bar b})(\tau ,\vec \sigma ) \Big(\delta_{r(a)}{\cal T}^u_{(a)v}+\delta
_{u(a)}{\cal T}^r_{(a)v}\Big) (\vec \sigma ,{\vec \sigma}_1,\tau |q,r_{\bar a}]
\Big] -\nonumber \\
&-&\sigma^r e^{-{1\over {\sqrt{3}}}\sum_{\bar a}\gamma_{\bar as}r_{\bar a}
(\tau ,\vec \sigma )}\nonumber \\
&&\Big[ (\partial_sq+{1\over {\sqrt{3}}}\sum_{\bar b}\gamma_{\bar bu}\partial_sr
_{\bar b}](\tau ,\vec \sigma )
\delta_{u(a)}{\cal T}^u_{(a)v}(\vec \sigma ,{\vec \sigma}_1,\tau
|q,r_{\bar a}] ]-\nonumber \\
&-&(\partial_uq+{1\over {\sqrt{3}}}\sum_{\bar b}\gamma_{\bar bs}\partial_ur
_{\bar b})(\tau ,\vec \sigma ) \Big(\delta_{s(a)}{\cal T}^u_{(a)v}+\delta
_{u(a)}{\cal T}^s_{(a)v}\Big) (\vec \sigma ,{\vec \sigma}_1,\tau |q,r_{\bar a}]
\Big] \Big] \nonumber \\
&&(e^{-q-{1\over {\sqrt{3}}}\sum_{\bar a}\gamma_{\bar av}r_{\bar a}} [{{\rho}
\over 3}+\sqrt{3}\sum_{\bar c}\gamma_{\bar cv}\pi_{\bar c}])(\tau ,{\vec
\sigma}_1)  ,\nonumber \\
{\hat J}^{\tau r}_{ADM,R}&=&\epsilon \int d^3\sigma  \, \sigma^r
\Big( k \Big[ e^q \sum_r
e^{-{1\over {\sqrt{3}}}\sum_{\bar a}\gamma_{\bar ar}r_{\bar a}} \times
\nonumber \\
&&\Big( {1\over 3} \sum_{\bar b\bar c}(2\gamma_{\bar br}\gamma_{\bar cr}+\delta
_{\bar br\bar cr})\partial_rr_{\bar b}\partial_rr_{\bar c}-{2\over {\sqrt{3}}}
(\sum_{\bar b}\gamma_{\bar br}\partial_rr_{\bar b})\partial_rq -(\partial_rq)^2
-\nonumber \\
&-&\sum_ue^{{2\over {\sqrt{3}}}\sum_{\bar a}(\gamma_{\bar ar}-\gamma_{\bar au})
r_{\bar a}} [{2\over 3}\sum_{\bar b\bar c}\gamma_{\bar br}\gamma_{\bar cr}
\partial_ur_{\bar b}\partial_ur_{\bar c}+\nonumber \\
&+&\sqrt{3}(\sum_{\bar b}\gamma_{\bar br}
\partial_ur_{\bar b})\partial_uq+(\partial_uq)^2] \Big) \Big] (\tau ,\vec
\sigma )-\nonumber \\
&-&{{e^{-q(\tau ,\vec \sigma )}}\over {8k}}\Big[ (e^{-2q}[6 \sum_{\bar a}\pi^2
_{\bar a}-{1\over 3}\rho^2])(\tau ,\vec \sigma )+\nonumber \\
&+&2 (e^{-q}\sum_ue^{{1\over {\sqrt{3}}}\sum_{\bar a}\gamma_{\bar au}r_{\bar a}}
[2\sqrt{3}\sum_{\bar b}\gamma_{\bar bu}\pi_{\bar b}-{1\over 3}\rho ])(\tau
,\vec \sigma )\times \nonumber \\
&&\int d^3\sigma_1 \sum_r \delta^u_{(a)} {\cal T}^u_{(a)r}(\vec \sigma ,{\vec
\sigma}_1,\tau |q,r_{\bar a}] \Big( e^{-q-{1\over {\sqrt{3}}}\sum_{\bar a}
\gamma_{\bar ar}r_{\bar a}}[{{\rho}\over 3}+\sqrt{3}\sum_{\bar b}\gamma_{\bar
br} \pi_{\bar b}]\Big) (\tau ,{\vec \sigma}_1)+\nonumber \\
&+&\int d^3\sigma_1d^3\sigma_2 \Big( \sum_u e^{{2\over {\sqrt{3}}}\sum_{\bar a}
\gamma_{\bar au}+r_{\bar a}(\tau ,\vec \sigma )} \times \nonumber \\
&&\sum_r{\cal T}^u_{(a)r}(\vec \sigma ,{\vec
\sigma}_1,\tau |q,r_{\bar a}] \Big( e^{-q-{1\over {\sqrt{3}}}\sum_{\bar a}
\gamma_{\bar ar}r_{\bar a}}[{{\rho}\over 3}+\sqrt{3}\sum_{\bar b}\gamma_{\bar
br} \pi_{\bar b}]\Big) (\tau ,{\vec \sigma}_1)\times \nonumber \\
&&\sum_s {\cal T}^u_{(a)s}(\vec \sigma ,{\vec
\sigma}_2,\tau |q,r_{\bar a}] \Big( e^{-q-{1\over {\sqrt{3}}}\sum_{\bar a}
\gamma_{\bar as}r_{\bar a}}[{{\rho}\over 3}+\sqrt{3}\sum_{\bar c}\gamma_{\bar
cs} \pi_{\bar c}]\Big) (\tau ,{\vec \sigma}_2)+\nonumber \\
&+&\sum_{uv} e^{{1\over {\sqrt{3}}}\sum_{\bar a}(\gamma_{\bar au}+\gamma_{\bar
av})r_{\bar a}(\tau ,\vec \sigma )} (\delta^u_{(b)}\delta^v_{(a)}-\delta^u_{(a)}
\delta^v_{(b)})\times \nonumber \\
&&\sum_r {\cal T}^u_{(a)r}(\vec \sigma ,{\vec
\sigma}_1,\tau |q,r_{\bar a}] \Big( e^{-q-{1\over {\sqrt{3}}}\sum_{\bar a}
\gamma_{\bar ar}r_{\bar a}}[{{\rho}\over 3}+\sqrt{3}\sum_{\bar b}\gamma_{\bar
br} \pi_{\bar b}]\Big) (\tau ,{\vec \sigma}_1)\nonumber \\
&&\sum_s {\cal T}^v_{(b)s}(\vec \sigma ,{\vec
\sigma}_2,\tau |q,r_{\bar a}] \Big( e^{-q-{1\over {\sqrt{3}}}\sum_{\bar a}
\gamma_{\bar as}r_{\bar a}}[{{\rho}\over 3}+\sqrt{3}\sum_{\bar c}\gamma_{\bar
cs} \pi_{\bar c}]\Big) (\tau ,{\vec \sigma}_2)\, \Big)\, \Big]\,
\Big)+\nonumber \\
&+&\epsilon k \int d^3\sigma \Big( e^{-q-{2\over {\sqrt{3}}}\sum_{\bar a}\gamma
_{\bar ar}r_{\bar a}}\Big) (\tau ,\vec \sigma ) \nonumber \\
&&\Big[ \sum_v e^{-{2\over {\sqrt{3}}}\sum_{\bar a}\gamma_{\bar av}r_{\bar a}}
\Big( e^{2q+{2\over {\sqrt{3}}}\sum_{\bar a}\gamma_{\bar av}r_{\bar a}}-1\Big)
(\partial_rq+{2\over {\sqrt{3}}}\sum_{\bar b}(\gamma_{\bar br}+\gamma_{\bar
bv})\partial_rr_{\bar b})-\nonumber \\
&-&e^{-{2\over {\sqrt{3}}}\sum_{\bar a}\gamma_{\bar ar}r_{\bar a}}\Big(
e^{2q+{2\over {\sqrt{3}}}\sum_{\bar a}\gamma_{\bar ar}r_{\bar a}}-1\Big)
(\partial_rq+{4\over {\sqrt{3}}}\sum_{\bar b}\gamma_{\bar br}\partial_rr_{\bar
b})\Big] (\tau ,\vec \sigma ),\nonumber \\
&&{}\nonumber \\
P^{\tau}_{ADM,R}&=&{\hat P}^{\tau}_{ADM,R}+\int d^3\sigma {\hat {\cal H}}
_R(\tau ,\vec \sigma )=\nonumber\\
&=&2 \epsilon k \sum_u \int_{S^2_{\tau ,\infty}} d^2\Sigma_u \Big( e^{q-{2\over
{\sqrt{3}}}\sum_{\bar a}\gamma_{\bar au}r_{\bar a}}[-2\partial_uq+{1\over
{\sqrt{3}}} \sum_{\bar b}\gamma_{\bar bu} \partial_ur_{\bar b}] \Big) (\tau
,\vec \sigma ),\nonumber \\
P^r_{ADM,R}&=&{\hat P}^r_{ADM,R}+\int d^3\sigma {\hat {\cal H}}^r(\tau ,\vec
\sigma )\equiv {\hat P}^r_{ADM,R}=\nonumber \\
&=&-\int_{S^2_{\tau ,\infty}} d^2\Sigma_r \Big[
e^{-q-{1\over {\sqrt{3}}}\sum_{\bar a}\gamma
_{\bar ar}r_{\bar a}}[{{\rho}\over 3}+\sqrt{3}\sum_{\bar c}\gamma_{\bar cr}
\pi_{\bar c}]\Big] (\tau ,\vec \sigma )-\nonumber \\
&-&{1\over 2}\sum_{u} \int_{S^2_{\tau ,\infty}} d^2\Sigma_u e^{-q(\tau ,\vec
\sigma )} \sum_v \int d^3\sigma_1 \Big( e^{-{1\over {\sqrt{3}}}\sum_{\bar a}
\gamma_{\bar ar}r_{\bar a}(\tau ,\vec \sigma )} \delta_{r(a)}{\cal T}^u_{(a)v}+
\nonumber \\
&+&e^{-{1\over {\sqrt{3}}}\sum_{\bar a}\gamma_{\bar au}r_{\bar a}(\tau ,\vec
\sigma )} \delta_{u(a)}{\cal T}^r_{(a)v}\Big) (\vec \sigma ,{\vec \sigma}_1,\tau
|q,r_{\bar a}]\nonumber \\
&&(e^{-q-{1\over {\sqrt{3}}}\sum_{\bar a}\gamma_{\bar av}r_{\bar a}}[{{\rho}
\over 3}+\sqrt{3}\sum_{\bar c}\gamma_{\bar cv}\pi_{\bar c}])(\tau ,{\vec
\sigma}_1) \} ,\nonumber \\
J^{rs}_{ADM,R}&=&{\hat J}^{rs}_{ADM,R}+{1\over 4}\int d^3\sigma [\sigma^s
{\hat {\cal H}}^r-\sigma^r {\hat {\cal H}}^s](\tau ,\vec \sigma )=\nonumber \\
&=&-{1\over 2} \sum_{u} \int_{S^2_{\tau ,\infty}} d^2\Sigma_u e^{-q(\tau ,\vec
\sigma )} \nonumber \\
&&\Big( \delta^r_u \sigma^s [e^{-{1\over {\sqrt{3}}}\sum_{\bar a}\gamma_{\bar
ar}r_{\bar a}}({{\rho}\over 3}+\sqrt{3}\sum_{\bar b}\gamma_{\bar br}\pi_{\bar b}
)]-\nonumber \\
&-&\delta^s_u \sigma^r [e^{-{1\over {\sqrt{3}}}\sum_{\bar a}\gamma_{\bar
as}r_{\bar a}}({{\rho}\over 3}+\sqrt{3}\sum_{\bar b}\gamma_{\bar bs}\pi_{\bar b}
)] \Big) (\tau ,\vec \sigma ) -\nonumber \\
&-&{1\over 4}\sum_u \int_{S^2_{\tau ,\infty}} d^2\Sigma_u e^{-q(\tau ,\vec
\sigma )} \sum_v \int d^3\sigma_1 \nonumber \\
&&\Big[ \sigma^s \Big( e^{-{1\over {\sqrt{3}}}\sum_{\bar a}\gamma_{\bar ar}
r_{\bar a}(\tau ,\vec \sigma )}\delta_{r(a)}{\cal T}^u_{(a)v}+
e^{-{1\over {\sqrt{3}}}\sum_{\bar a}\gamma_{\bar au}
r_{\bar a}(\tau ,\vec \sigma )}\delta_{u(a)}{\cal T}^r_{(a)v}\Big) -
\nonumber \\
&-& \sigma^r \Big( e^{-{1\over {\sqrt{3}}}\sum_{\bar a}\gamma_{\bar as}
r_{\bar a}(\tau ,\vec \sigma )}\delta_{s(a)}{\cal T}^u_{(a)v}+
e^{-{1\over {\sqrt{3}}}\sum_{\bar a}\gamma_{\bar au}
r_{\bar a}(\tau ,\vec \sigma )}\delta_{u(a)}{\cal T}^s_{(a)v}\Big) \Big]
\nonumber \\
&&(\vec \sigma ,{\vec \sigma}_1,\tau |q,r_{\bar a}]
(e^{-q-{1\over {\sqrt{3}}}\sum_{\bar a}\gamma_{\bar av}r_{\bar a}}[{{\rho}
\over 3}+\sqrt{3}\sum_{\bar c}\gamma_{\bar cv}\pi_{\bar c}](\tau ,{\vec
\sigma}_1) \} ,\nonumber \\
J^{\tau r}_{ADM,R}&=&{\hat J}^{\tau r}_{ADM,R}+{1\over 2} \int d^3\sigma
\sigma^r {\hat {\cal H}}_R(\tau ,\vec \sigma )=\nonumber \\
&=&2\epsilon k\sum_u \int_{S^2_{\tau ,\infty}} d^2\Sigma_u \sigma^r
\Big( e^{q-{2\over
{\sqrt{3}}}\sum_{\bar a}\gamma_{\bar au}r_{\bar a}}[-2\partial_uq+{1\over
{\sqrt{3}}} \sum_{\bar b}\gamma_{\bar bu} \partial_ur_{\bar b}] \Big) (\tau
,\vec \sigma )+\nonumber \\
&+&\epsilon k \int_{S^2_{\tau ,\infty}} d^2\Sigma_r \Big( e^{-q-{2\over
{\sqrt{3}}}\sum_{\bar a}\gamma_{\bar ar}r_{\bar a}}\Big[ \sum_ne^{-{2\over
{\sqrt{3}}}\sum_{\bar a}\gamma_{\bar an}r_{\bar a}}(e^{2q+{2\over {\sqrt{3}}}
\sum_{\bar a}\gamma_{\bar an}r_{\bar a}}-1)-\nonumber \\
&-&e^{-{2\over {\sqrt{3}}}\sum_{\bar a}\gamma_{\bar ar}r_{\bar a}}
(e^{2q+{2\over {\sqrt{3}}}\sum_{\bar a}\gamma_{\bar ar}r_{\bar a}}-1)\Big]
\Big) (\tau,\vec \sigma ).\nonumber \\
&&{}
\label{V5}
\end{eqnarray}

The reduced superhamiltonian constraint [see Eq.(104) of II] becomes

\begin{eqnarray}
{\hat {\cal H}}_R(\tau ,\vec \sigma )&=&-\epsilon
{{e^{-q(\tau ,\vec \sigma )}}\over {8k}}\Big[ (e^{-2q}[6 \sum_{\bar a}\pi^2
_{\bar a}-{1\over 3}\rho^2])(\tau ,\vec \sigma )+\nonumber \\
&+&2 (e^{-q}\sum_ue^{{1\over {\sqrt{3}}}\sum_{\bar a}\gamma_{\bar au}r_{\bar a}}
[2\sqrt{3}\sum_{\bar b}\gamma_{\bar bu}\pi_{\bar b}-{1\over 3}\rho ])(\tau
,\vec \sigma )\times \nonumber \\
&&\int d^3\sigma_1 \sum_r \delta^u_{(a)} {\cal T}^u_{(a)r}(\vec \sigma ,{\vec
\sigma}_1,\tau |q,r_{\bar a}] \Big( e^{-q-{1\over {\sqrt{3}}}\sum_{\bar a}
\gamma_{\bar ar}r_{\bar a}}[{{\rho}\over 3}+\sqrt{3}\sum_{\bar b}\gamma_{\bar
br} \pi_{\bar b}]\Big) (\tau ,{\vec \sigma}_1)+\nonumber \\
&+&\int d^3\sigma_1d^3\sigma_2 \Big( \sum_u e^{{2\over {\sqrt{3}}}\sum_{\bar a}
\gamma_{\bar au}+r_{\bar a}(\tau ,\vec \sigma )} \times \nonumber \\
&&\sum_r{\cal T}^u_{(a)r}(\vec \sigma ,{\vec
\sigma}_1,\tau |q,r_{\bar a}] \Big( e^{-q-{1\over {\sqrt{3}}}\sum_{\bar a}
\gamma_{\bar ar}r_{\bar a}}[{{\rho}\over 3}+\sqrt{3}\sum_{\bar b}\gamma_{\bar
br} \pi_{\bar b}]\Big) (\tau ,{\vec \sigma}_1)\times \nonumber \\
&&\sum_s {\cal T}^u_{(a)s}(\vec \sigma ,{\vec
\sigma}_2,\tau |q,r_{\bar a}] \Big( e^{-q-{1\over {\sqrt{3}}}\sum_{\bar a}
\gamma_{\bar as}r_{\bar a}}[{{\rho}\over 3}+\sqrt{3}\sum_{\bar c}\gamma_{\bar
cs} \pi_{\bar c}]\Big) (\tau ,{\vec \sigma}_2)+\nonumber \\
&+&\sum_{uv} e^{{1\over {\sqrt{3}}}\sum_{\bar a}(\gamma_{\bar au}+\gamma_{\bar
av})r_{\bar a}(\tau ,\vec \sigma )} (\delta^u_{(b)}\delta^v_{(a)}-\delta^u_{(a)}
\delta^v_{(b)})\times \nonumber \\
&&\sum_r {\cal T}^u_{(a)r}(\vec \sigma ,{\vec
\sigma}_1,\tau |q,r_{\bar a}] \Big( e^{-q-{1\over {\sqrt{3}}}\sum_{\bar a}
\gamma_{\bar ar}r_{\bar a}}[{{\rho}\over 3}+\sqrt{3}\sum_{\bar b}\gamma_{\bar
br} \pi_{\bar b}]\Big) (\tau ,{\vec \sigma}_1)\nonumber \\
&&\sum_s {\cal T}^v_{(b)s}(\vec \sigma ,{\vec
\sigma}_2,\tau |q,r_{\bar a}] \Big( e^{-q-{1\over {\sqrt{3}}}\sum_{\bar a}
\gamma_{\bar as}r_{\bar a}}[{{\rho}\over 3}+\sqrt{3}\sum_{\bar c}\gamma_{\bar
cs} \pi_{\bar c}]\Big) (\tau ,{\vec \sigma}_2)\, \Big)\, \Big]\, +\nonumber \\
&+&\epsilon k
\sum_{r,s} e^{q(\tau ,\vec \sigma )-{1\over {\sqrt{3}}}\sum_{\bar c}
(\gamma_{\bar cr}+\gamma_{\bar cs})r_{\bar c}(\tau ,\vec \sigma )} \epsilon
_{(a)(b)(c)}\delta_{(a)r}\delta_{(b)s}\, {}^3{\hat \Omega}_{rs(c)}[q,r_{\bar c}]
(\tau ,\vec \sigma ) \approx 0,\nonumber \\
&&{}\nonumber \\
with&& \epsilon k
\sum_{r,s} e^{q(\tau ,\vec \sigma )-{1\over {\sqrt{3}}}\sum_{\bar c}
(\gamma_{\bar cr}+\gamma_{\bar cs})r_{\bar c}(\tau ,\vec \sigma )} \epsilon
_{(a)(b)(c)}\delta_{(a)r}\delta_{(b)s}\, {}^3{\hat \Omega}_{rs(c)}[q,r_{\bar c}]
(\tau ,\vec \sigma )=\nonumber \\
&&=\epsilon k (e^{3q}\, {}^3{\hat R}[q,r_{\bar a}])(\tau ,\vec \sigma ).
\label{V6}
\end{eqnarray}

As already anticipated in II,
the gauge variable in which the superhamiltonian constraint
(or Lichnerowicz equation) has to be solved, is the conformal
factor $q(\tau ,\vec \sigma )$ [or better $\phi (\tau ,\vec \sigma )=e^{q(\tau
,\vec \sigma )/2}$], since, as shown in Appendix A, the surface integral giving
the ADM energy depends only on it and not on its conjugate momentum
$\rho (\tau ,\vec \sigma )$. In every Gauss law, the piece of the secondary
first class constraint corresponding to a divergence and giving the ``strong"
form of the conserved charge as the flux through the surface at infinity of a
corresponding density depends on the variable which has to be eliminated in the
canonical reduction by using the constraint (the conjugate variable is the
gauge variable): once the constraint is solved in this variable, it can be put
inside the volume expression of the ``weak" form of the conserved charge to
obtain its expression in the reduced phase space; the strong ADM energy is the
only known charge, associated with a constraint
bilinear in the momenta, depending only on the coordinates and not on the
momenta, so that $\phi$ and not $\rho$ is the unknown in the Lichnerowicz
equation.

It is the gauge-fixing constraint to the superhamiltonian one
which fixes the momentum $\rho (\tau ,\vec \sigma )$, the last gauge
variable. Starting from Lichnerowicz
\cite{conf}, usually one chooses the maximal slicing condition ${}^3K(\tau
,\vec \sigma ) \approx 0$ for this gauge-fixing; Lichnerowicz has shown that
with it the superhamiltonian and supermomentum constraints of metric gravity
form a system of 5 elliptic differential equations
which can be shown to have one and only one solution; moreover, with this
condition Schoen and Yau \cite{p17}
have shown that the ADM 4-momentum is timelike (i.e.
the ADM energy is positive or zero for Minkowski spacetime).

However Schoen-Yau have relaxed the maximal slicing condition in their last
proof of the positivity of the ADM energy. Therefore, if there is no
contradiction with the existence and unicity of the solution of the reduced
Lichnerowicz equation (\ref{V6})
(which is now an integro-differential equation for $\phi$),
one can replace ${}^3K(\tau ,\vec \sigma )\approx 0$ with the
gauge-fixing $\rho (\tau ,\vec \sigma )\approx 0$, which is natural in our
approach with 3-orthogonal coordinates, saving the positivity of the energy.

This  entails that at the level of the Dirac brackets associated with the
second class constraints ${\tilde {\cal H}}_R(\tau ,\vec \sigma )\approx 0$,
$\rho (\tau ,\vec \sigma )\approx 0$, the physical variables $r_{\bar a}(\tau
,\vec \sigma )$, $\pi_{\bar a}(\tau ,\vec \sigma )$, remain canonical and
describe the canonical physical degrees of freedom of the gravitational field
in this gauge. However, since a closed
form of the conformal factor in terms of $r_{\bar a}$, $\pi_{\bar a}$ as a
solution of the superhamiltonian constraint [after having put $\rho (\tau
,\vec \sigma )=0$ in it] is not known, the ADM energy (weakly coinciding with
the ADM invariant mass in the rest-frame instant form) cannot be explicitly
expressed in terms of the physical degrees of freedom of the gravitational
field in 3-orthogonal coordinates.

It seems difficult to be able to implement the last step of the programme,
namely to find the final Shanmugadhasan canonical transformation $q$, $\rho$,
$r_{\bar a}$, $\pi_{\bar a}$ $\mapsto$ ${\hat {\cal H}}^{'}_R$, $\rho^{'}$,
$r^{'}_{\bar a}$, $\pi^{'}_{\bar a}$  [${\hat {\cal H}}^{'}_R(\tau ,\vec \sigma
)\approx 0$ equivalent to ${\hat {\cal H}}_R(\tau ,\vec \sigma )\approx 0$ but
with $\{ {\hat {\cal H}}^{'}_R(\tau ,\vec \sigma),
{\hat {\cal H}}^{'}_R(\tau ,{\vec \sigma}^{'}) \} =0$], so that all the first
class constraints of tetrad gravity appear in the final canonical basis in
Abelianized form. Equally difficult is to find the analogue of the York
map\cite{yorkmap} (see also Appendix C of II) in the 3-orthogonal gauge:
$q$, $\rho$, $r_{\bar a}$, $\pi_{\bar a}$ $\mapsto$ ${}^3{\hat K}^{'}$,
$\rho^{(K)}$, $r^{(K)}_{\bar a}$, $\pi^{(K)}_{\bar a}$ [with ${}^3{\hat K}^{'}$
Abelianized version of ${}^3K$].

To transform the superhamiltonian constraint in the
reduced Lichnerowicz equation for
the conformal factor, let us use $\phi = e^{q/2}$ [$\, q, \rho \mapsto
\phi=e^{q/2}, \pi_{\phi}=2\phi^{-1} \rho$]. We get

\begin{eqnarray}
{}^3{\hat g}_{rs}&=&e^{2q+{2\over {\sqrt{3}}}\sum_{\bar a}\gamma_{\bar ar}r
_{\bar a}}\, \delta_{rs}\equiv e^{2q}\, {}^3{\tilde g}_{rs}=\phi^4\, {}^3{\tilde
g}_{rs},\quad\quad {}^3{\tilde g}_{rs}=e^{{2\over {\sqrt{3}}}\sum_{\bar a}
\gamma_{\bar ar}r_{\bar a}},\nonumber \\
&&\Rightarrow \sum_r \, {}^3{\tilde \Gamma}^r_{rs}=0,\nonumber \\
{}^3{\hat R}&=&e^{-2q}[-4\, {}^3{\tilde g}^{rs}\partial_r\partial_sq-2\,
{}^3{\tilde g}^{rs}\partial_rq \partial_sq-4 \partial_r\, {}^3{\tilde g}^{rs}
\partial_sq+{}^3{\tilde R}[r_{\bar a}]]=\nonumber \\
&=&e^{-2q}[e^{-q/2}(-8{\tilde \triangle}[r_{\bar a}]
e^{q/2}+8\, {}^3{\tilde \Gamma}^r_{rs}\, {}^3{\tilde g}^{su}
\partial_ue^{q/2})+{}^3{\tilde R}[r_{\bar a}]]=\nonumber \\
&=&\phi^{-5}[-8 {\tilde \triangle}[r_{\bar a}] \phi +{}^3{\tilde R}[r_{\bar a}]
 \phi ],
\label{V8}
\end{eqnarray}

\noindent where ${}^3\tilde R={}^3\tilde R[r_{\bar a}]$ and $\tilde \triangle
=\tilde \triangle [r_{\bar a}]$ are the scalar curvature and the
Laplace-Beltrami operator associated with the 3-metric ${}^3{\tilde g}_{rs}$
respectively [$\tilde \triangle -{1\over 8}{}^3\tilde
R$ is a conformally invariant operator \cite{conf}]. From Eq.(D1) of
Appendix D of II, we have [$\tilde \gamma = det\, |{}^3{\tilde g}_{rs}|=1$]

\begin{eqnarray}
{}^3\hat R [q,r_{\bar a}]&=&
-\sum_{uv}\{ (\partial_vq+{1\over {\sqrt{3}}}\sum_{\bar a}\gamma_{\bar au}
\partial_vr_{\bar a})(2\partial_vq-{1\over {\sqrt{3}}}\sum_{\bar b}\gamma
_{\bar bu}\partial_vr_{\bar b})+\nonumber \\
&&+e^{-2(q+{1\over {\sqrt{3}}}\sum_{\bar c}\gamma_{\bar cv}r_{\bar c})}
[\partial^2_vq+{1\over {\sqrt{3}}}\sum_{\bar a}\gamma_{\bar au}\partial_v^2r
_{\bar a}+\nonumber \\
&&+{2\over {\sqrt{3}}}(\partial_vq+{1\over {\sqrt{3}}}\sum_{\bar a}\gamma
_{\bar au}\partial_vr_{\bar a})\sum_{\bar b}(\gamma_{\bar bu}-\gamma_{\bar bv})
\partial_vr_{\bar b}-\nonumber \\
&&-(\partial_vq+{1\over {\sqrt{3}}}\sum_{\bar a}\gamma_{\bar av}\partial_vr
_{\bar a})(\partial_vq+{1\over {\sqrt{3}}}\sum_{\bar b}\gamma_{\bar bu}
\partial_vr_{\bar b}] \} +\nonumber \\
&&+\sum_ue^{-2(q+{1\over {\sqrt{3}}}\sum_{\bar c}\gamma_{\bar cu}r_{\bar c})}
[-\partial^2_uq+{2\over {\sqrt{3}}}\sum_{\bar a}\gamma_{\bar au} \partial^2_ur
_{\bar a}+\nonumber \\
&&+(\partial_uq+{1\over {\sqrt{3}}}\sum_{\bar a}\gamma_{\bar au}\partial_ur
_{\bar a})(\partial_uq-{2\over {\sqrt{3}}}\sum_{\bar b}\gamma_{\bar bu}
\partial_ur_{\bar b})]\nonumber \\
&&{\rightarrow}_{r_{\bar a}\, \rightarrow 0}\, -6\sum_u(\partial_uq)^2-
4e^{-2q}\sum_u[\partial_u^2q-(\partial_uq)^2]=\nonumber \\
&&-24 \sum_u(\partial_uln\, \phi )^2-8\phi^{-4}\sum_u[\partial^2_uln\, \phi -2
(\partial_uln\, \phi )^2]\, {\rightarrow}_{q\,
\rightarrow 0}\, 0,\nonumber \\
&&{}\nonumber \\
{\rightarrow}_{q\, \rightarrow 0}\,&& {}^3\tilde R[r_{\bar a}]=\nonumber \\
&&=-{1\over {\sqrt{3}}}\sum_{uv}\{
-{1\over {\sqrt{3}}}\sum_{\bar a\bar b}\gamma_{\bar au}\gamma_{\bar bu}
\partial_vr_{\bar a}\partial_vr_{\bar b}+e^{-{2\over {\sqrt{3}}}\sum_{\bar c}
\gamma_{\bar cv}r_{\bar c}}\sum_{\bar a}\gamma_{\bar au}\cdot \nonumber \\
&&[\partial^2_vr_{\bar a}+{2\over {\sqrt{3}}}\sum_{\bar b}(\gamma_{\bar bu}-
\gamma_{\bar bv})\partial_vr_{\bar a}\partial_vr_{\bar b}-{1\over {\sqrt{3}}}
\sum_{\bar b}\gamma_{\bar bv}\partial_vr_{\bar a}\partial_vr_{\bar b}]\}+
\nonumber \\
&&+{2\over {\sqrt{3}}}\sum_ue^{-{2\over {\sqrt{3}}}\sum_{\bar c}\gamma
_{\bar cu}t_{\bar c}}\sum_{\bar a}\gamma_{\bar au}[\partial^2_ur_{\bar a}+
{1\over {\sqrt{3}}}\sum_{\bar b}\gamma_{\bar bu}\partial_ur_{\bar a}
\partial_ur_{\bar b}]=\nonumber \\
&&={1\over 3}\sum_u (1-2e^{-{2\over {\sqrt{3}}}\sum_{\bar a}\gamma_{\bar au}
r_{\bar a}}) \sum_{\bar b}(\partial_ur_{\bar b})^2+\nonumber \\
&&+{2\over {\sqrt{3}}}\sum_ue^{-{2\over {\sqrt{3}}}\sum_{\bar c}\gamma
_{\bar cu}t_{\bar c}}\sum_{\bar a}\gamma_{\bar au}[\partial^2_ur_{\bar a}+
{1\over {\sqrt{3}}}\sum_{\bar b}\gamma_{\bar bu}\partial_ur_{\bar a}
\partial_ur_{\bar b}],\nonumber \\
&&{}\nonumber \\
\tilde \triangle [r_{\bar a}]&=& \partial_r [{}^3{\tilde g}^{rs}\, \partial_s]
={}^3{\tilde g}^{rs}\, {}^3{\tilde \nabla}_r\, {}^3{\tilde \nabla}_s=
\nonumber \\
&=&\sum_re^{-{2\over {\sqrt{3}}}\sum_{\bar a}\gamma_{\bar ar} r_{\bar a}}
[\partial_r^2-{2\over {\sqrt{3}}} \sum_{\bar b}\gamma_{\bar br}
\partial_rr_{\bar b} \partial_r].
\label{V9}
\end{eqnarray}

Using Eq.(\ref{V6}), the reduced superhamiltonian constraint becomes [$k
=c^3/16\pi G$]

\begin{eqnarray}
{\tilde {\cal H}}_R(\tau ,\vec \sigma )&=&\epsilon \{ k\phi^6\, {}^3{\hat R}-
{{\phi^{-6}}\over {8k}} {}^3G_{o(a)(b)(c)(d)}\, {}^3{\hat e}_{((a)r}\,
{}^3{\hat {\tilde \pi}}^r_{(b))}\, {}^3{\hat e}_{((c)s}\, {}^3{\hat {\tilde
\pi}}^s_{(d))} \} (\tau ,\vec \sigma )=\nonumber \\
&=&\epsilon \phi (\tau ,\vec \sigma ) \,
\{ \, k(-8{\tilde \triangle}[r_{\bar a}] +{}^3{\tilde R}[r_{\bar a}])\phi
-\nonumber \\
&-&{{\phi^{-7}}\over {8k}} {}^3G_{o(a)(b)(c)(d)}\, {}^3{\hat e}_{((a)r}\,
{}^3{\hat {\tilde \pi}}^r_{(b))}\, {}^3{\hat e}_{((c)s}\, {}^3{\hat {\tilde
\pi}}^s_{(d))} \} (\tau ,\vec \sigma )=\nonumber \\
&=&\epsilon \phi (\tau ,\vec \sigma )\, \{ \,
{{c^3}\over {16\pi G}} (-8{\tilde \triangle}[r_{\bar
a}] +{}^3{\tilde R}[r_{\bar a}])\phi -\nonumber \\
&-&{{2\pi G}\over {c^3}}
\Big[ (\phi^{-7} (6 \sum_{\bar a} \pi^2_{\bar a}-{1\over 3}\rho^2))(\tau ,\vec
\sigma )+\nonumber \\
&+&2(\phi^{-5}
\sum_ue^{{1\over {\sqrt{3}}}\sum_{\bar a}\gamma_{\bar au}r_{\bar a}}
[2\sqrt{3}\sum_{\bar b}\gamma_{\bar bu}\pi_{\bar b}-{1\over 3}\rho ])(\tau
,\vec \sigma )\times \nonumber \\
&&\int d^3\sigma_1 \sum_r \delta^u_{(a)} {\cal T}^u_{(a)r}(\vec \sigma ,{\vec
\sigma}_1,\tau |\phi ,r_{\bar a}] \Big( \phi^{-2}
e^{-{1\over {\sqrt{3}}}\sum_{\bar a}
\gamma_{\bar ar}r_{\bar a}}[{{\rho}\over 3}+\sqrt{3}\sum_{\bar b}\gamma_{\bar
br} \pi_{\bar b}]\Big) (\tau ,{\vec \sigma}_1)+\nonumber \\
&+&\phi^{-3}(\tau ,\vec \sigma )
\int d^3\sigma_1d^3\sigma_2 \Big( \sum_u e^{{2\over {\sqrt{3}}}\sum_{\bar a}
\gamma_{\bar au}+r_{\bar a}(\tau ,\vec \sigma )} \times \nonumber \\
&&\sum_r{\cal T}^u_{(a)r}(\vec \sigma ,{\vec
\sigma}_1,\tau |\phi ,r_{\bar a}] \Big( \phi^{-2}
e^{-{1\over {\sqrt{3}}}\sum_{\bar a}
\gamma_{\bar ar}r_{\bar a}}[{{\rho}\over 3}+\sqrt{3}\sum_{\bar b}\gamma_{\bar
br} \pi_{\bar b}]\Big) (\tau ,{\vec \sigma}_1)\times \nonumber \\
&&\sum_s {\cal T}^u_{(a)s}(\vec \sigma ,{\vec
\sigma}_2,\tau |\phi ,r_{\bar a}] \Big( \phi^{-2}
e^{-{1\over {\sqrt{3}}}\sum_{\bar a}
\gamma_{\bar as}r_{\bar a}}[{{\rho}\over 3}+\sqrt{3}\sum_{\bar c}\gamma_{\bar
cs} \pi_{\bar c}]\Big) (\tau ,{\vec \sigma}_2)+\nonumber \\
&+&\sum_{uv} e^{{1\over {\sqrt{3}}}\sum_{\bar a}(\gamma_{\bar au}+\gamma_{\bar
av})r_{\bar a}(\tau ,\vec \sigma )} (\delta^u_{(b)}\delta^v_{(a)}-\delta^u_{(a)}
\delta^v_{(b)})\times \nonumber \\
&&\sum_r {\cal T}^u_{(a)r}(\vec \sigma ,{\vec
\sigma}_1,\tau |\phi ,r_{\bar a}] \Big( \phi^{-2}
e^{-{1\over {\sqrt{3}}}\sum_{\bar a}
\gamma_{\bar ar}r_{\bar a}}[{{\rho}\over 3}+\sqrt{3}\sum_{\bar b}\gamma_{\bar
br} \pi_{\bar b}]\Big) (\tau ,{\vec \sigma}_1)\nonumber \\
&&\sum_s {\cal T}^v_{(b)s}(\vec \sigma ,{\vec
\sigma}_2,\tau |\phi ,r_{\bar a}] \Big( \phi^{-2}
e^{-{1\over {\sqrt{3}}}\sum_{\bar a}
\gamma_{\bar as}r_{\bar a}}[{{\rho}\over 3}+\sqrt{3}\sum_{\bar c}\gamma_{\bar
cs} \pi_{\bar c}]\Big) (\tau ,{\vec \sigma}_2)\, \Big)\, \Big]\,
\} \approx 0.\nonumber \\
&&{}
\label{V10}
\end{eqnarray}

This equation has to be implemented with a gauge-fixing to eliminate $\rho
(\tau ,\vec \sigma )$ [either ${}^3\hat K(\tau ,\vec \sigma ) \approx 0$ or
$\rho (\tau ,\vec \sigma )\approx 0$]. Let us remark that the maximal slicing
condition ${}^3\hat K(\tau ,\vec \sigma )\approx 0$ does not imply any real
simplification of this expression and has the extra complications that:
i) $r_{\bar a}$, $\pi_{\bar a}$ are no more canonical at the level of Dirac
brackets; ii) one has to solve ${}^3\hat K(\tau ,\vec \sigma )=0$ as an
integral equation in $\rho$, to eliminate this variable.

For $\rho(\tau ,\vec \sigma )\approx 0$ we get the
final reduced form of the Lichnerowicz equation

\begin{eqnarray}
(-{\tilde \triangle}[r_{\bar a}] &+&{1\over 8}{}^3{\tilde R}[r_{\bar a}])(\tau
,\vec \sigma ) \phi (\tau ,\vec \sigma )={{12 \pi^2 G^2}\over {c^6}}
\Big[ 2(\phi^{-7}  \sum_{\bar a} \pi^2_{\bar a})(\tau ,\vec
\sigma )+\nonumber \\
&+&4(\phi^{-5}
\sum_ue^{{1\over {\sqrt{3}}}\sum_{\bar a}\gamma_{\bar au}r_{\bar a}}
\sum_{\bar b}\gamma_{\bar bu}\pi_{\bar b})(\tau
,\vec \sigma )\times \nonumber \\
&&\int d^3\sigma_1 \sum_r \delta^u_{(a)} {\cal T}^u_{(a)r}(\vec \sigma ,{\vec
\sigma}_1,\tau |\phi ,r_{\bar a}] \Big( \phi^{-2}
e^{-{1\over {\sqrt{3}}}\sum_{\bar a}
\gamma_{\bar ar}r_{\bar a}}\sum_{\bar b}\gamma_{\bar
br} \pi_{\bar b}\Big) (\tau ,{\vec \sigma}_1)+\nonumber \\
&+&\phi^{-3}(\tau ,\vec \sigma )
\int d^3\sigma_1d^3\sigma_2 \Big( \sum_u e^{{2\over {\sqrt{3}}}\sum_{\bar a}
\gamma_{\bar au}r_{\bar a}(\tau ,\vec \sigma )} \times \nonumber \\
&&\sum_r{\cal T}^u_{(a)r}(\vec \sigma ,{\vec
\sigma}_1,\tau |\phi ,r_{\bar a}] \Big( \phi^{-2}
e^{-{1\over {\sqrt{3}}}\sum_{\bar a}
\gamma_{\bar ar}r_{\bar a}}\sum_{\bar b}\gamma_{\bar
br} \pi_{\bar b}\Big) (\tau ,{\vec \sigma}_1)\times \nonumber \\
&&\sum_s {\cal T}^u_{(a)s}(\vec \sigma ,{\vec
\sigma}_2,\tau |\phi ,r_{\bar a}] \Big( \phi^{-2}
e^{-{1\over {\sqrt{3}}}\sum_{\bar a}
\gamma_{\bar as}r_{\bar a}}\sum_{\bar c}\gamma_{\bar
cs} \pi_{\bar c}\Big) (\tau ,{\vec \sigma}_2)+\nonumber \\
&+&\sum_{uv} e^{{1\over {\sqrt{3}}}\sum_{\bar a}(\gamma_{\bar au}+\gamma_{\bar
av})r_{\bar a}(\tau ,\vec \sigma )} (\delta^u_{(b)}\delta^v_{(a)}-\delta^u_{(a)}
\delta^v_{(b)})\times \nonumber \\
&&\sum_r {\cal T}^u_{(a)r}(\vec \sigma ,{\vec
\sigma}_1,\tau |\phi ,r_{\bar a}] \Big( \phi^{-2}
e^{-{1\over {\sqrt{3}}}\sum_{\bar a}
\gamma_{\bar ar}r_{\bar a}}\sum_{\bar b}\gamma_{\bar
br} \pi_{\bar b}\Big) (\tau ,{\vec \sigma}_1)\nonumber \\
&&\sum_s {\cal T}^v_{(b)s}(\vec \sigma ,{\vec
\sigma}_2,\tau |\phi ,r_{\bar a}] \Big( \phi^{-2}
e^{-{1\over {\sqrt{3}}}\sum_{\bar a}
\gamma_{\bar as}r_{\bar a}}\sum_{\bar c}\gamma_{\bar
cs} \pi_{\bar c}\Big) (\tau ,{\vec \sigma}_2)\, \Big)\, \Big] .
\label{V11}
\end{eqnarray}

Let us remark that, if this integro-differential equation for $\phi (\tau
,\vec \sigma )= e^{{1\over 2}q(\tau ,\vec \sigma )}\, > 0$ admits different
solutions $\phi_1[r_{\bar a},\pi_{\bar a}]$, $\phi_2[r_{\bar a},\pi_{\bar a}]$
,..., they would correspond to inequivalent gravitational fields in vacuum
(there are no more gauge transformations for correlating them) evolving
according to the associated ADM energies.

However, if, as said in Section V of II, the existence and unicity of solutions
of the 5 equations of ADM metric gravity [the Lichnerowicz equation or
superhamiltonian constraint, the 3 supermomentum constraints and the gauge
fixing (maximal slicing condition) ${}^3K(\tau ,\vec \sigma )\approx 0$] remain
valid in tetrad gravity for the reduced Lichnerowicz equation [the original one
with the solution of the supermomentum constraints inserted into it] with the
gauge fixing ${}^3K(\tau ,\vec \sigma )\approx 0$ replaced by the natural one
$\rho (\tau ,\vec \sigma )\approx 0$ for the 3-orthogonal gauge, only one
solution $\phi \approx \phi [r_{\bar a},\pi_{\bar a}]$ will exist with the
boundary condition $\phi (\tau ,\vec \sigma )\, {\rightarrow}_{r\, \rightarrow
\infty}\, 1+{M\over {2r}}+o_4(r^{-3/2})$ of Eqs.(\ref{V3}).

Finally, let us restrict ourselves to WSW hypersurfaces with the gauge-fixing
procedure of Section III [$b^{(\mu )}_{(\infty )A}(\tau )\approx L^{(\mu )}{}
_{(\nu )=A}(p_{(\infty )},{\buildrel \circ \over p}_{(\infty )} )$], see
Eq.(\ref{I23}). The Dirac Hamiltonian is now [$\epsilon_{(\infty )}=
\sqrt{\epsilon p^2_{(\infty )}}$]

\begin{equation}
{\hat H}^{(WSW)}_{(D)ADM,R}= \int d^3\sigma [n {\hat {\cal H}}_R+\lambda_n
{\tilde \pi}^n](\tau ,\vec \sigma )-
{\tilde \lambda}_{\tau }(\tau ) [\epsilon_{(\infty )}
-{\hat P}^{\tau}_{ADM,R}]- {\tilde \lambda}_r(\tau ) {\hat P}^r_{ADM,R},
\label{V11a}
\end{equation}

\noindent and we have (see Eqs.(\ref{V4a}) for the expression of ${\hat n}_r$)
\hfill\break
\hfill\break
$N(\tau ,\vec \sigma )=-{\tilde \lambda}
_{\tau}(\tau )+n(\tau ,\vec \sigma )$, \hfill\break
$N_{(a)}(\tau ,\vec \sigma )
={}^3{\hat e}^r_{(a)}(\tau ,\vec \sigma ) \Big[ -{\tilde \lambda}
_r(\tau )+{\hat n}_r[\tau ,\vec \sigma |q, \rho ,r_{\bar a}, \pi_{\bar a},
{\tilde \lambda}_A] \Big] =$\hfill\break
$\quad\quad\quad\quad =-\phi^{-2}
(\tau ,\vec \sigma ) \delta^r_{(a)} \Big[ {\tilde \lambda}_r(\tau )-
{\hat n}_r[\tau ,\vec \sigma |q, \rho ,r_{\bar a}, \pi_{\bar a},
{\tilde \lambda}_A] \Big]$,\hfill\break
\hfill\break
If we add the natural gauge-fixing $\rho (\tau ,\vec \sigma )={1\over 2}\phi
(\tau ,\vec \sigma ) \pi_{\phi}(\tau ,\vec \sigma ) \approx 0$ to
${\hat {\cal H}}_R(\tau ,\vec \sigma )\approx 0$, its time
constancy implies

\begin{eqnarray}
\partial_{\tau} \rho (\tau ,\vec \sigma )\, &{\buildrel \circ \over =}\,&
\{ \rho (\tau ,\vec \sigma ), {\hat H}^{(WSW)}_{(D)ADM,R} \} = \int
d^3\sigma_1 n(\tau ,{\vec \sigma}_1) \{ \rho (\tau ,\vec \sigma ), {\hat {\cal
H}}_R(\tau ,{\vec \sigma}_1) \} +\nonumber \\
&+& {\tilde \lambda}_{\tau}(\tau ) \{ \rho (\tau ,\vec \sigma ),{\hat P}^{\tau}
_{ADM,R} \} + {\tilde \lambda}_r(\tau ) \{ \rho (\tau ,\vec \sigma ), {\hat P}^r
_{ADM,R} \} \approx \nonumber \\
&\approx& -{1\over 2}\phi (\tau ,\vec \sigma ) \Big[ \int d^3\sigma_1 n(\tau
,{\vec \sigma}_1) {{\delta {\hat {\cal H}}_R(\tau ,{\vec \sigma}_1)}\over
{\delta \phi (\tau ,\vec \sigma )}}+\nonumber \\
&+& {\tilde \lambda}_{\tau} {{\delta {\hat P}^{\tau}_{ADM,R}}\over
{\delta \phi (\tau ,\vec \sigma )}}+{\tilde \lambda}_r(\tau )
{{\delta {\hat P}^r_{ADM,R}}\over {\delta \phi (\tau ,\vec \sigma )}} \approx
0,\nonumber \\
&&{}\nonumber \\
\Rightarrow&& \quad n(\tau ,\vec \sigma ) - \hat n(\tau ,\vec \sigma |r_{\bar a}
,\pi_{\bar a},{\tilde \lambda}_A] \approx 0,\nonumber \\
&&{}\nonumber \\
\partial_{\tau}&& \Big[ n(\tau ,\vec \sigma ) - \hat n(\tau ,\vec \sigma
|r_{\bar a},\pi_{\bar a},{\tilde \lambda}_A] \Big] =\nonumber \\
&=&\lambda_n(\tau ,\vec \sigma )- \{ \hat n(\tau ,\vec \sigma
|r_{\bar a},\pi_{\bar a},{\tilde \lambda}_A] , {\hat H}^{(WSW)}_{(D)ADM,R} \}
\approx 0,\nonumber \\
&&{}\nonumber \\
\Rightarrow&& \quad \lambda_n(\tau ,\vec \sigma )\quad determined.
\label{V11b}
\end{eqnarray}

Therefore we find an integral equation for the lapse function $n(\tau ,\vec
\sigma )$ implying its being different from zero (this avoids a finite time
breakdown), even when ${\vec {\tilde \lambda}}(\tau )=0$ since ${\tilde
\lambda}_{\tau }(\tau )\not= 0$.

If we now go to the final Dirac brackets with respect to the second class
constraints $\rho (\tau ,\vec \sigma )\approx 0$, ${\hat {\cal H}}_R(\tau
,\vec \sigma )\approx 0$, $n(\tau ,\vec \sigma ) - \hat n(\tau ,\vec \sigma
|r_{\bar a},\pi_{\bar a},{\tilde \lambda}_A] \approx 0$, ${\tilde \pi}^n
(\tau ,\vec \sigma )\approx 0$, on the WSW hypersurfaces,
asymptotically orthogonal to ${\hat P}^{(\mu )}_{ADM,R}$ at spatial
infinity, we remain only with the canonical variables $r_{\bar a}$, $\pi_{\bar
a}$ from Eqs.(\ref{I23}) we get the following form of the
Dirac-Hamiltonian and of the remaining four first class constraints

\begin{eqnarray}
{\hat H}^{(WSW)}_{(D)ADM,R\rho =0}&=&
-{\tilde \lambda}_{\tau }(\tau ) \Big( \epsilon
_{(\infty )}-{\hat P}^{\tau}_{ADM,R}[r_{\bar a},\pi_{\bar a},\phi (r_{\bar a},
\pi_{\bar a})]\Big)- {\tilde \lambda}_r(\tau ) {\hat P}^r_{ADM,R}
[r_{\bar a},\pi_{\bar a},\phi (r_{\bar a},\pi_{\bar a})],\nonumber \\
&&{}\nonumber \\
&&\epsilon_{(\infty )}-{\hat P}^{\tau}_{ADM,R}
[r_{\bar a},\pi_{\bar a},\phi (r_{\bar a},\pi_{\bar a})] \approx 0,\nonumber \\
&&{\hat P}^r_{ADM,R}
[r_{\bar a},\pi_{\bar a},\phi (r_{\bar a},\pi_{\bar a})] \approx 0.
\label{V7a}
\end{eqnarray}

\noindent where $\phi (r_{\bar a},\pi_{\bar a})$ is the solution of the reduced
Lichnerowicz equation ${\hat {\cal H}}_R(\tau ,\vec \sigma ){|}_{\rho (\tau
,\vec \sigma )=0} =0$. The Dirac Hamiltonian takes the same form as in the
case of parametrized Minkowski theories restricted to
Wigner spacelike hyperplanes.

After the gauge-fixing $T_{(\infty )}-\tau \approx 0$, one gets
${\tilde \lambda}_{\tau}(\tau )=\epsilon$ and

\begin{eqnarray}
N(\tau ,\vec \sigma )&=&-\epsilon +\hat n(\tau ,\vec \sigma |r_{\bar a},\pi
_{\bar a},\phi (r_{\bar a},\pi_{\bar a}),{\tilde \lambda}_r]=\nonumber \\
&=&-\epsilon +\tilde n(\tau ,\vec \sigma |r_{\bar a}, \pi_{\bar a},{\tilde
\lambda}_r],\nonumber \\
N_r(\tau ,\vec \sigma )&=&-{\tilde \lambda}_r(\tau ) +{\hat n}_r
(\tau ,\vec \sigma |r_{\bar a},\pi
_{\bar a},\phi (r_{\bar a},\pi_{\bar a}),{\tilde \lambda}_r]=\nonumber \\
&=&-{\tilde \lambda}_r(\tau ) +{\tilde n}_r(\tau ,\vec \sigma |r_{\bar a},\pi
_{\bar a},{\tilde \lambda}_s],
\label{V7aa}
\end{eqnarray}

\begin{eqnarray}
&&{\hat H}^{(WSW)}_{(D)ADM}=\epsilon {\hat P}^{\tau}_{ADM,R}
[r_{\bar a},\pi_{\bar a},\phi (r_{\bar a},\pi_{\bar a})]-{\tilde \lambda}
_r(\tau ){\hat P}^r_{ADM,R}[r_{\bar a},\pi_{\bar a},\phi (r_{\bar a},
\pi_{\bar a})],\nonumber \\
&&{}\nonumber \\
&&{\hat P}^r_{ADM,R}[r_{\bar a},\pi_{\bar a},\phi (r_{\bar a},
\pi_{\bar a})] \approx 0.
\label{V7ab}
\end{eqnarray}

This is the asymptotic rest-frame instant form of dynamics for tetrad gravity.
\hfill\break
\hfill\break
In the gauge ${\vec {\tilde \lambda}}(\tau )=0$, implied by the gauge fixings
${\vec \sigma}_{ADM}[r_{\bar a},\pi_{\bar a}]\approx 0$ on the ``internal"
3-center-of-mass, we get the final Dirac Hamiltonian\hfill\break
\hfill\break
${\hat H}^{(WSW){'}}_{(D)ADM}=\epsilon {\hat P}^{\tau}_{ADM,R}$,\hfill\break
\hfill\break
and the following
normal form [namely solved in the accelerations] of the two dynamical
Einstein equations for the gravitational field Dirac observables
in the 3-orthogonal gauge with $\rho (\tau ,\vec \sigma )
\approx 0$ and in the rest frame

\begin{eqnarray}
\partial_{\tau} r_{\bar a}(\tau ,\vec \sigma )\, &{\buildrel \circ \over =}\,&
\{ r_{\bar a}(\tau ,\vec \sigma ),{\hat P}_{ADM,R}^{\tau}[r_{\bar b},\pi_{\bar
b},\phi (r_{\bar b},\pi_{\bar b})] \} ,
\nonumber \\
\partial_{\tau} \pi_{\bar a}(\tau ,\vec \sigma )\, &{\buildrel \circ \over =}\,&
\{ \pi_{\bar a}(\tau ,\vec \sigma ),{\hat P}_{ADM,R}^{\tau}[r_{\bar b},\pi
_{\bar b},\phi (r_{\bar b},\pi_{\bar b})] \} ,
\nonumber \\
&&{}\nonumber \\
{\hat P}^r_{ADM,R}[r_{\bar a},\pi_{\bar a},\phi (r_{\bar a},\pi_{\bar a})]
&\approx& 0.
\label{V13}
\end{eqnarray}

The 4-metric and the line element in adapted coordinates $\sigma^A$
on the WSW hypersurfaces are

\begin{eqnarray}
&&{}^4{\hat g}_{AB}=\nonumber \\
&&=\epsilon
\left( \begin{array}{c}  \Big(-\epsilon +\tilde n[r_{\bar a},\pi_{\bar a},
{\vec {\tilde \lambda}}]\Big)^2-\phi^{-4}[r_{\bar a},\pi_{\bar a}]\sum_r
e^{-{2\over {\sqrt{3}}}\sum_{\bar a}\gamma_{\bar ar}r_{\bar a}}
\Big({\tilde \lambda}^2_r(\tau )-{\tilde n}_r[r_{\bar a},\pi_{\bar a},
{\vec {\tilde \lambda}}\Big)^2 \\
{\tilde \lambda}_r(\tau )-{\tilde n}_r[r_{\bar a},
\pi_{\bar a},{\vec {\tilde \lambda}}]
\end{array} \right. \nonumber \\
&&\left. \begin{array}{c}
{\tilde \lambda}_s(\tau )-{\tilde n}_s[r_{\bar a},
\pi_{\bar a},{\vec {\tilde \lambda}}]\\
- [\phi^4[r_{\bar a},\pi_{\bar a}]e^{{2\over {\sqrt{3}}}
\sum_{\bar a}\gamma_{\bar ar}r_{\bar a}}]\delta_{rs}
\end{array} \right) \nonumber \\
&&{}\nonumber \\
&\rightarrow&_{{\vec {\tilde \lambda}}(\tau ,\vec \sigma )=0}\,\, \epsilon
\left( \begin{array}{c}
\Big(-\epsilon +\tilde n[r_{\bar a},\pi_{\bar a},0]\Big)^2-
\phi^{-4}[r_{\bar a},\pi_{\bar a}]\sum_r
e^{-{2\over {\sqrt{3}}}\sum_{\bar a}\gamma_{\bar ar}r_{\bar a}}
{\tilde n}^2_r[r_{\bar a},\pi_{\bar a},0]\\
-{\tilde n}_r[r_{\bar a},\pi_{\bar a},0]
\end{array} \right. \nonumber \\
&&\left. \begin{array}{c}
-{\tilde n}_s[r_{\bar a},\pi_{\bar a},0]\\
-\phi^4[r_{\bar a},\pi_{\bar a}]
e^{{2\over {\sqrt{3}}}\sum_{\bar a}\gamma_{\bar ar}r_{\bar a}}\delta_{rs}
\end{array} \right) ,\nonumber \\
&&{}\nonumber \\
&&{}\nonumber \\
ds^2&=&{}^4{\hat g}_{\tau\tau}(d\tau )^2 +2\, {}^4{\hat g}_{\tau r}
d\tau d\sigma^r +\sum_r {}^4{\hat g}_{rr}(d\sigma^r)^2.
\label{V7}
\end{eqnarray}

\noindent Even for ${\vec {\tilde \lambda}}(\tau )=0$ we do not get vanishing
shift functions [``synchronous" coordinates], like instead it is assumed by
Christodoulou and Klainermann for their singularity-free solutions.

Let us remark that Eqs.(\ref{V4a}) and (\ref{V11b}) imply that both $\tilde n$
and ${\tilde n}_r$ depend on $G/c^3$ and $c^3/G$ simultaneously, so that both
their post-Newtonian (expansion in $1/c$) and post-Minkowskian (formal expansion
in powers of G) may be non trivial after having done the gauge fixings.

Once we are on WSW hypersurfaces, each triad ${}^3e^r_{(a)}(\tau ,\vec \sigma )$
has an asymptotic limit, which is also the limit of one of
the Frauendiener triads ${}^3e^{(W) r}
_{(a)}(\tau ,\vec \sigma )$, solutions of Eqs.(\ref{IV2}).
In Eqs.(\ref{IV2}), $\, {}^3\hat K(\tau ,\vec \sigma )$ is
the function of $r_{\bar a}$, $\pi_{\bar a}$, $q$ [for $\rho (\tau ,\vec \sigma
)\approx 0$] determined by Eq.(\ref{V4}), so that also the triads ${}^3e
^{(W) r}_{(a)}(\tau ,\vec \sigma )$ become functionals of these same
variables.

From Eqs.(\ref{V5}) evaluated with the gauge fixing $\rho (\tau ,\vec \sigma )
\approx 0$, we get the weak ADM Poincar\'e charges in this gauge

\begin{eqnarray}
{\hat P}^{\tau}_{ADM,R}&=&\epsilon \int d^3\sigma \Big( {{c^3}\over {16\pi G}}
\Big[ \phi^2 \sum_r e^{-{1\over {\sqrt{3}}}\sum_{\bar a}\gamma_{\bar ar}r_{\bar
a}} \times \nonumber \\
&&\Big( {1\over 3}\sum_{\bar b\bar c}(2\gamma_{\bar br}\gamma_{\bar cr}+\delta
_{\bar b\bar c})\partial_rr_{\bar b}\partial_rr_{\bar c}-{4\over {\sqrt{3}}}
(\sum_{\bar b}\gamma_{\bar br}\partial_rr_{\bar b}) \partial_rln\, \phi -
4(\partial_rln\, \phi )^2-\nonumber \\
&-&\sum_ue^{{2\over {\sqrt{3}}}\sum_{\bar a}(\gamma_{\bar ar}-\gamma_{\bar au})
r_{\bar a}}[{2\over 3}\sum_{\bar b\bar c}\gamma_{\bar br}\gamma_{\bar cr}
\partial_ur_{\bar b}\partial_ur_{\bar c}+\nonumber \\
&+&2\sqrt{3}(\sum_{\bar b}\gamma_{\bar
br}\partial_ur_{\bar b})\partial_uln\, \phi +4(\partial_uln\, \phi )^2]\Big)
\Big] (\tau ,\vec \sigma )-\nonumber \\
&-&{{6\pi G}\over {c^3}} \phi^{-2}(\tau ,\vec \sigma )\Big[ (2\phi^{-4}\sum
_{\bar a} \pi^2_{\bar a})(\tau ,\vec \sigma )+\nonumber \\
&+&4(\phi^{-2}\sum_ue^{{1\over {\sqrt{3}}}\sum_{\bar a}\gamma_{\bar au}r_{\bar
a}} \sum_{\bar b}\gamma_{\bar bu}\pi_{\bar b})(\tau ,\vec \sigma )\times
\nonumber \\
&&\int d^3\sigma_1 \sum_r \delta^u_{(a)} {\cal T}^u_{(a)r}(\vec \sigma ,{\vec
\sigma}_1,\tau |\phi ,r_{\bar a}] \Big( \phi^{-2}
e^{-{1\over {\sqrt{3}}}\sum_{\bar a}
\gamma_{\bar ar}r_{\bar a}}\sum_{\bar b}\gamma_{\bar
br} \pi_{\bar b}\Big) (\tau ,{\vec \sigma}_1)+\nonumber \\
&+&
\int d^3\sigma_1d^3\sigma_2 \Big( \sum_u e^{{2\over {\sqrt{3}}}\sum_{\bar a}
\gamma_{\bar au}r_{\bar a}(\tau ,\vec \sigma )} \times \nonumber \\
&&\sum_r{\cal T}^u_{(a)r}(\vec \sigma ,{\vec
\sigma}_1,\tau |\phi ,r_{\bar a}] \Big( \phi^{-2}
e^{-{1\over {\sqrt{3}}}\sum_{\bar a}
\gamma_{\bar ar}r_{\bar a}}\sum_{\bar b}\gamma_{\bar
br} \pi_{\bar b}\Big) (\tau ,{\vec \sigma}_1)\times \nonumber \\
&&\sum_s {\cal T}^u_{(a)s}(\vec \sigma ,{\vec
\sigma}_2,\tau |\phi ,r_{\bar a}] \Big( \phi^{-2}
e^{-{1\over {\sqrt{3}}}\sum_{\bar a}
\gamma_{\bar as}r_{\bar a}}\sum_{\bar c}\gamma_{\bar
cs} \pi_{\bar c}\Big) (\tau ,{\vec \sigma}_2)+\nonumber \\
&+&\sum_{uv} e^{{1\over {\sqrt{3}}}\sum_{\bar a}(\gamma_{\bar au}+\gamma_{\bar
av})r_{\bar a}(\tau ,\vec \sigma )} (\delta^u_{(b)}\delta^v_{(a)}-\delta^u_{(a)}
\delta^v_{(b)})\times \nonumber \\
&&\sum_r {\cal T}^u_{(a)r}(\vec \sigma ,{\vec
\sigma}_1,\tau |\phi ,r_{\bar a}] \Big( \phi^{-2}
e^{-{1\over {\sqrt{3}}}\sum_{\bar a}
\gamma_{\bar ar}r_{\bar a}}\sum_{\bar b}\gamma_{\bar
br} \pi_{\bar b}\Big) (\tau ,{\vec \sigma}_1)\nonumber \\
&&\sum_s {\cal T}^v_{(b)s}(\vec \sigma ,{\vec
\sigma}_2,\tau |\phi ,r_{\bar a}] \Big( \phi^{-2}
e^{-{1\over {\sqrt{3}}}\sum_{\bar a}
\gamma_{\bar as}r_{\bar a}}\sum_{\bar c}\gamma_{\bar
cs} \pi_{\bar c}\Big) (\tau ,{\vec \sigma}_2)\, \Big)\, \Big]\, \Big)
,\nonumber \\
{\hat P}^r_{ADM,R}&=&\sqrt{3} \int d^3\sigma \phi^{-2}(\tau ,\vec \sigma )
e^{-{1\over {\sqrt{3}}}\sum_{\bar a}\gamma_{\bar ar}r_{\bar a}(\tau ,\vec
\sigma )}\nonumber \\
&&\Big( \phi^{-2}(\tau ,\vec \sigma ) \sum_{\bar c} \Big[ \sum_u \gamma_{\bar
cu} e^{-{1\over {\sqrt{3}}}\sum_{\bar a}\gamma_{\bar au}r_{\bar a}}
(2\partial_rln\, \phi +{1\over {\sqrt{3}}}\sum_{\bar b}\gamma_{\bar bu}
\partial_rr_{\bar b})-\nonumber \\
&-&2\gamma_{\bar cr} e^{-{1\over {\sqrt{3}}}\sum_{\bar a}\gamma_{\bar ar}r
_{\bar a}} (2\partial_rln\, \phi +{1\over {\sqrt{3}}}\sum_{\bar b}\gamma_{\bar
br}\partial_rr_{\bar b})\Big] (\tau ,\vec \sigma ) \pi_{\bar c}(\tau ,\vec
\sigma )+\nonumber \\
&+&\sum_{uv} \int d^3\sigma_1 \Big[ (2\partial_rln\, \phi +{1\over {\sqrt{3}}}
\sum_{\bar b}\gamma_{\bar bu}\partial_rr_{\bar b}
 )(\tau ,\vec \sigma )\delta_{u(a)}{\cal T}^u_{(a)v}(\vec \sigma ,{\vec
\sigma}_1,\tau |\phi ,r_{\bar a}]-\nonumber \\
&-&(2\partial_uln\, \phi +{1\over {\sqrt{3}}}\sum_{\bar b}\gamma_{\bar br}
\partial_ur_{\bar b})(\tau ,\vec \sigma )\Big( \delta_{r(a)}{\cal T}^u_{(a)v}+
\delta_{u(a)}{\cal T}^r_{(a)v}\Big) (\vec \sigma ,{\vec \sigma}_1,\tau |\phi
,r_{\bar a}]\Big] \nonumber \\
&&(\phi^{-2} e^{-{1\over {\sqrt{3}}}\sum_{\bar a}\gamma_{\bar av}r_{\bar a}}
\sum_{\bar c}\gamma_{\bar cv}\pi_{\bar c})(\tau ,{\vec
\sigma}_1) \},\nonumber \\
{\hat J}^{rs}_{ADM,R}&=&-{{\sqrt{3}}\over 2}\int d^3\sigma  \phi^{-4}(\tau
,\vec \sigma ) \sum_{\bar c} \pi_{\bar c}(\tau ,\vec \sigma )\nonumber \\
&&\Big[ \sigma^s e^{-{1\over {\sqrt{3}}}\sum_{\bar a}\gamma_{\bar ar}r_{\bar
a}}\Big( \sum_u \gamma_{\bar cu} e^{-{1\over {\sqrt{3}}}\sum_{\bar a}\gamma
_{\bar au}r_{\bar a}} (2\partial_rln\, \phi +{1\over {\sqrt{3}}}\sum_{\bar b}
\gamma_{\bar bu}\partial_rr_{\bar b})-\nonumber \\
&-&2\gamma_{\bar cr} e^{-{1\over {\sqrt{3}}}\sum_{\bar a}\gamma_{\bar ar}r
_{\bar a}} (2\partial_rln\, \phi +{1\over {\sqrt{3}}}\sum_{\bar b}\gamma_{\bar
br}\partial_rr_{\bar b})\Big)-\nonumber \\
&-&\sigma^r e^{-{1\over {\sqrt{3}}}\sum_{\bar a}\gamma_{\bar as}r_{\bar
a}}\Big( \sum_u \gamma_{\bar cu} e^{-{1\over {\sqrt{3}}}\sum_{\bar a}\gamma
_{\bar au}r_{\bar a}} (2\partial_sln\, \phi +{1\over {\sqrt{3}}}\sum_{\bar b}
\gamma_{\bar bu}\partial_sr_{\bar b})-\nonumber \\
&-&2\gamma_{\bar cs} e^{-{1\over {\sqrt{3}}}\sum_{\bar a}\gamma_{\bar as}r
_{\bar a}} (2\partial_sln\, \phi +{1\over {\sqrt{3}}}\sum_{\bar b}\gamma_{\bar
bs}\partial_sr_{\bar b})\Big) \Big] (\tau ,\vec \sigma )+\nonumber \\
&+&{{\sqrt{3}}\over 2}\sum_{uv} \int d^3\sigma d^3\sigma_1 \phi^{-2}(\tau
,\vec \sigma )\nonumber \\
&&\Big[ \sigma^s e^{-{1\over {\sqrt{3}}}\sum_{\bar a}\gamma_{\bar ar}r_{\bar
a}(\tau ,\vec \sigma )} \Big( (2\partial_rln\, \phi +{1\over {\sqrt{3}}}
\sum_{\bar b}\gamma_{\bar bu}\partial_rr_{\bar b})(\tau ,\vec \sigma )\delta
_{u(a)}{\cal T}^u_{(a)v} (\vec \sigma ,{\vec \sigma}_1,\tau |\phi
,r_{\bar a}]-\nonumber \\
&-&(2\partial_uln\, \phi +{1\over {\sqrt{3}}}\sum_{\bar b}\gamma_{\bar br}
\partial_ur_{\bar b})(\tau ,\vec \sigma )\Big( \delta_{r(a)}{\cal T}^u_{(a)v}+
\delta_{u(a)}{\cal T}^r_{(a)v}\Big) (\vec \sigma ,{\vec
\sigma}_1,\tau |\phi ,r_{\bar a}] \Big) -\nonumber \\
&-&\sigma^r e^{-{1\over {\sqrt{3}}}\sum_{\bar a}\gamma_{\bar as}r_{\bar
a}(\tau ,\vec \sigma )} \Big( (2\partial_sln\, \phi +{1\over {\sqrt{3}}}
\sum_{\bar b}\gamma_{\bar bu}\partial_sr_{\bar b})(\tau ,\vec \sigma )\delta
_{u(a)}{\cal T}^u_{(a)v} (\vec \sigma ,{\vec \sigma}_1,\tau |\phi
,r_{\bar a}]-\nonumber \\
&-&(2\partial_uln\, \phi +{1\over {\sqrt{3}}}\sum_{\bar b}\gamma_{\bar bs}
\partial_ur_{\bar b})(\tau ,\vec \sigma )\Big( \delta_{s(a)}{\cal T}^u_{(a)v}+
\delta_{u(a)}{\cal T}^s_{(a)v}\Big) (\vec \sigma ,{\vec
\sigma}_1,\tau |\phi ,r_{\bar a}] \Big) \Big] \nonumber \\
&&(\phi^{-2}e^{-{1\over {\sqrt{3}}}\sum_{\bar a}\gamma_{\bar av}r_{\bar a}}
\sum_{\bar c}\gamma_{\bar cv}\pi_{\bar c})(\tau ,{\vec
\sigma}_1) \} ,\nonumber \\
{\hat J}^{\tau r}_{ADM,R}&=&\epsilon \int d^3\sigma   \sigma^r
\Big( {{c^3}\over {16\pi G}}
\Big[ \phi^2 \sum_r e^{-{1\over {\sqrt{3}}}\sum_{\bar a}\gamma_{\bar ar}r_{\bar
a}} \times \nonumber \\
&&\Big( {1\over 3}\sum_{\bar b\bar c}(2\gamma_{\bar br}\gamma_{\bar cr}+\delta
_{\bar b\bar c})\partial_rr_{\bar b}\partial_rr_{\bar c}-{4\over {\sqrt{3}}}
(\sum_{\bar b}\gamma_{\bar br}\partial_rr_{\bar b}) \partial_rln\, \phi -
4(\partial_rln\, \phi )^2-\nonumber \\
&-&\sum_ue^{{2\over {\sqrt{3}}}\sum_{\bar a}(\gamma_{\bar ar}-\gamma_{\bar au})
r_{\bar a}}[{2\over 3}\sum_{\bar b\bar c}\gamma_{\bar br}\gamma_{\bar cr}
\partial_ur_{\bar b}\partial_ur_{\bar c}+\nonumber \\
&+&2\sqrt{3}(\sum_{\bar b}\gamma_{\bar
br}\partial_ur_{\bar b})\partial_uln\, \phi +4(\partial_uln\, \phi )^2]\Big)
\Big] (\tau ,\vec \sigma )-\nonumber \\
&-&{{6\pi G}\over {c^3}} \phi^{-2}(\tau ,\vec \sigma )\Big[ (2\phi^{-4}\sum
_{\bar a} \pi^2_{\bar a})(\tau ,\vec \sigma )+\nonumber \\
&+&4(\phi^{-2}\sum_ue^{{1\over {\sqrt{3}}}\sum_{\bar a}\gamma_{\bar au}r_{\bar
a}} \sum_{\bar b}\gamma_{\bar bu}\pi_{\bar b})(\tau ,\vec \sigma )\times
\nonumber \\
&&\int d^3\sigma_1 \sum_r \delta^u_{(a)} {\cal T}^u_{(a)r}(\vec \sigma ,{\vec
\sigma}_1,\tau |\phi ,r_{\bar a}] \Big( \phi^{-2}
e^{-{1\over {\sqrt{3}}}\sum_{\bar a}
\gamma_{\bar ar}r_{\bar a}}\sum_{\bar b}\gamma_{\bar
br} \pi_{\bar b}\Big) (\tau ,{\vec \sigma}_1)+\nonumber \\
&+&
\int d^3\sigma_1d^3\sigma_2 \Big( \sum_u e^{{2\over {\sqrt{3}}}\sum_{\bar a}
\gamma_{\bar au}r_{\bar a}(\tau ,\vec \sigma )} \times \nonumber \\
&&\sum_r{\cal T}^u_{(a)r}(\vec \sigma ,{\vec
\sigma}_1,\tau |\phi ,r_{\bar a}] \Big( \phi^{-2}
e^{-{1\over {\sqrt{3}}}\sum_{\bar a}
\gamma_{\bar ar}r_{\bar a}}\sum_{\bar b}\gamma_{\bar
br} \pi_{\bar b}\Big) (\tau ,{\vec \sigma}_1)\times \nonumber \\
&&\sum_s {\cal T}^u_{(a)s}(\vec \sigma ,{\vec
\sigma}_2,\tau |\phi ,r_{\bar a}] \Big( \phi^{-2}
e^{-{1\over {\sqrt{3}}}\sum_{\bar a}
\gamma_{\bar as}r_{\bar a}}\sum_{\bar c}\gamma_{\bar
cs} \pi_{\bar c}\Big) (\tau ,{\vec \sigma}_2)+\nonumber \\
&+&\sum_{uv} e^{{1\over {\sqrt{3}}}\sum_{\bar a}(\gamma_{\bar au}+\gamma_{\bar
av})r_{\bar a}(\tau ,\vec \sigma )} (\delta^u_{(b)}\delta^v_{(a)}-\delta^u_{(a)}
\delta^v_{(b)})\times \nonumber \\
&&\sum_r {\cal T}^u_{(a)r}(\vec \sigma ,{\vec
\sigma}_1,\tau |\phi ,r_{\bar a}] \Big( \phi^{-2}
e^{-{1\over {\sqrt{3}}}\sum_{\bar a}
\gamma_{\bar ar}r_{\bar a}}\sum_{\bar b}\gamma_{\bar
br} \pi_{\bar b}\Big) (\tau ,{\vec \sigma}_1)\nonumber \\
&&\sum_s {\cal T}^v_{(b)s}(\vec \sigma ,{\vec
\sigma}_2,\tau |\phi ,r_{\bar a}] \Big( \phi^{-2}
e^{-{1\over {\sqrt{3}}}\sum_{\bar a}
\gamma_{\bar as}r_{\bar a}}\sum_{\bar c}\gamma_{\bar
cs} \pi_{\bar c}\Big) (\tau ,{\vec \sigma}_2)\, \Big)\, \Big]\, \Big) +
\nonumber \\
&+&{{\epsilon c^3}\over {8\pi G}} \int d^3\sigma \phi^{-2}(\tau ,\vec \sigma )
e^{-{2\over {\sqrt{3}}}\sum_{\bar a}\gamma_{\bar ar}r_{\bar a}(\tau ,\vec
\sigma )}\nonumber \\
&&\Big[ \sum_v e^{-{2\over {\sqrt{3}}}\sum_{\bar a}\gamma_{\bar av}r_{\bar a}}
\Big( \phi^4 e^{{2\over {\sqrt{3}}}\sum_{\bar a}\gamma_{\bar av}r_{\bar a}}-1
\Big) (\partial_rln\, \phi +{1\over {\sqrt{3}}}\sum_{\bar b}(\gamma_{\bar br}+
\gamma_{\bar bv})\partial_rr_{\bar b})-\nonumber \\
&-&e^{-{2\over {\sqrt{3}}}\sum_{\bar a}\gamma_{\bar ar}r_{\bar a}}
\Big( \phi^4 e^{{2\over {\sqrt{3}}}\sum_{\bar a}\gamma_{\bar ar}r_{\bar a}}-1
\Big) (\partial_rln\, \phi +{2\over {\sqrt{3}}}\sum_{\bar b}\gamma_{\bar br}
\partial_rr_{\bar b}) \Big] (\tau ,\vec \sigma ).
\label{V12}
\end{eqnarray}

From Eq.(\ref{V11}) it is clear that $\phi [r_{\bar a},\pi_{\bar a}]$ depends on
$G^2/c^6$, while the previous equation implies that ${\hat H}^{(WSW) '}_{(D)ADM}
=\epsilon {\hat P}^{\tau}_{ADM,R}$ depends a priori on both $G/c^3$ and
$c^3/G$.

Let us remark that to try to make realistic calculations one needs normal
coordinates around a point (see Appendix A of II):
only in such a chart one can find the explicit expression of the Synge-DeWitt
bitensor $d^r_{\gamma_{PP_1}}(\vec \sigma ,{\vec \sigma}_1)$ (giving the
tangent in $P_1$ [${\vec \sigma}_1$] of the geodesic emanating from P [$\vec
\sigma$]).

A fundamental open problem, which will be studied in a future paper,
is to find a solution of a linearization of the
Lichnerowicz equation, which put inside a linearization of the weak ADM energy
will imply a linear equation for the physical canonical variables
$r_{\bar a}(\tau ,\vec \sigma )$ describing the gravitational field, so to make
contact with the theory of gravitational waves.

A connected open problem is to find the relation of our canonical variables
$r_{\bar a}$, $\pi_{\bar a}$ for the gravitational field in the special
3-orthogonal gauge with $\rho =0$ with the statement of Christodoulou-
Klainermann \cite{ckl} that the independent degrees of freedom of the
gravitational field are described by symmetric trace-free 2-tensors on 2-planes.
See the Conclusions and Appendix F for some comments on this point.

Finally in Appendix G there are some comments on the post-Newtonian
approximation.

\vfill\eject

\section{Void Spacetimes in the 3-orthogonal Gauge.}

As said in II it is interesting  to study the tiny subspace of the
reduced phase space in the 3-orthogonal gauges defined by putting by hand
equal to zero the Dirac's observables of the gravitational field
[$r_{\bar a}(\tau ,\vec \sigma )\approx 0$,
$\pi_{\bar a}(\tau ,\vec \sigma )\approx 0$] before
adding the gauge-fixing $\rho (\tau ,\vec \sigma )\approx 0$, since in
absence of matter this is consistent with Einstein equations. These
spacetimes may be named ``void spacetimes" and it will be shown in this
Section that they are gauge equivalent to Minkowski spacetime in Cartesian
coordinates by means of the Hamiltonian group of gauge transformations and that
they have vanishing Poincar\'e charges. They should
correspond to the relativistic generalization of the class of Galilean
non inertial frames (with their inertial forces) obtainable from an inertial
frame of the nonrelativistic Galileo spacetime [for example the (maybe
time-dependent) pseudo-diffeomorphisms in $Diff\, \Sigma_{\tau}$ replace the
Galilean coordinate transformations generating the inertial forces].

The concept of void spacetime implements the viewpoint of Synge\cite{synge}
that, due to tidal (i.e. curvature) effects, there is a difference between
true gravitational fields and accelerated motions, even if, as shown in
Ref.\cite{norton}, Einstein arrived at general relativity through the
intermediate step of showing the equivalence of uniform acceleration with
special homogeneous gravitational fields. It is only in the not generally
covariant Hamiltonian approach that one is able to identify the genuine
physical degrees of freedom of the gravitational field.

Since in void spacetimes without matter there are no physical degrees of freedom
of the gravitational field but only gauge degrees of freedom, we expect that
this equivalence class of spacetimes is not described by scenario b) of Section
III, but that it corresponds to scenario a) with vanishing Poincar\'e charges
(the exceptional Poincar\'e orbit).
Indeed, in this way Minkowski spacetime (and its gauge copies) would be
selected as the static background for special relativity with zero energy (in
contrast with the viewpoint of Ref.\cite{hh} of an infinite background energy),
starting point for parametrized Minkowski theories where the special
relativistic energy would be generated only by the added matter (and/or fields).
Tetrad gravity with matter would be described by scenario b) (with the
WSW hypersurfaces corresponding to Wigner hyperplanes; both of them are defined
only in presence of matter) and in the limit $G \rightarrow 0$ the weak ADM
energy would tend to the special relativistic energy of that matter system with
no trace left of the ``gravitational field energy" [present in scenario b) but
not in scenario a)].

To define ``void spacetimes" independently from the 3-orthogonal gauge, let us
remark that, since the conditions $r_{\bar a}(\tau ,\vec \sigma )=0$ imply the
vanishing of the 3-conformal Cotton-York tensor [see I after Eq.(9) for the
definition of this tensor and Eq.(D2) of Appendix D of II for its vanishing],
this means that void spacetimes should have the leaves $\Sigma_{\tau}$
conformally flat as Riemannian 3-manifolds but with the conformal factor
determined by the reduced Lichnerowicz equation (since the solution depends on
the gauge variable $\rho$, the conformal factor is gauge dependent). Therefore,
the general theory of void spacetimes
could be reformulated in arbitrary gauges by adding with Lagrange multipliers
the two independent components of the Cotton-York tensor ${}^3{\cal Y}
_{rs}(\tau ,\vec \sigma )$ [which is a function only of cotriads]
to the tetrad ADM Lagrangian of Eq.(50) of I for
tetrad gravity. In this way one should get two extra holonomic constraints
equivalent to $r_{\bar a}(\tau ,\vec \sigma )\approx 0$. Their time
constancy should produce two secondary (momentum dependent) constraints
equivalent to $\pi_{\bar a}(\tau ,\vec \sigma )\approx 0$.

Deferring to a future paper the study of the general case, let us explore
the properties of void spacetimes in the 3-orthogonal gauges.

To get void spacetimes starting from the 3-orthogonal gauge definition of the
gravitational field degrees of freedom, we add with Dirac's multipliers to the
tetrad gravity version of the Dirac Hamiltonian of Eq(\ref{III15})
the two pairs of
primary second class constraints $r_{\bar a}(\tau ,\vec \sigma )\approx 0$,
$\pi_{\bar a}(\tau ,\vec \sigma )\approx 0$:\hfill\break
\hfill\break
${\hat H}_{(D)ADM}^{(1)} \mapsto {\hat H}^{(1)}_{(D)ADM}+
\int d^3\sigma \sum_{\bar a} [\xi_{\bar a}r_{\bar a}+
\xi^{'}_{\bar a} \pi_{\bar a}](\tau \vec \sigma )$. \hfill\break
\hfill\break
The time constancy of these
constraints determines the Dirac multipliers $\xi_{\bar a}(\tau ,\vec \sigma )$,
$\xi^{'}_{\bar a}(\tau ,\vec \sigma )$. By going to new Dirac brackets
[$r_{\bar a}(\tau ,\vec \sigma )\equiv \pi_{\bar a}(\tau ,\vec \sigma ) \equiv
0$] we get the 3-metric ${}^3{\hat g}_{rs}=e^{2q}\delta_{rs}=\phi^4\delta
_{rs}$ and we remain only with the following variables:\hfill\break
i) $\phi(\tau ,\vec \sigma ) =e^{q(\tau ,\vec \sigma )/2}$, to be determined
by the reduced Lichnerowicz equation;\hfill\break
ii) $\pi_{\phi}(\tau ,\vec \sigma )=2\phi^{-1}(\tau ,\vec \sigma ) \rho(\tau
,\vec \sigma )$, the conjugate gauge variable.

In this way we have identified some members [they differ in the arbitrary value
of $\pi_{\phi}=2\phi^{-1}\rho$] of a Hamiltonian equivalence class,
which corresponds to the absence of the gravitational field. For it
we have

\begin{eqnarray}
&&r_{\bar a}=\pi_{\bar a}=0,\quad\quad \phi =e^{q/2}\quad
[or\, q=2 ln\, \phi],\nonumber \\
&&{}^3{\hat e}_{(a)r}=\phi^2 \delta_{(a)r},\quad\quad {}^3{\hat e}^r_{(a)}=
\phi^{-2} \delta^r_{(a)},\nonumber \\
&&{}^3{\hat g}_{rs}=\phi^4\delta_{rs},\quad\quad {}^3{\tilde g}_{rs}[r_{\bar
a}=0]=\delta_{rs},\nonumber \\
&&\tilde \triangle [r_{\bar a}=0]={\triangle}_{FLAT},\quad\quad
{}^3\tilde R[r_{\bar a}=0]=0.
\label{VI0}
\end{eqnarray}

From Eqs.(\ref{V3}), (\ref{V4}), one has in void spacetimes [before putting
$\rho ={1\over 2}\phi \pi_{\phi}=0$]

\begin{eqnarray}
{}^3{\hat g}_{rs}(\tau ,\vec \sigma )&=&=\phi^4(\tau ,\vec \sigma )\delta_{rs}
,\nonumber \\
{}^3{\hat {\tilde \pi}}^r_{(a)}(\tau ,\vec \sigma )&=&
{1\over 3}\int d^3\sigma_1 {\cal K}
^r_{(a)s}(\vec \sigma ,{\vec \sigma}_1;\tau |\phi ,0]
\rho (\tau ,{\vec \sigma}_1),\nonumber \\
&&{}\nonumber \\
{\cal K}^r_{(a)s}(\vec \sigma ,{\vec \sigma}_1,\tau |\phi ,0]&=&
\delta^r_{(a)}\delta^r_s\delta^3(\vec \sigma ,{\vec \sigma}_1)+{\cal T}^r
_{(a)s}(\vec \sigma ,{\vec \sigma}_1,\tau |\phi ,0],\nonumber \\
&&{}\nonumber \\
{\cal T}^r_{(a)s}(\vec \sigma ,{\vec \sigma}_1;\tau |\phi ,0]
&=&{1\over 2}[2 \sum_{w\not= s} \delta_{(k)w}
 {{\partial ln\, \phi (\tau ,{\vec \sigma}_1)}\over {\partial \sigma_1^w}}
\phi^{-2}(\tau ,\vec \sigma )
\delta^r_{(b)} T_{(b)(a)(k)}(\vec \sigma ,{\vec \sigma}_1;\tau )+\nonumber \\
&&+\delta_{(k)s} {{\partial}\over {\partial \sigma_1^s}} \phi^{-2}(\tau, \vec
\sigma ) \delta^r_{(b)} T_{(b)(a)(k)}(\vec \sigma ,{\vec \sigma}_1;\tau ) ],
\nonumber \\
&&{}\nonumber \\
\phi^{-2}(\tau ,\vec \sigma )&& \delta^r_{(b)} T_{(b)(a)(k)}(\vec \sigma ,{\vec
\sigma}_1;\tau ) =\nonumber \\
&=&d^r_{\gamma_{PP_1}} (P_{\gamma_{PP_1}}\, e^{\int
^{\vec \sigma}_{{\vec \sigma}_1}d\sigma_2^w\, {}^3{\hat \omega}_{w(c)}(\tau ,
{\vec \sigma}_2){\hat R}^{(c)} }\, )_{(a)(k)}+\nonumber \\
&+&\sum_u\delta_{(a)u} d^u_{\gamma_{PP_1}}(\vec \sigma ,{\vec \sigma}_1)
\nonumber \\
&&\delta^r_{(b)}(P_{\gamma_{PP_1}}\, e^{\int
^{\vec \sigma}_{{\vec \sigma}_1}d\sigma_2^w\, {}^3{\hat \omega}_{w(c)}(\tau ,
{\vec \sigma}_2){\hat R}^{(c)} }\, )_{(b)(k)},\nonumber \\
&&{}\nonumber \\
{}^3{\hat \omega}_{t(d)}(\tau ,\vec \sigma )
&=&2\epsilon_{(d)(m)(n)} \delta_{(m)t}\delta_{(n)u}
\partial_uln\, \phi (\tau ,\vec \sigma ).
\label{VI2}
\end{eqnarray}

\noindent and

\begin{eqnarray}
{}^3{\hat K}_{rs}(\tau ,\vec \sigma )&=&{{\epsilon}\over {4k}} [
\sum_u(\delta_{ru}\delta_{(a)s}
+\delta_{su}\delta_{(a)r}-\delta_{rs}\delta_{(a)u})\,
{}^3{\hat {\tilde \pi}}^u_{(a)}](\tau ,\vec \sigma ),\nonumber \\
{}^3{\hat K}(\tau ,\vec \sigma )
&=&-{{\epsilon}\over {4k}} [\phi^{-4} \sum_u \delta_{(a)u}
\, {}^3{\hat {\tilde \pi}}
^u_{(a)}](\tau ,\vec \sigma )=\nonumber \\
&=&-{{\epsilon}\over {4k}}\phi^{-6}(\tau ,\vec \sigma ) \{
\rho(\tau ,\vec \sigma )+{1\over 3}\sum_u \int d^3\sigma_1 \delta_{(a)u}
\nonumber \\
&&{\cal T}^u_{(a)s}(\vec \sigma ,{\vec \sigma}_1;\tau |\phi ,0]
\phi^{-2}(\tau,
{\vec \sigma}_1)\rho (\tau ,{\vec \sigma}_1) \} ,\nonumber \\
&&{}\nonumber \\
{\hat {\cal H}}^{'}_R(\tau ,\vec \sigma )&=&
{\hat {\cal H}}_R(\tau ,\vec \sigma ){|}_{r_{\bar a}=\pi_{\bar a}=0}=
\nonumber \\
&=&\epsilon \phi (\tau ,\vec \sigma ) \{ -{{c^3}\over {2\pi G}} \triangle_{FLAT}
\phi (\tau ,\vec \sigma )+{{2\pi G}\over {3c^3}}\Big[ {1\over 3}(\phi^{-7}
\rho )(\tau ,\vec \sigma )+\nonumber \\
&+&{2\over 3} (\phi^{-5} \rho )(\tau ,\vec \sigma ) \int d^3\sigma_1 \sum_r
\delta^u_{(a)}{\cal T}^u_{(a)r}(\vec \sigma ,{\vec \sigma}_1;\tau |\phi ,0]
(\phi^{-2} \rho (\tau ,{\vec \sigma}_1)-\nonumber \\
&-&{1\over 3} \phi^{-3}(\tau ,\vec \sigma ) \int d^3\sigma_1 d^3\sigma_2
\Big( \sum_u \nonumber \\
&& \sum_r{\cal T}^u_{(a)r}(\vec \sigma ,{\vec \sigma}_1;\tau |\phi
,0] (\phi^{-2} \rho )(\tau ,{\vec \sigma}_1)\nonumber \\
&&\sum_s {\cal T}^u_{(a)s}(\vec \sigma ,{\vec \sigma}_2;\tau |\phi ,0]
(\phi^{-2} \rho )(\tau ,{\vec \sigma}_2)+\nonumber \\
&+&\sum_{uv}(\delta^u_{(b)}\delta^v_{(a)}-\delta^u_{(a)}\delta^v_{(b)})
\nonumber \\
&&\sum_r{\cal T}^u_{(a)r}(\vec \sigma ,{\vec \sigma}_1;\tau |\phi
,0] (\phi^{-2} \rho )(\tau ,{\vec \sigma}_1)\nonumber \\
&&\sum_s {\cal T}^v_{(a)s}(\vec \sigma ,{\vec \sigma}_2;\tau |\phi ,0]
(\phi^{-2} \rho )(\tau ,{\vec \sigma}_2) \Big) \Big] \} \approx 0.
\label{VI3}
\end{eqnarray}

Therefore, the leaves $\Sigma_{\tau}$ of the 3+1 splittings of void
spacetimes are conformally flat 3-manifolds diffeomorphic to $R^3$, but with
the conformal factor $\phi$ determined by the reduced Lichnerowicz equation.

With the natural gauge $\rho (\tau ,\vec \sigma )\approx 0$, one has

\begin{eqnarray}
{}^3{\hat {\tilde \pi}}^r_{(a)}(\tau ,\vec \sigma )&\approx 0,\nonumber \\
{}^3{\hat K}_{rs}(\tau ,\vec \sigma )&\approx& 0,\quad\quad
\Rightarrow  {}^3\hat K(\tau ,\vec \sigma )\approx 0,
\label{VI4}
\end{eqnarray}

\noindent and
the reduced superhamiltonian constraint becomes the reduced Lichnerowicz
equation [$\triangle_{FLAT}={\vec \partial}^2$]

\begin{equation}
\triangle_{FLAT} \phi (\tau ,\vec \sigma )\approx 0,\quad \Rightarrow
\phi (\tau ,\vec \sigma ) = 1, \Rightarrow {}^3{\hat g}_{rs}=\delta_{rs},
\label{VI5}
\end{equation}

\noindent where we have shown the solution corresponding to the boundary
condition of Eq.(\ref{V3}).

Let us now delineate the sequence of natural gauge-fixings to get void
spacetimes in the 3-orthogonal gauges
from scenario a) [in which ${\tilde \lambda}_A(\tau )$, ${\tilde
\lambda}_{AB}(\tau )$ are gauge variables]:\hfill\break
\hfill\break
i) ${\tilde \lambda}_{\tau}(\tau )\approx \epsilon$, ${\tilde \lambda}_r(\tau )
\approx 0$, ${\tilde \lambda}_{AB}(\tau )\approx 0$ [so that $N_{(as)}(\tau
,\vec \sigma )\approx 0$, $N_{(as)r}(\tau ,\vec \sigma )\approx 0$]: these are
the gauge-fixings for the primary constraints ${\tilde \pi}^A(\tau )\approx 0$,
${\tilde \pi}^{AB}(\tau )\approx 0$ and imply $\zeta_A(\tau )\approx 0$,
$\zeta_{AB}(\tau )\approx 0$.\hfill\break
ii) $\alpha_{(a)}(\tau ,\vec \sigma )=\varphi_{(a)}(\tau ,\vec \sigma )=0$ [so
that $\lambda^{\vec \varphi}_{(a)}(\tau ,\vec \sigma )={\hat \mu}_{(a)}(\tau
,\vec \sigma )=0$ in the Dirac Hamiltonian];\hfill\break
iii) $\xi^r(\tau ,\vec \sigma )=\sigma^r$. Since, as we have seen,
 in void spacetimes
${}^3{\tilde \pi}^r_{(a)}$ and ${}^3{\tilde \Pi}^{rs}$ vanish for $\rho (\tau
,\vec \sigma )\approx 0$, also the Poisson bracket $\{ \xi^r(\tau ,\vec \sigma )
,{\hat P}^{\tau}_{ADM} \}$ vanishes for $\rho (\tau ,\vec \sigma )\approx 0$.
Therefore the requirement $\partial_{\tau}[\xi^r(\tau ,\vec \sigma )-\sigma^r]
\approx 0$ now implies $n_r(\tau ,\vec \sigma )-{\hat n}_r(\tau ,\vec \sigma |
q,\rho ]\approx 0$ but with ${\hat n}_r(\tau ,\vec \sigma |q,\rho ]{|}_{\rho
=0}\, =0$; the time constancy of these constraints determine $\lambda^{\vec n}_r
(\tau ,\vec \sigma )$. For $\rho (\tau ,\vec \sigma )\approx 0$ we get a
vanishing shift function $N_r(\tau ,\vec \sigma )\approx 0$
(synchronous coordinates).\hfill\break
iv) At this stage the lapse function is \hfill\break
\hfill\break
$N(\tau ,\vec \sigma )\approx -\epsilon +n(\tau ,\vec \sigma )$.\hfill\break
\hfill\break
v) Now we add the second class constraints $r_{\bar a}(\tau ,\vec \sigma )
\approx 0$, $\pi_{\bar a}(\tau ,\vec \sigma )\approx 0$ which imply the
previous results.\hfill\break
vi) The Dirac Hamiltonian becomes\hfill\break
\hfill\break
$H^{(1)}_{(D)ADM}=\int d^3\sigma [n{\hat {\cal H}}^{'}_R+\lambda_n{\tilde
\pi}^n](\tau ,\vec \sigma )+\epsilon {\hat P}^{\tau \, {'}}_{ADM}$\hfill\break
\hfill\break
with ${\hat {\cal H}}^{'}_R={\hat {\cal H}}_R{|}_{r_{\bar a}=\pi_{\bar a}=0}$,
${\hat P}^{\tau \, {'}}_{ADM}={\hat P}^{\tau}_{ADM}{|}_{r_{\bar a}=\pi_{\bar
a}=0}$.\hfill\break
vii) The natural gauge fixing $\rho (\tau ,\vec \sigma )\approx 0$ implies
\hfill\break
\hfill\break
$\partial_{\tau}\rho (\tau ,\vec \sigma )\, {\buildrel \circ \over =}\, \int
d^3\sigma_1 n(\tau ,{\vec \sigma}_1) \{ \rho (\tau ,\vec \sigma ),{\hat {\cal
H}}^{'}_R(\tau ,{\vec \sigma}_1) \} +\epsilon \{ \rho (\tau ,\vec \sigma ),
{\hat P}^{\tau \, {'}}_{ADN} \}$; \hfill\break
\hfill\break
but from Eq.(\ref{III13}) and from ${}^3{\tilde \Pi}^{rs}\approx 0$ we see that
only the term bilinear in the Christoffel symbols contributes to $\{ \rho (\tau
,\vec \sigma ), {\hat P}^{\tau \, {'}}_{ADN} \}$ for $\rho (\tau ,\vec \sigma )
\approx 0$. Now from Eq.(\ref{V4}) we get ${}^3{\hat \Gamma}^r_{uv}=2\phi^{-1}
[\delta_{uv}\partial_r\phi +\delta_{ru}\partial_v\phi +\delta_{rv}\partial_u
\phi ] \rightarrow_{\phi \rightarrow const.}\, 0$. Since $\phi (\tau ,\vec
\sigma )=1$ is the solution of the reduced Lichnerowicz equation for $\rho (\tau
,\vec \sigma )\approx 0$, we get $\{ \rho (\tau ,\vec \sigma ), {\hat P}^{\tau
\, {'}}_{ADN} \} \approx 0$ and then $n(\tau ,\vec \sigma )\approx 0$ and
$\lambda_n(\tau ,\vec \sigma )\approx 0$. Therefore, at the end the lapse
function is $N(\tau ,\vec \sigma ) \approx -\epsilon$.\hfill\break
viii) Since, as we shall see, ${\hat P}^{\tau \, {'}}_{ADN}$ vanishes for
$\rho (\tau ,\vec \sigma )\approx 0$, $\phi (\tau ,\vec \sigma )=1$, the
final Dirac Hamiltonian vanishes: $H^{(1)}_{(D)ADM}\approx 0$, and
the final 4-metric becomes \hfill\break
\hfill\break
${}^4{\hat g}_{AB}(\tau ,\vec \sigma )=\epsilon \,
\left( \begin{array}{cc} 1 & 0\\ 0 & -\delta_{rs} \end{array} \right)$.
\hfill\break
\hfill\break
ix) In void spacetimes the two gauge-fixings $\rho (\tau
,\vec \sigma )\approx 0$ and ${}^3\hat K(\tau ,\vec \sigma )\approx 0$
are equivalent and one chooses $\phi (\tau ,\vec \sigma )=1$ [i.e. $q(\tau
,\vec \sigma )=0$]; in this gauge one has ${}^3\hat R=0$
for the 3-hypersurfaces $\Sigma_{\tau}$
[they have both the scalar curvature and the trace of the extrinsic one
vanishing], but in other gauges the 3-curvature and the trace of the extrinsic
one may be not vanishing because the solution $\phi (\tau ,\vec \sigma )$ of the
reduced Lichnerowicz equation may become nontrivial.
From Eqs.(\ref{V5}), the weak and strong Poincar\'e charges are

\begin{eqnarray}
{\hat P}^{\tau}_{ADM,R}&=&-\epsilon \int d^3\sigma \Big( {{ \phi^{-6}(\tau
,\vec \sigma )}\over {24 k}} [-(\phi^{-4} \rho^2)(\tau ,\vec \sigma )-
\nonumber \\
&-&2(\phi^{-2} \rho )(\tau ,\vec \sigma )\delta_{r(a)}\int d^3\sigma_1 \sum_n
{\cal T}^r_{(a)n}(\vec \sigma ,{\vec \sigma}_1,\tau |\phi ,0] (\phi^{-2} \rho )
(\tau ,{\vec \sigma}_1)+\nonumber \\
&+&{1\over 3}\sum_{rs}(\delta_{rs}\delta_{(a)(b)}+\delta_{r(b)}\delta_{s(a)}-
\delta_{r(a)}\delta_{s(b)})\nonumber \\
&&\int d^3\sigma_1 \sum_m {\cal T}^r_{(a)m}(\vec \sigma ,{\vec \sigma}_1,\tau
|\phi ,0] (\phi^{-2} \rho )(\tau ,{\vec \sigma}_1)\nonumber \\
&&\int d^3\sigma_2 \sum_n {\cal T}^s_{(b)n}(\vec \sigma ,{\vec \sigma}_2,\tau
|\phi ,0] (\phi^{-2} \rho )(\tau ,{\vec \sigma}_2) ]-\nonumber \\
&-&8k [\phi^2 \sum_r(\partial_r ln\, \phi )^2](\tau ,\vec \sigma ) \Big) \,
{\rightarrow}_{\rho \, \rightarrow 0}\, 8\epsilon k \int d^3\sigma
[\phi^2\sum_r(\partial_rln\, \phi )^2](\tau ,\vec \sigma )\nonumber \\
&&{\rightarrow}_{\phi \rightarrow const.}\, 0,\nonumber \\
{\hat P}^r_{ADM,R}&=&{2\over 3} \int d^3\sigma \phi^{-2}(\tau ,\vec \sigma )
\Big( \Big( \rho \phi^{-2}\partial_rln\, \phi \Big) (\tau ,\vec \sigma )+
\nonumber \\
&+&\sum_{uv}\int d^3\sigma_1 \Big[ \partial_rln\, \phi (\tau ,\vec \sigma )
\delta_{u(a)}
{\cal T}^u_{(a)v}(\vec \sigma ,{\vec \sigma}_1,\tau |\phi ,0]-\nonumber \\
&-&\partial_uln\, \phi (\tau ,\vec \sigma )(\delta_{r(a)}{\cal T}^u_{(a)v}+
\delta_{u(a)}{\cal T}^r_{(a)v})(\vec \sigma ,{\vec \sigma}_1,\tau |\phi ,0]
\Big] (\phi^{-2}\rho )(\tau ,{\vec \sigma}_1) \Big) \nonumber \\
&&{\rightarrow}_{\rho \rightarrow 0}\, 0,\nonumber \\
{\hat J}^{rs}_{ADM,R}&=&{1\over 3}\int d^3\sigma [\phi^{-4} \rho ](\tau ,\vec
\sigma ) \Big[ \sigma_s\partial_rln\, \phi -\sigma_r\partial_sln\, \phi \Big]
(\tau ,\vec \sigma )+\nonumber \\
&+&{1\over 3}\sum_{uv} \int d^3\sigma d^3\sigma_1 \phi^{-2}(\tau ,\vec \sigma )
\nonumber \\
&&\Big[ (\sigma_s\partial_rln\, \phi -\sigma_r\partial_sln\, \phi )
(\tau ,\vec \sigma )\delta_{u(a)}{\cal T}^u_{(a)v}(\vec \sigma ,{\vec \sigma}
_1,\tau |\phi ,0]-\nonumber \\
&-&\partial_uln\, \phi (\tau ,\vec \sigma )[\sigma^s (\delta_{r(a)}{\cal T}^u
_{(a)v}+\delta_{u(a)}{\cal T}^r_{(a)v})-\nonumber \\
&-&\sigma^r (\delta_{s(a)}{\cal T}^u_{(a)v}+\delta_{u(a)}{\cal T}^s_{(a)v})
(\vec \sigma ,{\vec \sigma}_1,\tau |\phi ,0] \Big]
(\phi^{-2}\rho )(\tau ,{\vec \sigma}_1)\nonumber \\
&&{\rightarrow}_{\rho \rightarrow 0}\, 0,\nonumber \\
{\hat J}^{\tau r}_{ADM,R}&=&\epsilon \int d^3\sigma  \sigma^r
\Big( {{ \phi^{-6}(\tau
,\vec \sigma )}\over {24 k}} [-(\phi^{-4} \rho^2)(\tau ,\vec \sigma )-
\nonumber \\
&-&2(\phi^{-2} \rho )(\tau ,\vec \sigma )\delta_{r(a)}\int d^3\sigma_1 \sum_n
{\cal T}^r_{(a)n}(\vec \sigma ,{\vec \sigma}_1,\tau |\phi ,0] (\phi^{-2} \rho )
(\tau ,{\vec \sigma}_1)+\nonumber \\
&+&{1\over 3}\sum_{rs}(\delta_{rs}\delta_{(a)(b)}+\delta_{r(b)}\delta_{s(a)}-
\delta_{r(a)}\delta_{s(b)})\nonumber \\
&&\int d^3\sigma_1 \sum_m {\cal T}^r_{(a)m}(\vec \sigma ,{\vec \sigma}_1,\tau
|\phi ,0] (\phi^{-2} \rho )(\tau ,{\vec \sigma}_1)\nonumber \\
&&\int d^3\sigma_2 \sum_n {\cal T}^s_{(b)n}(\vec \sigma ,{\vec \sigma}_2,\tau
|\phi ,0] (\phi^{-2} \rho )(\tau ,{\vec \sigma}_2) ]-\nonumber \\
&-&8k [\phi^2 \sum_r(\partial_r ln\, \phi )^2](\tau ,\vec \sigma ) \Big) +
\nonumber \\
&+&4\epsilon k \int d^3\sigma \Big[ \phi^{-2} (\phi^4-1) \partial_rln\, \phi
\Big] (\tau ,\vec \sigma )\nonumber \\
&&{\rightarrow}_{\rho \rightarrow 0}\, \epsilon \int d^3\sigma \{ \sigma^r
[\phi^2\sum_u(\partial_uln \, \phi )^2](\tau ,\vec \sigma )-\nonumber \\
&-&2k\sum_u\delta^r_u\Big( \phi^{-2}\sum_s(\phi^4-1)(\delta_{us}-1)\partial_uln
\, \phi \Big) (\tau ,\vec \sigma ) \} \nonumber \\
&&{\rightarrow}_{\phi \rightarrow 1}\, 0,\nonumber \\
&&{}\nonumber \\
P^{\tau}_{ADM,R}&=&{\hat P}^{\tau}_{ADM,R}+\int d^3\sigma {\hat {\cal H}}
_R(\tau ,\vec \sigma )=
-8 \epsilon k \sum_u \int_{S^2_{\tau ,\infty}} d^2\Sigma_u \, \{
\phi \partial_u\phi \} (\tau ,\vec \sigma )\nonumber \\
&&{\rightarrow}_{\phi \rightarrow const.}\, 0,\nonumber \\
P^r_{ADM,R}&=&{\hat P}^r_{ADM,R}=-{1\over 3}\int_{S^2_{\tau ,\infty}}
d^2\Sigma_r [\phi^{-2} \rho ] (\tau ,\vec \sigma )-\nonumber \\
&-&{1\over 6}\sum_{uv} \int_{S^2_{\tau ,\infty}} d^2\Sigma_u \, \phi^{-2}(\tau
,\vec \sigma ) \int d^3\sigma_1 \nonumber \\
&&(\delta_{r(a)}{\cal T}^u_{(a)v}+\delta_{u(a)}{\cal
T}^r_{(a)v})(\vec \sigma ,{\vec \sigma}_1,\tau |\phi ,0] (\phi^{-2}\rho )
(\tau ,{\vec \sigma}_1) \nonumber \\
&&{\rightarrow}_{\rho \rightarrow 0}\, 0,\nonumber \\
J^{rs}_{ADM,R}&=&{\hat J}^{rs}_{ADM,R}=\nonumber \\
&=&{1\over 6} \sum_{u} \int_{S^2_{\tau ,\infty}} d^2\Sigma_u
(\sigma^r\delta_{u}^s-\sigma^s\delta_{u}^r)(\phi^{-4}\rho )
(\tau ,\vec \sigma )+\nonumber \\
&+&{1\over {12}}\sum_{uv}\int_{S^2_{\tau ,\infty}} d^2\Sigma_u
\int d^3\sigma_1\Big[\sigma^r(\delta_{s(a)}{\cal T}^u_{(a)v}+\delta_{u(a)}
{\cal T}^s_{(a)v})-\nonumber \\
&-&\sigma^s(\delta_{r(a)}{\cal T}^u_{(a)v}+\delta_{u(a)}
{\cal T}^r_{(a)v})\Big] (\vec \sigma ,{\vec \sigma}_1,\tau |\phi ,0]
(\phi^{-2}\rho )(\tau ,{\vec \sigma}_1) \nonumber \\
&&{\rightarrow}_{\rho \rightarrow 0}\, 0,\nonumber \\
J^{\tau r}_{ADM,R}&=&{\hat J}^{\tau r}_{ADM,R}+{1\over 2} \int d^3\sigma
\sigma^r {\hat {\cal H}}_R(\tau ,\vec \sigma )=\nonumber \\
&=&-16 \epsilon k \sum_u \int_{S^2_{\tau ,\infty}} d^2\Sigma_u
\sigma^r (\phi \partial_u\phi )(\tau ,\vec \sigma )+\nonumber \\
&+&2\epsilon k \int_{S^2_{\tau ,\infty}} d^2\Sigma_r [\phi^{-2}(\phi^4-1)]
(\tau ,\vec \sigma ) \nonumber \\
&&{\rightarrow}_{\phi \rightarrow 1}\, 0.
\label{VI8}
\end{eqnarray}

This shows that
at the level of Dirac brackets with respect to the natural gauge fixing
$\rho (\tau ,\vec \sigma )\approx 0$ [i.e. with respect to the pair of second
class constraints $\rho \approx 0$, $\phi - 1 \approx 0$]
the ten weak and strong Poincar\'e charges vanish for the solution
$\phi (\tau ,\vec \sigma )=1$ selected by the boundary conditions (\ref{V3})
[so that they must vanish in all the others
gauges connected with this solution, being conserved gauge invariant
quantities].

x) Since $r_{\bar a}(\tau ,\vec \sigma )=\pi_{\bar a}(\tau ,\vec \sigma )=0$
are solutions of the two independent dynamical equations contained in
Einstein's equations, void spacetimes are an equivalence class of ``Hamiltonian
dynamical gravitational fields" and not a kinematical one (see the Conclusions
of II and Appendix A).
For them the group manifold of the Hamiltonian gauge transformations
is restricted and only gauge transformations which are also dynamical symmetries
of Einstein's equations (i.e. solutions of the associated Jacobi equations) are
allowed. This implies that void spacetimes are described by the standard
equivalence class of ``Einstein or dynamical gravitational fields" corresponding
to the flat 4-geometry, one of whose representative is Minkowski spacetime in
rectangular coordinates. This is verified in Appendix E.

As we shall see in Ref.\cite{russo4}, in presence of matter, in the static
limit $|r_{\bar a}(\tau ,\vec \sigma )| << 1$, $|\pi_{\bar a}(\tau ,\vec
\sigma )| << 1$,  Eq.(\ref{VI5})  becomes the Poisson equation
$-\triangle_{FLAT} \phi =\rho_{matter}+O(r_{\bar a},\pi_{\bar a})$, showing
that $\phi (\tau ,\vec \sigma )$
is the general relativistic generalization of the Newton potential [see also
the term M in the boundary conditions (\ref{V3})]. But now the Poincar\'e
charges are not vanishing so that we cannot use void spacetimes as
approximations of spacetimes $M^4$ for extremely weak gravitational fields.
This is contrary to the expectation that for weak gravitational fields
a spacetime can be approximated by a void spacetime with ``test" matter, but
it is consistent with parametrized Minkowski theory for that ``test" matter:
its arbitrary spacelike hypersurfaces embedded in Minkowski spacetime
describe a family of accelerated observers much bigger of the one allowed in
tetrad gravity (namely the WSW hypersurfaces, which do not exist in void
spacetimes). The implications of this fact, which have still to be
investigated, are that most of the possible accelerated reference systems of
Minkowski spacetime (which are the starting point for the classical basis of the
Unruh effect) are unrelated with general relativity, at least with its
canonical ADM formulation presented in this paper. Also,
more study will be needed to clarify the conceptual difference between a
``test particle" following a geodesic of an external gravitational field and
a ``dynamical particle plus the gravitational field" (such a particle will
not follow a geodesic of the resulting dynamical gravitational field).

Given the previous interpretation of the conformal factor $\phi$,
let us remark that the definition of the so called
``gravitoelectric field" as minus the gradient of the Newton potential plus
the post-Newtonian gravitoelectric corrections \cite{mash} (see also Ref.
\cite{ciuf}) [describing effects
like gravitational redshift, perihelion precession of Mercury,
bending of light and Saphiro's radar time delay], becomes with our notation
${\vec E}_G=-\vec \partial \phi (\tau ,\vec \sigma )$: it goes to zero for
$\phi \rightarrow 1$ in void spacetimes.
Analogously, the ``gravitomagnetic vector potential" is
${\vec A}_G(\tau ,\vec \sigma )=\{ {}^4g_{\tau r}(\tau ,\vec \sigma )=
-\epsilon N_r(\tau ,\vec \sigma ) \}$, and the ``gravitomagnetic field"
is ${\vec B}_G(\tau ,\vec \sigma )=\vec \partial \times {\vec A}_G(\tau
,\vec \sigma )=c \vec \Omega (\tau ,\vec \sigma )$ [with $\vec \Omega$ the
angular velocity connected with the precessional effects like De Sitter
and Lense-Thirring effects and the dragging of inertial frames] go to zero
for ${\vec {\tilde \lambda}}(\tau )=0$ and $\rho (\tau ,\vec \sigma )\approx 0$.
The reformulation of gravitomagnetism in ADM tetrad gravity will be the
subject of a future paper.

Another use of the terms ``gravitoelectric" and ``gravitomagnetic" effects is
connected with the electric and magnetic parts of the Weyl tensor (see the
end of Appendix A of I and  Refs.\cite{maar}). By choosing the normals to
$\Sigma_{\tau}$ as privileged timelike 4-vectors, one has ${}^4E_{rs}={}^4
C_{r\tau s\tau}$, ${}^4H_{rs}={1\over 2}\epsilon_{s\tau}{}^{uv}\,
{}^4C_{r\tau uv}$ (see Appendix A of I), which go to zero for $\phi
\rightarrow 1$ and $\rho \rightarrow 0$ (see Appendix E) in void spacetimes.
While the electric part represents the tidal force of the curvature,
the magnetic part, which has no Newtonian analogue, is generated by
the vorticity, shear, .... of the congruence of  the timelike
worldlines used for the decomposition .

\vfill\eject

\section{Conclusions.}

In this  paper and in the previous two I and II we studied the canonical
reduction of tetrad gravity in absence of matter, we investigated its
Hamiltonian group of gauge transformations, we found a canonical basis of
Dirac's observables for the Hamiltonian kinematical and dynamical gravitational
fields (see the Conclusions of II) and we have solved many interpretational
problems at every level of the theory.

i)
A modification of ADM tetrad gravity along lines suggested by Dirac has been
proposed to give a solution to the
deparametrization problem of general relativity so to recover parametrized
Minkowski theories restricted to spacelike hyperplanes in the limit of absence
of gravity. The requirement of absence of supertranslations restricts the
allowed coordinate systems and the
boundary conditions of the fields and of the gauge transformations generated by
the first class constraints. As a consequence, the Hamiltonian formalism and the
Hamiltonian group of gauge transformations turns out to be well defined for the
family of spacetimes identified by Christodoulou and Klainermann. The allowed
3+1 splittings of spacetime are only the Wigner-Sen-Witten (WSW) family of
foliations, with the leaves tending to Minkowski spacelike hyperplanes in a
direction-independent way at spatial infinity,
having the asymptotic  normal parallel to the weak (timelike) ADM 4-momentum
and corresponding to the family of foliations of Minkowski spacetime, when
matter is present, with the Wigner hyperplanes intrinsically defined by the
isolated system. Therefore, when tetrad gravity with matter is restricted to
WSW spacelike hypersurfaces, one gets a generalized ``rest-frame
Wigner-covariant 1-time instant form" description of the dynamics of a globally
hyperbolic, asymptotically flat at spatial infinity,
spacetime $M^4$ with spacelike slices $\Sigma_{\tau}$ diffeomorphic to
$R^3$ and with the given matter content (matter will be treated in Ref.
\cite{russo4} starting with scalar particles).

The allowed 3+1 splittings of these spacetimes have the spacelike hypersurfaces
tending asymptotically to Minkowski hyperplanes in a direction-independent way
and at spatial infinity there are preferred (inertial in the rest frame)
observers, which however are not static but dynamically defined. They replace
static concepts like the ``fixed stars" in the study of the dragging of
inertial frames. Since the WSW hypersurfaces
and the 3-metric on them are dynamically determined (the solution of Einstein
equations is needed to find the physical 3-metric, the allowed WSW
hypersurfaces and the Sen connection), one has neither a static background
on system-independent hyperplanes like in parametrized Newton theories
nor a static one on the system-dependent Wigner hyperplanes like in
parametrized Minkowski theories. Now both the WSW hyperplanes and the metric on
it are system dependent.

The equivalence class of dynamical ``flat" spacetimes (i.e. without
gravitational field degrees of freedom), containing Minkowski spacetime in
Cartesian coordinates, turns out to be special, because its Poincar\'e charges
vanish, so that there are no WSW hypersurfaces in them. These ``void
spacetimes" can be defined only in absence of matter (consistently with
parametrized Minkowski theories, which exist only in presence of matter) and
describe pure acceleration effects without dynamical gravitational field
(no tidal effects) allowed in flat spacetimes as the relativistic
generalization of Galilean non inertial observers. Therefore, they cannot be
used to describe ``test matter in flat spacetimes" in some post-Minkowskian
approximation. Instead in Ref.\cite{russo4} we shall study the action-at-a-
distance instantaneous effects on scalar particles implied by Einstein theory in
the ideal limit of a negligible gravitational field (a more realistic situation
with tidal effects will be possible only after having studied the linearization
of tetrad gravity in 3-orthogonal gauges) and it will be shown that already
this kind of post-Minkowskian approximation lives in non trivial spacetimes
(with non trivial WSW hypersurfaces)
not gauge equivalent to Minkowski spacetime in Cartesian coordinates.

Moreover, we showed that
parametrized Minkowski theories on arbitrary spacelike
hypersurfaces are not connected to general relativity since there is not
agreement with the general relativistic lapse and shift functions: they seem
to be connected with the description of physics in a more general class of
accelerated frames, which are not well defined in asymptotically flat at
spatial infinity general relativity, so that one will have to re-examine the
classical background of the Unruh effect.

The final rest-frame instant form of tetrad gravity on WSW hypersurfaces
labelled by the time parameter $\tau \equiv T_{(\infty )}$ in the
special 3-orthogonal gauge with $\rho (\tau ,\vec \sigma )\approx 0$, assuming
to know the solution $\phi [r_{\bar a},\pi_{\bar a}](\tau ,\vec \sigma )$ of the
reduced Lichnerowicz equation (\ref{V11}), is based on the pair of canonical
variables $r_{\bar a}(\tau ,\vec \sigma )$, $\pi_{\bar a}(\tau ,\vec \sigma )$
satisfying the Hamilton equations (the only independent dynamical combinations
of Einstein's equations)

\begin{eqnarray}
\partial_{\tau} r_{\bar a}(\tau ,\vec \sigma )\, &{\buildrel \circ \over =}\,&
\{ r_{\bar a}(\tau ,\vec \sigma ),{\hat P}^{\tau}_{ADM,R} \},\nonumber \\
\partial_{\tau} \pi_{\bar a}(\tau ,\vec \sigma )\, &{\buildrel \circ \over =}\,&
\{ \pi_{\bar a}(\tau ,\vec \sigma ),{\hat P}^{\tau}_{ADM,R} \},
\label{co1}
\end{eqnarray}

\noindent where the final Dirac Hamiltonian is the weak ADM energy ${\hat P}
^{\tau}_{ADM,R}[r_{\bar a},\pi_{\bar a},\phi [r_{\bar a},\pi_{\bar a}]]$ of
Eqs.(\ref{V12}). In these equations there is also the final form of the weak
ADM 3-momentum, whose vanishing gives the three first class constraints
defining the rest frame

\begin{equation}
{\hat P}^r_{ADM,R}[r_{\bar a},\pi_{\bar a}, \phi [r_{\bar a},\pi_{\bar a}]]
\approx 0,
\label{co2}
\end{equation}

\noindent whose natural gauge fixing is [${\hat J}^{\tau r}_{ADM,R}$ is the
weak ADM boost]

\begin{equation}
\sigma^r_{ADM} \approx - {{ {\hat J}^{\tau r}_{ADM,R}}\over { {\hat P}^{\tau}
_{ADM,R}}} \approx 0.
\label{co3}
\end{equation}

In this gauge the cotriad and the 3-metric have the following form

\begin{eqnarray}
{}^3{\hat e}_{(a)r}(\tau ,\vec \sigma ) &=& \delta_{(a)r} \Big[ \phi^2[r_{\bar
a},\pi_{\bar a}] e^{ {1\over {\sqrt{3}}} \sum_{\bar a}\gamma_{\bar ar} r_{\bar
a}} \Big] (\tau ,\vec \sigma ),\nonumber \\
{}^3{\hat g}_{rs}(\tau ,\vec \sigma ) &=& \delta_{rs} \Big[ \phi^4[r_{\bar
a},\pi_{\bar a}] e^{ {2\over {\sqrt{3}}} \sum_{\bar a}\gamma_{\bar ar} r_{\bar
a}} \Big] (\tau ,\vec \sigma ),
\label{co4}
\end{eqnarray}

\noindent while the lapse and shift functions are given by [$N_{(a)}={}^3{\hat
e}^r_{(a)}N_r$]

\begin{eqnarray}
N(\tau ,\vec \sigma ) &=& -\epsilon +\hat n (\tau ,\vec \sigma | r_{\bar a},
\pi_{\bar a}, \phi [r_{\bar a},\pi_{\bar a}] ],\nonumber \\
N_r(\tau ,\vec \sigma ) &=& {\hat n}_r(\tau ,\vec \sigma | r_{\bar a},
\pi_{\bar a}, \phi [r_{\bar a},\pi_{\bar a}] ],
\label{co5}
\end{eqnarray}

\noindent with $\hat n$ and ${\hat n}_r$ determined by Eqs.(\ref{V11b}) and
(\ref{V4a}) respectively [in these equations we put ${\tilde \lambda}_{\tau}
(\tau )=\epsilon$, ${\tilde \lambda}_r(\tau )=0$].

The final form of the 4-metric in coordinates adapted to the WSW hypersurface is

\begin{equation}
{}^4{\hat g}_{AB}=\epsilon
\left( \begin{array}{cc}  \Big(-\epsilon +\hat n\Big)^2-\phi^{-4}\sum_r
e^{-{2\over {\sqrt{3}}}\sum_{\bar a}\gamma_{\bar ar}r_{\bar a}}{\hat n}_r^2&
-{\hat n}_s\\
-{\hat n}_r& -\delta_{rs} \phi^4 e^{{2\over {\sqrt{3}}}
\sum_{\bar a}\gamma_{\bar ar}r_{\bar a}}
\end{array} \right) .
\label{co6}
\end{equation}

ii) Our quasi-Shanmugadhasan canonical transformation, defined in II
starting from a multi-time formalism \cite{dc3},
and the choice of 3-orthogonal coordinates transforms
the superhamiltonian constraint in a nonlinear integro-differential
equation (the reduced Lichnerowicz equation) for the conformal factor
$\phi =e^{q/2}$ of the 3-metric, whose conjugate momentum $\pi_{\phi}=2\phi
^{-1} \rho$ plays the role of the last gauge variable.
Since this gauge variable describes nonlocal properties of the extrinsic
curvature of the Cauchy surfaces $\Sigma_{\tau}$, its variation describes the
allowed 3+1 splitting of spacetime (the ADM theory is independent
from the choice of anyone of them)  : this
is the meaning of the gauge transformations generated by the superhamiltonian
constraint, whose effect is to change $\rho$ and, therefore, the extrinsic
curvature of the leaves.

This fact leads to the distinction between Hamiltonian kinematical and dynamical
gravitational fields made in the Conclusions of II. As shown there (see also
Appendix A) on the solutions of Einstein's equations we must restrict the
parameters of the gauge transformations (and, therefore, the gauge variables)
in such a way that the allowed gauge transformations are also dynamical
symmetries of the Einstein's equations: this fact implies that the Hamiltonian
dynamical gravitational fields coincide with the standard ``Einstein or
dynamical gravitational fields", namely with a single 4-geometry whose 4-metrics
are solutions of Einstein's equations (this 4-geometry is parametrized in terms
of one conformal 3-geometry).

The knowledge of the
solution $\phi (\tau ,\vec \sigma )=e^{q(\tau ,\vec \sigma )/2}=
F[\rho ,r_{\bar a},\pi_{\bar a}](\tau ,\vec \sigma )$ of the reduced
Lichnerowicz equation in the 3-orthogonal gauge would
allow to start the search of the final  Shanmugadhasan canonical
transformation $q, \rho , r_{\bar a}, \pi_{\bar a} \mapsto q^{'}=q -
2 ln\, F, \rho ,
{\tilde r}_{\bar a}, {\tilde \pi}_{\bar a}$ which would implement Kuchar's
program defined in Refs.\cite{dc10,kuchar1} and would be an
alternative to the York map \cite{ise}, with ${\tilde r}_{\bar a}$,
${\tilde \pi}_{\bar a}$ describing the independent degrees of freedom of the
gravitational field. However, by adding the natural gauge
fixing $\rho (\tau ,\vec \sigma )\approx 0$ and by going to Dirac's brackets
[to eliminate $q, \rho$ even if one has not been able to solve explicitly the
Lichnerowicz equation], one finds that $r_{\bar a}$, $\pi_{\bar a}$ are the
natural canonical variables of the gravitational field in this gauge.

Regarding other approaches to the observables in general relativity
see also Refs.\cite{dc8}: the ``perennials" introduced
in this approach are essentially our Dirac observables.
See Ref.\cite{gr13} for the difficulties in observing perennials
experimentally at the classical and quantum levels and in their
quantization. See also Ref.\cite{torre} on the non existence of observables for
the vacuum gravitational field in a closed universe, built as spatial
integrals of local functions of Cauchy data and their first derivatives.

Our approach violates the
geometrical structure of general relativity breaking general covariance (but in
a way associated with the privileged presymplectic Darboux bases naturally
selected by the Shanmugadhasan canonical transformations) and avoiding the
``spacetime problem" with the choice of the privileged
WSW foliations, but it allows
the deparametrization of general relativity and a soldering with parametrized
Minkowski theories (and parametrized Newton theories for $c\, \rightarrow
\infty$) and to make contact with the kinematical framework, which will be
used\cite{bari} to find the Tomonaga-Schwinger
asymptotic states needed for relativistic bound states (the Fock asymptotic
states have no control on the relative times of the asymptotic particles).
The problem whether general covariance may be recovered at the quantum
level has to be attacked only after having seen if this minimal quantization
program can work.

iii) Let us add some comments on time in general relativity in our case of
globally hyperbolic asymptotically flat at spatial infinity spacetimes with
the modified ADM tetrad theory.

In general relativity, Isham\cite{ish} and Kuchar\cite{kuchar1}
have made a complete review of the problem of time [see Ref.\cite{butter} for a
recent contribution to the problem], showing that till now there is no
consistent quantization procedure for it.

We are in a scheme in which time is identified before quantization.
The unphysical mathematical
1-time of our rest-frame instant form of dynamics on WSW
hypersurfaces is the rest-frame ``global time" $T_{(\infty )}
=p_{(\infty )}\cdot {\tilde x}_{(\infty )}/\sqrt{\epsilon
p^2_{(\infty )}}={\hat P}_{ADM}\cdot x_{(\infty )}/\sqrt{\epsilon
{\hat P}^2_{ADM}}=\tau$ [let us note that this is
possible for globally hyperbolic, asymptotically flat at spatial infinity,
spacetimes; instead a global time does not exist, even with
a finite number of degrees of freedom, when the configuration space is compact;
see for instance Refs.\cite{dc8}] and not an internal time.
It is the gauge-fixing $T_{(\infty )}-\tau\approx 0$ to the extra Dirac
constraint $\epsilon_{(\infty )}-\sqrt{\epsilon
{\hat P}^2_{ADM}}\approx 0$
which identifies the foliation parameter with the rest-frame time.
The evolution in $T_{(\infty )}=\tau$ of the two canonical pairs of
gravitational degrees of freedom is governed by the weak ADM energy
${\hat P}^{\tau}_{ADM}$.

The positions of the non-covariant ``external"
center-of-mass variable ${\tilde x}^{(\mu )}
_{(\infty )}(\tau )$, replacing the arbitrary origin $x^{(\mu )}_{(\infty )}$
of the coordinates on the WSW hypersurfaces,
and of this origin are irrelevant, because,
as already said, at the end the 6 variables ${\vec z}_{(\infty )}$, ${\vec k}
_{(\infty )}$ are decoupled: they describe the
``external" center of mass of the isolated
universe or equivalently a decoupled external observer with his ``point
particle clock" [therefore one does not need ``matter clocks and reference
fluids"\cite{kuchar1,brk}].
They are not to be quantized because they can be said to
belong to the classical part of the Copenhagen interpretation,
but their non-covariance is fundamental in defining the classical M\"oller
radius $|{\hat {\vec S}}_{ADM}|/ {\hat P}^{\tau}_{ADM}$ [where, due to
${\hat {\vec P}}_{ADM}\approx 0$, we have $|{\hat {\vec S}}_{ADM}|=
\sqrt{-\epsilon W^2_{ADM}}/
{\hat P}^{\tau}_{ADM}$ with $W^A_{ADM}$ the asymptotic Pauli-Lubanski 4-vector]
to be used as a ultraviolet cutoff.

The ``internal" center-of-mass 3-variable ${\vec \sigma}_{ADM}[r_{\bar a},\pi
_{\bar a}]$ (built in terms of the weak Poincar\'e charges as it is done
for the Klein-Gordon field on the Wigner hyperplane in Ref.\cite{mate};
due to ${\hat P}^r_{ADM}\approx 0$ we have $\sigma^r_{ADM}\approx -{\hat J}
^{\tau r}_{ADM}/ {\hat P}^{\tau}_{ADM}$)
of the universe inside a WSW hypersurface identifies the 3 gauge-fixings
${\vec \sigma}
_{ADM}\approx 0$ to be added to ${\hat {\vec P}}_{ADM}[r_{\bar a},\pi_{\bar
a}]\approx 0$. With these gauge fixings this point
coincides with the arbitrary origin $x^{(\mu )}
_{(\infty )}(\tau )$. With ${\vec \sigma}_{ADM} \approx 0$ the origin
$x^{(\mu )}_{(\infty )}(\tau )$ becomes simultaneously
\cite{mate} the Fokker-Price
center of inertia, the Dixon center of mass and Pirani and Tulczjyew centroids
of the universe, while the non-covariant ``external"
center-of-mass variable ${\tilde x}^{(\mu )}
_{(\infty )}(\tau )$ is the analog of the Newton-Wigner position operator.

Our final picture of the reduced phase space has similarities
the frozen Jacobi picture of Barbour\cite{dc12} and his proposal to substitute
time with  the astronomical ephemeris time\cite{dc120} [in his timeless
and frameless approach based on Ref.\cite{dc11} the local ephemeris time
coincides with the local proper time] may be a starting point for correlating
local physical clocks with the mathematical time parameter $\tau =T_{(\infty )}$
of the foliation [and not for defining a timeless theory by using Jacobi's
principle]. We think that scenario a) of Section III, used for the
description of void spacetimes without matter, is a realization of the fully
Machian approach of Barbour which, however, seems possible only in absence of
matter. Instead the scenario b) with a decoupled free
``external" center-of-mass variable is
Machian only in the fact that there are only dynamical relative variables
left both in asymptotically flat general relativity and in parametrized
Minkowski theories.

Let us remark that the interpretation of the superhamiltonian constraint
as a generator of gauge transformations with natural gauge-fixing
$\rho (\tau ,\vec \sigma )\approx 0$ (at least in 3-orthogonal coordinates)
leads to the conclusion that neither York's internal extrinsic time nor
Misner's internal intrinsic time are to be used as time parameters: Misner's
time (the conformal factor) is determined by the Lichnerowicz equation while
York's time (the trace of the extrinsic curvature) by the natural
gauge-fixing $\rho \approx 0$. This implies a refusal of the standard
(interconnected)
interpretations  of the superhamiltonian constraint  either as a generator of
time evolution (being a time-dependent Hamiltonian) like in the
commonly accepted viewpoint based on the Klein-Gordon
interpretation of the quantized superhamiltonian constraint, i.e. the
Wheeler-DeWitt equation
[see Kuchar in Ref.\cite{ku1}, Wheeler's evolution of 3-geometries in
superspace in Ref.\cite{dc11,mtw} and the associated
``sandwich conjecture" (see for instance Refs.\cite{dc14}, \cite{dc11} and the
review in Ref.\cite{kuchar1}); see Ref.\cite{paren} for the
cosmological implications]
or as a quantum Hamilton-Jacobi  equation
without any time [one can introduce a concept of evolution,
somehow connected with an effective time, only in a WKB sense \cite{kiefer}].

Since the superhamiltonian  constraint is
quadratic in the momenta, one is naturally driven to make a comparison with
the free scalar relativistic particle with the first class constraint
$p^2-\epsilon m^2\approx 0$. As shown in Refs.\cite{lus1,dc1}, the constraint
manifold in phase space has 1-dimensional gauge orbits (the two disjointed
branches of the mass-hyperboloid); the $\tau$-evolution
generated by the Dirac Hamiltonian $H_D=\lambda (\tau ) (p^2-\epsilon m^2)$
gives the parametrized solution $x^{\mu}(\tau )$. Instead, if we go to the
reduced phase space by adding the non-covariant gauge-fixing $x^o-\tau \approx 0$
and eliminating the pair of canonical variables $x^o\approx \tau$, $p^o \approx
\pm \sqrt{{\vec p}^2+m^2}$, we get a frozen Jacobi
description in terms of independent Cauchy data, in which the same Minkowski
trajectory of the particle can be recovered in the non-covariant form
$\vec x(x^o)$ by introducing as Hamiltonian the energy generator
$\pm \sqrt{{\vec p}^2+m^2}$ of the Poincar\'e group [with the variables of
Ref.\cite{longhi}, one adds the covariant gauge-fixing $p\cdot x/\sqrt{p^2}
-\tau \approx 0$ and eliminates the pair $T=p\cdot x/
\sqrt{p^2}$, $\epsilon =\eta \sqrt{p^2} \approx \pm m$; now, since the
invariant mass is constant, $\pm m$, the non-covariant Jacobi data $\vec z=
\epsilon (\vec x-\vec p x^o/p^o)$, $\vec k=\vec p/\epsilon$ cannot be made to
evolve]. Instead the superhamiltonian constraint, being a secondary first
class constraint of a field theory, has an associated ``Gauss law" (see
Eq.(\ref{a5}) of
Appendix A) like the supermomentum constraints, even if it is quadratic in the
momenta and this fact is connected to the definition of the ADM energy.
This Gauss law (defining the strong ADM energy as a surface integral)
shows that the superhamiltonian constraint has to be solved in
the conformal factor q or in $\phi =e^{q/2}$. Therefore, its gauge orbit in
superspace is parametrized by $\rho$ and this is not a time evolution
(instead it is connected with the normal deformations of the spacelike
hypersurfaces so to change a 3+1 splitting in another one):
the solution of the superhamiltonian constraint do not define the weak
ADM energy, which, instead, is connected with an integral over
3-space of that part of the superhamiltonian constraint
dictated by the Gauss law. This does not contradict
Teitelboim's results\cite{tei}: the algebra of supermomentum and
superhamiltonian constraints reflects the embeddability of $\Sigma_{\tau}$
into $M^4$ (see also Ref.\cite{anton}): the theory is simply independent from
the allowed 3+1 splittings with embeddable leaves $\Sigma_{\tau}$.

Let us remember that in Ref.\cite{paur} the nonrelativistic limit of
the ADM action for metric gravity was considered: it allowed the identification
of a singular Lagrangian density with general Galileo covariance depending on 27
fields [coming from the development in series of powers of $1/c^2$ of $N$,
$N_r$, ${}^3g_{rs}$] describing Newton gravity in arbitrary coordinates. This
theory has first class constraints connected with inertial forces and second
class constraints, determining the static Newton potential in arbitrary frames
of reference when massive particles are present [see Ref.\cite{paur1} for
alternative nonrelativistic gravity theories]. This implies that it will be
possible to consider the nonrelativistic limit of our modified tetrad gravity
and establish its connections with the postNewtonian approximations
\cite{dc9}, in particular the recent one of Ref.\cite{dam,dam1}: see Appendix
G for some preliminary comments.

Now, at the nonrelativistic level there is an absolute time t and the evolution
in this numerical parameter of every system is described by the Hamilton
equations associated with a Hamiltonian function H describing the energy of
the system [it is a generator of the kinematical (extended) Galileo group when
the system is isolated]. Alternatively, one can use a parametrized
reformulation of the system by enlarging phase space with a canonical pair
$t,\, E$ [$\{ t,E \} =\epsilon =\pm 1$, if $\epsilon$ is the signature of the
time axis], by adding the first class constraint $\chi =E-H\approx 0$ (so that
the Dirac Hamiltonian is $H_D=\lambda (\tau ) \chi$) and by calling $\tau$
the scalar parameter associated with the canonical transformations generated by
$\chi$. The parameter $\tau$ labels the leaves of a foliation of Galilei
spacetime; the leaves are (rest-frame)
hyperplanes, which are the limit of Wigner
hyperplanes in parametrized Minkowski theories for $c \rightarrow \infty$.
One gets a parametric description of the same physics with t and the
solutions of the original Hamilton equations now expressed as functions of the
new time parameter $\tau$. If one adds the gauge-fixing $t-\tau \approx 0$
, one gets a frozen reduced phase space (equal to the
original one) like in the Jacobi theory, in which one reintroduce an evolution
by using the energy E=H for the evolution in $t=\tau$. However, with more
general gauge-fixings $t-f(\tau ,...)\approx 0$,
where dots mean other canonical
variables, the associated Hamiltonian is no more the energy
[see Ref.\cite{dc1}].

In the standard nonrelativistic quantization of the system one defines a
Hilbert space and writes a Schroedinger equation in which t is a parameter and
in which the t-evolution is governed by an operator obtained by quantizing the
Hamiltonian function corresponding to the energy [see Ref.\cite{dc2} for a
discussion of this point and of the associated ambiguities and problems].
Instead, in the parametrized theory, one should quantize also the pair $t, E$
(one introduces a unphysical Hilbert space in which the t-dependence of wave
functions is restricted to be square integrable) and
write a Schroedinger equation
in $\tau$ with the quantum Dirac Hamiltonian [see Ref.\cite{dc3} on this point
and on the problem of the unphysical and physical scalar products]
and then impose the constraint to identify the physical states. This procedure
is ambiguous, because in this way the energy operator has no lower bound for
its spectrum in the unphysical Hilbert space and it is delicate to recover the
physical Hilbert space from the quotient of the unphysical one with respect to
the quantum unitary gauge transformations generated by the quantum constraint.
In particular, physical states have infinite unphysical norm [usually the zero
eigenvalue belongs to the continuum spectrum of the constraint operators] and
the construction of the physical scalar product for physical states (without any
restriction on the t-dependence) depends on the form of the constraint
(see Ref.\cite{longhi} for a relativistic example).

Moreover, the absolute time t, which labels the Euclidean leaves of the
absolute foliation of Galileo spacetime, is unrelated to physical clocks. As
shown in Ref.\cite{dc4} (see also Ref.\cite{ish}), in the physical Hilbert space
there is no operator such that: i) it can be used as a ``perfect clock", in the
sense that, for some initial state, its observed values increase monotonically
with t; ii) is canonically conjugate to the Hamiltonian operator (this would
imply that this operator is not definite positive).
All this is also related with Rovelli's proposal\cite{dc5} of replacing t (in
the nonrelativistic case) with an ``evolving constant of the motion", i.e. a
t-dependent function of operators commuting with the Hamiltonian. This proposal
can be done either in the standard or in the parametrized version of the theory
(see also Ref.\cite{dc7}); among others\cite{dc6}, Kuchar\cite{kuchar1}
critics it for the ambiguities coming from the operator ordering problem. In
any case there are all the previously mentioned problems and also the fact
that the conjugate variables of these evolving constants of motion
generically have nothing to do with the energy
and can have spectra and symmetries of every type (see Ref.\cite{dc2}).

All the proposals of replacing the parameter t with some physical time function
(or operator) show that this is the main unsolved problem: how to identify (at
least locally, possibly globally) the leaves of the foliation of Galileo
spacetime with ``physical clocks", i.e. with an apparatus described in the given
either phase or Hilbert space. See again in this connection Barbour\cite{dc12}
who uses as local time functions special space coordinates (the astronomical
ephemeris time\cite{dc120} or some its relativistic extension).

Now, in the approach based on parametrized special relativistic theories in
Minkowski spacetime, the final result is that every isolated system (or better
all its configurations with a timelike total 4-momentum) identifies a Wigner
foliation of Minkowski spacetime. Its leaves (the Wigner hyperplanes) are
labelled by a scalar parameter $T_s=\tau$ (the center-of-mass time in the rest
frame) in the ``rest-frame Wigner-covariant 1-time instant form" with the
evolution in this parameter governed by the invariant mass of the system. There
is also a decoupled non-covariant center-of-mass point with free motion.
The quantization of this instant form produces a 1-time Schroedinger equation
as in the standard unparametrized nonrelativistic case with the Newtonian time t
replaced by the Lorentz-scalar rest frame time $T_s$.

In our modified tetrad gravity we have again the same picture in
the generalized rest-frame instant form with WSW foliations. Therefore, in our
unified approach to general relativity, special relativity and Newton-Galileo
theories we are never going to quantize any time variable and the problem of
time is replaced by the problem of how to correlate (locally) physical
clocks with the mathematical time parameter labelling the leaves of the 3+1
splitting of spacetime.

iv)
Let us remark that our ADM tetrad formulation assumed the existence of a
mathematical abstract 4-manifold, the spacetime $M^4$, to which we added 3+1
splittings with spacelike leaves $\Sigma_{\tau}\approx R^3$. The mathematical
points of $M^4$ have no physical meaning [invariance under $Diff\, M^4$, hole
argument \cite{stachel}; see the Conclusions of II] and are coordinatized with
$\Sigma_{\tau}$-adapted coordinates $(\tau ,\vec \sigma )$. All fields (also
matter fields when present) depend on these mathematical coordinates for $M^4$,
but till now there is no justification for saying that the points (or events)
of the spacetime have any physical meaning [instead in special relativity they
are physical points by hypothesis].

Is it possible to label the points of $M^4$ in terms of Dirac's observables
a posteriori by introducing ``physical points"? As said in the Conclusions of
II, once all gauge freedoms have been eliminated this can be done, in analogy
to what happens with the vector potential of electromagnetism which becomes
measurable in a completely fixed gauge like the Coulomb one. Indeed, one can
build a system of coordinates for a spacetime $M^4$ without
Killing vectors (at least in absence of
matter) in terms of the Dirac observables $r_{\bar a}$, $\pi_{\bar a}$,
describing the gravitational field in our complete 3-orthogonal gauge with the
natural gauge fixing $\rho (\tau ,\vec \sigma )\approx 0$ by restricting to
that gauge the 4 independent Komar-Bergmann ``individuating fields"
\cite{be,komar}
(see also Ref.\cite{dc11,ish}). These fields are bilinears
and trilinears in the Weyl tensor (independent from the lapse and shift
functions)
invariant under $Diff\, M^4$ but not under the Hamiltonian group of gauge
transformations. Using the results of Appendices D,E, of II plus the natural
gauge fixing the individuating fields can be expressed in terms of Dirac's
observables and there is a local 1-1 correspondence between them and the
mathematical coordinates $(\tau ,\vec \sigma )$.

v) In Appendix F there is a definition of null tetrads which is natural from
the Hamiltonian point of view. It will be used in a future paper to find
the phase space expressions of the Newman-Penrose 20 scalars carrying all the
information on the Riemann tensor (see for instance Ref.\cite{stewart}; they
are ${}^4R$, nine scalars for the trace-free Ricci tensor and five complex
scalars for the Weyl tensor). This will allow to find the dependence of these
scalars on the gauge variables of tetrad gravity and to study their asymptotic
expansions at null infinity. Moreover, the two spacelike vector fields in the
null tetrad will identify in the tangent space in each point of $M^4$ a 2-plane:
this will allow to connect our canonical variables $r_{\bar a}$, $\pi_{\bar a}$
in the special 3-orthogonal gauge with the symmetric trace-free 2-tensors on
2-planes of Ref.\cite{ckl}. The 2-planes identified in the tangent space in
each point of $M^4$ are orthogonal to the timelike normal $l^A(\tau ,\vec
\sigma )$ at $\Sigma_{\tau}$ in that point and to the spacelike unit vector
${\cal N}^A(\tau ,\vec \sigma )$, defined in Appendix F (the gauge direction
identified by the shift functions), in that point. In general (for instance in
the 3-orthogonal gauges) this vector field is not surface-forming, namely the
associated differential form ${\cal N}_A(\tau ,\vec \sigma ) d\sigma^A$ is not
proportional to a closed differential 1-form (non zero vorticity). As said in
Appendix F, it will be important to study those coordinate systems for $\Sigma
_{\tau}$ implying that ${\cal N}^A(\tau ,\vec \sigma )$ is surface forming,
because in them there is the possibility of making a 2+2 decomposition of $M^4$
with a conformal 2-structure \cite{inverno,inverno1}, of having variables in the
spirit of the Newman-Penrose formalism and of finding canonical variables
$r^{'}_{\bar a}$ for the gravitational field naturally defined on 2-surfaces.
Our approach opens the path to a systematic search of these 3-coordinates for
$\Sigma_{\tau}$, which should be investigated in the future like the normal
coordinates around a point of $M^4$.

vi) Let us now make some comments on the quantization of tetrad gravity in this
scheme in which general covariance is completely broken having
completely fixed all the gauges.
See Ref.\cite{gr13} for an updated discussion of quantization problems in
canonical gravity
(and Ref.\cite{ku1} for the quantization of parametrized theories).

The quantization of the rest-frame instant form of tetrad gravity in the
3-orthogonal gauge with the natural gauge fixing $\rho (\tau ,\vec \sigma )
\approx 0$ by using the mathematical time parameter $T_{(\infty )}\equiv \tau$
(the rest-frame time of the ``external" decoupled point particle clock) on the
Wigner-Sen-Witten hypersurfaces should be done with the following steps:
\hfill\break
a) Assume to have found either the exact or an approximate solution of the
classical reduced Lichnerowicz equation $\phi = \phi (r_{\bar a},\pi_{\bar a})$
and to have evaluated the associated weak ADM 4-momentum ${\hat P}^A_{ADM,R}=
{\hat P}^A_{ADM,R}[r_{\bar a},\pi_{\bar a},\phi (r_{\bar a},\pi_{\bar a})]$.
\hfill\break
b) On each WSW hypersurface $\Sigma_{\tau} \approx R^3$ replace the
Hamiltonian gravitational field physical degrees of freedom $r_{\bar a}(\tau
,\vec \sigma )$, $\pi_{\bar a}(\tau ,\vec \sigma )$ with operators ${\hat r}
_{\bar a}(\tau ,\vec \sigma )=r_{\bar a}(\tau ,\vec \sigma )$, ${\hat \pi}
_{\bar a}(\tau ,\vec \sigma )=i{{\delta}\over {\delta r_{\bar a}(\tau ,\vec
\sigma )}}$ (Schroedinger representation) on some Hilbert space.\hfill\break
c) Write the functional Schroedinger wave equation

\begin{equation}
i {{\partial}\over {\partial \tau}} \Psi (\tau ,\vec \sigma |r_{\bar a}] =
{\hat P}^{(op)\, \tau}_{ADM,R}[r_{\bar a},{\hat \pi}_{\bar a},\phi (r_{\bar a},
{\hat \pi}_{\bar a})] \Psi (\tau ,\vec \sigma |r_{\bar a}],
\label{VII1}
\end{equation}

\noindent plus the 3 conditions defining the rest frame

\begin{equation}
{\hat P}^{(op)\, r}_{ADM,R}[r_{\bar a},{\hat \pi}_{\bar a},\phi (r_{\bar a},
{\hat \pi}_{\bar a})] \Psi (\tau ,\vec \sigma |r_{\bar a}] =0,
\label{VII2}
\end{equation}

\noindent after having choosen (if possible!) an ordering such that $[ {\hat P}
^{(op)\, A}_{ADM,R}, {\hat P}^{(op)\, B}_{ADM,R} ] =0$.
Let us remark that at this
stage it could be useful the suggestion of Ref.\cite{dc13} that the
unphysical space of these functionals does not need to be a Hilbert space and
that, in it, the observables need not to be self-adjoint operators (these
properties must hold only in the physical space with the physical scalar
product). This Schroedinger equation has not
an``internal Schroedinger interpretation" since neither ``Misner internal
intrinsic time" nor ``York internal extrinsic time" nor any function like the
``Komar-Bergmann individuating fields" are the time: it does not
use the superhamiltonian constraint (like the Wheeler-DeWitt equation)
but the derived weak ADM energy.

The scalar product associated with this Schroedinger equation defines the
Hilbert space and the operators ${\hat P}^{(op)A}_{ADM,R}$
should be self-adjoint
with respect to it. Since there are the 3 conditions coming from the 3 first
class constraints defining the rest frame, the physical Hilbert space of the
wave functionals $\Psi_{phys}$ solution of Eq.(\ref{VII2}) will have an induced
physical scalar product which depends on the functional form of the constraints
${\hat P}^r_{ADM,R}\approx 0$ as it can be shown explicitly in
finite-dimensional examples \cite{lll,longhi}, so that it is not given by a
system-independent rule.

Another possibility is to add and quantize also the gauge fixings ${\vec
\sigma}_{ADM}\approx 0$. In this case one could impose the second class
constraints in the form
$< \Psi |\sigma^{(op)r}_{ADM}| \Psi > =0$, $< \Psi | {\hat P}^{(op)r}_{ADM,R}
| \Psi > =0$ and look whether it is possible to define a Gupta-Bleuler
procedure.

The best would be to be able to find the
canonical transformation $r_{\bar a}(\tau ,\vec \sigma )$, $\pi_{\bar a}(\tau
,\vec \sigma )$ $\mapsto$ ${\vec \sigma}_{ADM}$, ${\hat {\vec P}}_{ADM,R}$,
$R_{\bar a}(\tau ,\vec \sigma )$, $\Pi_{\bar a}(\tau ,\vec \sigma )$
[$R_{\bar a}$, $\Pi_{\bar a}$ being relative variables], since in this case
we would quantize only the final relative variables:

\begin{eqnarray}
\Psi_{phys}&=& \tilde \Psi (\tau ,\vec \sigma |R_{\bar a}],\nonumber \\
i{{\partial}\over {\partial \tau}} \Psi_{phys}&=& {\hat E}^{(op)}_{ADM}[R_
{\bar a},{\hat \Pi}_{\bar a}=i{{\delta}\over {\delta R_{\bar a}}}] \Psi_{phys},
\nonumber \\
&&with\quad {\hat E}_{ADM}={\hat P}^{\tau}_{ADM,R}[r_{\bar a},\pi_{\bar a},\phi
(r_{\bar a},\pi_{\bar a})] {|}_{{\vec \sigma}_{ADM}={\hat {\vec P}}_{ADM,R}=0}.
\label{VII3}
\end{eqnarray}

Let us remark that many aspects of the problem of time in quantum gravity
\cite{kuchar1} would be avoided: i) there would be no ``multiple choice
problem" since there is only one mathematical
time variable $T_{(\infty )}=\tau$;
ii) the problem of ``functional evolution" would be reduced to find an
ordering such that $[ {\hat P}^{(op)\, A}_{ADM,R},
{\hat P}^{(op)\, B}_{ADM,R} ] =0$
; iii) the ``Hilbert space problem" is not there because we do not have the
Wheeler-DeWitt equation but an ordinary Schroedinger equation; iv) there is
a physical ultraviolet cutoff (the M\"oller radius) like in parametrized
Minkowski theories which could help in regularization problems.

Naturally general covariance is completely broken and everything is defined only
on the Wigner-Sen-Witten foliation associated with the natural gauge fixing
$\rho (\tau ,\vec \sigma )\approx 0$. If we would do the same quantization
procedure in 3-normal coordinates on their WSW hypersurfaces associated with
the corresponding natural gauge fixing $\rho_{normal}(\tau ,\vec \sigma )
\approx 0$, we would get a different physical Hilbert space whose being
unitarily equivalent to the one in 3-orthogonal coordinates is a completely
open problem.

If this quantization can be done,  the completely gauge-fixed
4-metric ${}^4g_{AB}$ on the mathematical manifold $M^4$ would become an
operator ${}^4{\hat g}_{AB}(\tau ,\vec \sigma |R_{\bar a},{\hat \pi}_{\bar a}]$
with the implication of a quantization of the Dirac observables associated
with 3-volumes (the volume element Dirac observable is the solution $\phi$ of
the reduced Lichnerowicz equation for $\rho =0$), 2-areas and lengths. Let us
remark that these quantities would not a priori commute among themselves:
already at the classical level there is no reason that they should have
vanishing Dirac brackets (however, two quantities with compact disjoint supports
relatively spacelike would have vanishing Dirac brackets).

Since in the Dirac-Bergmann canonical reduction of tetrad gravity spin networks
do not show up (but they could be hidden in the non-tensorial character of the
Dirac observables $r_{\bar a}$, $\pi_{\bar a}$ still to be explored), it is not
clear which could be the overlap with Ashtekar-Rovelli-Smolin program
\cite{ars}, which is generally covariant but only after having fixed the
lapse and shift functions (so that it is not clear how one can rebuild the
spacetime from the 3-geometries) and replaces local variables of the type
$r_{\bar a}(\tau ,\vec \sigma )$ with global holonimies of the 3-spin
connection over closed 3-loops.

If the quantization can be made meaningful, the quantum Komar-Bergmann
individuating fields would lead to a quantization of the ``physical
coordinates" for the spacetime $M^4$. This will give a quantum spacetime
connected with non commutative geometry approaches.

Let us also remark that instead of using a solution of the classical reduced
Lichnerowicz equation with $\rho (\tau ,\vec \sigma )=0$, one could use
weak ADM 4-momentum ${\hat P}^{(op) A}_{ADM,R}[r_{\bar a},{\hat \pi}_{\bar a},
\phi^{(op)}]$ with $\phi^{(op)}$ an operatorial solution of a quantum
operatorial reduced Lichnerowicz equation (not a quantum constraint
on the states but the quantization of the classical Lichnerowicz equation
with $\rho =0$ after having gone to Dirac brackets).

Finally, let us observe that even if our approach is more complicated than
Ashtekar and string ones, it opens the possibility of a unified description of
the four interactions after having learned how to couple the standard
SU(3)xSU(2)xU(1) model to tetrad gravity and how to make the canonical
reduction of the complete theory. There is a unification of the mathematical
tools, namely one needs to learn the properties of special functions like the
Wilson lines along geodesics of either su(3)-valued connections or so(3)-valued
3-spin-connections. Moreover, the problem of which choice to make for the
function space of the fields associated with the four interactions will require
to understand whether the Gribov ambiguity is only a mathematical obstruction
to be avoided (in tetrad gravity this would eliminate the isometries and in
Yang-Mills theory the gauge symmetries and the gauge copies) or whether there
is some physics in it (in this case one should learn how to make the canonical
reduction in presence of gauge symmetries, gauge copies and isometries).

Even if it is too early to understand whether our approach can be useful either
from a computational point of view (like numerical gravity) or for the search
of exact solutions, we felt the necessity to revisit the Hamiltonian formulation
of tetrad gravity with its intrinsic naturalness for the search of the physical
degrees of freedom of any gauge theory and for the formulation of quantization
rules so that one can have a clear idea of the meaning of the gauge fixings and
the possibility to have an insight on the role of the gauge degrees of freedom
in the realm of exact solutions where traditionally one starts with suitable
parametrizations of the line element $ds^2$ and then uses symmetries to
simplify the mathematics.

But before attacking these problems, we have to study tetrad gravity in
3-normal coordinates, its linearization to make contact with the theory of
gravitational waves, its coupling to matter and how to define the analog of the
post-Minkowskian approximation (formal expansion in series of powers of G; see
for instance Ref.\cite{luc,dam}) starting from the rest-frame instant form on
WSW hypersurfaces. Instead in Appendix G there is a comparison of the rest-frame
instant form of tetrad gravity with the formulation of Ref.\cite{dam,dam1} of
the post-Newtonian approximation.

\vfill\eject

\appendix

\section{The Second Noether Theorem for ADM Metric Gravity.}

In Section V of I there was a review of ADM canonical metric gravity, whose
secondary first class constraints and Dirac Hamiltonian are
[the $\lambda (\tau ,\vec \sigma )$'s are arbitrary Dirac multipliers;
$k=c^3/16\pi G$]

\begin{eqnarray}
{\tilde {\cal H}}(\tau ,\vec \sigma )&=&\epsilon
[k\sqrt{\gamma}\, {}^3R-{1\over {2k
\sqrt{\gamma}}} {}^3G_{rsuv}\, {}^3{\tilde \Pi}^{rs}\, {}^3{\tilde \Pi}^{uv}]
(\tau ,\vec \sigma )=\nonumber \\
&=&\epsilon [\sqrt{\gamma}\, {}^3R-{1\over {k\sqrt{\gamma}}}({}^3{\tilde \Pi}
^{rs}\, {}^3{\tilde \Pi}_{rs}-{1\over 2}({}^3\tilde \Pi )^2)](\tau ,\vec
\sigma )=\nonumber \\
&=&\epsilon k \{ \sqrt{\gamma} [{}^3R-({}^3K_{rs}\, {}^3K^{rs}-({}^3K)^2 )]
(\tau ,\vec \sigma )\approx 0,
\nonumber \\
{}^3{\tilde {\cal H}}^r(\tau ,\vec \sigma )&=&-2\, {}^3{\tilde \Pi}^{rs}{}_{| s}
(\tau ,\vec \sigma )=-2[\partial_s\, {}^3{\tilde \Pi}^{rs}+{}^3\Gamma^r_{su}
{}^3{\tilde \Pi}^{su}](\tau ,\vec \sigma )=\nonumber \\
&=&-2\epsilon k \{ \partial_s[\sqrt{\gamma}({}^3K^{rs}-{}^3g^{rs}\, {}^3K)]+
\nonumber \\
&+&{}^3\Gamma^r_{su}\sqrt{\gamma}({}^3K^{su}-{}^3g^{su}\, {}^3K) \}
(\tau ,\vec \sigma )\approx 0,\nonumber \\
&&{}\nonumber \\
\lbrace {}^3{\tilde {\cal H}}_r(\tau ,\vec \sigma ),{}^3{\tilde {\cal H}}_s
(\tau ,{\vec \sigma}^{'})\rbrace &=&
{}^3{\tilde {\cal H}}_r(\tau ,{\vec
\sigma}^{'} )\, {{\partial \delta^3(\vec \sigma ,{\vec \sigma}^{'})}\over
{\partial \sigma^s}} + {}^3{\tilde {\cal H}}_s(\tau ,\vec \sigma ) {{\partial
\delta^3(\vec \sigma ,{\vec \sigma}^{'})}\over {\partial \sigma^r}},
\nonumber \\
\lbrace {\tilde {\cal H}}(\tau ,\vec \sigma ),{}^3{\tilde {\cal H}}_r(\tau ,
{\vec \sigma}^{'})\rbrace &=& {\tilde {\cal H}}(\tau ,\vec \sigma )
{{\partial \delta^3(\vec \sigma ,{\vec \sigma}^{'})}\over {\partial \sigma^r}},
\nonumber \\
\lbrace {\tilde {\cal H}}(\tau ,\vec \sigma ),{\tilde {\cal H}}(\tau ,{\vec
\sigma}^{'})\rbrace &=&[{}^3g^{rs}(\tau ,\vec \sigma ) {}^3{\tilde {\cal H}}_s
(\tau ,\vec \sigma )+\nonumber \\
&+&{}^3g^{rs}(\tau ,{\vec \sigma}^{'}){}^3{\tilde
{\cal H}}_s(\tau ,{\vec \sigma}^{'})]{{\partial \delta^3(\vec \sigma ,{\vec
\sigma}^{'})}\over {\partial \sigma^r}},
\label{a1}
\end{eqnarray}

\begin{eqnarray}
H_{(D)ADM}&=&H_{(c)ADM}+\int d^3\sigma \, [\lambda_N\, {\tilde \Pi}^N + \lambda
^{\vec N}_r\, {\tilde \Pi}^r_{\vec N}](\tau ,\vec \sigma ),\nonumber \\
H_{(c)ADM}&=& \int d^3\sigma [N\, {\tilde {\cal H}}+N_r\, {}^3{\tilde
{\cal H}}^r](\tau ,\vec \sigma )\approx 0.
\label{a2}
\end{eqnarray}

In Eqs.(25), (26) of I it was shown that the ADM action differs from the
Hilbert action by a surface term $\Sigma_{ADM}=
-{{\epsilon c^3}\over {8\pi G}} \int d^4x
\partial_{\mu} [\sqrt{{}^4g} ({}^3K l^{\mu}+{}^3a^{\mu})]$ (${}^3a^{\mu}=
l^{\beta} l^{\mu}{}_{;\beta}$ is the acceleration  of the observers
travelling along the normal $l^{\mu}$ at $\Sigma_{\tau}$), which becomes
$-\epsilon {{c^3}\over {8\pi G}} \int_{\partial U} d^3\Sigma \,
[N\sqrt{\gamma} \, {}^3K ]=
-\epsilon {{c^3}\over {8\pi G}} \int_{\tau_i}^{\tau_f} d\tau N_{(as)}(\tau )
\int_{S^2_{\tau ,\infty}} d^2\Sigma \sqrt{\gamma}\, {}^2K$ for suitable
boundary conditions. In this
case, following Ref.\cite{hh}, we assume that, given a subset $U\subset M^4$ of
spacetime, $\partial U$ consists of two
slices, $\Sigma_{\tau_i}$ (the initial one) and $\Sigma_{\tau_f}$ (the final
one) with outer normals $-l^{\mu}(\tau_i,\vec \sigma )$ and $l^{\mu}(\tau_f,
\vec \sigma )$ respectively, and of a surface $S_{\infty}$ near space infinity,
with outer unit (spacelike) normal $n^{\mu}(\tau ,\vec \sigma )$, orthogonal
to the slices [so that the normal $l^{\mu}(\tau ,\vec \sigma )$ to every slice
is asymptotically tangent to $S_{\infty}$]. The
3-surface $S_{\infty}$ is foliated by a family of 2-surfaces $S^2_{\tau ,
\infty}$ coming from its intersection with the slices $\Sigma_{\tau}$
[therefore, asymptotically $l^{\mu}(\tau ,\vec \sigma )$ is normal to the
corresponding $S^2_{\tau ,\infty}$]. The vector $b^{\mu}_{\tau}=z^{\mu}
_{\tau}=   N l^{\mu}+N^rb^{\mu}_r$ is not in general tangent to $S
_{\infty}$. In Ref.\cite{hh} it is shown that, if the lapse function
tends to an asymptotic limit $N_{(as)}(\tau )$, one gets  the trace ${}^2K$ of
the 2-dimensional extrinsic curvature of the 2-surface $S^2_{\tau ,\infty}=
S_{\infty}\cap \Sigma_{\tau}$, and  that for
$N_{(as)}(\tau )=1$, this surface term coincide with the ADM energy in
asymptotically flat spacetimes.

In Eq.(77) of I [see also Ref.\cite{witt,dew,hh}]
it was shown that in the Legendre transformation from the ADM
Lagrangian to the ADM canonical Hamiltonian there was a second surface term
$2 \int_{\partial S} d^2\Sigma_s [N_r\, {}^3{\tilde \Pi}^{rs}](\tau ,\vec
\sigma )=2 \int_{S^2_{\tau ,\infty}}\, d^2\Sigma [N_r\, {}^3{\tilde \Pi}^{rs}
n_s](\tau ,\vec \sigma )$ [$n_r={\hat b}^{\mu}_r n_{\mu}$]. Again it was
shown in Ref.\cite{hh} that, for constant asymptotic shifts, this surface
term reproduces the ADM momentum in asymptotically flat spacetimes.

In Refs,\cite{reg,reg1}, following
Refs.\cite{witt,dew}, it is shown that the differentiability of the ADM
canonical Hamiltonian $H_{(c)ADM}$ requires the introduction of the
following surface term $H_{(c)ADM} \mapsto H_{(c)ADM}+H_{\infty}$]

\begin{eqnarray}
H_{\infty}=-
\int_{S^2_{\tau ,\infty}} d^2\Sigma_u &&\{ \epsilon
k\sqrt{\gamma}\, {}^3g^{uv}\, {}^3g^{rs}
[N (\partial_r\, {}^3g_{vs}-\partial_v\, {}^3g_{rs})+ \nonumber \\
&+&\partial_uN ({}^3g_{rs}-\delta_{rs})
-\partial_rN ({}^3g_{sv}-\delta_{sv})] -2 N_r\, {}^3{\tilde \Pi}^{ru} \}
(\tau ,\vec \sigma ).
\label{a3}
\end{eqnarray}

\noindent Its use in Section III is connected with the ADM Poincar\'e
generators, so that it is equivalent to the two surfaces terms just discussed.

Let us show how this term arises from a suitable splitting of the
superhamiltonian and supermomentum constraints. By using
${}^3\Gamma^s_{sr}={1\over {\sqrt{\gamma}}} \partial_r \sqrt{\gamma},$
${}^3g^{uv}\, {}^3\Gamma^r_{uv}=-{1\over {\sqrt{\gamma}}} \partial_s
(\sqrt{\gamma}\, {}^3g^{rs}),$\hfill\break
${}^3R={}^3g^{uv} ({}^3\Gamma^r_{us}\, {}^3\Gamma^s_{vr}-
{}^3\Gamma^r_{uv}\, {}^3\Gamma^s_{sr})+
{1\over {\sqrt{\gamma}}} \partial_r [\sqrt{\gamma} ({}^3g^{uv}\,
{}^3\Gamma^r_{uv}-{}^3g^{ur}\, {}^3\Gamma^v_{vu})],$\hfill\break
and ${}^3g^{rs}\, {}^3\Gamma^u_{rs}-{}^3g^{uv}\,
{}^3\Gamma^s_{sv}={}^3g^{rs}\, {}^3g^{uv}(\partial_r\, {}^3g_{vs}-\partial_v\,
{}^3g_{rs})$, one gets

\begin{eqnarray}
&&\int d^3\sigma [ N {\tilde {\cal H}}+N_r\, {}^3{\tilde {\cal H}}^r]
(\tau ,\vec \sigma )=\nonumber \\
&&=\int d^3\sigma \{ \epsilon k N \partial_u[\sqrt{\gamma}({}^3g^{rs}\,
{}^3\Gamma^u_{rs}-{}^3g^{uv}\, {}^3\Gamma^s_{sv})]-2N_r\partial_u\,
{}^3{\tilde \Pi}^{ru}\} (\tau ,\vec \sigma )+\nonumber \\
&&+\int d^3\sigma \{ \epsilon kN[\sqrt{\gamma}\, {}^3g^{rs}({}^3\Gamma^u_{rv}\,
{}^3\Gamma^v_{su}-{}^3\Gamma^u_{rs}\, {}^3\Gamma^v_{vu})-\nonumber \\
&&-{1\over {2k\sqrt{\gamma}}}{}^3G_{rsuv}\, {}^3{\tilde \Pi}^{rs}\, {}^3{\tilde
\Pi}^{uv}]-2 N_r\, {}^3\Gamma^r_{su}\, {}^3{\tilde \Pi}^{su} \} (\tau ,\vec
\sigma )=\nonumber \\
&&=\int_{S^2_{\tau ,\infty}}d^2\Sigma_u
\{ \epsilon k N \sqrt{\gamma}\, {}^3g^{rs}\,
{}^3g^{uv} (\partial_r\, {}^3g_{vs}-\partial_v\, {}^3g_{rs})-2 N_r\,
{}^3{\tilde \Pi}^{ru} \} (\tau ,\vec \sigma )+\nonumber \\
&&+\int d^3\sigma \{ \epsilon
kN [\sqrt{\gamma}\, {}^3g^{rs}({}^3\Gamma^u_{rv}\,
{}^3\Gamma^v_{su}-{}^3\Gamma^u_{rs}\, {}^3\Gamma^v_{vu})-\nonumber \\
&&-{1\over {2k\sqrt{\gamma}}}{}^3G_{rsuv}\, {}^3{\tilde \Pi}^{rs}\, {}^3{\tilde
\Pi}^{uv} ]
-\epsilon k \partial_uN \sqrt{\gamma}\, {}^3g^{rs}\, {}^3g^{uv}(\partial_r\,
{}^3g_{vs}-\partial_v\, {}^3g_{rs})-\nonumber \\
&&-2 N_r\, {}^3\Gamma^r_{su}\, {}^3{\tilde \Pi}^{su}+2 \partial_uN_r\,
{}^3{\tilde \Pi}^{ru} \} (\tau ,\vec \sigma ).
\label{a4}
\end{eqnarray}

In Ref.\cite{reg1} it is noted that, with the boundary conditions of Refs.
\cite{reg,reg1} given in Section III, the term in $\partial_uN$ in the volume
integral diverges; the following (non-tensorial) regularization is
proposed: $\partial_r\, {}^3g_{vs}-\partial_v\, {}^3g_{rs}=\partial_r({}^3g
_{vs}-\delta_{vs})-\partial_v({}^3g_{rs}-\delta_{rs})$ [it may be thought as a
kind of subtraction of a static background metric in the spirit of Ref.
\cite{hh}; in this spirit one could think to use static
background metrics ${}^3f_{rs}$ different from $\delta_{rs}$: $\partial_r\,
{}^3g_{vs}-\partial_v\, {}^3g_{rs} \mapsto \partial_r({}^3g_{vs}-{}^3f_{vs})-
\partial_v({}^3g_{rs}-{}^3f_{rs})\not= \partial_r\, {}^3g_{vs}-\partial_v\,
{}^3g_{rs}$]. If we make a further integration by parts of the volume term
containing $\partial_uN$, we get the identity

\begin{eqnarray}
&&-\int_{S^2_{\tau ,\infty}} d^2\Sigma_u \{ \epsilon
k\sqrt{\gamma}\, {}^3g^{uv}\, {}^3g^{rs}
[N (\partial_r\, {}^3g_{vs}-\partial_v\, {}^3g_{rs})+ \partial_uN ({}^3g_{rs}-
\delta_{rs})-\nonumber \\
&&-\partial_rN ({}^3g_{sv}-\delta_{sv})] -2 N_r\, {}^3{\tilde \Pi}^{ru} \}
(\tau ,\vec \sigma )+\nonumber \\
&&+\int d^3\sigma [ N {\tilde {\cal H}}+N_r\, {}^3{\tilde {\cal H}}^r]
(\tau ,\vec \sigma )=\nonumber \\
&&{}\nonumber \\
&&=+\int d^3\sigma \{ \epsilon N [k \sqrt{\gamma}\, {}^3g^{rs} ({}^3\Gamma^u
_{rv}\, {}^3\Gamma^v_{su}-{}^3\Gamma^u_{rs}\, {}^3\Gamma^v_{vu})-{1\over
{2k\sqrt{\gamma}}}{}^3G_{rsuv}\, {}^3{\tilde \Pi}^{rs}\, {}^3{\tilde \Pi}^{uv} ]
+\nonumber \\
&&+\epsilon k ({}^3g_{vs}-\delta_{vs}) \partial_r[\sqrt{\gamma} \partial_uN
({}^3g^{rs}\, {}^3g^{uv}-{}^3g^{ru}\, {}^3g^{sv})]-\nonumber \\
&&-2 N_r\, {}^3\Gamma^r_{su}\, {}^3{\tilde \Pi}^{su} +2 \partial_uN_r\,
{}^3{\tilde \Pi}^{ru} \} (\tau ,\vec \sigma ).
\label{a5}
\end{eqnarray}

As shown in Eqs.(\ref{III10}), due to the splitting (\ref{III7}), which
explicitly separates
 the asymptotic part of the lapse and shift functions, the first member
of this equation is equal to \hfill\break
\hfill\break
${\tilde \lambda}_A(\tau ) P^A_{ADM}+{1\over 2}
{\tilde \lambda}_{AB}(\tau ) J^{AB}_{ADM}+\int d^3\sigma [N{\tilde {\cal H}}+
N_r\, {}^3{\tilde {\cal H}}^r](\tau ,\vec \sigma )$, \hfill\break
\hfill\break
with $P^A_{ADM}$ and
$J^{AB}_{ADM}$ being the ``strong improper Poincar\'e charges" given by the
surface integrals at spatial infinity of Eqs.(\ref{III11}).

This terminology derives from Ref.\cite{lusa} (see also Appendix D for the
treatment of the Hilbert action), where there is the definition
of the weak and strong improper conserved non-Abelian charges in Yang-Mills
theory and their derivation from the Noether identities implied by the second
Noether theorem. In this case one gets
(see Ref.\cite{sha} c) for the general theory):\hfill\break
 i) ``strong conserved improper currents" (their conservation is an identity
independent from the Euler-Lagrange equations), whose ``strong conserved
improper charges" are just surface integrals at spatial infinity like in
Eqs.(\ref{III11});\hfill\break
 ii) ``weak conserved improper
currents" (their conservation implies the Euler-Lagrange equations; it is a
form of first Noether theorem hidden in the second one), whose ``weak
conserved improper charges" are volume integrals, like the ones in the second
side of Eq.(\ref{a5}) and in Eqs.(\ref{III13});\hfill\break
 iii) the two kinds of charges differ by the Gauss law first class constraints
[like in Eq.(\ref{a5})] and, therefore, coincide when use is done of the
acceleration-independent Euler-Lagrange equations [i.e. the secondary first
class Gauss law constraints like it happens  in Eq.(\ref{a5})].

In ADM metric gravity it is difficult to check explicitly these statements,
because it is expressed in terms of lapse and shift functions and not in terms
of their splitted version of Eqs.(\ref{III7}). In this paper we shall adopt the
terminology ``strong" and ``weak" Poincar\'e charges to refer to surface and
volume integrals respectively, even if the strong charges are not strongly
conserved improper charges but only weakly conserved ones like the weak charges.

Let us now study the Noether identities \cite{sha,lusa} produced by the second
Noether theorem, which can be obtained from the quasi-invariance of the ADM
action $S_{ADM}$ under the gauge transformations generated by the first class
constraints, as it happens in the Yang-Mills case \cite{lusa}.

From Section V of I we know that the ADM Lagrangian density
is ${\cal L}_{ADM}(\tau
,\vec \sigma )=-\epsilon k (N \sqrt{\gamma} [{}^3R+{}^3K^{rs}\, {}^3K_{rs}-
({}^3K)^2])(\tau ,\vec \sigma )$. We shall use the notation \hfill\break
\hfill\break
$\delta_of(\sigma^A)={\bar f}(\sigma^A)-f(\sigma^A)$,
$\partial_B \delta_of(\sigma^A)=\delta_o \partial_Bf(\sigma^A)$, \hfill\break
\hfill\break
with $\sigma
^A=(\tau ;\vec \sigma )$ [instead one has $\delta f(\sigma^A)=\bar f({\bar
\sigma}^A)-f(\sigma^A)=\delta_o f(\sigma^A)+\partial_Bf(\sigma^A) \delta
\sigma^B$, if ${\bar \sigma}^B=\sigma^B+\delta \sigma^B(\sigma^A)$ with $\delta
\sigma^B\not= 0$].
The form of the general variation of the Lagrangian density is [
${\cal L}_{ADM}$ is considered a function of the following fields:
i) $N$ with dependence on $N$,
$\partial_{\tau}N$, $\partial_rN$;
ii) $N_s$ with dependence on $N_s$, $\partial_{\tau}N_s$,
$\partial_rN_s$; iii) ${}^3g_{rs}$ with dependence on ${}^3g_{rs}$, $\partial
_{\tau}\, {}^3g_{rs}$, $\partial_u\, {}^3g_{rs}$, $\partial_u\partial_v\,
{}^3g_{rs}$]

\begin{eqnarray}
&&\delta_o {\cal L}_{ADM}(\tau ,\vec \sigma )=\nonumber \\
&&={{\partial {\cal L}_{ADM}(\tau ,
\vec \sigma )}\over {\partial N(\tau ,\vec \sigma )}}\delta_o N(\tau ,\vec
\sigma )+{{\partial {\cal L}_{ADM}(\tau ,\vec \sigma )}\over {\partial \partial
_{\tau}N(\tau ,\vec \sigma )}}\delta_o \partial_{\tau}N(\tau ,\vec \sigma )+
{{\partial {\cal L}_{ADM}(\tau ,\vec \sigma )}\over {\partial \partial
_rN(\tau ,\vec \sigma )}}\delta_o \partial_rN(\tau ,\vec \sigma )+
\nonumber \\
&&+{{\partial {\cal L}_{ADM}(\tau ,\vec \sigma )}\over {\partial N_s(\tau ,\vec
\sigma )}}\delta_o N_s(\tau ,\vec \sigma )+{{\partial {\cal L}_{ADM}(\tau ,\vec
\sigma )}\over {\partial \partial_{\tau}N_s(\tau ,\vec \sigma )}}\delta_o
\partial_{\tau}N_s(\tau ,\vec \sigma )+{{\partial {\cal L}_{ADM}(\tau ,\vec
\sigma )}\over {\partial \partial_rN_s(\tau ,\vec \sigma )}}\delta_o \partial_r
N_s(\tau ,\vec \sigma )+\nonumber \\
&&+{{\partial {\cal L}_{ADM}(\tau ,\vec \sigma )}\over {\partial \, {}^3g_{uv}
(\tau ,\vec \sigma )}}\delta_o \, {}^3g_{uv}(\tau ,\vec \sigma )+{{\partial
{\cal L}_{ADM}(\tau ,\vec \sigma )}\over {\partial \partial_{\tau}\, {}^3g
_{uv}(\tau ,\vec \sigma )}}\delta_o \partial_{\tau}\, {}^3g_{uv}(\tau ,\vec
\sigma )+{{\partial {\cal L}_{ADM}(\tau ,\vec \sigma )}\over {\partial
\partial_r\, {}^3g_{uv}(\tau ,\vec \sigma )}}\delta_o \partial_r\, {}^3g_{uv}
(\tau ,\vec \sigma )+\nonumber \\
&&+{{\partial {\cal L}_{ADM}(\tau ,\vec \sigma )}\over {\partial \partial_r
\partial_s\, {}^3g_{uv}(\tau ,\vec \sigma )}}
 \delta_o\partial_r\partial_s\, {}^3g_{uv}(\tau ,\vec
\sigma )=\nonumber \\
&&=\delta_o N(\tau ,\vec \sigma ) \Big( {{\partial {\cal L}_{ADM}(\tau ,\vec
\sigma )}\over {\partial N(\tau ,\vec \sigma )}}-\partial_{\tau} {{\partial
{\cal L}_{ADM}(\tau ,\vec \sigma )}\over {\partial \partial_{\tau}N(\tau
,\vec \sigma )}} -\partial_r
{{\partial {\cal L}_{ADM}(\tau ,\vec \sigma )}\over {\partial \partial
_rN(\tau ,\vec \sigma )}}\Big) +\nonumber \\
&&+\delta_o N_s(\tau ,\vec \sigma )\Big(
{{\partial {\cal L}_{ADM}(\tau ,\vec \sigma )}\over {\partial N_s(\tau ,\vec
\sigma )}}-\partial_{\tau} {{\partial {\cal L}_{ADM}(\tau ,\vec \sigma )}\over
{\partial \partial_{\tau}N_s(\tau ,\vec \sigma )}}
-\partial_r{{\partial {\cal L}_{ADM}(\tau ,\vec \sigma )}\over
{\partial \partial_rN_s(\tau ,\vec \sigma )}}\Big) +\nonumber \\
&&+\delta_o \, {}^3g_{uv}(\tau ,\vec \sigma )[{{\partial {\cal L}_{ADM}(\tau ,
\vec \sigma )}\over {\partial \, {}^3g_{uv}(\tau ,\vec \sigma )}}-\partial
_{\tau}{{\partial {\cal L}_{ADM}(\tau ,\vec \sigma )}\over {\partial
\partial_{\tau}\, {}^3g_{uv}(\tau ,\vec \sigma )}}-\partial_r{{\partial
{\cal L}_{ADM}(\tau ,\vec \sigma )}\over {\partial \partial_r\, {}^3g
_{uv}(\tau ,\vec \sigma )}}+\nonumber \\
&&+\partial_r\partial_s
{{\partial {\cal L}_{ADM}(\tau ,\vec \sigma )}\over {\partial \partial_r
\partial_s\, {}^3g_{uv}(\tau ,\vec \sigma )}}]+
\partial_{\tau}[{{\partial {\cal L}_{ADM}(\tau ,\vec \sigma )}\over {\partial
\partial_{\tau}\, N(\tau ,\vec \sigma )}} \delta_oN(\tau ,\vec \sigma )+
\nonumber\\
&&+{{\partial {\cal L}_{ADM}(\tau ,\vec \sigma )}\over {\partial
\partial_{\tau}\, N_r(\tau ,\vec \sigma )}} \delta_oN_r(\tau ,\vec \sigma )+
{{\partial {\cal L}_{ADM}(\tau ,\vec \sigma )}\over {\partial
\partial_{\tau}\, {}^3g_{uv}(\tau ,\vec \sigma )}}\delta_o \, {}^3g_{uv}(\tau ,
\vec \sigma )]+\nonumber \\
&&+\partial_r\Big[{{\partial {\cal L}_{ADM}(\tau ,\vec \sigma )}\over {\partial
\partial_rN_s(\tau ,\vec \sigma )}}\delta_o N_s(\tau ,\vec \sigma )+\nonumber \\
&&+\Big(  {{\partial {\cal L}_{ADM}(\tau ,\vec \sigma )}\over {\partial
\partial_r\, {}^3g_{uv}(\tau ,\vec \sigma )}}-2\partial_s
{{\partial {\cal L}_{ADM}(\tau ,\vec \sigma )}\over {\partial \partial_r
\partial_s\, {}^3g_{uv}(\tau ,\vec \sigma )}}
\Big) \delta_o \, {}^3g_{uv}(\tau ,\vec \sigma )\Big] +\nonumber \\
&&+\partial_r\partial_s \Big(
{{\partial {\cal L}_{ADM}(\tau ,\vec \sigma )}\over {\partial \partial_r
\partial_s\, {}^3g_{uv}(\tau ,\vec \sigma )}} \delta_o\, {}^3g_{uv}(\tau
,\vec \sigma ) \Big) =\nonumber \\
&&=\delta_o N(\tau ,\vec \sigma ) L_N(\tau ,\vec \sigma )+\delta_o N_s(\tau
,\vec \sigma ) L^s_{\vec N}(\tau ,\vec \sigma )+\delta_o \, {}^3g_{uv}(\tau
,\vec \sigma ) L^{uv}_g(\tau ,\vec \sigma )+\nonumber \\
&&+\partial_{\tau} \Big[ {\tilde \pi}_N(\tau ,\vec \sigma )\delta_oN(\tau
,\vec \sigma )+{\tilde \pi}^r_{\vec N}(\tau ,\vec \sigma )\delta_oN_r+
{}^3{\tilde \Pi}^{uv}(\tau ,\vec \sigma ) \delta_o \,
{}^3g_{uv}(\tau ,\vec \sigma )\Big] +\nonumber \\
&&+\partial_r\Big[ {{\partial {\cal L}_{ADM}(\tau ,\vec \sigma )}\over {\partial
\partial_rN_s(\tau ,\vec \sigma )}}\delta_o N_s(\tau ,\vec \sigma )+
\Big( {{\partial
{\cal L}_{ADM}(\tau ,\vec \sigma )}\over {\partial \partial_r\, {}^3g_{uv}
(\tau ,\vec \sigma )}}-2\partial_s
{{\partial {\cal L}_{ADM}(\tau ,\vec \sigma )}\over {\partial \partial_r
\partial_s\, {}^3g_{uv}(\tau ,\vec \sigma )}}
\Big) \delta_o \, {}^3g_{uv}(\tau ,\vec \sigma )\Big] +\nonumber \\
&&+\partial_r\partial_s \Big(
{{\partial {\cal L}_{ADM}(\tau ,\vec \sigma )}\over {\partial \partial_r
\partial_s\, {}^3g_{uv}(\tau ,\vec \sigma )}} \delta_o\, {}^3g_{uv}(\tau
,\vec \sigma ) \Big) ,
\label{a7}
\end{eqnarray}

\noindent where $L_N(\tau ,\vec \sigma )\, {\buildrel \circ \over =}\, 0$,
$L^s_{\vec N}(\tau ,\vec \sigma )\, {\buildrel \circ \over =}\, 0$ [i.e.
${\tilde {\cal H}}(\tau ,\vec \sigma )\approx 0$, ${}^3{\tilde {\cal H}}
^r(\tau ,\vec \sigma )\approx 0$],
$L^{uv}_g(\tau ,\vec \sigma )\, {\buildrel \circ \over =}\, 0$, are the
Euler-Lagrange equations given in Eqs.(71) of I.

By using

\begin{eqnarray}
&&\delta_o \sqrt{\gamma} ={1\over 2} \sqrt{\gamma} \, {}^3g^{rs} \delta_o\,
{}^3g_{rs},\quad\quad
\delta_o {1\over {\sqrt{\gamma}}}=-{1\over {2\sqrt{\gamma}}} {}^3g^{rs}
\delta_o\, {}^3g_{rs},\nonumber \\
&&\delta_o\, {}^3g^{rs}=-{}^3g^{ru}\, {}^3g^{sv} \delta_o\, {}^3g_{uv},
\nonumber \\
&&\delta_o\, {}^3\Gamma^l_{uv}=-{}^3g^{lr}\, {}^3\Gamma^s_{uv} \delta_o\,
{}^3g_{rs}+{1\over 2} [{}^3g^{lr}(\delta^n_u\delta^s_v+\delta^n_v\delta^s_u)-
{}^3g^{ln}\delta^r_u\delta^s_v] \partial_n \delta_o\, {}^3g_{rs},\nonumber \\
&&\delta_o (\sqrt{\gamma}\, {}^3R)=-\sqrt{\gamma} ({}^3R^{rs}-{1\over 2}\,
{}^3g^{rs}\, {}^3R)\delta_o\, {}^3g_{rs}+\nonumber \\
&&+\sqrt{\gamma} [{}^3g_{rs}\delta_o\,
{}^3g^{rs|u}-\delta_o\, {}^3g^{ru}{}_{|r}]_{|u},
\nonumber \\
&&\delta_o\, {}^3K_{rs}=-{1\over N}\, {}^3K_{rs} \delta_oN+{1\over {2N}} \{
(\delta_o N_u)_{ |v}+(\delta_o N_v)_{ |u}+2N^r\, {}^3\Gamma^s_{uv} \delta_o\,
{}^3g_{rs}-\nonumber \\
&&-[N^r (\delta^n_u\delta^s_v+\delta^n_v\delta^s_u)-N^n\delta^r_u\delta^s_v]
\partial_n\delta_o\, {}^3g_{rs}-\partial_{\tau}\delta_o\, {}^3g_{rs} \} ,
\nonumber \\
&&\delta_o\, {}^3K=-{1\over N} {}^3K \delta_oN+{1\over N} {}^3g^{rs}(\delta_o
N_r)_{ |s}-[{}^3K^{rs}-{{N^r}\over N} {}^3g^{uv}\, {}^3\Gamma^s_{uv}] \delta_o\,
{}^3g_{rs}-\nonumber \\
&&-{1\over {2N}}(2N^r\, {}^3g^{ns}-N^n\, {}^3g^{rs})\partial_n\delta_o\, {}^3g
_{rs}-{1\over {2N}} {}^3g^{rs} \partial_{\tau}\delta_o\, {}^3g_{rs},
\label{a8}
\end{eqnarray}

\noindent we get the following form for explicit the variation of the ADM
Lagrangian density

\begin{eqnarray}
\delta_o {\cal L}_{ADM}(\tau ,\vec \sigma )&=&-\epsilon k \{ \sqrt{\gamma}
[{}^3R+{}^3K_{rs}\, {}^3K^{rs}-({}^3K)^2]\delta_o N +N\delta_o(\sqrt{\gamma}\,
{}^3R)+\nonumber \\
&+&N[{}^3K_{rs}\, {}^3K^{rs}-({}^3K)^2]\delta_o\sqrt{\gamma}+2N \sqrt{\gamma}
[{}^3K^{rs} \delta_o\, {}^3K_{rs}-{}^3K \delta_o\, {}^3K] \} =\nonumber \\
&=&-{\tilde {\cal H}} \delta_oN -2\, {}^3{\tilde \Pi}^{rs} (\delta_oN_r)_{ |s}
-N \{ \epsilon k \sqrt{\gamma} ({}^3R^{rs}-{1\over 2} {}^3g^{rs}\, {}^3R)+
\nonumber \\
&+&2(N^r\, {}^3\Gamma^s_{uv}\, {}^3{\tilde \Pi}^{uv}-{}^3K^r{}_m\, {}^3{\tilde
\Pi}^{ms})+{1\over 2}\epsilon k\sqrt{\gamma}({}^3K^{uv}\, {}^3K_{uv}-
({}^3K)^2){}^3g^{rs} \} \delta_o\, {}^3g_{rs}+\nonumber \\
&+&(N^n\, {}^3{\tilde \Pi}^{rs}-2N^r\, {}^3{\tilde \Pi}^{ns})\partial_n\delta_o
\, {}^3g_{rs}+{}^3{\tilde \Pi}^{rs} \partial_{\tau}\delta_o\, {}^3g_{rs}-
\nonumber \\
&&-\epsilon k N \partial_l \{ \sqrt{\gamma} [{1\over 2}\Big( {}^3g^{rs}\,
{}^3g^{uv}\, {}^3g^{lm}-2({}^3g^{ur}\, {}^3g^{vs}\, {}^3g^{lm}+{}^3g^{uv}\,
{}^3g^{lr}\, {}^3g^{ms})\Big) \cdot \nonumber \\
&&(\partial_v\, {}^3g_{um}-\partial_m\, {}^3g_{uv})\delta_o\, {}^3g_{rs}+
({}^3g^{rn}\, {}^3g^{ls}-{}^3g^{rs}\, {}^3g^{ln})\partial_n\delta_o\, {}^3g
_{rs}] \} ,
\label{a9}
\end{eqnarray}

We want to study the invariance properties of
$S_{ADM}=\int d\tau d^3\sigma {\cal L}_{ADM}(\tau ,\vec \sigma )$
under the most general gauge transformations generated by the first class
constraints, namely generated by \hfill\break
\hfill\break
$G[\alpha ,\alpha_r,\lambda_N,\lambda^{\vec N}
_r]=\int d^3\sigma [\alpha {\tilde {\cal H}}+
\alpha_r\, {}^3{\tilde {\cal H}}^r+\lambda_N {\tilde \pi}^N+\lambda_r^{\vec N}
{\tilde \pi}^r_{\vec N}](\tau ,\vec \sigma )$. \hfill\break
\hfill\break
They produce the following
variations of the Lagrangian variables

\begin{eqnarray}
&&\delta_o N(\tau ,\vec \sigma )=\lbrace N(\tau ,\vec \sigma ),G\rbrace
= \lambda_N(\tau ,\vec \sigma )\, {\buildrel \circ \over =}\, \partial
_{\tau}N(\tau ,\vec \sigma ),\nonumber \\
&&\delta_o N_r(\tau ,\vec \sigma )=\lbrace N_r(\tau ,\vec \sigma ),G
\rbrace = \lambda_r^{\vec N}(\tau ,\vec \sigma )\, {\buildrel \circ \over =}\,
\partial_{\tau}N_r(\tau ,\vec \sigma ),\nonumber \\
&&\delta_o \, {}^3g_{rs}(\tau ,\vec \sigma )=\lbrace {}^3g(\tau ,\vec \sigma ),
G\rbrace = [\alpha_{r|s}+\alpha_{s|r}-2\alpha \, {}^3K_{rs}](\tau
,\vec \sigma )=\nonumber \\
&&=\Big( \alpha_{r|s}+\alpha_{s|r}-{{\alpha}\over N} [N_{r|s}+N_{s|r}-
\partial_{\tau}\, {}^3g_{rs}] \Big) (\tau ,\vec \sigma ),
\label{a6}
\end{eqnarray}

\noindent  One sees that on the solutions of the Hamilton-Dirac
equations one has $\lambda_N\, {\buildrel \circ \over =}\, \partial_{\tau}N=
\partial_{\tau}\alpha$, $\lambda^{\vec N}_r\, {\buildrel \circ \over =}\,
\partial_{\tau}N_r=\partial_{\tau}\alpha_r$, if $\alpha =N$, $\alpha_r=N_r$, as
it happens if G is identified with the Dirac Hamiltonian.

Since it is difficult to find the associated quasi-invariances directly, let us
look at the various Noether transformations separately.

1) Under the variations generated by  \hfill\break
\hfill\break
$G_1[\lambda_N,\lambda^{\vec N}_r]=\int
d^3\sigma [\lambda_N {\tilde \pi}^N+\lambda^{\vec N}_r {\tilde \pi}^r_{\vec N}]
(\tau ,\vec \sigma )=G[0,0,\lambda_N,\lambda^{\vec N}_r]$ \hfill\break
\hfill\break
one gets the Noether identities

\begin{eqnarray}
\delta_o{\cal L}_{ADM}(\tau ,\vec \sigma )&=& \lambda_N(\tau ,\vec \sigma )
L_N(\tau ,\vec \sigma )+\lambda^{\vec N}_s(\tau ,\vec \sigma ) L^s_{\vec N}
(\tau ,\vec \sigma )+\nonumber \\
&+&\partial_{\tau} [{\tilde \pi}_N(\tau ,\vec \sigma )\lambda_N(\tau ,\vec
\sigma )+{\tilde \pi}^r_{\vec N}(\tau ,\vec \sigma )\lambda^{\vec N}_r(\tau
,\vec \sigma )]+\nonumber \\
&+&\partial_r [{{\partial {\cal L}_{ADM}(\tau ,\vec \sigma )}\over {\partial
\partial_rN_s(\tau ,\vec \sigma )}} \lambda^{\vec N}_s(\tau ,\vec \sigma )]
\equiv \nonumber \\
&\equiv& -
\Big( \lambda_N {\tilde {\cal H}} +\lambda^{\vec N}_r{\tilde {\cal H}}^r+
\partial_r[2\, {}^3{\tilde \Pi}^{rs} \lambda^{\vec N}_s]\Big) (\tau
,\vec \sigma )\, {\buildrel \circ \over =}\nonumber \\
&{\buildrel \circ \over =}&\, -\partial_r[2\, {}^3{\tilde \Pi}^{rs}
\lambda^{\vec N}_s](\tau ,\vec \sigma ),
\label{a10}
\end{eqnarray}

\noindent by using the definition of the ADM momentum and the acceleration
independent Euler-Lagrange equations corresponding to ${\tilde {\cal H}}
(\tau ,\vec \sigma )\approx 0$, ${\tilde {\cal H}}^r(\tau ,\vec \sigma )
\approx 0$ [see Eqs.(70) of I].

By equating the coefficients of $\partial_{\tau}\lambda_N$, $\lambda_N$ and of
$\partial_{\tau}\lambda_r^{\vec N}$, $\partial_s\lambda_r^{\vec N}$,
$\lambda_r^{\vec N}$ on the two sides of the previous identity one gets the
following Noether identities equivalent to the primary and secondary
constraints in the Hamiltonian formulation

\begin{eqnarray}
&&{\tilde \pi}_N \equiv 0,\nonumber \\
&&0 \equiv \partial_{\tau}{\tilde \pi}_N\equiv -{\tilde {\cal H}}-L_N\quad\quad
\Rightarrow {\tilde {\cal H}} \equiv -L_N\, {\buildrel \circ \over =}\, 0,
\label{a10a}
\end{eqnarray}

\begin{eqnarray}
&&{{\partial {\cal L}_{ADM}}\over {\partial \partial_rN_s}}\equiv -2\,
{}^3{\tilde \Pi}^{rs}=-2 {{\partial {\cal L}_{ADM}}\over {\partial \partial
_{\tau}\, {}^3g_{rs}}},\nonumber \\
&&{\tilde \pi}^r_{\vec N} \equiv 0,\nonumber \\
&&0 \equiv \partial_{\tau} {\tilde \pi}^r_{\vec N} \equiv -{\tilde {\cal H}}^r
-L^r_{\vec N}-\partial_s(2\, {}^3{\tilde \Pi}^{rs}+
{{\partial {\cal L}_{ADM}}\over {\partial \partial_rN_s}} )\equiv \nonumber \\
&&\equiv -{\tilde {\cal H}}^r-L^r_{\vec N}\quad\quad \Rightarrow {\tilde
{\cal H}}^r \equiv -L^r_{\vec N}\, {\buildrel \circ \over =}\, 0.
\label{a10b}
\end{eqnarray}

2) Since $S_{ADM}=\int d\tau d^3\sigma {\cal L}_{ADM}(\tau ,\vec \sigma )$
is invariant under 3-diffeomorphisms of $Diff\, \Sigma_{\tau}$ [$d^3\sigma
{\cal L}_{ADM}(\tau ,\vec \sigma )$ is a scalar under $\sigma^r
\mapsto {\bar \sigma}^r=\sigma^r+\delta \sigma^r(\vec \sigma )$; see Eqs.(31)
of I for the following formulas], under the variations

\begin{eqnarray}
\delta \sigma^r &=& \xi^r(\vec \sigma ),\nonumber \\
\delta^D_oN(\tau ,\vec \sigma )&=&-\xi^r(\vec \sigma )\partial_rN(\tau
,\vec \sigma ),\nonumber \\
\delta^D_oN^r(\tau ,\vec \sigma )&=&[\partial_s\xi^r(\vec \sigma )-
\delta^r_s\xi^u(\vec \sigma )\partial_u]N^s(\tau ,\vec \sigma )={\cal L}
_{-\xi^t\partial_t}N^r(\tau ,\vec \sigma ),\nonumber \\
\delta^D_oN_r(\tau ,\vec \sigma )&=&-[\partial_r\xi^s(\vec \sigma )+\delta^s_r
\xi^u(\vec \sigma )\partial_u]N_s(\tau ,\vec \sigma ),\nonumber \\
\delta^D_o\, {}^3g_{rs}(\tau ,\vec \sigma )&=&-[\delta^u_r\partial_s\xi
^v(\vec \sigma )+\delta^v_s\partial_r\xi^u(\vec \sigma )+\delta^u_r\delta^v
_s\xi^t(\vec \sigma )\partial_t]{}^3g_{uv}(\tau ,\vec \sigma )=\nonumber \\
&=&-[\xi_{r|s}+\xi_{s|r}](\tau ,\vec \sigma )={\cal L}_{-\xi^t\partial
_t} {}^3g_{rs}(\tau ,\vec \sigma )=\nonumber \\
&=&\lbrace {}^3g_{rs}(\tau ,\vec \sigma ), -\int d^3\sigma^{'} \xi
_u(\vec \sigma ) {\tilde {\cal H}}^u(\tau ,{\vec \sigma}^{'}) \rbrace ,
\nonumber \\
&&\Downarrow \nonumber \\
\delta^DN(\tau ,\vec \sigma )&=&\delta^D_oN(\tau ,\vec \sigma )+\xi^r(\vec
\sigma )\partial_rN(\tau ,\vec \sigma )=0,\nonumber \\
\delta^DN^r(\tau ,\vec \sigma )&=&\delta^D_oN^r(\tau ,\vec \sigma )+\xi^u
(\vec \sigma )\partial_uN^r(\tau ,\vec \sigma ),\nonumber \\
\delta^D {}^3g_{rs}(\tau ,\vec \sigma )&=&\delta^D_o {}^3g_{rs}(\tau ,\vec
\sigma )+\xi^u(\vec \sigma )\partial_u\, {}^3g_{rs}(\tau ,\vec \sigma ),
\label{a11}
\end{eqnarray}

\noindent we get the following Noether identity

\begin{eqnarray}
&&\delta^D_o {\cal L}_{ADM}(\tau ,\vec \sigma )=\nonumber \\
&&=\delta_o^D N(\tau ,\vec \sigma ) L_N(\tau ,\vec \sigma )+\delta^D_o N_s(\tau
,\vec \sigma ) L^s_{\vec N}(\tau ,\vec \sigma )+\delta_o^D \, {}^3g_{uv}(\tau
,\vec \sigma ) L^{uv}_g(\tau ,\vec \sigma )+\nonumber \\
&&+\partial_{\tau} \Big[ {\tilde \pi}^N(\tau ,\vec \sigma )\delta^D_oN(\tau
,\vec \sigma )+{\tilde \pi}^r_{\vec N}(\tau ,\vec \sigma )\delta^D_oN_r(\tau
,\vec \sigma )+{}^3{\tilde \Pi}^{uv}(\tau ,\vec \sigma ) \delta_o^D \,
{}^3g_{uv}(\tau ,\vec \sigma )\Big] +\nonumber \\
&&+\partial_r\Big[ {{\partial {\cal L}_{ADM}(\tau ,\vec \sigma )}\over {\partial
\partial_rN_s(\tau ,\vec \sigma )}}\delta_o^D N_s(\tau ,\vec \sigma )
+\Big( {{\partial
{\cal L}_{ADM}(\tau ,\vec \sigma )}\over {\partial \partial_r\, {}^3g_{uv}
(\tau ,\vec \sigma )}}-2\partial_s
{{\partial {\cal L}_{ADM}(\tau ,\vec \sigma )}\over {\partial \partial_r
\partial_s\, {}^3g_{uv}(\tau ,\vec \sigma )}}
\Big) \delta_o^D \, {}^3g_{uv}(\tau ,\vec \sigma )\Big] +\nonumber \\
&&+\partial_r\partial_s \Big(
{{\partial {\cal L}_{ADM}(\tau ,\vec \sigma )}\over {\partial \partial_r
\partial_s\, {}^3g_{uv}(\tau ,\vec \sigma )}} \delta_o^D\, {}^3g_{uv}(\tau
,\vec \sigma ) \Big) +\nonumber \\
&&+\partial_r[{\cal L}_{ADM} \xi^r](\tau ,\vec \sigma )\equiv 0.
\label{a12}
\end{eqnarray}

As shown in the paper c) in Ref.\cite{sha}, the same
Noether identities can be obtained with the transformations \hfill\break
\hfill\break
$\delta_o \sigma^r
=0$, $\delta_oN=\delta^D_oN$, $\delta_oN_r=\delta^D_oN_r$, $\delta_o\, {}^3g
_{rs}=\delta^D_o\, {}^3g_{rs}$, \hfill\break
\hfill\break
since now we get

\begin{eqnarray}
&&\delta_o{\cal L}_{ADM}(\tau ,\vec \sigma )=\nonumber \\
&&=\delta_o N(\tau ,\vec \sigma ) L_N(\tau ,\vec \sigma )+\delta_o N_s(\tau
,\vec \sigma ) L^s_{\vec N}(\tau ,\vec \sigma )+\delta_o \, {}^3g_{uv}(\tau
,\vec \sigma ) L^{uv}_g(\tau ,\vec \sigma )+\nonumber \\
&&+\partial_{\tau} \Big[ {\tilde \pi}^N(\tau ,\vec \sigma )\delta_oN(\tau
,\vec \sigma )+{\tilde \pi}^r_{\vec N}(\tau ,\vec \sigma )\delta_oN_r(\tau
,\vec \sigma )+{}^3{\tilde \Pi}^{uv}(\tau ,\vec \sigma ) \delta_o \,
{}^3g_{uv}(\tau ,\vec \sigma )\Big] +\nonumber \\
&&+\partial_r\Big[ {{\partial {\cal L}_{ADM}(\tau ,\vec \sigma )}\over {\partial
\partial_rN_s(\tau ,\vec \sigma )}}\delta_o N_s(\tau ,\vec \sigma )
+\Big( {{\partial
{\cal L}_{ADM}(\tau ,\vec \sigma )}\over {\partial \partial_r\, {}^3g_{uv}
(\tau ,\vec \sigma )}}-2\partial_s
{{\partial {\cal L}_{ADM}(\tau ,\vec \sigma )}\over {\partial \partial_r
\partial_s\, {}^3g_{uv}(\tau ,\vec \sigma )}}
\Big) \delta_o \, {}^3g_{uv}(\tau ,\vec \sigma )\Big] +\nonumber \\
&&+\partial_r\partial_s \Big(
{{\partial {\cal L}_{ADM}(\tau ,\vec \sigma )}\over {\partial \partial_r
\partial_s\, {}^3g_{uv}(\tau ,\vec \sigma )}} \delta_o\, {}^3g_{uv}(\tau
,\vec \sigma ) \Big) \equiv \nonumber \\
&&\equiv -\partial_r({\cal L}_{ADM}\xi^r)
(\tau ,\vec \sigma ).
\label{a13}
\end{eqnarray}

If the 3-diffeomorphisms are $\tau$-dependent, namely in Eqs.(\ref{a11})
the variations $\delta^D_o$ are replaced by $\delta^{D\tau}_o$
with $\delta \sigma^r=\xi^r(\tau ,\vec \sigma )$, one has

\begin{eqnarray}
\delta^{D\tau}_o\partial_{\tau}\, {}^3g_{rs}&=&\partial_{\tau} \delta^{D\tau}
_o\, {}^3g_{rs}=\delta^D_o\partial_{\tau}\, {}^3g_{rs}+\delta^{'} \partial
_{\tau}\, {}^3g_{rs}=\nonumber \\
&=&\delta^D_o \partial_{\tau}\, {}^3g_{rs}-[\delta^u_r\partial_s\partial
_{\tau}\xi^v+\delta^v_s\partial_r\partial_{\tau}\xi^u+\delta^u_r\delta^v_s
\partial_{\tau}\xi^t\partial_t]{}^3g_{uv},\nonumber \\
&&\Downarrow \nonumber \\
\delta^{D\tau}_o\, {}^3K_{rs}&=&\delta^D_o\, {}^3K_{rs}+\delta^{'}\,
{}^3K_{rs}=\nonumber \\
&=&\delta^D_o\, {}^3K_{rs}+{1\over {2N}}[\delta^u_r\partial_s\partial
_{\tau}\xi^v+\delta^v_s\partial_r\partial_{\tau}\xi^u+\delta^u_r\delta^v_s
\partial_{\tau}\xi^t\partial_t]{}^3g_{uv},\nonumber \\
&&\Downarrow \nonumber \\
\delta^{D\tau}_o {\cal L}_{ADM}(\tau ,\vec \sigma )&=&\delta_o{\cal L}
_{ADM}(\tau ,\vec \sigma )-[2\epsilon k N\sqrt{\gamma}({}^3K^{rs}-{}^3K\,
{}^3g^{rs})\delta^{'}\, {}^3K_{rs}](\tau ,\vec \sigma )\equiv \nonumber \\
&\equiv&-\partial_r({\cal L}_{ADM}\xi^r)(\tau ,\vec \sigma )
-[{}^3{\tilde \Pi}^{rs} \delta^{'}\, {}^3K_{rs}](\tau ,\vec \sigma )=
\nonumber \\
&=&-\partial_r({\cal L}_{ADM}\xi^r)(\tau ,\vec \sigma )-2[{}^3g
_{rv}\, {}^3{\tilde \Pi}^{rs}\partial_s\partial_{\tau}\xi^v](\tau ,\vec
\sigma )-\nonumber \\
&-&[{}^3{\tilde \Pi}^{rs}\partial_u\, {}^3g_{rs} \partial
_{\tau}\xi^u](\tau ,\vec \sigma )=\nonumber \\
&=&\delta^D_o{\cal L}_{ADM}(\tau ,\vec \sigma )-[\partial
_{\tau}\xi^u\, {}^3g_{ur}{\tilde {\cal H}}^r](\tau ,\vec \sigma )-\nonumber \\
&-&\partial_s[2 {}^3g_{ru}\, {}^3{\tilde \Pi}^{rs}\partial
_{\tau}\xi^u](\tau ,\vec \sigma ).
\label{a14}
\end{eqnarray}

Therefore, for \hfill\break
\hfill\break
$\delta \sigma^r=0$, $\delta_oN=\delta^{D\tau}_oN$, $\delta_oN_r
=\delta^{D\tau}_oN_r$, $\delta_o\, {}^3g_{rs}=\delta^{D\tau}_o\, {}^3g_{rs}$,
\hfill\break
\hfill\break
we get

\begin{eqnarray}
&&\delta_o{\cal L}_{ADM}(\tau ,\vec \sigma )=\nonumber \\
&&=\delta_o N(\tau ,\vec \sigma ) L_N(\tau ,\vec \sigma )+\delta_o N_s(\tau
,\vec \sigma ) L^s_{\vec N}(\tau ,\vec \sigma )+\delta_o \, {}^3g_{uv}(\tau
,\vec \sigma ) L^{uv}_g(\tau ,\vec \sigma )+\nonumber \\
&&+\partial_{\tau} \Big[ {\tilde \pi}^N(\tau ,\vec \sigma )
\delta_oN(\tau ,\vec \sigma )
+{\tilde \pi}^r_{\vec N}(\tau ,\vec \sigma )\delta_oN_r(\tau ,\vec \sigma )+
{}^3{\tilde \Pi}^{uv}(\tau ,\vec \sigma ) \delta_o \,
{}^3g_{uv}(\tau ,\vec \sigma )\Big] +\nonumber \\
&&+\partial_r\Big[ {{\partial {\cal L}_{ADM}(\tau ,\vec \sigma )}\over {\partial
\partial_rN_s(\tau ,\vec \sigma )}}\delta_o N_s(\tau ,\vec \sigma )+
\Big( {{\partial
{\cal L}_{ADM}(\tau ,\vec \sigma )}\over {\partial \partial_r\, {}^3g_{uv}
(\tau ,\vec \sigma )}}-2\partial_s
{{\partial {\cal L}_{ADM}(\tau ,\vec \sigma )}\over {\partial \partial_r
\partial_s\, {}^3g_{uv}(\tau ,\vec \sigma )}}
\Big) \delta_o \, {}^3g_{uv}(\tau ,\vec \sigma )\Big] +\nonumber \\
&&+\partial_r\partial_s \Big(
{{\partial {\cal L}_{ADM}(\tau ,\vec \sigma )}\over {\partial \partial_r
\partial_s\, {}^3g_{uv}(\tau ,\vec \sigma )}} \delta_o\, {}^3g_{uv}(\tau
,\vec \sigma ) \Big) \equiv \nonumber \\
&&\equiv -\partial_s[2\, {}^3g_{ru}\, {}^3{\tilde \Pi}
^{rs}\partial_{\tau}\xi^u+{\cal L}_{ADM} \xi^s](\tau ,\vec \sigma )-
[\partial_{\tau}\xi^u\, {}^3g_{ur}{\tilde {\cal H}}^r](\tau ,\vec
\sigma )\, {\buildrel \circ \over =}\nonumber \\
&&{\buildrel \circ \over =}\, -\partial_s[2\, {}^3g_{ru}\,
{}^3{\tilde \Pi}^{rs}\partial_{\tau}\xi^u+{\cal L}_{ADM} \xi^s](\tau ,\vec
\sigma )],
\label{a15}
\end{eqnarray}

\noindent by using the acceleration independent Euler-Lagrange equations
corresponding to ${\tilde {\cal H}}^r(\tau ,\vec \sigma )\approx 0$.

Therefore, if we choose $\alpha_r(\tau ,\vec \sigma )=-\xi_r(\tau ,\vec
\sigma )$ [$\alpha^r={}^3g^{rs}\alpha_s$], the generator

\begin{eqnarray}
G_2[\alpha_r]
&=&\int d^3\sigma [\alpha_r{\tilde {\cal H}}^r+\alpha^r\partial_rN {\tilde
\pi}^N+[\partial_r\alpha^s+\delta^s_r\alpha^u\partial_u]N_s {\tilde \pi}^r
_{\vec N}](\tau ,\vec \sigma )=\nonumber \\
&=&G[0,\alpha_r, \alpha^r\partial_rN, (\partial_r\alpha^s+\delta^s_r\alpha^u
\partial_u)N_s],
\label{a16}
\end{eqnarray}

\noindent yields the previous transformations.

By putting together Eqs.(\ref{a10}) and (\ref{a14}), we get that the gauge
transformations with generator

\begin{eqnarray}
G_3[\alpha_r,\lambda_N,\lambda_r^{\vec N}]
&=&\int d^3\sigma [\alpha_r{\tilde {\cal H}}^r+\lambda_N {\tilde \pi}^N
+\lambda^{\vec N}_r {\tilde \pi}^r_{\vec N}](\tau ,\vec \sigma )=\nonumber \\
&=&G[0,\alpha_r,\lambda_N, \lambda^{\vec N}_r]
=G_2[\alpha_r]+\nonumber \\
&+&G_1[\lambda_N-\alpha^r\partial_rN,
\lambda^{\vec N}_r-(\partial_r\alpha^s+\delta^s_r\alpha^u\partial_u)N_s],
\nonumber \\
&&\Downarrow \nonumber \\
&&\delta_oN=\lambda_N,\quad\quad \delta_oN_r=
\lambda^{\vec N}_r,\quad\quad \delta_o\, {}^3g_{rs}=\alpha_{r|s}+\alpha_{s|r},
\label{a17}
\end{eqnarray}

\noindent yields the Noether identities

\begin{eqnarray}
&&\delta_o{\cal L}_{ADM}(\tau ,\vec \sigma )=\nonumber \\
&&=\lambda_N(\tau ,\vec \sigma ) L_N(\tau ,\vec \sigma )+\lambda^{\vec N}
_s(\tau ,\vec \sigma ) L^s_{\vec N}(\tau ,\vec \sigma )+(\alpha_{u|v}+
\alpha_{v|u})(\tau
,\vec \sigma ) L^{uv}_g(\tau ,\vec \sigma )+\nonumber \\
&&+\partial_{\tau} \Big[ {\tilde \pi}^N(\tau ,\vec \sigma )\lambda_N(\tau
,\vec \sigma )+{\tilde \pi}^r_{\vec N}(\tau ,\vec \sigma )\lambda^{\vec N}
_r(\tau ,\vec \sigma )+{}^3{\tilde \Pi}^{uv}(\tau ,\vec \sigma )(\alpha_{u|v}+
\alpha_{v|u})(\tau ,\vec \sigma )\Big] +\nonumber \\
&&+\partial_r\Big[ {{\partial {\cal L}_{ADM}(\tau ,\vec \sigma )}\over {\partial
\partial_rN_s(\tau ,\vec \sigma )}}\lambda^{\vec N}_s(\tau ,\vec \sigma )+
\nonumber \\
&&+\Big( {{\partial
{\cal L}_{ADM}(\tau ,\vec \sigma )}\over {\partial \partial_r\, {}^3g_{uv}
(\tau ,\vec \sigma )}}-2\partial_s
{{\partial {\cal L}_{ADM}(\tau ,\vec \sigma )}\over {\partial \partial_r
\partial_s\, {}^3g_{uv}(\tau ,\vec \sigma )}}
\Big) (\alpha_{u|v}+\alpha_{v|u})(\tau ,\vec \sigma )\Big] +\nonumber \\
&&+\partial_r\partial_s \Big(
{{\partial {\cal L}_{ADM}(\tau ,\vec \sigma )}\over {\partial \partial_r
\partial_s\, {}^3g_{uv}(\tau ,\vec \sigma )}} (\alpha_{u|v}+\alpha_{v|u})(\tau
,\vec \sigma ) \Big) \equiv \nonumber \\
&&\equiv  -\Big( [\lambda_N-\alpha^r\partial_rN] {\tilde {\cal H}} \Big)
(\tau ,\vec \sigma )-\nonumber \\
&&-\Big( [\lambda^{\vec N}_r-(\partial_r\alpha^s+\delta^s_r\alpha^u\partial_u)
N_s -\, {}^3g_{ru}\partial_{\tau}\alpha^u] {\tilde {\cal
H}}^r \Big) (\tau ,\vec \sigma )-\nonumber \\
&&-\partial_r\Big( 2 \, {}^3{\tilde \Pi}^{rs}[\lambda^{\vec N}_s-
(\partial_s\alpha^v+\delta^v_s\alpha^u\partial_u)N_v]-\nonumber \\
&&-[2\, {}^3{\tilde \Pi}^{rs}\, {}^3g_{su} \partial_{\tau}
\alpha^u+{\cal L}_{ADM} \alpha^r]\Big) (\tau ,\vec \sigma ) .
\label{a18}
\end{eqnarray}

By putting $\lambda_N=0$ and $\lambda_r^{\vec N}=\partial_{\tau} \alpha_r\,
{\buildrel \circ \over =}\, \partial_{\tau}N_r$ (so that the generator
is $G[0,\alpha_r,0,\partial_{\tau}\alpha_r]$ and one has $\alpha_r\,
{\buildrel \circ \over =}\, N_r$),
one recovers the Noether identities of Eqs.(\ref{a10b}) and, in addition, new
identities.
Indeed, by equating the coefficients of $\partial_r\partial_{\tau}\alpha_s$,
$\partial^2_{\tau}\alpha_r$, $\partial_{\tau}\alpha_r$, $\partial_r\partial
_s\partial_u\alpha_v$, $\partial_r\partial_s\alpha_t$, $\partial_r\alpha_s$,
$\alpha_t$ on both sides we get 7 Noether identities [the first three
reproduce Eqs.(\ref{a10b})]

\begin{eqnarray}
&&{{\partial {\cal L}_{ADM}}\over {\partial \partial_rN_s}}+2\, {}^3{\tilde
\Pi}^{rs} \equiv 0,\nonumber \\
&&{\tilde \pi}^r_{\vec N} \equiv 0, \nonumber \\
&&\partial_{\tau}{\tilde \pi}^r_{\vec N}+{\tilde {\cal H}}^r+L^r_{\vec N} \equiv
0,\quad \Rightarrow \quad {\tilde {\cal H}}^r\equiv -L^r_{\vec N},\nonumber \\
&&{{\partial {\cal L}_{ADM}}\over {\partial \partial_{(r}\partial_s\, {}^3g
_{u)v}}} \equiv 0,\nonumber \\
&&{{\partial {\cal L}_{ADM}}\over {\partial \partial_{(r}\, {}^3g_{s)t}}}-
{}^3\Gamma^t_{uv} {{\partial {\cal L}_{ADM}}\over {\partial \partial_r\partial
_s\, {}^3g_{uv}}}-\, {}^3{\tilde \Pi}^{rs}\, {}^3g^{tu} N_u \equiv 0,
\nonumber \\
&&2L^{rs}_g-N_v\, {}^3g^{v(r}{\tilde {\cal H}}^{s)}
+2[\partial_{\tau}{}^3{\tilde \Pi}^{rs}-{}^3g_{tu}\, {}^3{\tilde
\Pi}^{t(r} \partial_{\tau}\, {}^3g^{s)u}]-\nonumber \\
&&-2\, {}^3g^{v(s} \partial_t({}^3{\tilde \Pi}^{r)t}N_v)-4N_v\, {}^3{\tilde
\pi}^{t(r} \partial_t\, {}^3g^{s)v}-2\, {}^3{\tilde \Pi}^{v(r}\, {}^3g^{s)u}
\partial_uN_v -{}^3g^{rs}{\cal L}_{ADM}+\nonumber \\
&&+2\partial_t ({{\partial {\cal L}_{ADM}}\over {\partial \partial_t\,
{}^3g_{rs}}}-\partial_u {{\partial {\cal L}_{ADM}}\over {\partial \partial_u
\partial_t\, {}^3g_{rs}}})-2\, {}^3\Gamma^{(r}_{uv} {{\partial {\cal L}_{ADM}}
\over {\partial \partial_{s)}\, {}^3g_{uv}}}+2 (\partial_t\, {}^3\Gamma^{(r}
_{uv}) {{\partial {\cal L}_{ADM}}\over {\partial \partial_{s)}\partial_t\,
{}^3g_{uv}}}\equiv 0,\nonumber \\
&&-2{}^3\Gamma^t_{uv} L^{uv}_g-{}^3g^{tr}\partial_r{\tilde {\cal H}} -N_s
\partial_r\, {}^3g^{st} {\tilde {\cal H}}^r-{}^3g^{tu}\partial_uN_s {\tilde
{\cal H}}^s+\nonumber \\
&&+\partial_{\tau}\, {}^3g^{ut}\, {}^3g_{ur}{\tilde {\cal H}}^r
-2\partial_{\tau}({}^3\Gamma^t_{uv}\, {}^3{\tilde \Pi}^{uv})-2\partial_r
({}^3{\tilde \Pi}^{rs}\, {}^3g_{su})\partial_{\tau}\, {}^3g^{ut}-2\,
{}^3{\tilde \Pi}^{rs}\, {}^3g_{su}\partial_r\partial_{\tau}\, {}^3g^{ut}-
\nonumber \\
&&-2\partial_r[{}^3{\tilde \Pi}^{rs}(N_v \partial_s\, {}^3g^{vt}+{}^3g^{vt}
\partial_vN_s)]-\partial_r({}^3g^{rt}{\cal L}_{ADM})-\nonumber \\
&&-2\partial_r({}^3\Gamma^t_{uv} [{{\partial {\cal L}_{ADM}}\over {\partial
\partial_r\, {}^3g_{uv}}}-2\partial_s{{\partial {\cal L}_{ADM}}\over {\partial
\partial_r\partial_s\, {}^3g_{uv}}}])-2\partial_r\partial_s({}^3\Gamma^t_{uv}
{{\partial {\cal L}_{ADM}}\over {\partial \partial_r\partial_s\,
{}^3g_{uv}}})\equiv 0.
\label{a18a}
\end{eqnarray}

One can verify that the last two identities are actually $0 \equiv 0$. Suitable
combinations of these identities should rebuild three of the contracted
Bianchi identities quoted after Eq.(10) of I and allow to rewrite the part of
Eq.(\ref{a5}) connected with the supermomentum constraints as a weak charge
(a Noether constant expressed as a volume integral of a charge density) equal
(modulo the constraints) to a strong charge (a surface integral of
a charge density expressible in terms of a superpotential like in Appendix D).
This will be studied elsewhere.

3) To preserve the 3+1 splitting of $M^4$, the ADM action (like the actions of
parametrized theories on spacelike hypersurfaces in Minkowski spacetime) is
invariant under $\vec \sigma$-independent $\tau$-reparametrizations $\tau
\mapsto \tau^{'}(\tau )=\tau +\delta \tau (\tau )$ [instead the
diffeomorphisms in $Diff\, M^4$ mix all the coordinates $x^{\mu}$]. Since the
lapse and shift functions transform like ${d\over {d\tau}}$, one has $\delta
_{\tau}S_{ADM}\equiv 0$ [namely $d\tau {\cal L}_{ADM}(\tau ,\vec \sigma )=
d\tau N(\tau ,\vec \sigma ) {\cal L}^{'}_{ADM}(\tau ,\vec \sigma )$ is a
scalar] under the ``non Lagrangian" $\tau$-reparametrizations

\begin{eqnarray}
\delta \tau (\tau )&=&\bar \tau (\tau )-\tau ,\nonumber \\
\delta_{\tau}N(\tau ,\vec \sigma )&=&-{{d\delta \tau}\over {d\tau}} N(\tau
,\vec \sigma ),\quad\quad \Rightarrow \, \delta_{\tau}[d\tau N(\tau ,\vec
\sigma )]=0,\nonumber \\
\delta_{\tau}N_r(\tau ,\vec \sigma )&=&-{{d\delta \tau}\over {d\tau}} N_r
(\tau ,\vec \sigma ),\nonumber \\
\delta_{\tau}\, {}^3g_{rs}(\tau ,\vec \sigma )&=&0,\quad\quad \delta_{\tau}
\partial_u\, {}^3g_{rs}(\tau ,\vec \sigma )=0,...\nonumber \\
\delta_{\tau} \partial_{\tau}\, {}^3g_{rs}(\tau ,\vec \sigma )&=&-{{d\delta
\tau}\over {d\tau}} \partial_{\tau}\, {}^3g_{rs}(\tau ,\vec \sigma ),
\nonumber \\
\Rightarrow && \delta_{\tau}\, {}^3K_{rs}(\tau ,\vec \sigma )=0,\nonumber \\
&&\Downarrow \nonumber \\
\delta_{\tau} [d\tau {\cal L}_{ADM}(\tau ,\vec \sigma )]&=&\delta_{\tau}[d\tau
N(\tau ,\vec \sigma ) {\cal L}^{'}_{ADM}(\tau ,\vec \sigma )]=\nonumber \\
&=&d\tau N(\tau ,\vec \sigma ) \delta_{\tau} {\cal L}^{'}_{ADM}(\tau ,\vec
\sigma ) \equiv 0.
\label{a19}
\end{eqnarray}

Following the paper c) in Ref.\cite{sha}, one
reconstructs a real Noether Lagrangian transformation

\begin{eqnarray}
\delta \tau (\tau )&&\nonumber \\
\delta_{o\tau}N(\tau ,\vec \sigma )&=&\delta_{\tau}N(\tau ,\vec \sigma )=
-{{d\delta \tau (\tau )}\over {d\tau}} N(\tau ,\vec \sigma ),\nonumber \\
\delta_{o\tau}N_r(\tau ,\vec \sigma )&=&\delta_{\tau}N_r(\tau ,\vec \sigma )=
-{{d\delta \tau (\tau )}\over {d\tau}} N_r(\tau ,\vec \sigma ),\nonumber \\
\delta_{o\tau}\, {}^3g_{rs}(\tau ,\vec \sigma )&=&-\delta \tau (\tau )\,
\partial_{\tau}\, {}^3g_{rs}(\tau ,\vec \sigma ),\nonumber \\
\delta_{o\tau}\partial_{\tau}\, {}^3g_{rs}(\tau ,\vec \sigma )&=&-{{d\delta
\tau (\tau )}\over {d\tau}} \partial_{\tau}\, {}^3g_{rs}(\tau ,\vec \sigma )-
\delta \tau (\tau )\, \partial^2_{\tau}\, {}^3g_{rs}(\tau ,\vec \sigma )=
\nonumber \\
&=&\delta_{\tau}\partial_{\tau}\, {}^3g_{rs}(\tau ,\vec \sigma )+\delta^{'}
\partial_{\tau}\, {}^3g_{rs}(\tau ,\vec \sigma ),
\label{a20}
\end{eqnarray}

Since, with ${\cal L}^{'}_{ADM}={\cal L}_{ADM}/N$, we get ${{\partial}\over
{\partial N}} {\cal L}^{'}_{ADM}={1\over N} ({{\partial {\cal L}_{ADM}}\over
{\partial N}}-{\cal L}^{'}_{ADM})$, then by using $\delta_{\tau} {\cal L}^{'}
_{ADM}={{\partial {\cal L}^{'}_{ADM}}\over {\partial N}}\delta_{\tau}N+
{{\partial {\cal L}^{'}_{ADM}}\over {\partial N_r}}\delta_{\tau}N_r+{{\partial
{\cal L}^{'}_{ADM}}\over {\partial_sN_r}}\partial_s\delta_{\tau}N_r+
{{\partial {\cal L}^{'}_{ADM}}\over {\partial \partial_{\tau}\, {}^3g_{rs}}}
\delta_{\tau}\partial_{\tau}\, {}^3g_{rs} \equiv 0$, we obtain

\begin{eqnarray}
\delta_{o\tau} {\cal L}^{'}_{ADM}&=&{1\over N}\Big[ \delta_{o\tau}{\cal L}
_{ADM}-{{ {\cal L}_{ADM}}\over N} \delta_{o\tau}N\Big] =\nonumber \\
&=&{1\over N}(L_N-
{\cal L}^{'}_{ADM})\delta_{o\tau}N+{1\over N}L^r_{\vec N}\delta_{o\tau}N_r+
{1\over N}L^{rs}_g\delta_{o\tau}\, {}^3g_{rs}+\nonumber \\
&+&{1\over N}
\partial_{\tau}[ {\tilde \pi}^N\delta_{o\tau}N+{\tilde \pi}^r
_{\vec N}\delta_{o\tau}N_r+{}^3{\tilde \Pi}^{rs}\delta_{o\tau}\, {}^3g_{rs}]+
\nonumber \\
&+&{1\over N}
\partial_r[{{\partial {\cal L}_{ADM}}\over {\partial \partial_rN
_s}}\delta_{o\tau}N_s+
({{\partial {\cal L}_{ADM}}\over {\partial \partial_r\, {}^3g
_{uv}}}-2\partial_s{{\partial {\cal L}_{ADM}}\over {\partial \partial_r\partial
_s\, {}^3g_{uv}}}) \delta_{o\tau}\, {}^3g_{uv}]+\nonumber\\
&+&{1\over N}\partial_r\partial_s(
{{\partial {\cal L}_{ADM}}\over {\partial \partial_r\partial_s\, {}^3g_{uv}}}
\delta_{o\tau}\, {}^3g_{uv}) \equiv \nonumber \\
&\equiv&{{\partial {\cal L}^{'}_{ADM}}\over {\partial \, {}^3g_{uv}}}\delta
_{o\tau}\, {}^3g_{uv}+{{\partial {\cal L}^{'}_{ADM}}\over {\partial \partial
_{\tau}\, {}^3g_{uv}}} \delta^{'}\partial_{\tau}\, {}^3g_{uv}+\nonumber \\
&+&{{\partial {\cal L}^{'}_{ADM}}\over {\partial \partial_r\, {}^3g_{uv}}}
\delta_{o\tau}\partial_r\, {}^3g_{uv}+{{\partial {\cal L}^{'}_{ADM}}\over
{\partial \partial_r\partial_s\, {}^3g_{uv}}}\delta_{o\tau}\partial_r\partial_s
\, {}^3g_{uv},\nonumber \\
&&\Downarrow \quad\quad by\, putting\, \beta (\tau )=-\delta \tau (\tau )
\nonumber \\
{{{\ddot \beta}(\tau )}\over N} && [{\tilde \pi}^N N+{\tilde \pi}^r_{\vec N}
N_r]+\nonumber \\
+{{\dot \beta (\tau )}\over N}&& [\partial_{\tau}({\tilde \pi}^N N+{\tilde \pi}
^r_{\vec N} N_r)+L_N N +L^r_{\vec N} N_r-{\cal L}_{ADM}+\nonumber \\
&+& {}^3{\tilde \Pi}^{rs}\partial_{\tau}\, {}^3g_{rs}+\partial_r
({{\partial {\cal L}_{ADM}}\over {\partial \partial_rN_s}} N_s)]+
\nonumber \\
+{{\beta (\tau )}\over N}&& [L^{rs}_g \partial_{\tau}\, {}^3g_{rs}+\partial
_{\tau}({}^3{\tilde \Pi}^{rs}\partial_{\tau}\, {}^3g_{rs})-{}^3{\tilde \Pi}^{rs}
\partial^2_{\tau}\, {}^3g_{rs}+\nonumber \\
&+&\partial_r\Big[ ({{\partial {\cal L}_{ADM}}\over {\partial \partial_r\,
{}^3g_{uv}}}-2\partial_s{{\partial
{\cal L}_{ADM}}\over {\partial \partial_r\partial_s\, {}^3g_{uv}}})
\partial_{\tau}\, {}^3g_{uv}\Big] +\partial_r\partial_s({{\partial {\cal L}
_{ADM}}\over {\partial \partial_r\partial_s\, {}^3g_{uv}}}\partial_{\tau}\,
{}^3g_{uv})-\nonumber \\
&-&{{\partial {\cal L}_{ADM}}\over {\partial \, {}^3g_{uv}}}\partial_{\tau}\,
{}^3g_{rs}-{{\partial {\cal L}_{ADM}}\over {\partial \partial_r\, {}^3g_{uv}}}
\partial_r\partial_{\tau}\, {}^3g_{rs}-{{\partial {\cal L}
_{ADM}}\over {\partial \partial_r\partial_s\, {}^3g_{uv}}}\partial_r\partial_s
\partial_{\tau}\, {}^3g_{rs} \equiv 0.
\label{a21}
\end{eqnarray}

As one can check the three identities given by the vanishing of the
coefficients of $\ddot \beta (\tau )$, $\dot \beta (\tau )$, $\beta (\tau )$
[due to Eqs.(77), (80) of I, the second
identity is $N{\tilde {\cal H}}+N_r\, {}^3{\tilde {\cal H}}^r\equiv -NL_N-N_r
L^r_{\vec N}$] are satisfied and contain implicitly the
fourth contracted Bianchi identity and the reformulation of the part of
Eq.(\ref{a5}) connected with the superhamiltonian constraint as a weak charge
weakly equal to a strong charge. Again this will be studied elsewhere.

Under the equivalent (see the paper c) in Ref.\cite{sha})
Noether transformation \hfill\break
\hfill\break
$\delta^{'}_{o\tau}N=\dot \beta N$, $\delta^{'}_{o\tau}
N_r=\dot \beta N_r$, $\delta^{'}_{o\tau}\, {}^3g_{rs}=\beta \partial_{\tau}\,
{}^3g_{rs}$, \hfill\break
\hfill\break
but without transforming
$\tau$ [$\delta \tau (\tau )=0$], this Noether identity is rewritten as

\begin{eqnarray}
&&\delta^{'}_{o\tau}{\cal L}_{ADM}(\tau ,\vec \sigma )=\nonumber \\
&&=L_N(\tau ,\vec \sigma )\delta^{'}_{o\tau}N(\tau ,\vec \sigma )+
L^s_{\vec N}(\tau ,\vec \sigma )\delta^{'}_{o\tau}N_s(\tau ,\vec \sigma ) +
L^{uv}_g(\tau ,\vec \sigma )\delta^{'}_{o\tau}\,
{}^3g_{uv}(\tau ,\vec \sigma )+\nonumber \\
&&+\partial_{\tau} \Big[ {\tilde \pi}^N(\tau ,\vec \sigma )\delta^{'}_{o\tau}
N(\tau ,\vec \sigma )+{\tilde \pi}^r_{\vec N}(\tau ,\vec \sigma )\delta^{'}
_{o\tau}N_r(\tau ,\vec \sigma )+{}^3{\tilde \Pi}^{uv}(\tau ,\vec \sigma )
\delta^{'}_{o\tau}\, {}^3g_{uv}(\tau ,\vec \sigma )\Big] +\nonumber \\
&&+\partial_r\Big[ {{\partial {\cal L}_{ADM}(\tau ,\vec \sigma )}\over {\partial
\partial_rN_s(\tau ,\vec \sigma )}} \delta^{'}_{o\tau}N_s(\tau ,\vec \sigma )
+\nonumber \\
&&+\Big( {{\partial
{\cal L}_{ADM}(\tau ,\vec \sigma )}\over {\partial \partial_r\, {}^3g_{uv}
(\tau ,\vec \sigma )}}-2\partial_s
{{\partial {\cal L}_{ADM}(\tau ,\vec \sigma )}\over {\partial \partial_r
\partial_s\, {}^3g_{uv}(\tau ,\vec \sigma )}}
\Big) \delta^{'}_{o\tau}\, {}^3g_{uv}(\tau ,\vec \sigma )\Big] +\nonumber \\
&&+\partial_r\partial_s \Big(
{{\partial {\cal L}_{ADM}(\tau ,\vec \sigma )}\over {\partial \partial_r
\partial_s\, {}^3g_{uv}(\tau ,\vec \sigma )}} \delta^{'}_{o\tau}\, {}^3g
_{uv}(\tau ,\vec \sigma ) \Big) \equiv \nonumber \\
&&\equiv -\dot \beta (\tau ) \Big( {}^3{\tilde \Pi}^{rs}\partial_{\tau}\,
{}^3g_{rs}-{\cal L}_{ADM}\Big) (\tau ,\vec \sigma )+{}^3{\tilde \Pi}^{rs}(\tau
,\vec \sigma )\delta^{'}_{o\tau}\, {}^3g_{rs}(\tau ,\vec \sigma )+\nonumber \\
&&+{{\partial {\cal L}_{ADM}(\tau ,\vec \sigma )}
\over {\partial \, {}^3g_{rs}(\tau ,\vec \sigma )}}\, \delta^{'}_{o\tau}\, {}^3g
_{rs}(\tau ,\vec \sigma )+{{\partial {\cal L}_{ADM}(\tau ,\vec \sigma )}\over
{\partial \partial_u\, {}^3g_{rs}(\tau ,\vec \sigma )}} \delta^{'}_{o\tau}
\partial_u\, {}^3g_{rs}(\tau ,\vec \sigma )+\nonumber \\
&&+{{\partial {\cal L}_{ADM}(\tau ,\vec \sigma )}
\over {\partial \partial_u\partial_v\, {}^3g_{rs}(\tau ,\vec \sigma )}}
\delta^{'}_{o\tau} \partial_u\partial_v\, {}^3g_{rs}(\tau ,\vec \sigma ).
\label{a22}
\end{eqnarray}

Since the generator $\int d^3\sigma \alpha (\tau ,\vec \sigma ) {\tilde {\cal
H}}(\tau ,\vec \sigma )$ gives \hfill\break
\hfill\break
$\delta_oN(\tau ,\vec \sigma )=0$, $\delta_oN_r
(\tau ,\vec \sigma )=0$, $\delta_o\, {}^3g_{rs}(\tau ,\vec \sigma )=-\big(
{{\alpha}\over N} [N_{r|s}+N_{s|r}-\partial_{\tau}\, {}^3g_{rs}]\Big)
(\tau ,\vec \sigma )$, \hfill\break
\hfill\break
to get the previous transformations $\delta^{'}_{o\tau}
q^i$ one has to put \hfill\break
\hfill\break
$\alpha (\tau ,\vec \sigma )= \beta (\tau ) N(\tau
,\vec \sigma )$ with $\beta(\tau )\rightarrow \,-\delta \tau (\tau )$ for
infinitesimal gauge transformations. \hfill\break
\hfill\break
Then, these transformations  are generated by

\begin{eqnarray}
G_{\tau}[\beta (\tau ),\dot \beta (\tau )]
&=& \int d^3\sigma \Big( \beta (\tau )[N {\tilde {\cal H}}+N_r {\tilde
{\cal H}}^r] +\dot \beta (\tau )[N {\tilde \pi}^N +N_r{\tilde \pi}^r_{\vec N}]
\Big) (\tau ,\vec \sigma )=\nonumber \\
&=&G[\beta N, \beta N_r, \dot \beta N, \dot \beta N_r].
\label{a23}
\end{eqnarray}

Therefore, the most general gauge transformation of the ADM action has the
generator

\begin{eqnarray}
G[\beta N,\alpha_r,\lambda_N,\lambda_r^{\vec N}]&=&\int d^3\sigma [\beta (\tau )
N {\tilde {\cal H}}+\alpha_r{\tilde {\cal H}}^r+\lambda_N{\tilde \pi}^N+
\lambda_r^{\vec N}{\tilde \pi}^r_{\vec N}](\tau ,\vec \sigma )=\nonumber \\
&=&G_{\tau}[\beta ,\dot \beta]+G_2[\alpha_r- \beta N_r]+\nonumber \\
&+&G_1[\lambda_N-\dot \beta (\tau )N-(\alpha^r- \beta (\tau )N^r)
\partial_rN,\nonumber \\
&&\lambda_r^{\vec N}-\dot \beta (\tau )N_r-[\partial_r(\alpha^s-\beta
(\tau )N^s)+\delta^s_r(\alpha^u-\beta (\tau )N^u)\partial_u]N_s]=
\nonumber \\
&=&G_{\tau}[\beta ,\dot \beta ]+G_3[\alpha_r-\beta (\tau )N_r,\lambda_N-\dot
\beta (\tau )N, \lambda_r^{\vec N}-\dot \beta (\tau )N_r],\nonumber \\
&&\Downarrow \nonumber \\
\delta_oN(\tau ,\vec \sigma )&=&\lambda_N(\tau ,\vec \sigma ),\nonumber \\
\delta_oN_r(\tau ,\vec \sigma )&=&\lambda_r^{\vec N}(\tau ,\vec \sigma ),
\nonumber \\
\delta_o\, {}^3g_{rs}(\tau ,\vec \sigma )&=&[(\alpha_r- \beta (\tau )N_r)
_{|s}+(\alpha_s-\beta (\tau )N_s)_{|r}+\beta (\tau ) \partial_{\tau}
\, {}^3g_{rs} ](\tau ,\vec \sigma ).\nonumber \\
&&{}
\label{a24}
\end{eqnarray}

Maybe there is also a local invariance with arbitrary $\alpha (\tau ,\vec
\sigma ) \not= \beta (\tau ) N(\tau ,\vec \sigma )$, but we have not found it.

The transformation generated by $G_{\tau}$ is what remains of the invariance
of the Hilbert action under the diffeomorphisms in $M^4$
given by Eq.(31) of I, which have the infinitesimal form\hfill\break
\hfill\break
$\delta_ox^{\mu}=\xi^{\mu}(x)$ \hfill\break
\hfill\break
[$0=\delta_{diff}S_H=\delta_{diff}S_{ADM}+
\delta_{diff}\Sigma_{ADM}$]. Since the diffeomorphisms imply\hfill\break
\hfill\break
$\delta_ob^{\mu}_A=\partial_A\delta_oz^{\mu}(\tau ,\vec \sigma )=
\partial_A\xi^{\mu}(z(\tau ,\vec \sigma ))=
\partial_{\rho}\xi^{\mu}(z(\tau ,\vec \sigma ))b^{\rho}_A(\tau ,\vec \sigma )$,
\hfill\break
\hfill\break
their effect on the ADM variables in $\Sigma_{\tau}$-adapted
coordinates is the transformations \hfill\break
\hfill\break
$\delta_o\, {}^4g_{AB}=\delta_o
(b^{\mu}_A\, {}^4g_{\mu\nu} b^{\nu}_B)=b^{\mu}_Ab^{\nu}_B\delta_o\, {}^4g
_{\mu\nu}+({}^4g_{CB}b^{\rho}_A+{}^4g_{AC}b^{\rho}_B)b^C_{\sigma}\partial
_{\rho}\xi^{\sigma}=-{}^4g_{AB}\partial_{\rho}\xi^{\rho}$, \hfill\break
\hfill\break
so that from Eq.(6) of I we get \hfill\break
\hfill\break
$\delta_o\, {}^3g_{rs}=-{}^3g_{rs}\partial_{\rho}\xi
^{\rho}$, $\delta_oN^r=0$, $\delta_oN_r=-N_r\partial_{\rho}\xi^{\rho}$,
$\delta_oN=-{1\over 2}N\partial_{\rho}\xi^{\rho}$. \hfill\break
\hfill\break
But this is a unique mixture
of Eqs.(\ref{a20}) and of a transformations generated by ${\tilde \pi}^N$.
If we assume the form \hfill\break
\hfill\break
$\eta =-\partial_{\rho}\xi^{\rho}=\beta (\tau ) N$ \hfill\break
for the infinitesimal parameter, the associated Hamiltonian generator is
\hfill\break
\hfill\break
$G_{\tau}[\beta]+G_1[-{1\over 2}\beta N, 0]$. \hfill\break
\hfill\break
However, for a generic $\eta =-
\partial_{\rho} \xi^{\rho}$ we get a 3-conformal infinitesimal transformation
$\delta_o\, {}^3g_{rs}=\eta \, {}^3g_{rs}$ enlarged with $\delta_o N={1\over 2}
\eta N$, $\delta_oN_r= \eta N_r$. From Appendix B of II [see after Eqs.(B1)]
with $\phi =1+{1\over 4}\eta$ we get \hfill\break
\hfill\break
$\delta_o\sqrt{\gamma}={3\over 2}\eta \sqrt{\gamma}$,\hfill\break
$\delta_o\, {}^3\Gamma^u_{rs}={1\over 2}(\delta^u_r\partial
_s\eta +\delta^u_s\partial_r\eta -{}^3g_{rs}\, {}^3g^{uv}\partial_v\eta )$,
\hfill\break
$\delta_o\, {}^3R=-\eta \, {}^3R-2 \triangle \eta$, \hfill\break
$\delta_o\, {}^3K_{rs}=
{1\over 2}\eta \, {}^3K_{rs}+{{{}^3g_{rs}}\over {2N}} (N_u\, {}^3g^{uv}
\partial_v\eta -\partial_{\tau}\eta )$, \hfill\break
\hfill\break
so that \hfill\break
\hfill\break
$\delta_o{\cal L}_{ADM}=-\eta
{\cal L}_{ADM}+2\epsilon k \sqrt{\gamma} [N \triangle \eta +{}^3K (N_u\,
{}^3g^{uv}\partial_v\eta -\partial_{\tau}\eta )]$. \hfill\break
\hfill\break
Namely, the ADM Lagrangian
density is not quasi-invariant under the diffeomorphisms of $M^4$ rewritten in
$\Sigma_{\tau}$-adapted coordinates, but only under the gauge
transformations generated by the first class constraints.

See Appendix D for a review of the second Noether theorem in the case of the
Hilbert action and for the consequences
of its 4-diffeomorphism invariance like the Komar
superpotential and the energy-momentum pseudotensors. More work is needed
to see whether it is possible to define an ADM superpotential and an associated
ADM energy-momentum pseudo-tensor deriving from the second Noether theorem
applied to the ADM action.

As said in the Conclusions of II, the consequences of this difference in the
invariance properties of the ADM and Hilbert actions (even if they generate
the same Einstein's equations) are the following:\hfill\break
\hfill\break
i) a ``Hamiltonian kinematical gravitational field", defined as an equivalence
class of spacetimes modulo the Hamiltonian group of gauge transformations
(whose group manifold is not well understood in the large), in general contains
many 4-geometries (the elements of $Riem\, M^4/Diff\, M^4$, which are the
standard ``kinematical gravitational fields") connected by arbitrary gauge
transformations;\hfill\break
ii) a ``Hamiltonian Einstein or dynamical gravitational field" is a kinematical
one satisfying the Hamilton-Dirac equations (namely all Einstein's equations):
it coincides with a standard ``Einstein or dynamical gravitational field" (a
4-geometry whose 4-metrics satisfy Einstein's equations), because on the space
of solutions of Einstein equations the spacetime diffeomorphisms are solutions
of the Jacobi equation associated with Einstein's equations, so that they are
dynamical symmetries of Einstein's equations. But this implies that on the
space of solutions of Einstein's equations the group manifold of the
Hamiltonian gauge transformations is constrained to contain only those gauge
transformations which are also dynamical symmetries of the Hamilton-Dirac
equations (and, therefore, also of Einstein's equations). The allowed gauge
transformations are the subset of spacetime diffeomorphisms under which the ADM
action is quasi-invariant; the other spacetime diffeomorphisms are dynamical
symmetries of the equations of motion but not Noether symmetries of the ADM
action.

\vfill\eject

\section{Proposals for the Reduced Phase Space of Metric Gravity.}

Besides our approach in I and II based on tetrad gravity,
the four main proposals to find a copy of the reduced phase space of
metric gravity present in the literature are:

1) The (non canonical) one of ADM\cite{adm} [see the review in Ref.\cite{reg};
in Ref.\cite{yoyo} it is called the `isotropic radiation gauge']. The
canonical variables ${}^3g_{\check r\check s}(\tau ,\vec \sigma )$, ${}^3{\tilde
\Pi}^{\check r\check s}(\tau ,\vec \sigma )$ are divided (not in a canonical
way) in a gauge sector of four pairs ${}^3g_T$, ${}^3{\tilde \Pi}_T$, ${}^4g
_{L\, \check r}$, ${}^3{\tilde \Pi}_L^{\check r}$, and in a physical sector
${}^3g_{TT\, \check r\check s}$, ${}^3{\tilde \Pi}_{TT}^{\check r\check s}$,
by means of a ``flat transverse-traceless" decomposition (see Appendix C of II
for the general TT decomposition) of ${}^3g_{\check r\check s}-\delta_{\check
r\check s}$ and of ${}^3{\tilde \Pi}^{\check r\check s}$ in the spirit of the
linearized theory of metric gravity. Namely, one uses the following splitting
of symmetric tensors [$\triangle
=-{\vec \partial}^2$ is the flat Laplacian; $\triangle^{-1}$ is assumed to exist
and to vanish at spatial infinity]

\begin{eqnarray}
f_{rs}&=&f_{sr}=f_{TT\, rs}+f_{T\, rs}+\partial_rf_{L\, s}+\partial_sf_{L\,
r},\nonumber \\
&&{}\nonumber \\
&&f_{L\, r}=-{1\over {\triangle}} (\partial_sf_{rs}+{1\over {2\triangle}}
\partial_r\partial_s\partial_uf_{us}),\nonumber \\
&&f_{T\, rs}={1\over 2} (\delta_{rs} f_T+{1\over {\triangle}}
\partial_r\partial_sf_T),\nonumber \\
&&f_T=f_{rr}+{1\over {\triangle}} \partial_r\partial_sf_{rs}.
\label{b1}
\end{eqnarray}

The four gauge-fixings, giving a fixation of the coordinates to asymptotic
Minkowski rectangular ones, are

\begin{eqnarray}
&&{}^3{\tilde \Pi}_T(\tau ,\vec \sigma ) \approx 0,\nonumber \\
&&{}^3g_{L\, \check r}(\tau ,\vec \sigma )-\sigma^{\check r} \approx 0,
\label{b2}
\end{eqnarray}

\noindent and one assumes that the constraints ${\tilde {\cal H}}(\tau ,\vec
\sigma ) \approx 0$, ${}^3{\tilde {\cal H}}^{\check r}(\tau ,\vec \sigma )
\approx 0$, may be solved in the form \hfill\break
\hfill\break
${}^3g_T-F[{}^3g_{TT}, {}^3{\tilde \Pi}
_{TT}]\approx 0\quad\quad\quad$,
${}^3{\tilde \Pi}_L^{\check u}-G^{\check u}[{}^3g_{TT},
{}^3{\tilde \Pi}_{TT}]\approx 0$. \hfill\break
\hfill\break
The non-canonical variables have the Poisson
brackets [the Dirac brackets with respect to the gauge-fixings could be
evaluated]

\begin{eqnarray}
&&\lbrace {}^3g_T(\tau ,\vec \sigma ),{}^3{\tilde \Pi}_T(\tau ,{\vec \sigma}
^{'}) \rbrace = 2 \delta^3(\vec \sigma ,{\vec \sigma}^{'}),\nonumber \\
&&\lbrace {}^3g_{L\, \check r}(\tau ,\vec \sigma ),{}^3{\tilde \Pi}_L^{\check s}
(\tau ,{\vec \sigma}^{'}) \rbrace = {1\over {2\triangle}} (\delta_{\check
r\check s}+{1\over {2\triangle}}\partial_{\check r}\partial_{\check s})
\delta^3(\vec \sigma ,{\vec \sigma}^{'}),\nonumber \\
&&\lbrace {}^3g_{TT\, \check r\check s}(\tau ,\vec \sigma ),{}^3{\tilde \Pi}
_{TT}^{\check u\check v}(\tau ,{\vec \sigma}^{'}) \rbrace =\delta
_{TT}{}^{\check u\check v}_{\check r\check s} \delta^3(\vec \sigma ,{\vec
\sigma}^{'}),\nonumber \\
&&{}\nonumber \\
&&\delta_{TT}{}^{\check u\check v}_{\check r\check s}={1\over 2}[(\delta_{\check
r\check v}+{{\partial_{\check r}\partial_{\check v}}\over {\triangle}})
(\delta_{\check
s\check u}+{{\partial_{\check s}\partial_{\check u}}\over {\triangle}})+
\nonumber \\
&&+(\delta_{\check
r\check u}+{{\partial_{\check r}\partial_{\check u}}\over {\triangle}})
(\delta_{\check
s\check v}+{{\partial_{\check s}\partial_{\check v}}\over {\triangle}})+
(\delta_{\check
r\check s}+{{\partial_{\check r}\partial_{\check s}}\over {\triangle}})
(\delta_{\check
u\check v}+{{\partial_{\check u}\partial_{\check v}}\over {\triangle}})],
\nonumber \\
&&\delta_{TT}{}^{\check r\check s}_{\check r\check s}=2,\quad
\delta_{TT}{}^{\check u\check v}_{\check r\check s}=\delta_{TT}{}^{\check
r\check s}_{\check u\check v},\quad \delta_{TT}{}^{\check u\check v}_{\check
r\check r}=\partial_{\check r}\delta_{TT}{}^{\check u\check v}_{\check r\check
s}=0,\nonumber \\
&&\delta_{TT}{}^{\check u\check v}_{\check r\check s}\delta_{TT}{}^{\check
m\check n}_{\check u\check v}=\delta_{TT}{}^{\check m\check n}_{\check r\check
s}, \quad \delta_{TT}{}^{\check u\check v}_{\check r\check s}f_{TT}{}^{\check
r\check s}=f_{TT}{}^{\check u\check v}.
\label{b3}
\end{eqnarray}

Without a study of the time constancy of the gauge-fixings and, therefore, of
the allowed behaviour of the lapse and shift functions, ADM take into account
only the ADM energy with asymptotic Minkowski rectangular coordinates $P^{\tau}
_{ADM}=\int_{S^2_{\tau ,\infty}}d^2\Sigma_{\check u}(\partial_{\check r}\,
{}^3g_{\check u\check r}-\partial_{\check u}\, {}^3g_{\check r\check r})
(\tau ,\vec \sigma )\approx P^{\tau}_{ADM}[{}^3g_{TT},{}^3{\tilde \Pi}_{TT}]$
and assume that the Hamiltonian in the reduced phase space is $P^{\tau}_{ADM}$
in its volume form ${\hat P}_{ADM}^{\tau}$.

2) The one of Dirac\cite{ppp} [see the review in Ref.\cite{reg}; in Ref.
\cite{yoyo} it is called ``radiation gauge"] of fixing only the one
parameter family of spacelike hypersurfaces $\Sigma_{\tau}$ by the ``maximal
slicing condition" ${}^3{\tilde \Pi}(\tau ,\vec \sigma )\approx 0$ [i.e.
${}^3K(\tau ,\vec \sigma )\approx 0$; as noted in Ref.\cite{yoyo}, Lichnerowicz
realized that this condition produces an effective anti-focusing condition,
which cures the coordinate singularities (created by the focusing of geodesics
in $M^4$ normal to $\Sigma_{\tau}$) of the `synchronous' reference systems
($N=1$, $N_r=0$); for closed universes it is better to use ${}^3K(\tau ,\vec
\sigma )\approx const.$]. Here the canonical variables are divided into one
gauge pair (see Appendix C of II for the notation) \hfill\break
\hfill\break
$\varphi =log\, \phi^4={2\over 3}log (-{\cal P}_{\cal T})={1\over 3} log\,
\gamma \quad\quad\quad$, ${}^3{\tilde \Pi}=-2
\epsilon k\sqrt{\gamma}\, {}^3K={3\over 2}\sqrt{\gamma} {\cal T}$, \hfill\break
\hfill\break
and into five other pairs \hfill\break
\hfill\break
${}^3\sigma_{\check r\check s}=\gamma^{-1/3}\, {}^3g_{\check
r\check s}$ [${}^3\sigma =det\, |{}^3\sigma_{\check r\check s}|=1\quad\quad$],
${}^3{\tilde \Pi}_B^{\check r\check s}=\gamma^{1/3}\, {}^3{\tilde \Pi}_A
^{\check r\check s}=\gamma^{1/3}({}^3{\tilde \Pi}^{\check r\check s}-{1\over 3}
{}^3g^{\check r\check s}\, {}^3{\tilde \Pi})$ \hfill\break
\hfill\break
with Poisson brackets

\begin{eqnarray}
&&\lbrace \varphi (\tau ,\vec \sigma ),{}^3{\tilde \Pi}(\tau ,{\vec \sigma}^{'})
\rbrace =\delta^3(\vec \sigma ,{\vec \sigma}^{'}),\nonumber \\
&&\lbrace {}^3\sigma_{\check r\check s}(\tau ,\vec \sigma ),{}^3{\tilde \Pi}_B
^{\check u\check v}(\tau ,{\vec \sigma}^{'}) \rbrace ={\tilde \delta}^{\check
u\check v}_{\check r\check s}(\vec \sigma )\delta^3(\vec \sigma
,{\vec \sigma}^{'}),\nonumber \\
&&\lbrace {}^3{\tilde \Pi}^{\check r\check s}_B(\tau ,\vec \sigma ),{}^3{\tilde
\Pi}_B^{\check u\check v}(\tau ,{\vec \sigma}^{'}) \rbrace ={1\over 3}
[{}^3{\tilde \Pi}_B^{\check r\check s}\, {}^3\sigma^{\check u\check v}-
{}^3{\tilde \Pi}^{\check u\check v}_B\, {}^3\sigma^{\check r\check s}]
(\tau ,\vec \sigma )\delta^3(\vec \sigma ,{\vec \sigma}^{'}),\nonumber \\
&&{}\nonumber \\
&&{\tilde \delta}^{\check u\check v}_{\check r\check s}(\vec \sigma )=
{1\over 2}(\delta^{\check u}_{\check r}\delta^{\check v}_{\check s}+\delta
^{\check v}_{\check r}\delta^{\check u}_{\check s})-{1\over 3}[{}^3\sigma
_{\check r\check s}\, {}^3\sigma^{\check r\check s}](\tau ,\vec \sigma ),
\nonumber \\
&&{\tilde \delta}^{\check r\check s}_{\check r\check s}(\vec \sigma )=5,\quad
{\tilde \delta}^{\check r\check s}_{\check u\check v}(\vec \sigma ){\tilde
\delta}^{\check u\check v}_{\check m\check n}(\vec \sigma )={\tilde \delta}
^{\check r\check s}_{\check m\check n}(\vec \sigma ),\nonumber \\
&&{\tilde \delta}^{\check u\check v}_{\check r\check s}(\vec \sigma )\,
{}^3\sigma_{\check u\check v}(\tau ,\vec \sigma )={\tilde \delta}^{\check
u\check v}_{\check r\check s}(\vec \sigma )\, {}^3\sigma^{\check r\check s}
(\tau ,\vec \sigma )=0,\quad {\tilde \delta}^{\check u\check v}_{\check
r\check s}(\vec \sigma )\, {}^3{\tilde \Pi}_B^{\check r\check s}(\tau ,\vec
\sigma )={}^3{\tilde \Pi}_B^{\check u\check v}(\tau ,\vec \sigma ).
\label{b4}
\end{eqnarray}

The gauge-fixing ${}^3{\tilde \Pi}(\tau ,\vec \sigma )\approx 0$ implies that
the constraint ${\tilde {\cal H}}(\tau ,\vec \sigma )\approx 0$ has to be
solved in $\varphi (\tau ,\vec \sigma )$ [see the Lichnerowicz equation in
Appendix C of II] and one can evaluate the Dirac brackets. It is assumed that
$P^{\tau}_{ADM}$ is the real Hamiltonian and, by restricting the asymptotic
behaviour of $\partial_{\check u}\, {}^3g_{\check r\check s}(\tau ,\vec \sigma )
=O(r^{-2})$ to $\partial_{\check u}\, {}^3\sigma_{\check r\check s}(\tau ,\vec
\sigma )=O(r^{-(2+\epsilon )})$ one has\hfill\break
\hfill\break
$P^{\tau}_{ADM}=-2 \int_{S^2_{\tau ,\infty}}d^2\Sigma_{\check u} \partial
_{\check u} \gamma^{1/3}$,\hfill\break
\hfill\break
because the change from rectangular coordinates $\sigma^{\check r}$ to ${\bar
\sigma}^{\check r} {\approx}_{r\, \rightarrow \infty}\, \sigma^{\check r}(1-
{M\over {16\pi r}})$ gives \hfill\break
\hfill\break
$ds^2\, {\rightarrow}_{r\, \rightarrow \infty}\,
\epsilon (1-{M\over {8\pi r}})d\tau^2+(1+{M\over {8\pi r}})\delta_{\check
r\check s}d{\bar \sigma}^{\check r}d{\bar \sigma}^{\check s}$.\hfill\break
\hfill\break
 Since the 3-coordinates are not fixed, one has still
the supermomentum constraints and their gauge freedom.

3) This state of affairs was further developed by York\cite{yoyo} (see also Ref.
\cite{reg2}) with the so called ``quasi-isotropic" (QI) coordinate conditions,
which are not phase space gauge-fixings [in Ref.\cite{yoyo} there is no
discussion of supertranslations]. He parametrizes the canonical variables
${}^3g_{\check r\check s}$, ${}^3{\tilde \Pi}^{\check r\check s}$, as in Eq.
(C5) of Appendix C of II
 using a TT-decomposition and then adds the following
quasi-isotropic coordinate conditions [the first one becomes the maximal slicing
condition if put =0; the second one is equivalent to $\partial_{\check s}\,
{}^3\sigma_{\check r\check s}(\tau ,\vec \sigma )=O(r^{-3})$ in asymptotic
rectangular coordinates]

\begin{eqnarray}
&&{\cal T}(\tau ,\vec \sigma )={2\over {3\sqrt{\gamma (\tau ,\vec \sigma )}}}
{}^3{\tilde \Pi}(\tau ,\vec \sigma )=-{4\over 3}\epsilon k\, {}^3K(\tau ,\vec
\sigma )= O(r^{-3}),\nonumber \\
&&[k^{\check r}_{(u)}\, {}^3\nabla^{\check s}\, {}^3\psi_{\check r\check s}]
(\tau ,\vec \sigma ) = O(r^{-3}),
\label{b5}
\end{eqnarray}

\noindent with ${}^3\psi_{\check r\check s}={}^3\psi_{TT\, \check r\check s}+
{}^3\psi_{L\, \check r\check s}={}^3h_{\check r\check s}-{1\over 3} {}^3h\,
{}^3f_{\check r\check s}$ [${}^3h={}^3f^{\check r\check s}\, {}^3h_{\check
r\check s}$], if one puts ${}^3g_{\check r\check s}={}^3f_{\check r\check s}+
{}^3h_{\check r\check s}$ with ${}^3f_{\check r\check s}$ being a flat
asymptotic 3-metric; $k^{\check r}_{(u)}$ are three unit orthogonal
translational Killing vectors of ${}^3f_{\check r\check s}$. York assumes
${\cal L}_{{\vec k}_{(u)}}\, {}^3h_{\check r\check s}=0$ and ${}^3K^{\check
r}{}_{\check s}=O(r^{-2})$ [sometimes the tidal forces are restricted by
`curvature conditions'  ${\cal L}_{{\vec k}_{(u)}}{\cal L}_{{\vec k}_{(v)}}\,
{}^3h^{\check r}{}_{\check s}=O(r^{-3})$ and ${\cal L}_{{\vec k}_{(u)}}\,
{}^3K^{\check r}{}_{\check s}=O(r^{-3})$].

After having posed these quasi-isotropic coordinate conditions (replacing the
gauge-fixings to the secondary constraints, see the end of Section V of I),
York puts the following gauge-fixings on N and $N_{\check r}$ (replacing the
time constancy of the gauge-fixings to the secondary constraints, which
produce the gauge-fixings to the primary ones, i.e. which give equations for N
and $N_{\check r}$)

\begin{eqnarray}
&&\partial_{\tau}\, {}^3K(\tau ,\vec \sigma ) =0,\nonumber \\
&&[{}^3\nabla^{\check s} \, {}^3\Sigma_{\check r\check s}](\tau
,\vec \sigma )=[{}^3\nabla^{\check s} \gamma^{1/3}\, \partial_{\tau}\,
{}^3\sigma_{\check r\check s}](\tau ,\vec \sigma )=\nonumber \\
&&={}^3\nabla^{\check s} [-2N({}^3K_{\check r\check s}-{1\over 3} {}^3g
_{\check r\check s}\, {}^3K)+(L\vec N)_{\check r\check s}](\tau ,\vec \sigma )
=0,
\label{b6}
\end{eqnarray}

\noindent where ${}^3\Sigma$ is called the ``distortion tensor" [it measures
the change of shape of a small spheroid dragged along $\tau$ from $\Sigma
_{\tau}$ to $\Sigma_{\tau +d\tau}$] and $(L\vec N)_{\check r\check s}=
N_{\check r|\check s}+N_{\check s|\check r}-{2\over 3} {}^3g_{\check r\check
s} N_{\check u|\check u}$  [see Eq.(C2) of Appendix C of II].
The equation $\partial
_{\tau}\, {}^3K(\tau ,\vec \sigma )=0$ becomes an elliptic equation for the
lapse function $N(\tau ,\vec \sigma )$ by using the Einstein equations for
$\partial_{\tau}\, {}^3g_{\check r\check s}$ and $\partial_{\tau}\, {}^3K
_{\check r\check s}$; in Ref.\cite{yoyo} it is shown that this equation, after
having used the superhamiltonian constraint, is $-{}^3\nabla_{\check u}\,
{}^3\nabla^{\check u} N+N\, {}^3K_{\check r\check s}\, {}^3K^{\check r\check
s}+N^{\check r}\, {}^3\nabla_{\check r}\, {}^3K\, {\buildrel \circ \over =}\,
0$. Then, by assuming $N(\tau ,\vec \sigma )=1+O(r^{-1})$, $N_{\check r}(\tau
,\vec \sigma )=O(r^{-1})$, York linearizes these equations with respect to the
flat 3-metric ${}^3f_{\check r\check s}$ and obtains the following (not phase
space) gauge-fixings

\begin{eqnarray}
&&{}^3f^{\check r\check s}\, {}^3K_{\check r\check s}(\tau ,\vec \sigma )=0,
\nonumber \\
&&\partial_{\tau} [{}^3\nabla^{(f) \check s}\, {}^3\psi_{\check r\check s}]
(\tau ,\vec \sigma )=0.
\label{b7}
\end{eqnarray}

\noindent It is shown in Ref.\cite{yoyo} that the ADM `isotropic radiation
gauge' is equivalent to the special case ${}^3f^{\check r\check s}\, {}^3K
_{\check r\check s}={}^3\nabla^{(f)\check s}\, {}^3\psi_{\check r\check s}=0$,
while Dirac's coordinate conditions ${}^3f^{\check r\check s}\, {}^3K_{\check
r\check s}=\partial^{\check s}\, {}^3\sigma_{\check r\check s}=0$ are a
non-covariant approximation of the ADM gauge.

Therefore, in some sense, York's construction can be rephrased in phase space
as first fixing N and $N_{\check r}$ (i.e.  giving first the gauge-fixings to
the primary constraints) and then deducing gauge-fixings for the secondary
constraints, a method which is not natural in constraint theory.

In Ref.\cite{reg2}, as already said,
it is shown that the quasi-isotropic coordinate conditions
kill the supertranslations and define a unique asymptotic Poincar\'e group.

4) Instead, in Ref.\cite{reg1} one imposes the following gauge-fixings to the
secondary constraints

\begin{eqnarray}
&&{}^3{\tilde \Pi}(\tau ,\vec \sigma ) \approx 0,\nonumber \\
&&{}^3g^{\check r\check s}(\tau ,\vec \sigma )\, {}^3\Gamma^{\check u}
_{\check r\check s}(\tau ,\vec \sigma )\approx 0\quad or\quad \partial_{\check
r}[\sqrt{\gamma}\, {}^3g^{\check r\check s}](\tau ,\vec \sigma )\approx 0,
\label{b8}
\end{eqnarray}

\noindent corresponding to the maximal slicing condition and to the request of
using 3-harmonic coordinates. Asking their time constancy [$\partial_{\tau}
\, {}^3{\tilde \Pi}
\approx 0$, $\partial_{\tau}({}^3g^{\check r\check s}\, {}^3\Gamma^{\check u}
_{\check r\check s}) \approx 0$], one gets a homogeneous system for N and
$N_{\check r}$

\begin{eqnarray}
&&{}^3\nabla_{\check u}\, {}^3\nabla^{\check u} N +{}^3K^{\check r\check s}\,
{}^3K_{\check r\check s} N \approx 0,\nonumber \\
&&{}^3\nabla_{\check u}\, {}^3\nabla^{\check u} N^{\check r}+2\, {}^3K^{\check
r\check s}\, {}^3\nabla_{\check s} N-2 N\, {}^3K^{\check u\check v}\,
{}^3\Gamma^{\check r}_{\check u\check v} \approx 0.
\label{b9}
\end{eqnarray}

\noindent If $N, N_{\check r}$ are requested to be parameters of ``proper" gauge
transformations [$N=m$, $N_{\check r}=m_{\check r}$], then the only solution
is $N=N_{\check r}=0$; in the case of improper gauge transformations there is
a unique solution of these elliptic equations with both N and $N_{\check r}$
growing linearly with $\vec \sigma$.

In this way, it is shown in Ref.\cite{reg1} that one can construct a copy of
the reduced phase space (the gauge-fixings intersect all the gauge orbits
once modulo the non studied problem of Gribov ambiguity), so that, by going to
Dirac brackets, a unique asymptotic Poincar\'e group is selected and there are
no more supertranslations. Then, in Ref.\cite{reg1} there is a comparison of
these results with the SPI formalism in the coordinate-dependent formulation
of Ref.\cite{p8} with an extension of the results of Ref.\cite{p11}.

All the formulations agree on the form of the ADM energy-momentum $P^{\tau}
_{ADM}$, $P^{\check r}_{ADM}$, which represents the conserved energy-momentum
of the $\tau$-slice $\Sigma_{\tau}$ of the asymptotically flat spacetime
$M^4$. It can be shown that $P^{(\mu )}_{ADM}$ is a four-vector under the
asymptotic Poincar\'e group, when this one is uniquely defined.

\vfill\eject

\section{Spinors on $M^4$ and on $\Sigma_{\tau}$.}

Let us add some notions about SL(2,C) spinors on $M^4$ and SU(2) spinors on
$\Sigma_{\tau}$ \cite{spinor,spinor1}. Our pseudoRiemannian spacetime $(M^4,
{}^4g_{\mu\nu})$ is assumed to admit a spin [or spinorial or principal
SL(2,C)-bundle] structure. Therefore, we can
consider an associated bundle over $M^4$ with structure group SL(2,C) and
standard fiber a complex 2-dimensional vector space V equipped with a
non-degenerate symplectic 2-form $\epsilon$; elements of V are denoted $\xi
^{\tilde A}$ [they are SL(2,C) spinor fields] and the symplectic 2-form has
components $\epsilon_{\tilde A\tilde B}=-\epsilon_{\tilde B\tilde A}$ with
inverse $\epsilon^{\tilde A\tilde B}$ [$\, \epsilon^{\tilde A\tilde B}
\epsilon_{\tilde B\tilde C}=\delta^{\tilde A}_{\tilde B}$]. If $L\in SL(2,C)$,
then $\epsilon_{\tilde A\tilde B}L^{\tilde A}{}_{\tilde C} L^{\tilde B}{}
_{\tilde D}=\epsilon_{\tilde C\tilde D}$. If $V^{*}$ is the dual space of V,
with elements $\xi_{\tilde A}$, $\epsilon$ gives an isomorphism between it and
V, namely $\epsilon$ can be used to lower and raise indices: $\xi^{\tilde A}=
\epsilon^{\tilde A\tilde B}\xi_{\tilde B}$, $\xi_{\tilde B}=\xi^{\tilde A}
\epsilon_{\tilde A\tilde B}$. The complex conjugate vector space $\bar V$, with
elements ${\bar \xi}^{{\tilde A}^{'}}$ (the primed spinors), and its dual
${\bar V}^{*}$, with elements ${\bar \xi}_{{\tilde A}^{'}}$, are also isomorphic
(namely $\bar \epsilon$ and its inverse can be used to lower and raise primed
indices). The SL(2,C) spinors in V and $\bar V$ correspond to the
$({1\over 2},0)$ and $(0,{1\over 2})$ representations of SL(2,C) [when
representing massive particles, they satisfy the Klein-Gordon equation, while
in the massless case they satisfy Weyl spinorial equations]. Given SL(2,C)
spinor fields on $M^4$, we can restrict them to SL(2,C) spinor fields over a
spacelike hypersurface in $M^4$ like $\Sigma_{\tau}$. To relate SL(2,C) spinors
to 4-tensors over $M^4$ in each point $p\in M^4$, one considers the
4-dimensional real vector space $W_p$ of all objects $\xi^{\tilde A{\tilde A}
^{'}}=\epsilon {\bar \xi}^{\tilde A{\tilde A}^{'}}$ [$\epsilon (+---)$ is the
signature of $M^4$, not to be confused with the symplectic 2-form], which
is equipped with the natural metric $\epsilon_{\tilde A\tilde B}{\bar \epsilon}
_{{\tilde A}^{'}{\tilde B}^{'}}$ with signature $\epsilon (+---)$. Therefore,
it is natural to identify $W_p$, at each point $p\in M^4$, with the tangent
space $T_pM^4$ at that point through an isomorphism $\sigma^{\mu}_{\tilde
A{\tilde A}^{'}}$ [called SL(2,C) ``soldering form"; it is unique and globally
defined if $M^4$ admits a spinor structure] with inverse $\sigma_{\mu}{}
^{\tilde A{\tilde A}^{'}}$: ${}^4V^{\mu}=\sigma^{\mu}{}_{\tilde A{\tilde A}
^{'}} \xi^{\tilde A{\tilde A}^{'}}$. Since tangent vectors are real, one has
${\bar \sigma}_{\mu}{}^{\tilde A{\tilde A}^{'}}=-\sigma_{\mu}{}^{\tilde
A{\tilde A}^{'}}$. The real spinor ${\bar \xi}^{{\tilde A}^{'}}\xi^{\tilde A}$
corresponds to a null vector $t^{\mu}=\sigma^{\mu}{}_{\tilde A{\tilde A}^{'}}
{\bar \xi}^{{\tilde A}^{'}}\xi^{\tilde A}$, $t^2=0$. Then, one has ${}^4g
^{\mu\nu}=\sigma^{\mu}{}_{\tilde A{\tilde A}^{'}} \sigma^{\nu}{}_{\tilde
B{\tilde B}^{'}} \epsilon^{\tilde A\tilde B}{\bar \epsilon}^{{\tilde A}
^{'}{\tilde B}^{'}}$ and there is a 2-1 homomorphism from the local SL(2,C)
transformation group on spinors to the local proper Lorentz group of $(M^4,
{}^4g_{\mu\nu})$: $L^{\mu}{}_{\nu}=\sigma^{\mu}{}_{\tilde A{\tilde A}^{'}}
\sigma_{\nu}{}^{\tilde B{\tilde B}^{'}} L^{\tilde A}{}_{\tilde B}{\bar L}
^{{\tilde A}^{'}}{}_{{\tilde B}^{'}}$. The torsion-free covariant derivative
${}^4\nabla_{\mu}$ on $M^4$ with ${}^4\nabla_{\mu}\, {}^4g_{\alpha\beta}=0$ is
uniquely extended to a covariant derivative ${}^4{\tilde \nabla}_{\mu}$
acting on spinors with ${}^4{\tilde \nabla}_{\mu}\, \epsilon_{\tilde A\tilde
B}=0$ through the requirement to be compatible with the soldering form:
${}^4{\tilde \nabla}_{\mu}\, \sigma^{\nu}{}_{\tilde A{\tilde A}^{'}}=0$
[see Ref.\cite{spinor1} for the relation between its curvature tensor and the
Riemann tensor of $M^4$].

As shown for instance in Ref.\cite{stewart} given a ``spin basis"
$o^{\tilde A}$, $i^{\tilde A}$ [$\epsilon_{\tilde A\tilde B}o^{\tilde a}
i^{\tilde B}=1$] for V, it induces a null tetrad in $M^4$ (like the one of
Appendix F):\hfill\break
\hfill\break
$L^{\mu}=\sigma^{\mu}{}_{\tilde A{\tilde A}^{'}} o^{\tilde A}{\bar o}
^{{\tilde A}^{'}}$, ${\cal N}^{\mu}=\sigma^{\mu}{}_{\tilde A{\tilde A}^{'}}
i^{\tilde A}{\bar i}^{{\tilde A}^{'}}$, $M^{\mu}=
\sigma^{\mu}{}_{\tilde A{\tilde A}^{'}} o^{\tilde A}{\bar i}^{{\tilde A}^{'}}$,
${\bar M}^{\mu}=
\sigma^{\mu}{}_{\tilde A{\tilde A}^{'}} i^{\tilde A}{\bar o}^{{\tilde A}^{'}}$.

Let now $({}^3\Sigma ,{}^3g_{rs})$ be an abstract Riemannian 3-manifold with
torsion-free derivative ${}^3\nabla_u$ [${}^3\nabla_u\, {}^3g_{rs}=0$]. Since
3-manifolds are parallelizable, they always admit a spin [principal
SU(2)-bundle] structure. As before consider the associated bundle over
${}^3\Sigma$ with standard fiber the 2-dimensional complex space V with
symplectic 2-form $\epsilon_{\tilde A\tilde B}$. One needs an extra structure
on each fiber, namely a positive definite Hermitian inner product $< .,. >$,
\hfill\break
$< \xi ,\eta >={\bar \xi}^{{\tilde A}^{'}} G_{{\tilde A}^{'}\tilde A} \eta
^{\tilde A}$, which is equivalent to a positive definite Hermitian metric
$G_{{\tilde A}^{'}\tilde B}={\bar G}_{{\tilde A}^{'}\tilde B}$ with inverse
$G^{{\tilde A}^{'}\tilde B}$ [either $\epsilon_{\tilde A\tilde B}$ or
$G_{{\tilde A}^{'}\tilde B}$ is assumed so normalized to get ${\bar \epsilon}
^{{\tilde A}^{'}{\tilde B}^{'}} G_{{\tilde A}^{'}\tilde A}G_{{\tilde B}^{'}
\tilde B}=\epsilon_{\tilde A\tilde B}$, so that $G_{{\tilde A}^{'}\tilde A}
G^{{\tilde A}^{'}\tilde B}=\delta^{\tilde B}_{\tilde A}$]. This metric allows
to convert the primed indices to the unprimed ones, so that we can restrict
ourselves to unprimed SU(2) spinors. The SU(2) transformations are the ones
that preserve both $\epsilon$ and $G$ structure [the Hermitian conjugate of a
transformation $U^{\tilde A}{}_{\tilde B}$ is $(U^{\dagger})^{\tilde C}{}
_{\tilde D}=G^{{\tilde B}^{'}\tilde C}{\bar U}^{{\tilde A}^{'}}{}_{{\tilde B}
^{'}}G_{{\tilde A}^{'}\tilde D}$]. The tangent space $T_p\, {}^3\Sigma$ at
$p\in {}^3\Sigma$ is always globally isomorphic, through the SU(2) ``soldering
form" $\sigma^r{}_{\tilde A}{}^{\tilde B}$, to the 3-dimensional real vector
space H of all objects at $p\in {}^3\Sigma$ of the form $\alpha^{\tilde A}{}
_{\tilde B}$ satisfying $\alpha^{\tilde A}{}_{\tilde A}=0$, $(\alpha
^{\dagger})^{\tilde A}{}_{\tilde B}=-\alpha^{\tilde A}{}_{\tilde B}$ [H is
equipped with a natural positive definite metric $(\alpha ,\beta )=-\alpha
^{\tilde A}{}_{\tilde B}\beta^{\tilde B}{}_{\tilde A}$, so that $\sigma^r{}
_{\tilde A}{}^{\tilde A}=0$, $(\sigma^r{}_{\tilde A}{}^{\tilde B})^{\dagger}=-
\sigma^r{}_{\tilde A}{}^{\tilde B}$]. One has ${}^3g^{rs}=-\sigma^r{}_{\tilde A}
{}^{\tilde B} \sigma^s{}_{\tilde B}{}^{\tilde A}=-Tr(\sigma^r\sigma^s)$.
The objects $\psi^{\tilde A}$ are SU(2) spinor fields on $({}^3\Sigma ,{}^3g
_{rs})$ and SU(2) transformations on spinors, $U^{\tilde A}{}_{\tilde B}$, are
tied to SO(3) transformations on the tangent spaces of ${}^3\Sigma$ by a 2-1
homomorphism $U^r{}_s=\sigma^{r\tilde A}{}_{\tilde B} U^{\tilde B}{}_{\tilde
C} \sigma^{s\tilde C}{}_{\tilde D} (U^{\dagger})^{\tilde D}{}_{\tilde A}$.
There is a unique extension of ${}^3\nabla_u$ to a covariant derivative
${}^3{\tilde \nabla}_u$ acting on SU(2) spinors such that ${}^3{\tilde \nabla}
_u \sigma^r{}_{\tilde A}{}^{\tilde B}={}^3{\tilde \nabla}_u\, \epsilon_{\tilde
A\tilde B}={}^3{\tilde \nabla}_u\, G_{{\tilde A}^{'}\tilde B}=0$.

Let us now consider the Riemannian spacelike hypersurface $(\Sigma_{\tau},
{}^3g_{rs})$ embedded in the pseudo-Riemannian spacetime $M^4$ and let us define
SU(2) spinors on $\Sigma_{\tau}$ starting from SL(2,C) spinors on $M^4$, just as
spatial 3-tensors on $\Sigma_{\tau}$ are identified with 4-tensors on $M^4$
which are tangent to $\Sigma_{\tau}$. As at each point of $\Sigma_{\tau}$
one can identify SO(3) transformations with proper Lorentz transformations
which preserve the future-directed unit normal $l^{\mu}(\tau ,\vec \sigma )$ to
$\Sigma_{\tau}$, similarly SU(2) transformations can be identified with
SL(2,C) transformations preserving $l^{\tilde A{\tilde A}^{'}}=l^{\mu} \sigma
_{\mu}{}^{\tilde A{\tilde A}^{'}}$ [one has $l_{\tilde A{\tilde A}^{'}}=l
_{\mu} \sigma^{\mu}{}_{\tilde A{\tilde A}^{'}}$ and $l_{\mu}l^{\mu}=\epsilon$
becomes $l^{\tilde A{\tilde A}^{'}}l_{\tilde B{\tilde A}^{'}}={{\epsilon}\over
2}\delta^{\tilde A}_{\tilde B}$], where $\sigma_{\mu}{}^{\tilde A{\tilde A}
^{'}}$ is the SL(2,C) soldering form. To reduce the structure group SL(2,C) of
the spin bundle to SU(2), one needs the extra structure of a positive definite
Hermitian inner product or,
equivalently, of a positive definite Hermitian metric\hfill\break
\hfill\break
$G_{{\tilde A}^{'}\tilde A}=\sqrt{2} l_{\tilde A{\tilde A}^{'}}$ \hfill\break
\hfill\break
[hermiticity
follows from ${\bar \sigma}_{\mu}^{\tilde B{\tilde A}^{'}}=-\sigma_{\mu}{}
^{\tilde A{\tilde B}^{'}}$]. The SL(2,C) transformations preserving $l_{\tilde
A{\tilde A}^{'}}$ can be identified with SU(2) transformations. The horizontal
subspace of the associated bundle over $M^4$ with standard fiber V with respect
to the normal $l_{\tilde A{\tilde A}^{'}}$ consists of all the elements
$\alpha^{\tilde A{\tilde A}^{'}}$ of V satisfying $\alpha^{\tilde A{\tilde A}
^{'}}l_{\tilde A{\tilde A}^{'}}=0$. This is a 3-dimensional real vector space
[with a positive definite metric $\epsilon_{\tilde A\tilde B}{\bar \epsilon}
_{{\tilde A}^{'}{\tilde B}^{'}}-\epsilon \, l_{\tilde A{\tilde A}^{'}}l
_{\tilde B{\tilde B}^{'}}=\epsilon_{\tilde A\tilde B}{\bar \epsilon}_{{\tilde
A}^{'}{\tilde B}^{'}}+{{\epsilon}\over 2} G_{{\tilde A}^{'}\tilde A}G_{{\tilde
B}^{'}\tilde B}$ induced by the metric $\epsilon_{\tilde A\tilde B}{\bar
\epsilon}_{{\tilde A}^{'}{\tilde B}^{'}}$ on V] isometric to the space H used
in the theory of SU(2) spinors, through the isometry \hfill\break
\hfill\break
$\alpha^{\tilde A{\tilde
A}^{'}} \mapsto \alpha^{\tilde A}{}_{\tilde B}=\alpha^{\tilde A{\tilde A}^{'}}
G_{{\tilde A}^{'}\tilde B}$ \hfill\break
\hfill\break
[the metric in H is $G^{{\tilde A}^{'}\tilde C}G
^{{\tilde B}^{'}\tilde D}(\epsilon_{\tilde A\tilde B}{\bar \epsilon}_{{\tilde
A}^{'}{\tilde B}^{'}}+{{\epsilon}\over 2}G_{{\tilde A}^{'}\tilde A}G_{{\tilde
B}^{'}\tilde B})=\epsilon^{\tilde C\tilde D}\epsilon_{\tilde A\tilde B}+
{{\epsilon}\over 2}\delta^{\tilde C}_{\tilde A}\delta^{\tilde D}_{\tilde B}$].
The SU(2) soldering form is \hfill\break
\hfill\break
$\sigma_{\mu}{}^{\tilde A}{}_{\tilde B}={}^3g
_{\mu}{}^{\nu}\sigma_{\nu}^{\tilde A{\tilde A}^{'}} G_{{\tilde A}^{'}\tilde
B}$ \hfill\break
\hfill\break
[${}^3g_{\mu\nu}=-\epsilon \, {}^3h_{\mu\nu}=-\epsilon ({}^4g_{\mu\nu}-
\epsilon l_{\mu}l_{\nu})$ so that ${}^3g_{\tilde
A\tilde C\tilde B\tilde D}=-{1\over 2}(\epsilon_{\tilde A\tilde D}\epsilon
_{\tilde B\tilde C}+\epsilon_{\tilde A\tilde B}\epsilon_{\tilde D\tilde C})$].
Since the spinor field ${}^4\alpha_{\tilde A{\tilde A}^{'}}$ on $M^4$ is
``spatial" if \hfill\break
\hfill\break
${}^4\alpha_{\tilde A{\tilde A}^{'}}l^{\tilde A{\tilde A}^{'}}
=0$, \hfill\break
\hfill\break
i.e. if ${}^4\alpha_{\tilde A\tilde B}={}^4\alpha_{\tilde A{\tilde A}
^{'}}l^{{\tilde A}^{'}}{}_{\tilde B}={}^4\alpha_{\tilde B\tilde A}={}^4\alpha
_{(\tilde A\tilde B)}$ [$(\tilde A\tilde B)$ means symmetrization],
each spatial 3-tensor ${}^3T^{\mu ...\nu}_{\alpha
...\beta}$ defines a spatial unprimed spinor field ${}^3T^{({\tilde A}_1{\tilde
A}_2)...({\tilde B}_1{\tilde B}_2)}_{({\tilde C}_1{\tilde C}_2)...({\tilde D}
_1{\tilde D}_2)}$. If $\xi_{\tilde A} \mapsto \xi^{\dagger}_{\tilde A}={\bar
\xi}_{{\tilde A}^{'}} l^{{\tilde A}^{'}}{}_{\tilde A}$ is the conjugation map,
a ``spatial spinor" field represents a ``real" spatial 3-tensor field if and
only if $({}^3T^{({\tilde A}_1{\tilde
A}_2)...({\tilde B}_1{\tilde B}_2)}_{({\tilde C}_1{\tilde C}_2)...({\tilde D}
_1{\tilde D}_2)})^{\dagger}=(-)^k\, {}^3T^{({\tilde A}_1{\tilde
A}_2)...({\tilde B}_1{\tilde B}_2)}_{({\tilde C}_1{\tilde C}_2)...({\tilde D}
_1{\tilde D}_2)}$ with 2k being the number of indices.

The torsion-free covariant derivative ${}^4\nabla_{\mu}$ [${}^4\nabla_{\mu}\,
{}^4g_{\alpha\beta}=0$] on $M^4$ goes down to the torsion-free ${}^3\nabla
_{\mu}$ [${}^3\nabla_{\mu}\, {}^3g_{\alpha\beta}=0$] acting on spatial
3-tensors on $\Sigma_{\tau}$ [see after Eq.(8) of I]. They are extended
to ${}^4{\tilde \nabla}_{\mu}$ [or ${}^4{\tilde \nabla}_{\tilde A{\tilde A}
^{'}}=\sigma^{\mu}_{\tilde A{\tilde A}^{'}}{}^4{\tilde \nabla}_{\mu}$;
${}^4{\tilde \nabla}_{\mu} \epsilon_{\tilde A\tilde B}={}^4{\tilde \nabla}_{\mu}
\sigma^{\nu}{}_{\tilde A{\tilde A}^{'}}=0$] acting on SL(2,C) spinors on $M^4$
and to ${}^3{\tilde \nabla}_{\tilde A\tilde B}=\sqrt{2} l_{(\, \tilde A}{}
^{{\tilde A}^{'}}\, {}^4\nabla_{\tilde B)\, {\tilde A}^{'}}$ (the
torsion-free Levi-Civita connection as we have seen) acting on spatial spinors
on $\Sigma_{\tau}$ [${}^3{\tilde \nabla}_u \sigma^r{}_{\tilde A}{}^{\tilde B}=0$
and ${}^3{\tilde \nabla}_u(\epsilon^{\tilde C\tilde D}\epsilon_{\tilde A\tilde
B}+{{\epsilon}\over 2}\delta^{\tilde C}_{\tilde A}\delta^{\tilde D}_{\tilde
B})=0$; it is the Levi-Civita connection of ${}^3g$].
One has the splitting \hfill\break
\hfill\break
${}^4{\tilde \nabla}_{\tilde A\tilde B}=
l^{{\tilde A}^{'}}{}_{\tilde B}\, {}^4{\tilde \nabla}_{\tilde A{\tilde A}^{'}}=
{1\over 2}\epsilon_{\tilde A\tilde B} T+l_{(\, \tilde B}{}^{{\tilde A}^{'}}\,
{}^4{\tilde \nabla}_{\tilde A)\, {\tilde A}^{'}}$. \hfill\break
\hfill\break
The first term $T=l^{\tilde
C{\tilde D}^{'}}\, {}^4{\tilde \nabla}_{\tilde C{\tilde D}^{'}}$ represents
the ``time derivative". The second term, which depends only on the intrinsic
geometry of $\Sigma_{\tau}$, represents the ``spatial" derivative ``only" when
$\Sigma_{\tau}$ has zero extrinsic curvature ${}^3K_{\mu\nu}=0$.

The true ``spatial" derivative , called the ``Sen connection", is given by real
operators ${}^3{\cal D}_{\tilde A\tilde B}={}^3{\cal D}_{\tilde B\tilde A}$
[acting on real spinor fields they produce real spinor fields] which are not
only the pullback to $\Sigma_{\tau}$ of ${}^4\nabla_{\mu}$ but also an
extension [depending also from the extrinsic geometry of $\Sigma_{\tau}$] of
${}^3\nabla_{\mu}$ from spatial tensors to SU(2) spinors [the Sen
connection is torsion-free, satisfies ${}^3{\cal D}_{\tilde A\tilde B}\,
{}^3g_{\tilde C\tilde D\tilde E\tilde F}=0$ but is not the Levi-Civita
connection  of ${}^3g$]. On scalars one has ${}^3{\cal D}_{\tilde A\tilde B}
\phi={}^3{\tilde \nabla}_{\tilde A\tilde B}\phi$. Instead its action on SU(2)
spinors is

\begin{equation}
{}^3{\cal D}_{\tilde A\tilde B}\, \psi_{\tilde C}={}^3{\tilde \nabla}_{\tilde
A\tilde B}\, \psi_{\tilde C}+{1\over {\sqrt{2}}}\, {}^3K_{\tilde A\tilde
B\tilde C}{}^{\tilde D}\, \psi_{\tilde D},
\label{c1}
\end{equation}

\noindent so that one has

\begin{equation}
{}^3{\cal D}_{\tilde A\tilde B}\, \psi^{\tilde B} ={}^3{\tilde \nabla}
_{\tilde A\tilde B} \psi^{\tilde B}+{1\over {2\sqrt{2}}}\, {}^3K \psi_{\tilde
A}.
\label{c2}
\end{equation}

\vfill\eject

\section{Komar Superpotential and Energy-Momentum Pseudotensors from the
Hilbert Action.}

For the sake of completeness let us add the standard derivation of the Noether
identities, the determination of the Komar superpotential and of the
energy-momentum pseudotensors in metric gravity starting from the Hilbert action
and from its invariance under 4-diffeomorphisms, since it is in this way that
one usually defines the weak Poincar\'e charges (even if with open problems
for the ADM angular momentum).

The invariance under $Diff\, M^4$ of the Hilbert action
 may be used to set the 4-metric tensor equal to the Minkowski 4-metric
and the affine connection to zero at any point of $M^4$. Indeed, these
conditions may be satisfied along an arbitrary geodesic of $M^4$. It is in this
way that the principle of equivalence, the equality of inertial and
gravitational mass, is described in general relativity: to first order in their
separation, all bodies moving on parallel geodesics move at the same rate. Just
this property is also responsible for the inability to define a ``local energy
density" for the gravitational field. Minkowski spacetime describes a
spacetime with no gravitational field, so that its energy density must be zero.
But a general spacetime can be made to appear Minkowskian along an arbitrary
geodesic. As a result, any non-tensorial (even if covariant) ``energy density"
can be made to be zero along an arbitrary geodesic and, therefore, has no
invariant meaning. To define a tensorial quantity requires the introduction of
an auxiliary vector field $\xi^{\mu}$, which is an element of arbitrariness.
It follows that only the global energy-momentum and angular momentum may be
given a meaning in general relativity.

Conservation laws in general relativity were first formulated by Einstein in
1916 \cite{lc2} [Noether's theorems appeared in 1918], who found a canonical
pseudo-tensor of the gravitational field homogeneous quadratic in the first
derivatives of the metric tensor. Due to its non-tensoriality, the local
energy density does not have a covariant significance and was criticized by
Schroedinger\cite{lc3} [he found a coordinate system in which all the
components of the pseudo-tensor vanished for the Schwarzschild metric outside
the Schwarzschild radius]. Bauer \cite{lc4} showed that simply by transforming
the description of flat space from Cartesian coordinates to spherical
coordinates an apparent nonzero ``energy density" results which yields an
infinite total ``energy". This
criticism was answered only when Einstein \cite{lc5,lc6} showed that the total
energy and momentum, the only meaningful quantities, are constants of the
motion and transform as a ``free-vector" [an affine tensor: a set of quantities
which are not defined at a particular point in space] under linear
coordinate transformations. This is the so-called Einstein\cite{lc5}-
Klein\cite{lc7} theorem: it assumes the existence of ``asymptotically flat"
coordinates such that\cite{lc8} ${}^4g_{\mu\nu}={}^4\eta_{\mu\nu}+O(r^{-1})$,
$\partial_{\alpha}\, {}^4g_{\mu\nu}=O(r^{-2})$ [$r$ is the distance measured
along geodesics from a point on a spacelike asymptotically flat hypersurface]
and, by writing the Schwarzschild line element in coordinates which are
Cartesian at infinity, one gets $p^o=m$, $p^i=0$. Einstein showed that under
certain conditions (essentially, no radiation) the total energy and momentum in
a closed domain of space (outside one uses Minkowski coordinates) is
independent of the choice of the coordinates within the domain [this was called
the ``flux theorem" by Pauli\cite{lc9};
see also Refs.\cite{lc10,lc10a,lc10b} based on the
work of P.von Freud\cite{lc11}, who was the first to find a superpotential for
the Einstein pseudo-tensor]. Trautman\cite{lc13} added conditions for extending
the Einstein-Klein results from the case of asymptotically flat isolated
non-radiating systems to that of radiating systems [see also the discussion in
Ref.\cite{lc14}, where a background Minkowski metric is used to covariantize
the treatment].

The pseudo-tensor was named a ``complex" by Lorentz\cite{lc12}, who gave a
different, non satisfactory, definition of energy and momentum of the
gravitational field. In order to discuss angular momentum, a symmetric
energy-momentum tensor is desirable\cite{lc9}, although not necessary
\cite{lc15}. The Einstein canonical pseudo-tensor has mixed indices, and raising
one with the metric tensor does not yield a symmetric quantity. Landau and
Lifshitz\cite{ll} succeeded in constructing a symmetric pseudo-tensor [see
Ref.\cite{lc14} for the associated angular momentum tensor in presence of a
flat background], but the associated total energy and momentum transform as a
vector density rather than as a vector as in the case of particles. Bergmann
\cite{lc16a} started the investigation on the local invariances of the action
in general relativity (second
Noether theorem) and of their Hamiltonian generators.
In the study of the relationship between Landau-Lifshitz and Einstein canonical
pseudotensors, Goldberg\cite{lc16} discovered a whole family of mixed and
symmetric pseudotensors with different weights. All the mixed tensors have the
same physical content (total energy and momentum), whereas the symmetric ones
are all different in their physical content. Of the symmetric quantities, only
the Landau-Lifshitz pseudo-tensor has the same total energy and momentum as the
Einstein canonical one (but has the wrong transformation properties).

Komar\cite{lc17}, trying to generalize an earlier theory due to M$\o$ller
\cite{lc18}, looked for a superpotential depending on a vector field $\xi^{\mu}
$, such that, when $\xi^{\mu}$ is the timelike Killing vector of the
Schwarzschild solution, the constant of the motion is the mass. In this way he
gets a Noether (weak) conservation law $t^{\nu}{}_{,\nu}{\buildrel \circ \over
=}\, 0$ in which $t^{\nu}$ is a tensor. However, this coordinate-free expression
depends on the choice of the vector field $\xi^{\mu}$. In asymptotically flat
spacetimes, one takes for $\xi^{\mu}$ asymptotic Killing vectors to flat
spacetime [it works for translations, but there are problems with rotations
\cite{lc19}].

M$\o$ller \cite{lc20} proposed a theory designed to provide a definition of
energy invariant under time independent spatial transformations without any
restriction on the asymptotic form of the metric. M$\o$ller's theory is based on
the Hilbert variational principle with the 4-metric reexpressed in terms of
orthogonal tetrads. Due to the extra 6 gauge degrees of freedom (there are
constraints generating the local Lorentz transformations in the tangent planes),
he proposed 6 gauge-fixings. The main point is that, when one assumes that the
universe is asymptotically flat [instead of a spatially closed (without
boundaries) one], one is introducing an absolute family of privileged
observers, namely those whose associated tetrads tend to Minkowski tetrads at
space infinity [against the philosophy of Mach's principle, which, to avoid
absolute motions, requires a spatially closed universe; see Section IV].
Then M$\o$ller\cite{lc21} applied his theory to the
axis-symmetric solution and found that the total energy agrees with the
expression of the Bondi mass. M$\o$ller's theory gives no information concerning
the linear momentum of the system.

contrastedWhile for regular Lagrangian systems proper conservation laws are related to
its invariances under global symmetries according to the first Noether
theorem, with singular Lagrangians one has improper [either weak (i.e.
implying the use of the equation of motion) or strong (i.e. independent from
the equations of motion)] conservation laws
\cite{lc22,lc31}, which are hidden in the Noether
identities implied by local symmetries (either inner gauge groups or
diffeomorphisms of time and/or space and/or spacetime) according to the second
Noether theorem [see the original Refs.\cite{lc10a,lc16a,lc16,lc32}
and Refs.\cite{sha} b), c), d)
for a systematic and complete treatment for singular
Lagrangians depending upon the first derivatives of the fields]. In general
relativity\cite{lc31} the local symmetry
transformations of the theory are the diffeomorphisms of the spacetime $M^4$.

In Refs.\cite{lc55} there is a complete bibliography on the modern variational
techniques based on jet bundles for treating Noether's theorems also for
Lagrangians depending on higher derivatives of the fields. This is the
Lagrangian approach to be  with either the use of the equations of
motion and Bianchi identities or the non-covariant Hamiltonian formalism using
momentum mapping methods in symplectic or presymplectic manifolds. In
particular, in the papers of Refs.\cite{lc55} one makes use of the formulation
based on the globally defined Poincar\'e-Cartan one-form (uniquely defined for
first and second order Lagrangians) applied to general relativity.

Let ${\cal L}(x,\phi_A,\phi_{A,\mu}, \phi_{A,\mu\nu})\, {\buildrel {def}
\over =}\, {\cal L}(x;\phi (x))$ [$S=\int d^4x\, {\cal
L}$] be a singular Lagrangian density depending on a set of fields $\phi_A(x)$
[A=1..N; for Einstein general relativity $\phi_A={}^4g_{\mu\nu}$ and
${\cal L}={\cal L}_H$, not explicitly depending on $x^{\mu}$, and $S=S_H=\int
d^4x\, {\cal L}_H$ of Eq.(22) of I] and their first and second derivatives (in
Einstein general relativity the dependence on the second derivatives is linear,
due to Eq.(3) of I, and can be reabsorbed in a 4-divergence:
${\cal L}_H={\cal L}_E+{{c^3}\over {16\pi G}} \partial_{\lambda} [
\sqrt{{}^4g} ({}^4g^{\lambda\alpha}\, {}^4g_{\mu\nu}\partial_{\alpha}\,
{}^4g^{\mu\nu}-\partial_{\mu}\, {}^4g^{\lambda\mu})]\,\,$ and this defines the
(not general covariant) Einstein action of Eq.(24) of I ). The
Euler-Lagrange equations are \hfill\break
\hfill\break
$L^A(x,\phi (x))={{\delta S}\over {\delta \phi_A
(x)}} = {{\partial {\cal L}}\over {\partial \phi_A}}(x)-\partial_{\mu}\,
{{\partial {\cal L}}\over {\partial \partial_{\mu}\phi_A}}(x)+\partial_{\mu}
\partial_{\nu}\, {{\partial {\cal L}}\over {\partial \partial_{\mu}
\partial_{\nu}\phi_A}}(x)\, {\buildrel \circ \over =}\, 0$. \hfill\break
\hfill\break
A given physical situation can be described in different frames of reference
(systems of coordinates) and by means of different sets of variables $\phi$,
or, for short, in different gauges. The class of finite gauge transformations
to be considered is given by \hfill\break
\hfill\break
$x^{{'}\, \mu}=X^{\mu}(x)$, $\phi^{'}_A(x^{'})=Y_A(x;\phi )$ \hfill\break
\hfill\break
[for metric general relativity ${}^4g^{'}_{\mu\nu}(x^{'}(x))=
{{\partial x^{\alpha}}\over {\partial x^{{'}\, \mu}}}{{\partial x^{\beta}}
\over {\partial x^{{'}\, \nu}}}\, {}^4g_{\alpha\beta}(x)$ ]. It contains
Lorentz and general coordinate transformations (diffeomorphisms) and symmetry
transformations of classical mechanics as well as electromagnetic gauge
transformations.

Given $S=\int_Ud^4x\, {\cal L}(x;\phi (x))$, a sufficient condition for the new
equations of motion for $\phi^{'}_A(x^{'})$ to be equivalent to the old ones
is that the new action $S^{'}$ is given by \hfill\break
\hfill\break
$S^{'}=\int_{U^{'}}d^4x^{'}\,
{\cal L}^{'}(x^{'};\phi^{'}(x^{'}))=\int_Ud^4x\, [{\cal L}(x;\phi (x))-
\partial_{\mu}Q^{\mu}(x;\phi (x))]$ \hfill\break
\hfill\break
for some 4 functions $Q^{\mu}$. Here,
${\cal L}^{'}$ is the new Lagrangian density and $U^{'}$ denotes the image of
U by the transformation $x^{\mu} \mapsto x^{{'}\, \mu}=X^{\mu}(x)$. The
functions $Q^{\mu}$ are arbitrary except for the condition that they must vanish
whenever the $\phi$'s and their derivatives vanish. This condition on $S^{'}$
does not ensure numerical invariance of the action ($S\not= S^{'}$), but one
has $\delta S=\delta S^{'}$ provided the $\delta \phi$'s vanish with their
derivatives on $\partial U$, the boundary of U. From this it follows that the
new Lagrangian density ${\cal L}^{'}$ is not uniquely determined by ${\cal L}$,
and, generally, there is an arbitrariness in the choice of the Lagrangian.
The Lagrangian density ${\cal L}^{'}$ as function of its arguments is different
from ${\cal L}$. The corresponding new equations of motion $L^{{'}\, A}(x;
\phi (x))=\delta S^{'}/\delta \phi^{'}_A(x)\, {\buildrel \circ \over =}\, 0$
will differ in form from the old ones.

But let us assume that the gauge transformation $x^{{'}\, \mu}=X^{\mu}(x)$,
$\phi^{'}_A(x^{'})+Y_A(x;\phi )$, is a
``symmetry transformation": this means that
it leaves the form of the equations of motion unaltered, namely $L^{{'}\, A}
(x;\phi (x))=L^A(x;\phi (x))$. It will be so if, for a certain choice of the
functions $Q^{\mu}$, the form of the Lagrangian density is not changed:
${\cal L}^{'}(x;\phi (x))={\cal L}(x;\phi (x))$.

Let us now assume that these gauge symmetries form a continuous group. Then,
they can be characterized by the infinitesimal transformations\hfill\break
\hfill\break
$x^{{'}\, \mu}=
x^{\mu}+\delta x^{\mu}(x)$, $\phi^{'}_A(x^{'})=\phi_A(x)+\delta \phi_A(x)=
\phi_A(x)+\delta_o\phi_A(x)+\delta x^{\mu}(x)\partial_{\mu} \phi_A(x)$
\hfill\break
\hfill\break
with
$\delta_o\phi_A(x)=\phi^{'}_A(x)-\phi_A(x)$ so that ${\cal L}(x;\phi^{'}(x))-
{\cal L}(x;\phi (x))=\delta_o{\cal L}$ [$\delta_o$ commutes with differentiation
]. Then one gets

\begin{eqnarray}
&&d^4x^{'}{\cal L}^{'}(x^{'};\phi^{'}(x^{'}))-d^4x {\cal L}(x;\phi (x))= d^4x
[(1+\partial_{\mu}\delta x^{\mu}){\cal L}(x^{'};\phi^{'}(x^{'}))-{\cal L}(x;
\phi (x))]=\nonumber \\
&&=d^4x [{\cal L}(x;\phi (x)) \partial_{\mu}\delta x^{\mu}+\delta
{\cal L}]=d^4x [{\cal L} \partial_{\mu} \delta x^{\mu}+\delta_o{\cal L}+
\delta x^{\mu} {\cal L}] \equiv -d^4x \partial_{\mu} Q^{\mu}_{INF},\nonumber \\
{}&&\nonumber \\
&&\Rightarrow \delta {\cal L}+{\cal L} \partial_{\mu} \delta x^{\mu}=
\delta_o{\cal L}+\partial_{\mu} ({\cal L} \delta x^{\mu}) \equiv \partial_{\mu}
F^{\mu},
\label{i10}
\end{eqnarray}

\noindent where $F^{\mu}=-Q^{\mu}_{INF}$ denotes the functions $Q^{\mu}$
corresponding to the infinitesimal transformations. This is the statement of
``quasi-invariance", which becomes ``invariance" when $F^{\mu}\equiv 0$. Since
\hfill\break
\hfill\break
$\delta_o{\cal L}={{\partial {\cal L}}\over {\partial \phi_A}}\delta_o\phi_A+
{{\partial {\cal L}}\over {\partial \partial_{\mu}\phi_A}}\delta_o\partial
_{\mu}\phi_A+{{\partial {\cal L}}\over {\partial \partial_{\mu}\partial_{\nu}
\phi_A}}\delta_o\partial_{\mu}\partial_{\nu}\phi_A=L^A\delta_o\phi_A+\partial
_{\mu}[({{\partial {\cal L}}\over {\partial \partial_{\mu}\phi_A}}-\partial
_{\nu}{{\partial {\cal L}}\over {\partial \partial_{\mu}\partial_{\nu}\phi_A}})
\delta_o\phi_A+{{\partial {\cal L}}\over {\partial \partial_{\mu}\partial_{\nu}
\phi_A}} \delta_o\partial_{\nu}\phi_A]$, \hfill\break
\hfill\break
one gets the Noether identity

\begin{equation}
\partial_{\mu}\theta^{\mu}_{(W)}\, {\buildrel {def} \over =}\, \partial_{\mu}
[({{\partial {\cal L}}\over {\partial \partial_{\mu}\phi_A}}-\partial
_{\nu}{{\partial {\cal L}}\over {\partial \partial_{\mu}\partial_{\nu}\phi_A}})
\delta_o\phi_A+{{\partial {\cal L}}\over {\partial \partial_{\mu}\partial_{\nu}
\phi_A}} \delta_o\partial_{\nu}\phi_A+{\cal L} \delta x^{\mu}-F^{\mu}]\equiv
-\delta_o\phi_A L^A\, {\buildrel \circ \over =}\, 0.
\label{i11}
\end{equation}

If the theory is quasi-invariant under a general group $G_{\infty q}$, that is
to say under transformations which depend upon q arbitrary functions
$\epsilon^a(x)$, a=1,..,q. of the $x^{\mu}$'s [in metric general relativity
one has the group $G_{\infty 4}$ of diffeomorphisms with arbitrary functions
$\epsilon^{\mu}(x)$], assumed for simplicity of the form \hfill\break
\hfill\break
$\delta x^{\mu}=
\epsilon^a(x)\xi^{\mu}_a(x)$, $\delta_o\phi_A(x)=\epsilon^a(x) \eta_{Aa}(x,\phi
,\partial_{\mu}\phi )+\partial_{\nu}\epsilon^a(x) \eta^{\nu}_{Aa}(x,\phi ,
\partial_{\mu}\phi )$,\hfill\break
\hfill\break
so that \hfill\break
\hfill\break
$F^{\mu}=\epsilon^a F^{\mu}_a+\partial_{\nu}
\epsilon^a F^{\mu\nu}_a$,\hfill\break
\hfill\break
then one gets the following Noether identities
from the vanishing of the coefficients of the arbitrary functions $\partial
_{\mu}\partial_{\nu}\partial_{\rho}\epsilon^a$, $\partial_{\mu}\partial_{\rho}
\epsilon^a$, $\partial_{\rho}\epsilon^a$, $\epsilon^a$ [indices inside round
brackets are completely symmetrized $t^{(\mu\nu )}={1\over 2}(t^{\mu\nu}+
t^{\nu\mu})$]

\begin{eqnarray}
&&\partial_{\mu}\theta^{\mu}_{(W)}[\epsilon^a]=\partial_{\mu}[\epsilon^at^{\mu}
_a+\partial_{\rho}\epsilon^a t^{\mu\rho}_a+\partial_{\nu}\partial_{\rho}
\epsilon^a {{\partial {\cal L}}\over {\partial \partial_{\mu}\partial_{\nu}
\phi_A}} \eta^{\rho}_{Aa}]\equiv \nonumber \\
&&\equiv -(\epsilon^a\eta_{Aa}+\partial_{\rho}\epsilon^a \eta^{\rho}_{Aa}) L^A\,
{\buildrel \circ \over =}\, 0,\nonumber \\
{}&&\nonumber \\
&&{{\partial {\cal L}}\over {\partial_{(\mu}\partial_{\nu}\phi_A}}\, \eta
^{\rho )}_{Aa} \equiv 0,\nonumber \\
{}&&\nonumber \\
&&\partial_{\nu}({{\partial {\cal L}}\over {\partial \partial_{\nu}\partial
_{(\mu}\phi_A}} \eta^{\rho )}_{Aa}) \equiv -t^{(\mu\rho )}_a\, {\buildrel
{def} \over =}\nonumber \\
&&{\buildrel {def} \over =}
\, -[({{\partial {\cal L}}\over {\partial \partial_{(\mu}\phi_A}}
-\partial_{\nu}{{\partial {\cal L}}\over {\partial \partial_{\nu}
\partial_{(\mu}\phi_A}}) \eta^{\rho )}_{Aa}+{{\partial {\cal L}}\over
{\partial \partial_{\nu}\partial_{(\mu}\phi_A}} (\delta^{\rho )}_{\nu}
\eta_{Aa}+\partial_{\nu}\eta^{\rho )}_{Aa})-F^{(\mu\rho )}_a],\nonumber \\
{}&&\nonumber \\
&&\partial_{\mu} t^{\mu\rho}_a+t^{\rho}_a\, {\buildrel {def} \over =}\, \partial
_{\mu} [({{\partial {\cal L}}\over {\partial \partial_{\mu}\phi_A}}-\partial
_{\nu} {{\partial {\cal L}}\over {\partial \partial_{\mu}\partial_{\nu}\phi_A}})
\eta^{\rho}_{Aa} +{{\partial {\cal L}}\over {\partial \partial_{\mu}\partial
_{\nu}\phi_A}} (\delta^{\rho}_{\nu} \eta_{Aa}+\partial_{\nu} \eta^{\rho}_{Aa})-
F^{\mu\rho}_a]+ \nonumber \\
&&+[({{\partial {\cal L}}\over {\partial \partial_{\rho}\phi_A}}-
\partial_{\nu} {{\partial {\cal L}}\over {\partial \partial_{\rho}\partial_{\nu}
\phi_A}}) \eta_{Aa}+{{\partial {\cal L}}\over {\partial \partial_{\rho}\partial
_{\nu}\phi_A}} \partial_{\nu} \eta^{\rho}_{Aa}+{\cal L} \xi^{\rho}_a-F^{\rho}_a]
\equiv -\eta^{\rho}_{Aa} L^A\, {\buildrel \circ \over =}\, 0,\nonumber \\
{}&&\nonumber \\
&&\partial_{\mu} t^{\mu}_a\, {\buildrel {def} \over =}\, \partial_{\mu} [(
{{\partial {\cal L}}\over {\partial \partial_{\mu}\phi_A}}-\partial_{\nu}
{{\partial {\cal L}}\over {\partial \partial_{\mu}\partial_{\nu}\phi_A}})
\eta_{Aa}+{{\partial {\cal L}}\over {\partial \partial_{\mu}\partial_{\nu}
\phi_A}} \partial_{\nu} \eta_{Aa}+ {\cal L} \xi^{\mu}_a-F^{\mu}_a]\equiv -
\eta_{Aa} L^A\, {\buildrel \circ \over =}\, 0.
\label{i12}
\end{eqnarray}

These Noether identities imply the ``contracted Bianchi identities"

\begin{equation}
\eta_{Aa} L^A - \partial_{\rho} (\eta^{\rho}_{Aa} L^A) \equiv 0,
\label{i13}
\end{equation}

\noindent and, since $\partial_{\mu}\partial_{\rho} t^{\mu\rho}_a=\partial_{\mu}
\partial_{\rho} t^{(\mu\rho )}_a\equiv \partial_{\rho}\partial_{\mu}\partial
_{\nu} ({{\partial {\cal L}}\over {\partial \partial_{\mu}\partial_{(\nu}
\phi_A}} \eta^{\rho )}_{Aa}) \equiv \partial_{\rho}\partial_{\mu}\partial_{\nu}
({{\partial {\cal L}}\over {\partial \partial_{(\mu}\partial_{\nu}\phi_A}}
\eta^{\rho )}_{Aa})\equiv 0$, one also gets the
strong (i.e. independent from the equations of motion) conservation laws

\begin{equation}
\partial_{\mu} \theta^{\mu}_{(S)a}\, {\buildrel {def} \over =}\, \partial
_{\mu} (t^{\mu}_a + \eta^{\mu}_{Aa} L^A) \equiv 0.
\label{i14}
\end{equation}

The original Noether identities together with the contracted Bianchi identities
allow to get \hfill\break
\hfill\break
$\partial_{\mu} \theta^{\mu}_{(W)}[\epsilon^a]\equiv -\delta_o
\phi_A L^A=-(\epsilon^a\eta_{Aa}+\partial_{\rho}\epsilon^a \eta^{\rho}_{Aa})
L^A\equiv -[\partial_{\rho}\epsilon^a \eta^{\rho}_{Aa} L^A+\epsilon^a\partial
_{\rho}(\eta^{\rho}_{Aa} L^A)]=-\partial_{\rho}(\epsilon^a \eta^{\rho}_{Aa} L^A)
$,\hfill\break
\hfill\break
which is equivalent to the generalized Trautman strong conservation law

\begin{eqnarray}
&&\partial_{\mu} \theta^{\mu}_{(S)}[\epsilon^a]=\partial_{\mu} [\theta^{\mu}
_{(W)}[\epsilon^a] + \epsilon^a \eta^{\mu}_{Aa} L^A ] \equiv 0,\nonumber \\
{}&&\nonumber \\
&&\Downarrow \nonumber \\
&&{}\nonumber \\
&&\theta^{\mu}_{(S)}[\epsilon^a]=\partial_{\nu}\, U^{[\mu\nu ]}
[\epsilon^a],\nonumber \\
&&\theta^{\mu}_{(W)}[\epsilon^a]=\epsilon^a t^{\mu}_a+\partial_{\rho}\epsilon^a
t_a^{\mu\rho}+\partial_{\nu}\partial_{\rho}\epsilon^a {{\partial {\cal L}}
\over {\partial \partial_{\mu}\partial_{\nu}\phi_A}} \eta^{\rho}_{Aa}+
\partial_{\nu} V^{[\mu\nu ]}=\nonumber \\
&&=\theta^{\mu}_{(S)}[\epsilon^a]-\epsilon^a \eta^{\mu}_{Aa} L^A
=\partial_{\nu}U^{[\mu\nu ]}[\epsilon^a]-\epsilon^a \eta^{\mu}_{Aa}L^A,
\label{i15}
\end{eqnarray}

\noindent where we introduced the superpotential $U^{[\mu\nu ]}[\epsilon^a]=-
U^{[\nu\mu ]}[\epsilon^a]$ and the arbitrariness $V^{[\mu\nu ]}$ implicit
in the first line of Eq.(\ref{i12}).

Let us now assume that we have a subgroup $G_p$ of the general group $G_{\infty
q}$ so that one can write $\epsilon^a(x)=\epsilon^{\bar a}\zeta^a_{\bar a}(x)$,
with $\epsilon^{\bar a}$, $\bar a$=1,..,p, the constant parameters of $G_p$.
Then one has the following restrictions to $G_p$: \hfill\break
\hfill\break
$\delta x^{\mu}=\epsilon
^{\bar a}\zeta^a_{\bar a}(x)\xi^{\mu}_a(x)=\epsilon^{\bar a} {\hat \xi}^{\mu}
_{\bar a}(x)$, $\delta_o\phi_A=\epsilon^{\bar a}(\zeta^a_{\bar a}(x)\eta_{Aa}
(x)+\partial_{\nu}\zeta^a_{\bar a}(x)\eta^{\nu}_{Aa}(x))=\epsilon^{\bar a}
{\hat \eta}_{A\bar a}(x)$, $F^{\mu}=\epsilon^{\bar a}(\zeta^a_{\bar a}F^{\mu}
_a+\partial_{\nu}\zeta^a_{\bar a}F^{\mu\nu}_a)=
\epsilon^{\bar a}{\hat F}^{\mu}_{\bar a}$.\hfill\break
\hfill\break
Then, one gets\cite{lc16a} the
following improper weak conservation laws

\begin{equation}
\partial_{\mu} t^{\mu}_{\bar a}\, {\buildrel {def} \over =}\, \partial_{\mu}
[({{\partial {\cal L}}\over {\partial \partial_{\mu}\phi_A}}-\partial_{\nu}
{{\partial {\cal L}}\over {\partial \partial_{\mu}\partial_{\nu}\phi_A}})
{\hat \eta}_{A\bar a}+{{\partial {\cal L}}\over {\partial \partial_{\mu}
\partial_{\nu}\phi_A}} \partial_{\nu}{\hat \eta}_{A\bar a}+{\cal L} {\hat
\xi}^{\mu}_{\bar a}-{\hat F}^{\mu}_{\bar a}]\equiv -{\hat \eta}_{A\bar a} L^A\,
{\buildrel \circ \over =}\, 0,
\label{i16}
\end{equation}

\noindent and, by using the contracted Bianchi identities, also the strong ones
\hfill\break
\hfill\break
$\partial_{\mu}(t^{\mu}_{\bar a}-\zeta^a_{\bar a}\eta^{\mu}_{A\bar a} L^A)
\equiv 0$.\hfill\break
\hfill\break
Let us now apply the second Noether theorem to the Lagrangian densities
${\cal L}_H=\sqrt{{}^4g}\, {}^4R$ with ${}^4g_{\mu\nu}$ as independent
variables and with equations of motion $L^{\mu\nu}=\sqrt{{}^4g}\, {}^4G^{\mu\nu}
\, {\buildrel \circ \over =}\, 0$. One has strict invariance $\delta_o{\cal L}
_H\equiv 0$, i.e. $F^{\mu}_H\equiv 0$, under the infinitesimal diffeomorphisms
[see Eqs.(31) of I]

\begin{eqnarray}
&&\delta x^{\mu}(x)=\epsilon^{\mu}(x),\nonumber \\
{}&&\nonumber \\
&&\delta_o\, {}^4g_{\mu\nu}(x)=-\epsilon^{\rho}\partial_{\rho}\, {}^4g_{\mu\nu}
(x)-\partial_{\sigma}\epsilon^{\rho}(x) [\delta^{\sigma}_{\nu}\, {}^4g_{\mu\rho}
(x)+\delta^{\sigma}_{\mu}\, {}^4g_{\rho\nu}(x)]=\nonumber \\
&&+\delta \, {}^4g_{\mu\nu}(x)-
\epsilon^{\rho}(x)\partial_{\rho}\, {}^4g_{\mu\nu}(x)={\cal L}_{-\epsilon
^{\alpha}\partial_{\alpha}}\, {}^4g_{\mu\nu}(x).
\label{i17}
\end{eqnarray}

From Eqs.(3) of I and from $\partial \, {}^4\Gamma^{\lambda}
_{\rho\sigma}/\partial \partial_{\mu}\, {}^4g_{\alpha\beta}={1\over 2}\, {}^4g
^{\lambda\tau} (\delta^{\mu}_{\rho} \delta^{\alpha\beta}_{(\tau\sigma )}+
\delta^{\mu}_{\sigma} \delta^{\alpha\beta}_{(\tau\rho )}-\delta^{\mu}_{\tau}
\delta^{\alpha\beta}_{(\rho\sigma )})$ with $\delta^{\alpha\beta}_{(\mu\nu )}=
{1\over 2}(\delta^{\alpha}_{\mu}\delta^{\beta}_{\nu}+\delta^{\alpha}_{\nu}
\delta^{\beta}_{\mu})$, one gets

\begin{eqnarray}
&&{{\partial {\cal L}_H}\over {\partial \partial_{\mu}\partial_{\nu}\, {}^4g
_{\alpha\beta}}} = \sqrt{{}^4g} [{1\over 2}({}^4g^{\mu\alpha}\, {}^4g^{\nu\beta}
+{}^4g^{\mu\beta}\, {}^4g^{\nu\alpha})-{}^4g^{\mu\nu}\, {}^4g^{\alpha\beta}],
\nonumber \\
&&{}\nonumber \\
&&{{\partial {\cal L}_H}\over {\partial \partial_{\mu}\, {}^4g_{\alpha\beta}}}=
\sqrt{{}^4g}[{}^4g^{\mu\rho}({}^4g^{\alpha\sigma}\, {}^4\Gamma^{\beta}
_{\rho\sigma}+{}^4g^{\beta\sigma}\, {}^4\Gamma^{\alpha}_{\rho\sigma})-
{}^4g^{\alpha\rho}\, {}^4g^{\beta\sigma}\, {}^4\Gamma^{\mu}_{\rho\sigma}]+
\nonumber \\
&&+{}^4g^{\mu\alpha} \partial_{\rho}(\sqrt{{}^4g}\, {}^4g^{\rho\beta})+{}^4g
^{\mu\beta} \partial_{\rho}(\sqrt{{}^4g}\, {}^4g^{\rho\alpha})-{}^4g^{\alpha
\beta} \partial_{\rho} (\sqrt{{}^4g}\, {}^4g^{\rho\mu}),\nonumber \\
&&{}\nonumber \\
&&{{\partial {\cal L}_H}\over {\partial \, {}^4g_{\alpha\beta}}}=\sqrt{{}^4g}
[-{}^4G^{\alpha\beta}+\partial_{\rho}\, {}^4g_{\eta\tau}\, {}^4g^{\mu\nu}\,
{}^4\Gamma^{\sigma}_{\mu\nu} ({}^4g^{\rho\eta}\, \delta^{(\alpha}_{\sigma}\,
{}^4g^{\beta )\tau}-\delta^{\tau}_{\sigma}\, {}^4g^{\rho (\alpha}\, {}^4g
^{\beta )\eta})-\nonumber \\
&&-{}^4g^{\rho (\alpha}\, {}^4g^{\beta )\delta}\, {}^4g^{\mu\nu} \partial_{\rho}
({}^4g_{\delta\sigma}\, {}^4\Gamma^{\sigma}_{\mu\nu})-{1\over 4} \partial_{\rho}
\, {}^4g_{\gamma\delta}\, {}^4g^{\mu\nu}\, {}^4\Gamma^{\eta}_{\mu\nu} ({}^4g
^{\gamma\delta}\, {}^4g^{\rho (\alpha}\, \delta^{\beta )}_{\eta}+\delta^{\rho}
_{\eta}\, {}^4g^{\gamma (\alpha}\, {}^4g^{\beta )\delta})+\nonumber \\
&&+{1\over 2} {}^4g^{\mu\nu} ({}^4g^{\alpha\gamma}\, {}^4g^{\beta\delta}
\partial_{\mu}\partial_{\nu}\, {}^4g_{\gamma\delta}-2\, {}^4g^{\eta (\alpha}\,
{}^4g^{\beta )\gamma}\, {}^4g^{\tau\delta} \partial_{\mu}\, {}^4g_{\eta\tau}
\partial_{\nu}\, {}^4g_{\gamma\delta})+\nonumber \\
&&+{}^4g^{\mu\nu}\, {}^4\Gamma^{\eta}_{\mu\sigma}\, {}^4\Gamma^{\tau}_{\nu\rho}
({}^4g^{\rho (\alpha}\, \delta^{\beta )}_{\eta}\delta^{\sigma}_{\tau}+\delta
^{\rho}_{\eta}\, {}^4g^{\sigma (\alpha}\, \delta^{\beta )}_{\tau})].
\label{i18}
\end{eqnarray}

The weak and strong  conservation laws are respectively

\begin{eqnarray}
&&\partial_{\mu}\, {}_H\theta^{\mu}_{(W)}[\epsilon^{\rho}] \equiv [\epsilon
^{\rho}\partial_{\rho}\, {}^4g_{\mu\nu}+\partial_{\sigma}\epsilon^{\rho}
(\delta^{\sigma}_{\nu}\, {}^4g_{\mu\rho}+\delta^{\sigma}_{\mu}\, {}^4g_{\rho\nu}
)] \sqrt{{}^4g}\, {}^4G^{\mu\nu}\, {\buildrel \circ \over =}\, 0,\nonumber \\
&&\partial_{\mu}\, {}_H\theta^{\mu}_{(S)\rho}[\epsilon^{\rho}]=\partial_{\mu}
[{}_H\theta^{\mu}_{(W)}[\epsilon^{\rho}]-\epsilon^{\rho}(\delta^{\mu}_{\alpha}
\, {}^4g_{\rho\beta}+\delta^{\mu}_{\beta}\, {}^4g_{\rho\alpha})\sqrt{{}^4g}
\, {}^4G^{\alpha\beta}] \equiv 0,\nonumber \\
{}&&\nonumber \\
&&{}_{H}\theta^{\mu}_{(S)}[\epsilon^{\rho}]=\partial_{\nu}U^{[\mu\nu ]}
[\epsilon^{\rho}],\nonumber \\
&&{}_H\theta^{\mu}_{(W)}[\epsilon^{\rho}]=\epsilon^{\rho} t^{\mu}_{\rho}+
\partial_{\sigma}\epsilon^{\rho} t^{\mu\sigma}_{\rho}-\partial_{\nu}\partial
_{\sigma}\epsilon^{\rho}{{\partial {\cal L}_H}\over {\partial \partial_{\mu}
\partial_{\nu}\, {}^4g_{\alpha\beta}}}(\delta^{\sigma}_{\alpha}\, {}^4g
_{\beta\rho}+\delta^{\sigma}_{\beta}\, {}^4g_{\alpha\rho})+\partial_{\nu}
V^{[\mu\nu ]}=\nonumber \\
&&=-[\epsilon^{\rho}\partial_{\rho}\,
{}^4g_{\alpha\beta}+\partial_{\sigma}\epsilon^{\rho}(\delta^{\sigma}_{\alpha}\,
{}^4g_{\beta\rho}+\delta^{\sigma}_{\beta}\, {}^4g_{\alpha\rho})] ({{\partial
{\cal L}_H}\over {\partial \partial_{\mu}\, {}^4g_{\alpha\beta}}}-\partial_{\nu}
{{\partial {\cal L}_H}\over {\partial \partial_{\mu}\partial_{\nu}\, {}^4g
_{\alpha\beta}}})-\nonumber \\
&&-\partial_{\nu}[\epsilon^{\rho}\partial_{\rho}\, {}^4g_{\alpha\beta}+\partial
_{\sigma}\epsilon^{\rho}(\delta^{\sigma}_{\alpha}\, {}^4g_{\beta\rho}+\delta
^{\sigma}_{\beta}\, {}^4g_{\alpha\rho})]{{\partial {\cal L}_H}\over {\partial
\partial_{\mu}\partial_{\nu}\, {}^4g_{\alpha\beta}}}+\epsilon^{\mu}
\sqrt{{}^4g}\, {}^4R=\nonumber \\
&&=\partial_{\nu} U^{[\mu\nu ]}+\epsilon^{\rho} (\delta^{\mu}_{\alpha}\,
{}^4g_{\rho\beta}+\delta^{\mu}_{\beta}\, {}^4g_{\rho\alpha}) \sqrt{{}^4g}\,
{}^4G^{\alpha\beta}.
\label{i19}
\end{eqnarray}

\noindent Here $t^{\mu}_{\rho}$ is the candidate for the energy-momentum
pseudo-tensor once one has done a choice for $U^{[\mu\nu ]}$ and $V^{[\mu\nu ]}$.

The Noether identities (\ref{i12}) are satisfied with

\begin{eqnarray}
&&{}_Ht^{\mu\nu}_{\rho}=-(\delta^{\nu}_{\alpha}\, {}^4g_{\beta\rho}+\delta^{\nu}
_{\beta}\, {}^4g_{\alpha\rho})({{\partial {\cal L}_H}\over {\partial \partial
_{\mu}\, {}^4g_{\alpha\beta}}}-\partial_{\nu}{{\partial {\cal L}_H}\over
{\partial \partial_{\mu}\partial_{\nu}\, {}^4g_{\alpha\beta}}})-\nonumber \\
&&-[\delta^{\nu}_{\sigma}\partial_{\rho}\, {}^4g_{\alpha\beta}+\partial_{\sigma}
(\delta^{\nu}_{\alpha}\, {}^4g_{\beta\rho}+\delta^{\nu}_{\beta}\, {}^4g_{\alpha
\rho})]{{\partial {\cal L}_H}\over {\partial \partial_{\mu}\partial_{\nu}\,
{}^4g_{\alpha\beta}}},\nonumber \\
{}&&\nonumber \\
&&{}_Ht^{\mu}_{\rho}=-\partial_{\rho}\, {}^4g_{\alpha\beta}
({{\partial {\cal L}_H}\over {\partial \partial
_{\mu}\, {}^4g_{\alpha\beta}}}-\partial_{\nu}{{\partial {\cal L}_H}\over
{\partial \partial_{\mu}\partial_{\nu}\, {}^4g_{\alpha\beta}}})-\nonumber \\
&&-\partial_{\nu}\partial_{\rho}\, {}^4g_{\alpha\beta}
{{\partial {\cal L}_H}\over {\partial \partial_{\mu}\partial_{\nu}\,
{}^4g_{\alpha\beta}}}+\delta^{\mu}_{\rho} \sqrt{{}^4g}\, {}^4R,\nonumber \\
{}&&\nonumber \\
&&\partial_{\mu}\, {}_Ht^{\mu\nu}_{\rho}+{}_Ht^{\nu}_{\rho}\equiv (\delta^{\mu}
_{\alpha}\, {}^4g_{\rho\beta}+\delta^{\mu}_{\beta}\, {}^4g_{\rho\alpha})
\sqrt{{}^4g}\, {}^4G^{\alpha\beta}\, {\buildrel \circ \over =}\, 0,
\nonumber \\
&&\partial_{\mu}\, {}_Ht^{\mu}_{\rho}\equiv \partial_{\rho}\, {}^4g_{\alpha
\beta} \sqrt{{}^4g}\, {}^4G^{\alpha\beta}\, {\buildrel \circ \over =}\, 0,
\label{i20}
\end{eqnarray}

\noindent and the contracted Bianchi identities are

\begin{equation}
{}^4\nabla_{\alpha}\, {}^4G^{\alpha\beta}={1\over {\sqrt{{}^4g}}}\partial
_{\alpha}(\sqrt{{}^4g}\, {}^4G^{\alpha\beta})+{}^4\Gamma^{\beta}_{\alpha\gamma}
\, {}^4G^{\alpha\gamma} \equiv 0.
\label{i21}
\end{equation}

The conclusion of this discussion based on the second Noether theorem for the
generally covariant Hilbert action is that any vector field \hfill\break
\hfill\break
$\xi^{\mu}=\sqrt{{}^4g}\, \epsilon^{\mu}$ \hfill\break
\hfill\break
(it is more convenient to use $\xi^{\mu}$
rather than $\epsilon^{\mu}$ as a parameter) generates a
one-parameter group of diffeomorphisms, which
gives rise \hfill\break
\hfill\break
to a ``weak" [$\partial_{\mu}\, \theta^{\mu}_{(W)}[\xi ]\, {\buildrel
\circ \over =}\, 0$] and a ``strong" [$\partial_{\mu}\, \theta^{\mu}_{(S)}[\xi
]\, \equiv 0$] conservation law. \hfill\break
\hfill\break
Any strongly conserved quantity $\theta^{\mu}
_{(S)}[\xi ]$ can be written in the form $\theta^{\mu}_{(S)}[\xi ] = \partial
_{\nu} U^{[\mu\nu ]}[\xi ]$ [whatever the transformation properties of
$\xi^{\mu}$ are; $\xi^{\mu}$ may also not be a 4-vector], where
$U^{[\mu\nu ]}=-U^{[\nu\mu ]}$ is a ``superpotential" defined up to another
antisymmetric quantity $V^{[\mu\nu ]}
=-V^{[\nu\mu ]}$. One can show that \hfill\break
\hfill\break
$\theta^{\mu}_{(S)}[\xi ] =\partial_{\nu}
U^{[\mu\nu ]}[\xi ] \equiv \theta^{\mu}_{(W)}[\xi ] +{}^4G^{\mu}{}_{\nu} \xi
^{\nu}\, {\buildrel \circ \over =}\, \theta^{\mu}_{(W)}[\xi ] +{}^4T^{\mu}{}
_{\nu}\xi^{\nu}$ \hfill\break
\hfill\break
so that by changing the superpotential, $U^{[\mu\nu ]}\mapsto
U^{{'}[\mu\nu ]}=U^{[\mu\nu ]}+V^{[\mu\nu ]}$, one has $\theta^{\mu}_{(W)}
\mapsto \theta^{{'}\mu}_{(W)}=\theta^{\mu}_{(W)}+\partial_{\nu}V^{[\mu\nu ]}$.
Two superpotentials are especially important

A) ``Komar\cite{lc17} covariant superpotential" ($\theta^{\mu}_{(S)}[\xi ]$ is a
vector density):

\begin{equation}
{}_{(K)}U^{[\mu\nu ]}[\xi ]\, = {{c^3}\over {8\pi G}} \sqrt{{}^4g} ({}^4\nabla
^{\mu}\, \xi^{\nu}-{}^4\nabla^{\nu}\, \xi^{\mu}).
\label{i27}
\end{equation}

In this case the associated weak conservation law $\partial_{\mu}\, {}_{(K)}
\theta^{\mu}_{(W)}\, {\buildrel \circ \over =}\, 0$ is the divergence of a
tensor density [$\partial_{\mu}\, {}_{(K)}\theta^{\mu}_{(W)}=\sqrt{{}^4g}\,
{}^4\nabla_{\mu}\, {}_{(K)}{\hat \theta}^{\mu}_{(W)}\, {\buildrel \circ \over
=}\, 0$]. Therefore, this is a coordinate-free expression which depends,
however, on the choice of the vector field $\xi^{\mu}$, which, in
asymptotically flat spacetimes, is chosen to tend to asymptotic Killing vectors
for the evaluation of the global conserved quantities. Let us remark that
 ${}_{(K)}\theta^{\mu}_{(W)}$ contains the second derivatives of the metric.

B) ``Bergmann\cite{lc23} superpotential"

\begin{eqnarray}
&&{}_{(B)}U^{[\mu\nu ]}[\xi ] = {}_{(F)}U_{\lambda}{}^{[\mu\nu ]} \xi^{\lambda},
\nonumber \\
&&{}_{(F)}U_{\lambda}{}^{[\mu\nu ]}=-{{c^3}\over {16\pi G}}{1\over
{\sqrt{{}^4g}}}\, {}^4g_{\lambda\rho}\partial_{\gamma}[{}^4g ({}^4g^{\mu\gamma}
\, {}^4g^{\nu\rho}-{}^4g^{\nu\gamma}\, {}^4g^{\mu\rho})],\quad or
\nonumber \\
&&{}_{(F)}U^{\lambda [\mu\nu ]}={}^4g^{\lambda\rho}\, {}_{(F)}U_{\rho}{}
^{[\mu\nu ]}={{c^3}\over {16\pi G}} {1\over {\sqrt{{}^4g}}} \partial_{\beta}
({}^4{\hat g}^{\nu\beta}\, {}^4{\hat g}^{\mu\lambda}-{}^4{\hat g}^{\nu\lambda}
\, {}^4{\hat g}^{\mu\beta})=\nonumber \\
&&={{c^3}\over {16\pi G}} {1\over {\sqrt{{}^4g}}} \partial_{\beta}
{\cal T}^{\nu\lambda\beta\mu},
\label{i28}
\end{eqnarray}

\noindent where ${}_{(F)}U_{\lambda}{}^{[\mu\nu ]}$ is the Freud superpotential
\cite{lc11} (the notation of Eqs.(\ref{i24})-(\ref{i26}) is used).

All the known gravitational pseudotensors or complexes can be derived from the
superpotentials ${}_{(K)}U^{[\mu\nu ]}[\xi ]$ and ${}_{(B)}U^{[\mu\nu ]}[\xi ]$.

1) ``Einstein canonical pseudo-tensor" - If in ${}_{(B)}U^{[\mu\nu ]}[\xi ]$ one
chooses $\xi^{\mu}$ to be an object with constant components in any coordinate
system, one gets [see also Eq.(\ref{i25})]

\begin{eqnarray}
&&\theta^{\mu}_{(S)}[\xi ] =\xi^{\lambda} \partial_{\nu}\, {}_{(F)}U_{\lambda}
{}^{[\mu\nu ]} =\xi^{\lambda}\, {}_{(E)}\theta_{\lambda}{}^{\mu}\, {\buildrel
\circ \over =}\, \theta^{\mu}_{(W)}[\xi ]+{}^4T^{\mu}{}_{\lambda} \xi^{\lambda}
=\xi^{\lambda} ({}_{(E)}t_{\lambda}{}^{\mu}+{}^4T_{\lambda}{}^{\mu}),
\nonumber \\
&&{}_{(E)}\theta_{\nu}{}^{\mu}\, {\buildrel \circ \over =}\, {}_{(E)}t_{\nu}{}
^{\mu}+{}^4T_{\nu}{}^{\mu},\nonumber \\
&&\partial_{\mu}\, {}_{(E)}\theta_{\nu}{}^{\mu}\equiv 0,\quad\quad {}_{(E)}
\theta^{\nu\mu}={}^4g^{\nu\alpha}\, {}_{(E)}\theta_{\alpha}{}^{\mu}\not=
{}_{(E)}\theta^{\mu\nu},\nonumber \\
&&{}_{(E)}t_{\nu}{}^{\mu}=-\delta^{\mu}_{\nu}\, {}^4Z+\partial_{\nu}\, {}^4g
_{\alpha\beta} {{\partial \, {}^4Z}\over {\partial \partial_{\mu}\, {}^4g
_{\alpha\beta} }},\quad\quad {}^4Z={}^4R-\partial_{\rho}(\partial_{\mu}\,
{}^4g_{\alpha\beta} {{\partial \, {}^4R}\over {\partial \partial_{\rho}
\partial_{\mu}\, {}^4g_{\alpha\beta} }} ),\nonumber \\
&&{}_{(E)}t_{\nu}{}^{\mu}=2\sqrt{{}^4g}\, {}^4G_{\nu}{}^{\mu}+{{c^3}\over
{16\pi G}} \partial_{\alpha} ({{{}^4g_{\mu\sigma}}\over {\sqrt{{}^4g}}}
{\cal T}^{\alpha\nu\beta\sigma}),
\label{i29}
\end{eqnarray}

\noindent where ${}_{(E)}t_{\nu}{}^{\mu}$ is Einstein's energy-momentum
pseudo-tensor\cite{lc24} and ${}_{(E)}\theta_{\nu}{}^{\mu}$ is a tensor density
of weight 1.

2) ``Landau-Lifschitz\cite{ll} symmetric pseudo-tensor" - It is obtained from
${}_{(B)}U^{[\mu\nu ]}[\xi ]$ by choosing the ``covariant" components $\xi_{\mu}
/ \sqrt{{}^4g}$ to be constant [see also Eq.(\ref{i24})]:

\begin{eqnarray}
&&{}_{(L)}\theta^{\mu\nu} \equiv \partial_{\gamma}(\sqrt{{}^4g}\, {}^4g
^{\mu\delta}\, {}_{(F)}U_{\delta}{}^{[\nu\gamma ]})\, {\buildrel \circ \over =}
\, {}_{(L)}t^{\mu\nu} +\sqrt{{}^4g}\, {}^4T^{\mu\nu},\nonumber \\
&&{}_{(L)}\theta^{\mu\nu}={}_{(L)}\theta^{\nu\mu},\quad\quad \partial_{\nu}
\, {}_{(L)}\theta^{\mu\nu} \equiv 0,\nonumber \\
&&{}_{(L)}t^{\mu\nu}=2 \, {}^4g\, {}^4G^{\mu\nu}+{{c^3}\over {16\pi G}}
\partial_{\rho}\partial_{\sigma} {\cal T}^{\alpha\nu\beta\mu}.
\label{i30}
\end{eqnarray}

\noindent ${}_{(L)}\theta^{\mu\nu}$ is a tensor density of weight 2 and
${}_{(L)}t^{\mu\nu}$ is the Landau-Lifschitz pseudo-tensor [see Refs.
\cite{ll1} for some of its applications], which contains no second derivatives
of the metric and gives meaningful results only in an asymptotically flat
Cartesian coordinate system. It was found trying to rewrite
${}^4\nabla_{\mu}\, {}^4T^{\mu\nu}\, {\buildrel \circ \over =}\, 0$ in the
form $\partial_{\mu} [{}_{(L)}t^{\mu\nu}+\sqrt{{}^4g}\, {}^4T^{\mu\nu}]\,
{\buildrel \circ \over =}\, 0$. Starting from Einstein's equations
${}^4T^{\mu\nu}\, {\buildrel \circ \over =}\, {{c^3}\over {8\pi G}}({}^4R
^{\mu\nu}-{1\over 2}{}^4g^{\mu\nu}\, {}^4R)$, one rewrites them as\hfill\break
\hfill\break
${}_{(L)}t^{\mu\nu}+\sqrt{{}^4g}\, {}^4T^{\mu\nu}\, {\buildrel \circ \over =}\,
\partial_{\rho} h^{\mu [\nu\rho ]}$ with $h^{\mu [\nu\rho ]}=-h^{\mu [\rho\nu
]}={{c^3}\over {16\pi G}} \partial_{\sigma} [\sqrt{{}^4g} ({}^4g^{\mu\nu}\,
{}^4g^{\rho\sigma}-{}^4g^{\mu\rho}\, {}^4g^{\nu\sigma})]$, \hfill\break
\hfill\break
which imply
$\partial_{\nu}\partial_{\rho} h^{\mu [\nu\rho ]}=0$. Then one gets

\begin{eqnarray}
&&{}_{(L)}t^{\mu\nu}={{c^3}\over {16\pi G}} \sqrt{{}^4g} [(2\, {}^4\Gamma^{\rho}
_{\tau\sigma}\, {}^4\Gamma^{\gamma}_{\rho\gamma}-{}^4\Gamma^{\rho}_{\tau\gamma}
\, {}^4\Gamma^{\gamma}_{\sigma\rho}-{}^4\Gamma^{\rho}_{\tau\rho}\,
{}^4\Gamma^{\gamma}_{\sigma\gamma})({}^4g^{\mu\tau}\, {}^4g^{\nu\sigma}-
{}^4g^{\mu\nu}\, {}^4g^{\tau\sigma})+\nonumber \\
&&+{}^4g^{\mu\tau}\, {}^4g^{\sigma\rho}
({}^4\Gamma^{\nu}_{\tau\gamma}\, {}^4\Gamma^{\gamma}_{\sigma\rho}+{}^4\Gamma
^{\nu}_{\sigma\rho}\, {}^4\Gamma^{\gamma}_{\tau\gamma}-{}^4\Gamma^{\nu}
_{\rho\gamma}\, {}^4\Gamma^{\gamma}_{\tau\sigma}-{}^4\Gamma^{\nu}_{\tau\sigma}
\, {}^4\Gamma^{\gamma}_{\rho\gamma})+\nonumber \\
&&+{}^4g^{\nu\tau}\, {}^4g^{\sigma\rho}
({}^4\Gamma^{\mu}_{\tau\gamma}\, {}^4\Gamma^{\gamma}_{\sigma\rho}+
{}^4\Gamma^{\mu}_{\sigma\rho}\, {}^4\Gamma^{\gamma}_{\tau\gamma}-
{}^4\Gamma^{\mu}_{\rho\sigma}\, {}^4\Gamma^{\gamma}_{\tau\sigma}-
{}^4\Gamma^{\mu}_{\tau\sigma}\, {}^4\Gamma^{\gamma}_{\rho\gamma})+\nonumber \\
&&+{}^4g^{\tau
\sigma}\, {}^4g^{\rho\gamma} ({}^4\Gamma^{\mu}_{\tau\rho}\, {}^4\Gamma^{\nu}
_{\sigma\gamma}-{}^4\Gamma^{\mu}_{\tau\sigma}\, {}^4\Gamma^{\nu}_{\rho
\gamma})]=\nonumber \\
&&={{c^3}\over {16\pi G}} [\partial_{\rho}\, {}^4{\hat g}^{\mu\nu}\,
\partial_{\gamma}\, {}^4{\hat g}^{\gamma\rho}-\partial_{\rho}\, {}^4{\hat g}
^{\mu\rho}\partial_{\gamma}\, {}^4{\hat g}^{\nu\gamma}+{1\over 2}{}^4g^{\mu\nu}
\, {}^4g_{\rho\sigma}\, \partial_{\sigma}\, {}^4{\hat g}^{\rho\delta}\,
\partial_{\delta}\, {}^4{\hat g}^{\sigma\gamma}-\nonumber \\
&&-({}^4g^{\mu\rho}\, {}^4g_{\gamma
\delta} \partial_{\sigma}\, {}^4{\hat g}^{\nu\delta}\, \partial_{\rho}\,
{}^4{\hat g}^{\gamma\sigma}+{}^4g^{\nu\rho}\, {}^4g_{\gamma\delta} \partial
_{\sigma}\, {}^4{\hat g}^{\mu\delta} \, \partial_{\rho}\, {}^4{\hat g}^{\gamma
\sigma})+{}^4g_{\rho\sigma}\, {}^4g^{\gamma\delta} \partial_{\gamma}\,
{}^4{\hat g}^{\mu\rho} \partial_{\delta}\, {}^4{\hat g}^{\nu\sigma}+\nonumber \\
&&+{1\over 8}
(2\, {}^4g^{\mu\rho}\, {}^4g^{\nu\sigma}-{}^4g^{\mu\nu}\, {}^4g^{\rho\sigma})
(2\, {}^4g_{\gamma\delta}\, {}^4g_{\alpha\beta}-{}^4g_{\delta\alpha}\,
{}^4g_{\gamma\beta}) \partial_{\rho}\, {}^4{\hat g}^{\gamma\beta} \partial
_{\sigma}\, {}^4{\hat g}^{\delta\alpha}].
\label{i31}
\end{eqnarray}

The Einstein pseudo-tensor and Landau-Lifschitz complex are the only complexes
that are homogeneous quadratic in first derivatives of the metric tensor.
Therefore, both complexes can be made to vanish along any one geodesic by a
suitable choice of coordinates. For a spacetime that is asymptotically flat
either in spacelike or null directions, the Landau-Lifschitz and Einstein
superpotentials give exactly the same definition of energy-momentum. Both
require an asymptotically rectangular coordinate system to be meaningful. Once
defined, both transform as free vectors under the asymptotic symmetry group.

3) ``M$\o$ller's energy-momentum complex"\cite{lc18} -
It is obtained by ${}_{(K)}
U^{[\mu\nu ]}[\xi]$ by putting $\xi^{\mu}=const.$ [Komar found his
superpotential trying to generalize M$\o$ller's one]:

\begin{equation}
{}_{(M)}\theta_{\mu}{}^{\nu}=\partial_{\gamma}\, {}_{(M)}U_{\mu}{}
^{[\nu\gamma ]},\quad\quad {}_{(M)}U_{\mu}{}^{[\nu\gamma ]}={{c^3}\over
{8\pi G}}\sqrt{{}^4g}\,
{}^4g^{\nu\alpha}\, {}^4g^{\gamma\beta} (\partial_{\alpha}\, {}^4g_{\mu\beta}-
\partial_{\beta}\, {}^4g_{\mu\alpha}).
\label{i32}
\end{equation}

While all the previous pseudotensors depend on the metric and its first
derivatives, M$\o$ller complex also depends on the second derivatives of the
metric.

4) Lorentz\cite{lc12} chose $\theta^{\mu}_{(S)}[\xi ]=0$ and identified
${{c^3}\over {8\pi G}} {}^4G^{\mu\nu}\, {\buildrel \circ \over =}\, 0$ with
the energy momentum of the gravitational field in absence of matter
\cite{lc5}. This also was the point of view of Levi-Civita\cite{lc12a}.

Let us remark that in Ref.\cite{lc14} there is the following derivation of
three pseudotensors based only on the use of the Einstein equations.
Given Einstein's equations of motion ${{\partial {\cal L}_E}\over {\partial
\, {}^4g_{\mu\nu}}}-\partial_{\alpha}{{\partial {\cal L}_E}\over {\partial
\partial_{\alpha}\, {}^4g_{\mu\nu}}}\, {\buildrel \circ \over =}\, -{{8\pi G}
\over {c^3}}\, {}^4T^{\mu\nu}$ and defined the tensor density of weight +2,
\hfill\break
\hfill\break
${\cal T}^{\alpha\nu\beta\mu}={}^4{\hat g}^{\alpha\beta}\, {}^4{\hat g}
^{\mu\nu}-{}^4{\hat g}^{\alpha\mu}\, {}^4{\hat g}^{\beta\nu}=-{\cal T}
^{\nu\alpha\beta\mu}=-{\cal T}^{\alpha\nu\mu\beta}={\cal T}^{\beta\mu\alpha\nu}$
\hfill\break
\hfill\break
[${\cal T}^{\alpha\nu\beta\mu}+{\cal T}^{\alpha\beta\mu\nu}+{\cal T}^{\alpha
\mu\nu\beta}=0$], one has the following three ways of rewriting the
equations of motion:

1) the Landau-Lifschitz\cite{ll} weak conservation law

\begin{eqnarray}
&&-{1\over {2\sqrt{{}^4g}}} \partial_{\alpha}\partial_{\beta}{\cal T}^{\alpha\nu
\beta\mu}+(terms\, homogeneous\, quadratic\, in\, \partial \, {}^4g )\,
{\buildrel \circ \over =}\, -{{8\pi G}\over {c^3}}\, {}^4T^{\mu\nu},\nonumber \\
&&{{c^3}\over {16\pi G}} \partial_{\alpha}\partial_{\beta} {\cal T}^{\alpha\nu
\beta\mu}\, {\buildrel \circ \over =}\, {}_{(L)}t^{\mu\nu}+\sqrt{{}^4g}\,
{}^4T^{\mu\nu},\quad with\quad {}_{(L)}t^{\mu\nu}={}_{(L)}t^{\nu\mu},
\nonumber \\
&&\partial_{\nu} [{}_{(L)}t^{\mu\nu}+\sqrt{{}^4g}\, {}^4T^{\mu\nu}]\,
{\buildrel \circ \over =}\, 0;
\label{i24}
\end{eqnarray}

2) the Einstein weak conservation law: by taking the factor ${}^4g_{\mu\sigma}/
\sqrt{{}^4g}$ through $\partial_{\alpha}$ one gets

\begin{eqnarray}
&&-{1\over 2}\partial_{\alpha} [{{{}^4g_{\mu\sigma}}\over {\sqrt{{}^4g}}}
\partial_{\beta} {\cal T}^{\alpha\nu\beta\sigma}]
+(terms\, homogeneous\, quadratic\, in\, \partial \, {}^4g )\,
{\buildrel \circ \over =}\, -{{8\pi G}\over {c^3}}\, {}^4T^{\nu}_{\mu},
\nonumber \\
&&{{c^3}\over {16\pi G}} \partial_{\alpha}[{{{}^4g_{\mu\sigma}}\over
{\sqrt{{}^4g}}} \partial_{\beta} {\cal T}^{\alpha\nu\beta\sigma}]\,
{\buildrel \circ \over =}\, {}_{(E)}t^{\nu}_{\mu}+{}^4T^{\nu}_{\mu},
\nonumber \\
&&\partial_{\nu} [{}_{(E)}t^{\nu}_{\mu}+{}^4T^{\nu}_{\mu}]\, {\buildrel \circ
\over =}\, 0;
\label{i25}
\end{eqnarray}

3) the Bergmann-Thompson\cite{lc15} weak conservation law: by taking the
factor $1/\sqrt{{}^4g}$ through $\partial_{\alpha}$ one gets

\begin{eqnarray}
&&-{1\over 2}\partial_{\alpha}[{1\over {\sqrt{{}^4g}}} \partial_{\beta}
{\cal T}^{\alpha\nu\beta\mu}]
+(terms\, homogeneous\, quadratic\, in\, \partial \, {}^4g )\,
{\buildrel \circ \over =}\, -{{8\pi G}\over {c^3}} {}^4T^{\mu\nu},
\nonumber \\
&&{{c^3}\over {16\pi G}} \partial_{\alpha} [{1\over {\sqrt{{}^4g}}} \partial
_{\beta} {\cal T}^{\alpha\nu\beta\mu}]\, {\buildrel \circ \over =}\,
{}_{(BT)}t^{\mu\nu}+{}^4T^{\mu\nu},\quad\quad {}_{(BT)}t^{\mu\nu}\not=
{}_{(BT)}t^{\nu\mu},\nonumber \\
&&\partial_{\nu}({}_{(BT)}t^{\mu\nu}+{}^4T^{\mu\nu})\, {\buildrel \circ \over
=}\, 0,\quad\quad \partial_{\mu}({}_{(BT)}t^{\mu\nu}+{}^4T^{\mu\nu})\not= 0.
\label{i26}
\end{eqnarray}

A whole class of weak conservation equations may be obtained by taking the
factor $(\sqrt{{}^4g})^m$, for any integer m, with or without a factor
${}^4g_{\mu\sigma}$ through the $\partial_{\alpha}$. Some of these conservation
equations are among those considered by Goldberg\cite{lc16}. All the
conserved quantities in the class contain the source term ${}^4T^{\mu\nu}$
added to a homogeneous quadratic function of the first derivatives of the
metric; in a sense all the conservation equations in the class are equivalent
to one another.

The total energy contained in a finite or infinite 3-volume V with
2-dimensional boundary $\partial V$ is [$d^3\Sigma_{\mu}$ and $d^2\Sigma
_{\mu\nu}$ denote the volume and area elements of V and $\partial V$
respectively]

\begin{eqnarray}
&&\psi [\xi ]=\int_Vd^3\Sigma_{\mu}\, \theta^{\mu}_{(S)}[\xi ] =\int_{\partial
V}d^2\Sigma_{\mu\nu}\, U^{[\mu\nu ]}[\xi ]\equiv \nonumber \\
&&\equiv \int_Vd^3\Sigma_{\mu}\, [\theta^{\mu}_{(W)}[\xi ]+{}^4G^{\mu}{}_{\nu}\,
\xi^{\nu}]\, {\buildrel \circ \over =}\, \int_Vd^3\Sigma_{\mu}\, [\theta^{\mu}
_{(W)}[\xi ]+{}^4T^{\mu}{}_{\nu} \xi^{\nu}].
\label{i33}
\end{eqnarray}

\noindent Here $\psi [\xi ]$ is a scalar if $\theta^{\mu}_{(S)}[\xi ]$ is a
vector density as in Komar's case. However, this scalar does not characterize
intrinsically the region V unless we find a way of picking out a ``privileged"
$\xi^{\mu}$. Since the action $S_H$ is invariant under all the
one-parameter groups generated by a vector filed $\xi^{\mu}$, one gets an
infinity of weak conservation laws\cite{lc23}, as well as of strong ones.

If the spacetime is asymptotically flat, one can take for $\xi^{\mu}$ a vector
field which coincides with a Killing vector at spatial infinity (if the flat
spacetime is Minkowski one, one has 10 asymptotic Killing vectors
satisfying $\xi_{\mu ;\nu}+\xi_{\nu ;\mu}=\partial_{\mu}\xi_{\nu}+\partial_{\nu}
\xi_{\mu}=0$) and one can
show that $\psi [\xi ]=\int_{\partial V \rightarrow \infty}\, d^2\Sigma
_{\mu\nu}\, U^{[\mu\nu ]}[\xi ]$ does not depend on the choice of $\xi^{\mu}$ at
finite distances, so that 10 global quantities can be constructed
if $\xi^{\mu}(\vec x,x^o){\rightarrow}_{|\vec x|
\rightarrow \infty}\, a^{\mu}+\omega^{\mu}{}_{\nu}x^{\nu}$ [$\omega_{\mu\nu}=-
\omega_{\nu\mu}$]. Since $\xi^{\mu}$ generates an one-parameter group of
diffeomorphisms, here are hidden the restrictions that one has to impose on
the global structure of the diffeomorphism group.

For $\xi^{\mu}(x)\rightarrow a^{\mu}$ one gets $\psi [\xi]=a^{\mu}P_{\mu}$
with $P_{\mu}=\int_{\partial V}d^2\Sigma_{\alpha\beta}\, {}_{(B)}
U^{[\alpha\beta ]}
{}_{\mu}\, {\buildrel \circ \over =}\, \int_V d^3\Sigma_{\alpha}\, [\theta
^{\alpha}_{(W)}{}_{\mu}+{}^4T^{\alpha}{}_{\mu}]$ [if $\theta^{\alpha}_{(W)}=
\theta^{\alpha}_{(W)}{}_{\mu} \xi^{\mu}$] in the case of the Bergmann
superpotential. This formula holds in particular for the Einstein canonical
superpotential: ${}_{(E)}P_{\mu}[V]=\int_{\partial V}d^2\Sigma_{\alpha\beta}\,
{}_{(F)}U^{[\alpha\beta ]}{}_{\mu}\, {\buildrel \circ \over =}\, \int_V
d^3\Sigma_{\alpha}\, {}_{(E)}\theta_{\mu}{}^{\alpha}$, and the Einstein-Klein
theorem says that the surface integral is convergent in asymptotically flat
coordinates ${}^4g_{\mu\nu}={}^4\eta_{\mu\nu}+O(r^{-1})$, $\partial_{\gamma}\,
{}^4g_{\mu\nu}=O(r^{-2})$ [isolated nonradiating system], is independent
from the spacelike hypersurface $\Sigma$ containing the volume
V [the conservation law gives $\int_{V^{'}}
d^3\Sigma_{\alpha}\, {}_{(E)}\theta_{\mu}{}^{\alpha} -\int_V d^3\Sigma_{\alpha}
\, {}_{(E)}\theta_{\mu}{}^{\alpha}=\int_{S_{\infty}} d^3\Sigma_{\alpha}\,
{}_{(E)}t_{\mu}{}^{\alpha}=0$, because the matter is confined and,
with the assumed boundary conditions on the 4-metric, the Einstein pseudo-tensor
${}_{(E)}t_{\nu}{}^{\mu}$ is of order $O(r^{-4})$ and does not contribute with
a gravitational field 4-momentum; $S_{\infty}$ is a timelike 3-surface at
space infinity connecting $\Sigma^{'}$ and $\Sigma$ limits of $V^{'}$ and $V$
respectively], is unaltered by coordinate transformations
with the same asymptotic limit and that ${}_{(E)}P_{\mu}$ transforms as a
covariant vector under linear transformations. By using Cartesian coordinates
at infinity, one gets $P_o=m$ and $P_i=0$ for the Schwarzschild solution.

The previous asymptotic boundary conditions  on the 4-metric are satisfied by
the static fields produced by matter confined to a finite volume, but in general
they exclude the possibility of radiation. Comparison with electrodynamics
suggests that radiation fields in general relativity should be characterized by
$\partial_{\gamma}\, {}^4g_{\mu\nu}=O(r^{-1})$ rather that by $\partial_{\gamma}
\, {}^4g_{\mu\nu}=O(r^{-2})$. However, if the integral of ${}_{(E)}t_{\nu}{}
^{\mu}$ over $S_{\infty}$ does not vanish, the argument used to prove the
Einstein-Klein theorem is no longer valid and the meaning of $P_{\mu}[\Sigma ]$
becomes obscure. If the radiation goes on with a finite rate from $x^o=-\infty$
, one cannot even expect the integrals $P_{\mu}[\Sigma ]$ to be convergent
[the space is filled with an infinite amount of energy in the form of
radiation]. Nevertheless, if the system remains quiescent till, say, $x^o=0$,
then radiates for a while and again quiets down, one can give a reasonable
prescription for calculating the total energy and its rate of change with the
same procedure used in the linearized theory\cite{lc22}: one calculates the
energy in coordinates systems which asymptotically satisfy the harmonic
condition $\partial_{\beta} (\sqrt{{}^4g}\, {}^4g^{\alpha\beta})=0$
\cite{lc25}. Let us assume that the gravitational filed in question defines a
scalar field u(x) whose gradient, $k_{\mu}=\partial_{\mu} u$, is null and
diverging. This allows us to introduce a ``luminosity distance" r through
${}^4\nabla_{\mu} (r^{-2}\, k^{\mu})=0$ [see Bondi and Sachs\cite{lc26,lc26a}].
Now we can formulate the boundary conditions which constitute a generalization
of Sommerfeld's radiation conditions to gravitational fields: there exist
coordinate systems and functions $i_{\mu\nu}=O(r^{-1})$ such that ${}^
4g_{\mu\nu}={}^4\eta_{\mu\nu}+O(r^{-1})$, $\partial_{\gamma}\, {}^4g_{\mu\nu}=
i_{\mu\nu} k_{\gamma}+O(r^{-2})$, $(i_{\mu\nu}-{1\over 2} {}^4\eta_{\mu\nu}\,
{}^4\eta^{\gamma\delta}i_{\gamma\delta})=O(r^{-2})$. The expression $P_{\mu}(u)=
lim_{u=const.,r\rightarrow \infty}\, \int_{\partial V} d^2\Sigma_{\alpha\beta}
\, {}_{(F)}U_{\mu}{}^{\alpha\beta}$, if they converge, give the total energy
and momentum of the system as function of the retarded time u; $P_o$ is called
the Bondi mass [see Refs.\cite{lc26b}, where after a reformulation of the
Landau-Lifschitz complex in a manifest covariant way with a background metric
following Refs.\cite{lc26c,lc14}, its covariant formulation at null infinity in
asymptotically flat spacetimes is given and the Bondi 4-momentum is recovered].
Our boundary conditions ensure that the 1/r terms in the
integrand cancel out and so make plausible the existence of $P_{\mu}(u)$.
Moreover, the $P_{\mu}(u)$ are invariant under coordinate transformations which
preserve the boundary conditions and reduce to the identity for $r\rightarrow
\infty$. If radiation goes on only for a finite interval of time, $P_{\mu}
[\Sigma ]$ is well defined and equal to $P_{\mu}(u=-\infty)$. The pseudo-tensor
in the wave zone is given by the same expression as the canonical tensor in
the linear theory\cite{lc22}. Under the further assumption $\partial_{\gamma}
\partial_{\delta}\, {}^4g_{\mu\nu}=O(r^{-1})$, the previous boundary conditions
allow to prove that the 1/r part of the curvature tensor is of Petrov's type
II null.

Formally, the Komar superpotential differs from the Einstein pseudo-tensor: it
gives the same results at spatial infinity, but at null infinity it must be
modified in order to give the Bondi mass\cite{lc33,lc34} in the case of the Kerr
metric\cite{lc35}. As noted in Ref.\cite{lc36}, the conserved Komar quantities
$K=\int_{\Sigma}d^3\Sigma_{\mu}\, \partial_{\nu}\, {}_{(K)}U^{[\mu\nu ]}[\xi ]=
{{c^3}\over {8\pi G}} \int_{\partial \Sigma}d^2\Sigma_{\mu\nu} \sqrt{{}^4g}
[{}^4\nabla_{\mu} \xi^{\nu}-{}^4\nabla_{\nu} \xi^{\mu}]$ [$\Sigma$ is a
spacelike hypersurface] have the following properties: i) K is equal to the
mass M for Schwarzschild and Kerr black holes if $\xi^{\mu}=(1,0,0,0)$ in
coordinates $(x^o, r,\theta,\varphi )$; ii) for Kerr black holes and
$\xi^{\mu}=(0,0,0,1)$ in Boyer-Linquist coordinates, K is ``twice" the angular
momentum J [this is the ``anomalous factor 2"]; iii) for the radiating
asymptotic solution of Ref.\cite{lc27}, if $\xi^{\mu}$ is asymptotically
(1,0,0,0) in radiative coordinates ($u=x^o-r, r,\theta ,\varphi$), K is not
Bondi's mass $M(u,\theta )$ but rather $M+{1\over 2}c\dot c$ in Bondi's notation
and the correction to Komar's integral to get rid of the $c\dot c$ term has
been found in Ref.\cite{lc33}.

In Ref.\cite{lc14} by using a Minkowski background metric it is shown that, if
Trautman's boundary conditions are satisfied, the same 4-momentum $P_{\mu}
[\Sigma ]$ is obtained starting from the Einstein, the Landau-Lifschits and
Bergmann-Thomson\cite{lc15} pseudotensors [a system is defined as non-radiative
if the 4-momentum is the same for all spacelike hypersurfaces $\Sigma$]. Then,
in the second of Refs.\cite{lc14}, it is analyzed the case of the solution
corresponding to an isolated axi-symmetric system generating gravitational
waves found by Bondi, van der Burg and Metzner\cite{lc27}. These authors were
able to show that the behaviour of the system was fully determined by a single
function, the ``news" function, and initial conditions specified on the
light cone. They used a definition of mass of the system such that in the
static case the definition led to the correct quantity; their mass is the
Bondi mass. In the approach with the background metric one recovers the same
value of the mass [ due to symmetry, in this case there are only the energy
and the momentum along the symmetry axis]. One uses expansions such that an
outgoing radiation condition of Sommerfeld type {probably equivalent to the
Trautmann conditions] is automatically satisfied.

All this discussion of the conservation laws in the generally covariant approach
based on the Hilbert action is not directly connected with the weak and strong
ADM charges, as can be seen in the more recent review given in Ref.\cite{wald}
of the definitions of energy-momentum at spatial and null infinity in
asymptotically flat spacetimes. Essentially one has that the variation in time
of the Bondi energy at null infinity may be interpreted as defining the flux of
energy carried away to infinity by gravitational radiation and that this agrees
with the energy flux computed from the Landau-Lifschitz pseudotensor with an
appropriately chosen background and, then, with the energy flux of the
linearized theory.

Instead, in the case of a timelike Killing field $\xi^{\mu}$
asymptotically orthogonal to the Cauchy surface the strong ADM energy-moemtum
4-vector gives the same definition for the total mass of all stationary
asymptotically flat at spatial infinity spacetimes as one would get from the
Komar superpotential. More work will be needed to get a consistent picture
englobing all these properties for tetrad gravity. In particular, the results
of this paper show that one has to study in detail the effect of gravitational
radiation in the tetrad gravity reformulation of the Christodoulou and
Klainermann spacetimes and to find a bridge to the Bondi results at null
infinity. Since the natural formalism for discussing null infinity is the
Newman-Penrose one \cite{stewart}, in Appendix F we give the definition
of a set of null tetrads natural from the Hamiltonian point of view to be used
as a starting point to find the Hamiltonian version of the Newman-Penrose
formalism.

\vfill\eject

\section{4-tensors in void spacetimes.}

From Eqs.(103) and Appendices D and E of II, we get the following expression
for various 3- and 4-tensors in void spacetimes in the 3-orthogonal gauges
[$N(\tau ,\vec \sigma )\approx -\epsilon +n(\tau ,\vec \sigma )\approx -
\epsilon$, $N_r(\tau ,\vec \sigma )\approx {\hat n}_r(\tau ,\vec \sigma |q,
\rho ] \approx 0$ for $\rho (\tau ,\vec \sigma )\approx 0$; see Section VI for
a complete gauge fixing;
we use either $q$, $\rho$ or $\phi =e^{q/2}$, $\pi_{\phi}=2
\phi^{-1} \rho$ with $q=0$ ($\phi =1$) and $\rho =\pi_{\phi}=0$ as the
realization of the flat limit; when $\rho =0$ we have $n=0$]

\begin{eqnarray}
{}^3{\hat {\tilde \pi}}^r_{(a)}(\tau ,\vec \sigma )&=&{1\over 3} \int
d^3\sigma_1 {\cal K}^r_{(a)s}(\vec \sigma ,{\vec \sigma}_1;\tau |q,0] \rho
(\tau ,{\vec \sigma}_1)\, \rightarrow_{\rho \rightarrow 0}\, 0,\nonumber \\
{}^3{\hat K}_{rs}&=&{{\epsilon}\over {4k}} \sum_u(\delta_{rs}\delta_{(a)s}+
\delta_{su}\delta_{(a)r}-\delta_{rs}\delta_{(a)u}) {}^3{\hat {\tilde \pi}}^u
_{(a)}\, \rightarrow_{\rho \rightarrow 0}\, 0,\nonumber \\
{}^3{\hat \omega}_{r(a)}&=&\epsilon_{(a)(b)(c)} \delta_{(b)r} \delta_{(c)u}
\partial_uq\, \rightarrow_{q \rightarrow 0}\, 0,\nonumber \\
{}^3{\hat \Omega}_{rs(a)}&=& \epsilon_{(a)(b)(c)}\sum_u\delta
_{(c)u}[\delta_{(b)s} \partial_u\partial_rq-\delta_{(b)r} \partial_u\partial
_sq]+\nonumber \\
&+&{1\over 2}[\delta_{(a)(b)}\epsilon_{(c)(d)(e)}-\delta_{(a)(c)}\epsilon
_{(b)(d)(e)}+\delta_{(a)(d)}\epsilon_{(e)(c)(b)}-\delta_{(a)(e)}\epsilon
_{(d)(c)(b)}]\partial_uq\partial_vq\, {\rightarrow}_{q\, \rightarrow 0}\, 0,
\nonumber \\
{}^3{\hat R}_{rusv}&=&
 (\delta_{rv}\delta_{su}-\delta
_{rs}\delta_{uv}) e^{4q} \sum_n (\partial_nq)^2+\nonumber \\
&+&e^{2q}\{ \delta_{rv}[\partial_s\partial_uq-\partial_sq \partial_uq]-
\delta_{rs}[\partial_v\partial_uq-\partial_vq \partial_uq]+\nonumber \\
&+&\delta_{su}[\partial_v\partial_rq-\partial_vq \partial_rq]-\delta_{uv}
[\partial_s\partial_rq-\partial_sq \partial_rq] \} {\rightarrow}_{q\,
\rightarrow 0}\, 0,\nonumber \\
{}^3{\hat R}_{uv}&=& -\partial_u\partial_vq+
\partial_uq \partial_vq-\delta_{uv}e^{2q}\sum_n[2e^{2q}(\partial_nq)^2+
\partial^2_nq-(\partial_nq)^2]\, {\rightarrow}_{q\, \rightarrow 0}\, 0,
\nonumber \\
{}^3\hat R&=& -6\sum_u(\partial_uq)^2-
4e^{-2q}\sum_u[\partial_u^2q-(\partial_uq)^2]\, {\rightarrow}_{q\,
\rightarrow 0}\, 0,
\label{e1}
\end{eqnarray}

\begin{eqnarray}
{}^4{\hat \Gamma}^{\tau}_{\tau\tau}\, &{\buildrel \circ \over =}\,&
{1\over {N}} \partial_{\tau}N\, \rightarrow_{n \rightarrow 0, {\tilde \lambda}
_{\tau}\rightarrow \epsilon}\, 0 ,\nonumber \\
{}^4{\hat \Gamma}^{\tau}_{r\tau}&=&{}^4{\hat \Gamma}^{\tau}_{\tau r}={1\over
{N}} \partial_rN\, \rightarrow_{n \rightarrow 0}\, 0,\nonumber \\
{}^4{\hat \Gamma}^{\tau}_{rs}&=&{}^4{\hat \Gamma}^{\tau}_{sr}=-{{\epsilon}\over
{4kN}} \sum_u {}^3G_{o(a)(b)(c)(d)}\delta_{(a)r}
\delta_{(b)s}\delta_{(c)u}\, {}^3{\hat {\tilde \pi}}^u_{(d)}
\, \rightarrow_{\rho \rightarrow 0}\, 0,\nonumber \\
{}^4{\hat \Gamma}^u_{\tau\tau}\, &{\buildrel \circ \over =}\,&
\phi^{-4} N\partial_uN\, \rightarrow_{n \rightarrow 0}\, 0,\nonumber \\
{}^4{\hat \Gamma}^u_{r\tau}&=&{}^4{\hat \Gamma}^u_{\tau r}=-
{{\epsilon N}\over {4k}}\sum_{uv} \phi^{-4}
{}^3G_{o(a)(b)(c)(d)}\delta^u_{(a)}\delta
_{(b)r}\delta_{(c)v}\, {}^3{\hat {\tilde \pi}}^v_{(d)}
\, \rightarrow_{\rho \rightarrow 0}\, 0,\nonumber \\
{}^4{\hat \Gamma}^u_{rs}&=&{}^3{\hat \Gamma}^u_{rs}=-\delta_{uv}\sum_v\delta^u_v
\partial_vq+\delta^u_r\partial_sq+\delta^u_s\partial_rq
\, \rightarrow_{q \rightarrow 0}\, 0,\nonumber \\
&&{}\nonumber \\
{}^4{\hat {\buildrel \circ \over {\omega}}}_{\tau (o)(a)}&=&-{}^4{\hat
{\buildrel \circ \over {\omega}}}_{\tau (a)(o)}=-\epsilon \phi^{-2}
\sum_r\delta_{(a)r} \partial_rN\, \rightarrow_{n \rightarrow 0}\, 0
,\nonumber \\
{}^4{\hat {\buildrel \circ \over {\omega}}}_{\tau (a)(b)}&=&-{}^4{\hat
{\buildrel \circ \over {\omega}}}_{\tau (b)(a)}\,
{\buildrel \circ \over =}\, 0,\nonumber \\
{}^4{\hat {\buildrel \circ \over {\omega}}}_{r(o)(a)}&=&-{}^4{\hat
{\buildrel \circ \over {\omega}}}_{r(a)(o)}=-{1\over {4k}} \phi^{-2}\sum_u
{}^3G_{o(a)(b)(c)(d)}\delta_{(b)r}\delta_{(c)u}\, {}^3{\hat
{\tilde \pi}}^u_{(d)}\, \rightarrow_{\rho \rightarrow 0}\, 0,\nonumber \\
{}^4{\hat {\buildrel \circ \over {\omega}}}_{r(a)(b)}&=&-{}^4{\hat
{\buildrel \circ \over {\omega}}}_{r(b)(a)}=-\epsilon
{}^3{\hat \omega}_{r(a)(b)}\, \rightarrow_{q \rightarrow 0}\, 0,\nonumber \\
&&{}\nonumber \\
{}^4{\hat {\buildrel \circ \over {\Omega}}}_{rs(a)(b)}&=&-\epsilon
\, [{}^3{\hat \Omega}_{rs(a)(b)}+
{{\phi^{-4}}\over {4k}}\, {}^3G_{o(a)(c)(d)(e)}\, {}^3G_{o(b)(f)(g)(h)}\cdot
\nonumber \\
&&\sum_{uv}
(\delta_{(c)r}\delta_{(f)s}-\delta_{(c)s}\delta_{(f)r}) \delta_{(d)u}
\, {}^3{\hat {\tilde \pi}}^u_{(e)}\, \delta_{(g)v}\, {}^3{\hat {\tilde \pi}}^v
_{(h)}]\, \rightarrow_{q,\rho \rightarrow 0}\, 0,\nonumber \\
{}^4{\hat {\buildrel \circ \over {\Omega}}}_{rs(o)(a)}&=&{1\over N}\phi^{-4}
\sum_v \delta
_{(a)v}({}^4{\hat R}_{\tau vrs}-N_{(b)}\sum_u
\delta_{(b)u}\, {}^4{\hat R}_{uvrs})=\nonumber \\
&=&{1\over {4k}} \phi^{-2}\sum_u \delta_{(a)u} [({}^3G_{o(b)(c)(d)(e)}\,
\delta_{(b)r}\delta_{(c)u}\delta_{(d)v}\, {}^3{\hat {\tilde \pi}}^v_{(e)})_{|s}-
\nonumber \\
&-&({}^3G_{o(b)(c)(d)(e)}\,
\delta_{(b)s}\delta_{(c)u}\delta_{(d)v}\, {}^3{\hat {\tilde \pi}}^v_{(e)})
_{|r}]\, \rightarrow_{\rho \rightarrow 0}\, 0,\nonumber \\
{}^4{\hat {\buildrel \circ \over {\Omega}}}_{\tau r(a)(b)}\, &=&\phi^{-4}
\sum_{uv}\delta_{(a)u}\delta_{(b)v)}\, {}^4{\hat R}_{uv\tau r}
{\buildrel \circ \over =}\, \nonumber \\
&&{\buildrel \circ \over =}\, -\epsilon \{ \partial_{\tau}\, {}^3{\hat \omega}
_{r(a)(b)}+{1\over 2}(\epsilon_{(a)(b)(c)}\epsilon_{(d)(e)(f)}-\epsilon
_{(a)(b)(d)}\epsilon_{(c)(e)(f)})\cdot \nonumber \\
&&\sum_s \phi^{-2}\delta_{(c)s}
[{{\epsilon N}\over {4k}} \phi^{-2}\sum_v\, {}^3G_{o(d)(l)(m)(n)}\,
\delta_{(l)s}\delta_{(m)v}\, {}^3{\hat {\tilde \pi}}^v_{(n)}+\nonumber \\
&+&N_{(l)}\phi^2\sum_u
\delta_{(l)u}\partial_u(\delta_{(d)s}\phi^2)+\phi^2\sum_u
\delta_{(d)u}\partial_s(N_{(l)}\delta_{(l)u}\phi^{-2})+\nonumber \\
&+&\phi^2 \epsilon_{(d)(m)(n)}{\hat \mu}_{(m)}\delta_{(n)s}-
N_{(g)}\phi^{-2}\sum_u
\delta_{(g)u} \partial_u(\delta_{(d)s}\phi^2)-\nonumber \\
&-&\phi^2\sum_u
\delta_{(d)u}\partial_s(N_{(g)}\delta_{(g)u}\phi^{-2}
)] {}^3{\hat \omega}_{r(e)(f)}+N_{(c)}\phi^{-2}\delta_{(c)s}
\, [{}^3{\hat \omega}_s,{}^3{\hat \omega}_r]_{(a)(b)}+\nonumber \\
&+&{{\epsilon}\over {4k}} \phi^{-4}\sum_u
\, {}^3G_{o(c)(d)(e)(f)}\delta_{(c)r}\delta_{(e)u}\,
{}^3{\hat {\tilde \pi}}^u_{(f)}\nonumber \\
&&(\delta_{(a)(d)}\delta_{(b)u}-\delta_{(b)(d)}\delta_{(a)u}) \partial_uN+
\nonumber \\
&+&{1\over {(4k)^2}}(\delta_{(a)(l)}\delta_{(b)(d)}-\delta_{(a)(d)}
\delta_{(b)(l)}){}^3G_{o(d)(e)(f)(g)}\, {}^3G_{o(h)(l)(m)(n)}\cdot
\nonumber \\
&\cdot& \phi^{-6}\sum_{wv}\delta_{(h)r}
N_{(e)}\delta_{(f)w}\, {}^3{\hat {\tilde \pi}}^w_{(g)}\delta_{(m)v}\,
{}^3{\hat {\tilde \pi}}^v_{(n)} \}\, \rightarrow_{q,\rho \rightarrow 0}\, 0
,\nonumber \\
{}^4{\hat {\buildrel \circ \over {\Omega}}}_{\tau r(o)(a)}\, &{\buildrel \circ
\over =}\,& {1\over N}\, \phi^{-2}\sum_u
\delta_{(a)u}[{}^4{\hat R}_{\tau u\tau r}-N_{(b)}\phi^{-2}\sum_s
\delta_{(b)s}\, {}^4{\hat R}_{su\tau r}]\, {\buildrel \circ \over
=}\nonumber \\
&{\buildrel \circ \over =}&\, -\epsilon \, \phi^{-2} \sum_s
\delta_{(a)s}[\partial_{\tau}\,
{}^3{\hat K}_{rs} + N_{|s|r}-\nonumber \\
&-&{{\epsilon}\over {4k}}\sum_{uw}
\,{}^3G_{o(c)(d)(e)(f)}\delta_{(d)u}
\delta_{(e)w}\, {}^3{\hat {\tilde \pi}}^w_{(f)}\nonumber \\
&&\{ \delta_{(c)r}(N_{(b)}\phi^{-2}\delta_{(b)u})_{|s} + \delta_{(c)s}
(N_{(b)}\phi^{-2}\delta_{(b)u})_{|r} \} -\nonumber \\
&-&{{\epsilon}\over {4k}}\phi^2\sum_{usw} N_{(b)}\delta_{(b)u} \nonumber \\
&&({}^3G_{o(c)(d)(e)(f)}\,
\delta_{(c)s}\delta_{(d)u}\delta_{(e)w}\, {}^3{\hat {\tilde \pi}}^w_{(f)})
_{|r}\, ]\, \rightarrow_{q,\rho ,n \rightarrow 0}\, 0,\nonumber \\
&&{}\nonumber \\
{}^4{\hat R}_{rsuv}&=&\phi^4\delta_{(a)r}\delta_{(b)s}
\, {}^4{\hat {\buildrel \circ \over {\Omega}}}_{uv(a)(b)}=\nonumber \\
&&=-{}^3{\hat R}_{rsuv}+{{N^2}\over {16k^2}}\sum_{tw}
{}^3G_{o(a)(b)(c)(d)}\, {}^3G_{o(e)(f)(g)(h)}\nonumber \\
&&\cdot \delta_{(a)r}\delta_{(e)s}(\delta_{(b)u}\delta_{(f)v}-\delta_{(b)v}
\delta_{(f)u})\delta_{(c)t}\delta_{(g)w}\, {}^3{\hat {\tilde \pi}}^t_{(d)}\,
{}^3{\hat {\tilde \pi}}^w_{(h)}\, \rightarrow_{q,\rho \rightarrow 0}\,
0,\nonumber \\
{}^4{\hat R}_{\tau ruv}&=&N \phi^2\delta_{(a)r}\, {}^4{\hat {\buildrel \circ
\over {\Omega}}}_{uv(o)(a)}\, \rightarrow_{q,\rho \rightarrow 0}\,
0,\nonumber \\
{}^4{\hat R}_{\tau r\tau s}&=&N \phi^2\delta_{(a)r}\, {}^4{\hat {\buildrel
\circ \over {\Omega}}}_{\tau s(o)(a)}\, \rightarrow_{q,\rho \rightarrow 0}\,
0, \nonumber \\
&&{}\nonumber \\
{}^4{\hat R}_{\tau\tau}&=&-\epsilon \phi^{-4}\sum_r\, {}^4{\hat R}
_{r\tau r\tau}\, \rightarrow_{q,\rho \rightarrow 0}\, 0,\nonumber \\
{}^4{\hat R}_{\tau r}&=&{}^4{\hat R}_{r\tau}=-\epsilon \phi^{-4}\sum_u
\, {}^4{\hat R}_{u\tau ur}\, \rightarrow_{q,\rho \rightarrow 0}\, 0,\nonumber \\
{}^4{\hat R}_{rs}&=&{}^4{\hat R}_{sr}={{\epsilon}\over {N^2}}{}^4{\hat R}
_{\tau r\tau s}-\epsilon \phi^{-4}\sum_u\, {}^4{\hat R}_{urus}\, \rightarrow
_{q,\rho \rightarrow 0}\, 0,\nonumber \\
{}^4{\hat R}&=&{{\epsilon}\over {N^2}}{}^4{\hat R}_{\tau\tau}-\epsilon
\phi^{-4}\sum_r\, {}^4{\hat R}_{rr}\, \rightarrow_{q,\rho \rightarrow 0}\,
0,\nonumber \\
&&{}\nonumber \\
{}^4{\hat C}_{rsuv}&=&{}^4{\hat R}_{rsuv}+{{\epsilon}\over 2}[\phi^4
(\delta_{rv}\, {}^4{\hat R}_{su}-\delta_{ru}\, {}^4{\hat R}_{sv})+\nonumber \\
&&+\phi^4(\delta_{su}\,
{}^4{\hat R}_{rv}-\delta_{sv}\, {}^4{\hat R}_{ru})]+\nonumber \\
&&+{1\over 6}\phi^8
(\delta_{ru}\delta_{sv}-\delta_{rv}\delta_{su})
{}^4{\hat R}\, \rightarrow_{q,\rho \rightarrow 0}\, 0,\nonumber \\
{}^4{\hat C}_{\tau ruv}&=&{}^4{\hat R}_{\tau ruv}+{{\epsilon}\over 2}
\phi^4(\delta_{ru}\, {}^4{\hat R}_{\tau v}-\delta_{rv}\, {}^4{\hat R}_{\tau u})
\, \rightarrow_{q,\rho \rightarrow 0}\, 0,\nonumber \\
{}^4{\hat C}_{\tau r\tau s}&=&{}^4{\hat R}_{\tau r\tau s}+{1\over 2}(N^2\,
{}^4{\hat R}_{rs}-\epsilon \phi^4
\delta_{rs}\, {}^4{\hat R}_{\tau\tau})-\nonumber \\
&&-{1\over 6}N^2\phi^4\delta_{rs}\, {}^4\hat R\, \rightarrow_{q,\rho
\rightarrow 0}\, 0.
\label{e2}
\end{eqnarray}

Using Eqs.(69) and (70) of I, Ashtekar's variables become [we give the
limits for $\phi =e^{q/2}=1$ and $\rho =0$]

\begin{eqnarray}
&&{}^3{\tilde h}^r_{(a)}(\tau ,\vec \sigma )
\mapsto {}^3{\hat {\tilde h}}^r_{(a)}(\tau ,\vec \sigma )=\delta^r_{(a)}
\phi^4 (\tau ,\vec \sigma ) \rightarrow \delta^r_{(a)},\nonumber \\
&&{}\nonumber \\
&&{}^3A_{(a)r}(\tau ,\vec \sigma )
\mapsto {}^3{\hat A}_{(a)r}(\tau ,\vec \sigma )=
{1\over {6k}}\phi^4(\tau ,\vec \sigma )\nonumber \\
&&\int d^3\sigma_1\, {\cal K}^r_{(a)s}(\vec \sigma ,{\vec \sigma}_1,\tau |
\phi ,0]\, [\phi^{-2}\rho](\tau ,{\vec \sigma}_1)+\nonumber \\
&+&2i\epsilon_{(a)(b)(c)}\delta_{(b)r}\delta_{(c)u}
\partial_uln\, \phi (\tau ,\vec \sigma )\rightarrow {1\over {6k}}.
\label{e3}
\end{eqnarray}

\vfill\eject

\section{$\Sigma_{\tau}$-adapted Null Tetrads.}

The 3+1 splitting of $M^4$ with spacelike slices $\Sigma_{\tau}$ not only
identifies a timelike vector field, namely the unit normal to $\Sigma_{\tau}$
[see Eqs.(40) of I; $l^A=\epsilon \, {}^4_{(\Sigma )}{\check {\tilde E}}^A
_{(o)}={{\epsilon}\over N} (1; -{}^3e^r_{(a)} N_{(a)} )$, ${}^4g_{AB}l^Al^B=
\epsilon$, $l_A={}^4_{(\Sigma )}{\check {\tilde E}}^{(o)}_A=(N; \vec 0)$; by
construction it is surface forming: ${1\over N} l_Ad\sigma^A$ is a closed
differential 1-form], but
also a spacelike vector field ${\cal N}^A$ tangent to $\Sigma_{\tau}$ by
means of the shift functions:\hfill\break
\hfill\break
${\cal N}^A={{N_{(a)}}\over {\sqrt{\sum_{(c)}N^2_{(c)}}}}\, {}^4_{(\Sigma
)}{\check {\tilde E}}^A_{(a)} ={1\over {\sqrt{\sum_{(c)}N^2_{(c)}}}}\, (0;
{}^3e^r_{(a)}N_{(a)})\quad\quad$, ${}^4g_{AB}{\cal N}^A{\cal N}^B=-\epsilon$,
\hfill\break
${\cal N}_A=-\epsilon N_{(a)}\, {}^4_{(\Sigma )}{\check {\tilde E}}^{(a)}_A=
-\epsilon (\sqrt{\sum_{(c)}N^2_{(c)}} ; {}^3e_{(a)r} {{N_{(a)}}\over
{\sqrt{\sum_{(c)}N^2_{(c)}}}} )$.\hfill\break
\hfill\break
The two directions $l^A$, ${\cal N}^A$ are intrinsically selected by the gauge
nature of the lapse and shift functions in every 3+1 splitting.

Let us remark that the vector field ${\cal N}^A(\tau ,\vec \sigma )$ is not in
general surface forming, namely the associated differential 1-form ${\cal
N}(\tau ,\vec \sigma )={\cal N}
_A(\tau ,\vec \sigma ) d\sigma^A$ is not proportional to a closed 1-form. Since
we have \hfill\break
\hfill\break
$d\Big[ -{{\epsilon {\cal N}}\over {\sqrt{\sum_{(c)}N^2_{(c)}} }} \Big] =
\partial_r \Big[ {{{}^3e_{(a)s} N_{(a)}}\over { \sum_{(c)}N_{(c)}^2}} \Big]
d\sigma^r\wedge d\sigma^s$,\hfill\break
\hfill\break
its vanishing implies\hfill\break
\hfill\break
$\partial_{\tau} {{N_r}\over {{\vec N}^2}} =0\quad\quad$, $\partial_r {{N_s}
\over {{\vec N}^2}} = \partial_s {{N_r}\over {{\vec N}^2}}$,\hfill\break
\hfill\break
with $N_r={}^3e_{(a)r} N_{(a)}$, ${\vec N}^2=\sum_{(c)}N_{(c)}^2={}^3g^{rs}
N_rN_s$. Therefore, the condition for having ${\cal N}^A(\tau ,\vec \sigma )$
surface-forming (zero vorticity) is the choice of a coordinate system on
$\Sigma_{\tau}$ such that Eq.(\ref{V2a}) [for $N_r=n_r$, $N_{(as)r}=0$] implies
\hfill\break
\hfill\break
$N_r={\vec N}^2 \partial_r f$ with $\partial_{\tau} f=0$.\hfill\break
\hfill\break
3-orthogonal coordinates do not imply this property of ${\cal N}^A$. In the
coordinate systems for $\Sigma_{\tau}$ in which  ${\cal N}^A$ is surface
forming there is a second foliation of $M^4$ with timelike hypersurfaces
$\Xi_{\zeta}$ (if $\zeta$ is the parameter labelling the leaves: ${\cal N}=k
d\zeta$), and the intersection $\Sigma_{\tau} \cap \Xi_{\zeta}=S_{\tau \zeta}$
is a 2-surface whose tangent space in each point is a 2-plane spanned by the two
spacelike vectors perpendicular to $l^A$ and ${\cal N}^A$ in that point (see
later on the vector field $M^A$, ${\bar M}^A$). Therefore, in these special
coordinate systems for $\Sigma_{\tau}$ one could perform a 2+2 decomposition
of $M^4$ along the lines of Refs.\cite{inverno1}. The study of the
Shanmugadhasan canonical transformation and of the
superhamiltonian constraint in these coordinates should allow to identify the
analog of the natural gauge fixing $\rho \approx 0$ in the 3-orthogonal gauges.

Therefore, in each point of $\Sigma_{\tau}$ we can select two orthogonal vectors
${\buildrel \circ \over V}^{(\alpha )}$ and $S^{(\alpha )}$ in the tangent
plane:\hfill\break
\hfill\break
one timelike ${\buildrel \circ \over V}^{(\alpha )}=l^A(\tau ,\vec \sigma )\,
{}^4_{(\Sigma )}{\check {\tilde E}}^{(\alpha )}_A(\tau ,\vec \sigma )=(1; \vec
0)\quad\quad\quad$,
${}^4\eta_{(\alpha )(\beta )} {\buildrel \circ \over V}^{(\alpha )}
{\buildrel \circ \over V}^{(\beta )}=\epsilon$,\hfill\break
\hfill\break
and the other spacelike $S^{(\alpha )}(\tau ,\vec \sigma )={\cal N}^A(\tau
,\vec \sigma )\, {}^4_{(\Sigma )}{\check {\tilde E}}^{(\alpha )}_A(\tau ,\vec
\sigma )=(0; {{N_{(a)}}\over {\sqrt{\sum_{(c)}N^2_{(c)}}}})(\tau ,\vec \sigma
),$
${}^4\eta_{(\alpha )(\beta )}[S^{(\alpha )}S^{(\beta )}](\tau ,\vec \sigma )=
-\epsilon$, ${}^4\eta_{(\alpha )(\beta )} {\buildrel \circ \over V}^{(\alpha )}
S^{(\beta )}(\tau ,\vec \sigma )=0$.\hfill\break
\hfill\break
Then in each point of $\Sigma_{\tau}$ we can define the following two null
tangent vectors

\begin{eqnarray}
{\cal K}^{(\alpha )}(\tau ,\vec \sigma )&=&\sqrt{{1\over 2}\sum_{(c)}N^2
_{(c)}(\tau ,\vec \sigma )}\Big[ {\buildrel \circ \over V}^{(\alpha )}+
S^{(\alpha )}(\tau ,\vec \sigma )\Big]={1\over {\sqrt{2}}}(\sqrt{\sum_{(c)}
N^2_{(c)}}; N_{(a)})(\tau ,\vec \sigma )=\nonumber \\
&=&\sqrt{{1\over 2}\sum_{(c)}N^2_{(c)}(\tau ,\vec \sigma )}\Big( l^A+{\cal N}^A
\Big)(\tau ,\vec \sigma )\, {}^4_{(\Sigma )}{\check {\tilde E}}^{(\alpha )}_A
(\tau ,\vec \sigma ),\nonumber \\
{\cal L}^{(\alpha )}(\tau ,\vec \sigma )&=&{1\over {\sqrt{2} \sum_{(c)}N^2
_{(c)}(\tau ,\vec \sigma )}}\Big[{\buildrel \circ \over V}^{(\alpha )}-
S^{(\alpha )}(\tau ,\vec \sigma )\Big]=\nonumber \\
&=&{1\over {\sqrt{2} \sum_{(c)}N^2
_{(c)}(\tau ,\vec \sigma )}}(\sqrt{\sum_{(c)}
N^2_{(c)}}; -N_{(a)})(\tau ,\vec \sigma )=\nonumber \\
&=&{1\over {\sqrt{2} \sum_{(c)}N^2_{(c)}(\tau ,\vec \sigma )}}
\Big( l^A-{\cal N}^A
\Big)(\tau ,\vec \sigma )\, {}^4_{(\Sigma )}{\check {\tilde E}}^{(\alpha )}_A
(\tau ,\vec \sigma ),\nonumber \\
&&{}\nonumber \\
&&{}^4\eta_{(\alpha )(\beta )}[{\cal K}^{(\alpha )}{\cal K}^{(\beta )}](\tau
,\vec \sigma )={}^4\eta_{(\alpha )(\beta )}[{\cal L}^{(\alpha )}{\cal L}
^{(\beta )}](\tau ,\vec \sigma )=0,\nonumber \\
&&{}^4\eta_{(\alpha )(\beta )}[{\cal K}^{(\alpha )}{\cal L}^{(\beta )}](\tau
,\vec \sigma )=\epsilon.
\label{f1}
\end{eqnarray}

The null vector ${\cal K}^{(\alpha )}(\tau ,\vec \sigma )$ may be obtained
from the reference vector ${\buildrel {(o)} \over {\cal K}}^{(\alpha )}=
\omega (1;001)$ [$\omega$ is a constant with the dimensions of the shift
functions] by means of the standard Wigner helicity boost\cite{photon}
[$(\lambda )=(1),(2)$]

\begin{eqnarray}
{}_HL^{(\alpha )}{}_{(\beta )}({\cal K},{\buildrel {(o)} \over {\cal K}})&=&
\left( \begin{array}{c}
{1\over 2}({{\sqrt{\sum_{(c)}N^2_{(c)}} }\over {\omega}}+{{\omega}\over
{\sqrt{\sum_{(c)}N^2_{(c)}}}})\\
{1\over 2}({{\sqrt{\sum_{(c)}N^2_{(c)}}}\over {\omega}}-{{\omega}\over
{\sqrt{\sum_{(c)}N^2_{(c)}}}})
{{N_{(\lambda )}}\over {\sqrt{\sum_{(c)}N^2_{(c)}}}}\\
{1\over 2}({{\sqrt{\sum_{(c)}N^2_{(c)}}}\over {\omega}}-{{\omega}\over
{\sqrt{\sum_{(c)}N^2_{(c)}}}})
{{N_{(3)}}\over {\sqrt{\sum_{(c)}N^2_{(c)}}}}
\end{array} \right.\nonumber \\
&&\left. \begin{array}{cc}
0& {1\over 2}({{\sqrt{\sum_{(c)}N^2_{(c)}}}\over {\omega}}-{{\omega}\over
{\sqrt{\sum_{(c)}N^2_{(c)}}}})\\
\delta^{(\lambda )}_{(\lambda^{'})}-{{ N_{(\lambda )}N_{(\lambda^{'})} }\over
{\sqrt{\sum_{(c)}N^2_{(c)}}(\sqrt{\sum_{(c)}N^2_{(c)}}+N_{(3)})}}&
{1\over 2}({{\sqrt{\sum_{(c)}N^2_{(c)}}}\over {\omega}}+{{\omega}\over
{\sqrt{\sum_{(c)}N^2_{(c)}}}})
{{N_{(\lambda )}}\over {\sqrt{\sum_{(c)}N^2_{(c)}}}}\\
-{{N_{(\lambda^{'})}}\over {\sqrt{\sum_{(c)}N^2_{(c)}}}}&
{1\over 2}({{\sqrt{\sum_{(c)}N^2_{(c)}}}\over {\omega}}+{{\omega}\over
{\sqrt{\sum_{(c)}N^2_{(c)}}}})
{{N_{(3)}}\over {\sqrt{\sum_{(c)}N^2_{(c)}}}}
\end{array} \right) ,\nonumber \\
&&{}\nonumber \\
&&{}\nonumber \\
&&{\cal K}^{(\alpha )}=
{}_HL^{(\alpha )}{}_{(\beta )}({\cal K},{\buildrel {(o)} \over {\cal K}})
{\buildrel {(o)} \over {\cal K}}^{(\beta )}.
\label{f2}
\end{eqnarray}

The columns of ${}_HL^{(\alpha )}{}_{(\beta )}({\cal K},{\buildrel {(o)} \over
{\cal K}})$ define a flat helicity tetrad\hfill\break
\hfill\break
 ${}_H\epsilon^{(\alpha )}_{(\tilde
\alpha )}(N_{(c)}(\tau ,\vec \sigma ))={}_HL^{(\alpha )}{}_{(\beta )=(\tilde
\alpha )}({\cal K},{\buildrel {(o)} \over {\cal K}})(\tau ,\vec \sigma )$,
\hfill\break
\hfill\break
${}^4\eta^{(\alpha )(\beta )}={}_H\epsilon^{(\alpha )}_{(\tilde \alpha )}\,
{}^4\eta^{(\tilde \alpha )(\tilde \beta )}\, {}_H\epsilon^{(\beta )}_{(\tilde
\beta )}$,\hfill\break
 such that

\begin{eqnarray}
{\cal K}^{(\alpha )}(\tau ,\vec \sigma )&=&{{\omega}\over {\sqrt{2}}}\Big[
{}_H\epsilon^{(\alpha )}_{(\tilde o)}+ {}_H\epsilon^{(\alpha )}_{(\tilde 3)}
\Big] (\tau ,\vec \sigma ),\nonumber \\
{\cal L}^{(\alpha )}(\tau ,\vec \sigma )&=&{1\over {\sqrt{2}\omega}}\Big[
{}_H\epsilon^{(\alpha )}_{(\tilde o)}- {}_H\epsilon^{(\alpha )}_{(\tilde 3)}
\Big] (\tau ,\vec \sigma ),\nonumber \\
&&{}\nonumber \\
{}^4\eta^{(\alpha )(\beta )}&=& 2\Big[ {\cal K}^{(\alpha )}{\cal L}^{(\beta )}+
{\cal K}^{(\beta )}{\cal L}^{(\alpha )}\Big] (\tau ,\vec \sigma ) -\sum
^{(\tilde 2)}_{(\tilde \lambda )=(\tilde 1)}\Big[ {}_H\epsilon^{(\alpha )}
_{(\tilde \lambda )}\, {}_H\epsilon^{(\beta )}_{(\tilde \lambda )}\Big]
(\tau ,\vec \sigma ).
\label{f3}
\end{eqnarray}

With the transverse helicity polarization vectors ${}_H\epsilon^{(\alpha )}
_{(\tilde \lambda )}(\tau ,\vec \sigma )$, in each point we can build
circular complex polarization vectors and then a null tetrad

\begin{eqnarray}
{\cal M}^{(\alpha )}(\tau ,\vec \sigma )&=& {}_H\epsilon^{(\alpha )}
_{(-)}(\tau ,\vec \sigma )={1\over {\sqrt{2}}}\Big[ {}_H\epsilon^{(\alpha )}
_{(\tilde 1)}-i\, {}_H\epsilon^{(\alpha )}_{(\tilde 2)}\Big] (\tau ,\vec
\sigma ),\nonumber \\
{\bar {\cal M}}^{(\alpha )}(\tau ,\vec \sigma )&=& {}_H\epsilon^{(\alpha )}
_{(+)}(\tau ,\vec \sigma )={1\over {\sqrt{2}}}\Big[ {}_H\epsilon^{(\alpha )}
_{(\tilde 1)}+i\, {}_H\epsilon^{(\alpha )}_{(\tilde 2)}\Big] (\tau ,\vec
\sigma ),\nonumber \\
&&{}\nonumber\\
{}^4\eta^{(\alpha )(\beta )}&=&2 \Big[ {\cal K}^{(\alpha )}{\cal L}^{(\beta )}+
{\cal K}^{(\beta )}{\cal L}^{(\alpha )}-\Big( {\cal M}^{(\alpha )}{\bar {\cal
M}}^{(\beta )}+{\cal M}^{(\beta )}{\bar {\cal M}}^{(\alpha )}\Big)
\Big] (\tau ,\vec \sigma ) .
\label{f4}
\end{eqnarray}

See Ref.\cite{photon} for the covariance properties of the polarization
vectors ${}_H\epsilon^{(\alpha )}_{(\tilde \lambda )}(\tau ,\vec \sigma )$
under Lorentz transformations and Ref.\cite{stewart} for the associated
transformation properties of the null tetrad.

Now we can build a helicity $\Sigma_{\tau}$-adapted tetrad in $\Sigma_{\tau}$-
adapted coordinates in each point of $M^4$

\begin{eqnarray}
{}^4_{H(\Sigma )}{\check {\tilde E}}^A_{(\tilde \alpha )}(\tau ,\vec \sigma )&=&
{}_H\epsilon^{(\alpha )}_{(\tilde \alpha )}(\tau ,\vec \sigma )\,\, {}^4
_{(\Sigma )}{\check {\tilde E}}^A_{(\alpha )},\nonumber \\
&&{}\nonumber \\
{}^4g^{AB}(\tau ,\vec \sigma )&=&{}^4\eta^{(\tilde \alpha )(\tilde \beta )}
\Big[ {}^4_{H(\Sigma )}{\check {\tilde E}}^A_{(\tilde \alpha )}\, \,
{}^4_{H(\Sigma )}{\check {\tilde E}}^B_{(\tilde \beta )}\Big] (\tau ,\vec
\sigma ),\nonumber \\
&&{}\nonumber \\
{}^4_{H(\Sigma )}{\check {\tilde E}}^A_{(\tilde o )}(\tau ,\vec \sigma )&=&
{1\over 2}\Big[ \Big( {{ \sqrt{\sum_{(c)}N^2_{(c)}} }\over {\omega}}+
{{\omega}\over {\sqrt{\sum_{(c)}N^2_{(c)}}}}\Big) \epsilon l^A+\nonumber \\
&+&\Big(
{{\sqrt{\sum_{(c)}N^2_{(c)}}}\over {\omega}}-{{\omega}\over {\sqrt{\sum_{(c)}
N^2_{(c)}}}}\Big) {\cal N}^A\Big] (\tau ,\vec \sigma )\nonumber \\
&&{\rightarrow}_{\omega =\sqrt{\sum_{(c)}N^2_{(c)}}}\quad\quad \epsilon l^A
(\tau ,\vec \sigma ),\nonumber \\
{}^4_{H(\Sigma )}{\check {\tilde E}}^A_{(\tilde 3 )}(\tau ,\vec \sigma )&=&
{1\over 2}\Big[ \Big( {{ \sqrt{\sum_{(c)}N^2_{(c)}} }\over {\omega}}-
{{\omega}\over {\sqrt{\sum_{(c)}N^2_{(c)}}}}\Big) \epsilon l^A+\nonumber \\
&+&\Big(
{{\sqrt{\sum_{(c)}N^2_{(c)}}}\over {\omega}}+{{\omega}\over {\sqrt{\sum_{(c)}
N^2_{(c)}}}}\Big) {\cal N}^A\Big] (\tau ,\vec \sigma )\nonumber \\
&&{\rightarrow}_{\omega =\sqrt{\sum_{(c)}N^2_{(c)}}}\quad\quad {\cal N}^A
(\tau ,\vec \sigma ),\nonumber \\
{}^4_{H(\Sigma )}{\check {\tilde E}}^A_{(\tilde \lambda )}(\tau ,\vec \sigma )
&=& {}^4_{(\Sigma )}{\check {\tilde E}}^A_{(a)=(\tilde \lambda )}(\tau ,\vec
\sigma )-{{ N_{(\tilde \lambda )}[{\cal N}^A+{}^4_{(\Sigma )}{\check {\tilde E}}
^A_{(3)} }\over {\sqrt{\sum_{(c)}N^2_{(c)}}+N_{(3)} }}
(\tau ,\vec \sigma ),\nonumber \\
&&{}\nonumber \\
&&{}\nonumber \\
l^A(\tau ,\vec \sigma )&=&
{{\epsilon}\over 2}\Big[ \Big( {{ \sqrt{\sum_{(c)}N^2_{(c)}} }\over {\omega}}+
{{\omega}\over {\sqrt{\sum_{(c)}N^2_{(c)}}}}\Big)
{}^4_{H(\Sigma )}{\check {\tilde E}}^A_{(\tilde o )}+\nonumber \\
&+&\Big(
{{\sqrt{\sum_{(c)}N^2_{(c)}}}\over {\omega}}-{{\omega}\over {\sqrt{\sum_{(c)}
N^2_{(c)}}}}\Big) {}^4_{H(\Sigma )}{\check {\tilde E}}^A_{(\tilde 3 )}
\Big] (\tau ,\vec \sigma )\nonumber \\
&&{\rightarrow}_{\omega =\sqrt{\sum_{(c)}N^2_{(c)}}}\quad\quad \epsilon \, {}^4
_{H(\Sigma )}{\check {\tilde E}}^A_{(\tilde o)}
(\tau ,\vec \sigma ),\nonumber \\
{\cal N}^A(\tau ,\vec \sigma )&=&
{1\over 2}\Big[ \Big( {{ \sqrt{\sum_{(c)}N^2_{(c)}} }\over {\omega}}-
{{\omega}\over {\sqrt{\sum_{(c)}N^2_{(c)}}}}\Big)
{}^4_{H(\Sigma )}{\check {\tilde E}}^A_{(\tilde o )}+\nonumber \\
&+&\Big(
{{\sqrt{\sum_{(c)}N^2_{(c)}}}\over {\omega}}+{{\omega}\over {\sqrt{\sum_{(c)}
N^2_{(c)}}}}\Big) {}^4_{H(\Sigma )}{\check {\tilde E}}^A_{(\tilde 3 )}
\Big] (\tau ,\vec \sigma )\nonumber \\
&&{\rightarrow}_{\omega =\sqrt{\sum_{(c)}N^2_{(c)}}}\quad\quad {}^4_{H(\Sigma )}
{\check {\tilde E}}^A_{(\tilde 3)}(\tau ,\vec \sigma ),
\label{f5}
\end{eqnarray}

\noindent and then a natural intrinsic [i.e. dictated by canonical tetrad
gravity itself] null tetrad (to be used for doing the transition to the
Newman-Penrose formalism\cite{stewart})

\begin{eqnarray}
L^A(\tau ,\vec \sigma )&=&{1\over {\sqrt{2}}}
\Big[ {}^4_{H(\Sigma )}{\check {\tilde E}}^A_{(\tilde o )}+
{}^4_{H(\Sigma )}{\check {\tilde E}}^A_{(\tilde 3 )}\Big] (\tau ,\vec \sigma )=
{{ \sqrt{\sum_{(c)}N^2_{(c)}}}\over {\sqrt{2}\omega}} (\epsilon l^A+
{\cal N}^A)(\tau ,\vec \sigma )=\nonumber \\
&=&{1\over {\sqrt{2} \omega N}} \Big( \sqrt{\sum_{(c)}N^2_{(c)}} ;
(N-\sqrt{\sum_{(c)}N^2_{(c)}}) {}^3e^r_{(a)}N_{(a)} \Big) (\tau ,\vec \sigma )
\nonumber \\
&&{\rightarrow}_{\omega =\sqrt{\sum_{(c)}N^2_{(c)}}}\quad\quad {1\over
{\sqrt{2}}}[\epsilon l^A+{\cal N}^A](\tau ,\vec \sigma )=\nonumber \\
&=&{1\over {\sqrt{2} N \sqrt{\sum_{(c)}N^2_{(c)}} }} \Big( \sqrt{\sum_{(c)}N^2
_{(c)}} ; (N-\sqrt{\sum_{(c)}N^2_{(c)}}) {}^3e^r_{(a)}N_{(a)} \Big) (\tau
,\vec \sigma ),\nonumber \\
K^A(\tau ,\vec \sigma )&=&{1\over {\sqrt{2}}}
\Big[ {}^4_{H(\Sigma )}{\check {\tilde E}}^A_{(\tilde o )}-
{}^4_{H(\Sigma )}{\check {\tilde E}}^A_{(\tilde 3 )}\Big] (\tau ,\vec \sigma )=
{{\omega} \over {\sqrt{2\sum_{(c)}N^2_{(c)}}}}(\epsilon l^A-{\cal N}^A)
(\tau ,\vec \sigma )=\nonumber \\
&=&{{\omega}\over {\sqrt{2} N \sum_{(c)}N^2_{(c)}}} \Big( \sqrt{\sum_{(c)}N^2
_{(c)}} ; -(N+\sqrt{\sum_{(c)}N^2_{(c)}}) {}^3e^r_{(a)}N_{(a)} \Big) (\tau
\vec \sigma ) \nonumber \\
&&{\rightarrow}_{\omega =\sqrt{\sum_{(c)}N^2_{(c)}}}\quad\quad {1\over
{\sqrt{2}}}[\epsilon l^A-{\cal N}^A](\tau ,\vec \sigma )=\nonumber \\
&=&{1\over {\sqrt{2} N \sqrt{\sum_{(c)}N^2_{(c)}}}} \Big( \sqrt{\sum_{(c)}N^2
_{(c)}} ; -(N+\sqrt{\sum_{(c)}N^2_{(c)}}) {}^3e^r_{(a)}N_{(a)} \Big) (\tau
\vec \sigma ),\nonumber \\
M^A(\tau ,\vec \sigma )&=&{}^4_{H(\Sigma )}{\check {\tilde E}}^A_{(-)}(\tau
,\vec \sigma )={1\over {\sqrt{2}}} \Big[{}^4_{H(\Sigma )}{\check {\tilde E}}
^A_{(\tilde 1 )}-i\, {}^4_{H(\Sigma )}{\check {\tilde E}}^A_{(\tilde 2 )}
\Big] (\tau ,\vec \sigma )=\nonumber \\
&=&(0; {}^3e^r_{(-)}-{{N_{(-)}}\over {\sqrt{\sum_{(c)}N^2_{(c)}}+N_{(3)}}}
[{}^3e^r_{(a)}{{N_{(a)}}\over {\sqrt{\sum_{(c)}N^2_{(c)}}}}+{}^3e^r_{(3)}])
(\tau ,\vec \sigma ),\nonumber \\
{\bar M}^A(\tau ,\vec \sigma )&=&
{}^4_{H(\Sigma )}{\check {\tilde E}}^A_{(+)}(\tau
,\vec \sigma )={1\over {\sqrt{2}}} \Big[{}^4_{H(\Sigma )}{\check {\tilde E}}
^A_{(\tilde 1 )}+i\, {}^4_{H(\Sigma )}{\check {\tilde E}}^A_{(\tilde 2 )}
\Big] (\tau ,\vec \sigma )=\nonumber \\
&=&(0; {}^3e^r_{(+)}-{{N_{(+)}}\over {\sqrt{\sum_{(c)}N^2_{(c)}}+N_{(3)}}}
[{}^3e^r_{(a)}{{N_{(a)}}\over {\sqrt{\sum_{(c)}N^2_{(c)}}}}+{}^3e^r_{(3)}])
(\tau ,\vec \sigma ),\nonumber \\
&&{}\nonumber \\
&&{}\nonumber \\
{}^4g^{AB}(\tau ,\vec \sigma )&=&2 \Big[ L^AK^B+L^BK^A-(M^A{\bar M}^B+M^B{\bar
M}^A)\Big] (\tau ,\vec \sigma ),\nonumber \\
&&{}\nonumber \\
&&{}\nonumber \\
L_A(\tau ,\vec \sigma )&=&
{{ \sqrt{\sum_{(c)}N^2_{(c)}}}\over {\sqrt{2}\omega}} (\epsilon l_A+
{\cal N}_A)(\tau ,\vec \sigma )=\nonumber \\
&=&{{\epsilon \sqrt{\sum_{(c)}N^2_{(c)}}}\over {\sqrt{2}\omega}} (N-\sqrt{\sum
_{(c)}N^2_{(c)}}; -{}^3e_{(3)r})(\tau ,\vec \sigma ),\nonumber \\
K_A(\tau ,\vec \sigma )&=&
{{\omega} \over {\sqrt{2\sum_{(c)}N^2_{(c)}}}}(\epsilon l_A-{\cal N}_A)
(\tau ,\vec \sigma )=\nonumber \\
&=&{{\epsilon \sqrt{\sum_{(c)}N^2_{(c)}}}\over {\sqrt{2}\omega}} (N+\sqrt{\sum
_{(c)}N^2_{(c)}}; +{}^3e_{(3)r})(\tau ,\vec \sigma ),\nonumber \\
M_A(\tau ,\vec \sigma )&=&-\epsilon (0; {}^3e_{(-)r}-
{{N_{(-)}}\over {\sqrt{\sum_{(c)}N^2_{(c)}}+N_{(3)}}}
[{}^3e_{(a)r}{{N_{(a)}}\over {\sqrt{\sum_{(c)}N^2_{(c)}}}}+{}^3e_{(3)r}])
(\tau ,\vec \sigma ),\nonumber \\
{\bar M}_A(\tau ,\vec \sigma )&=&-\epsilon (0; {}^3e_{(+)r}-
{{N_{(+)}}\over {\sqrt{\sum_{(c)}N^2_{(c)}}+N_{(3)}}}
[{}^3e_{(a)r}{{N_{(a)}}\over {\sqrt{\sum_{(c)}N^2_{(c)}}}}+{}^3e_{(3)r}])
(\tau ,\vec \sigma ),\nonumber \\
&&{}\nonumber \\
{}^4g_{AB}(\tau ,\vec \sigma )&=&2[L_AK_B+L_BK_A-(M_A{\bar M}_B+M_B{\bar M}_A)
](\tau ,\vec \sigma ),\nonumber \\
&&{}\nonumber \\
ds^2&=&2 [ \theta^{(L)}\otimes \theta^{(K)}+\theta^{(K)}\otimes \theta^{(L)}-
(\theta^{(M)}\otimes \theta^{(\bar M)}+\theta^{(\bar M)}\otimes
\theta^{(M)})],\nonumber \\
&&\theta^{(L)}=L_Ad\sigma^A,\quad \theta^{(K)}=K_Ad\sigma^A,\quad
\theta^{(M)}=M_Ad\sigma^A,\quad  \theta^{(\bar M)}={\bar M}_Ad\sigma^A,
\label{f6}
\end{eqnarray}

Then, for $\omega =\sqrt{\sum_{(c)}N^2_{(c)}}$ we get

\begin{eqnarray}
{}^3g_{rs}&=&-2[ L_rK_s+L_sK_r -(M_r{\bar M}_s+M_s{\bar M}_s)]=\nonumber \\
&=&-2[l_rl_s-({\cal N}_r{\cal N}_s+M_r{\bar M}_s+M_s{\bar M}_r)]=\nonumber \\
&=&2[{\cal N}_r{\cal N}_s+M_r{\bar M}_s+M_s{\bar M}_r],\nonumber \\
&&{}\nonumber \\
ds^2&=&\epsilon [ \theta^{(\tau ) }\otimes \theta^{(\tau )}-2(\theta
^{({\cal N})}\otimes \theta^{({\cal N})}+\theta^{(M)}\otimes
\theta^{(\bar M)}+\theta^{(\bar M)}\otimes \theta^{(M)}) ],\nonumber \\
&&\theta^{(\tau )}=Nd\tau ,\quad\quad \theta^{({\cal N})}={\cal N}_r(d\sigma^r+
N^rd\tau ),\quad\quad \theta^{(M)}=M_r(d\sigma^r+N^rd\tau ).
\label{f7}
\end{eqnarray}

In the 3-orthogonal gauges we have

\begin{eqnarray}
M_r&=& -\epsilon \phi^2 e^{{1\over {\sqrt{3}}}\sum_{\bar a}\gamma_{\bar ar}
r_{\bar a}} \Big[ {1\over {\sqrt{2}}} (\delta_{(1)r}-i\delta_{(2)r})-
\nonumber \\
&-&{{N_{(-)}}\over {\sqrt{\sum_{(c)}N^2_{(c)}}+N_{(3)}}}(\delta_{(a)r}
{{N_{(a)}}\over {\sqrt{\sum_{(c)}N^2_{(c)}}}}+\delta_{(3)r}) \Big] .
\label{f8}
\end{eqnarray}

Given the previous null tetrad one could find an associated spin basis (see
Appendix C) $o^{\tilde A}$, $i^{\tilde A}$ for the Newman-Penrose formalism,
 so to visualize its 3+1 decomposition and its dependence on
the Hamiltonian gauge variables.

Then, from the line element $-2(\theta^{(M)}\otimes \theta^{(\bar M)}+\theta
^{(\bar M)}\otimes \theta^{(M)})$, modulo a 2-conformal factor, one should
identify a 2-metric ${}^2g_{\bar a\bar b}$, which in suitable adapted
coordinates should depend only on the two new canonical variables $r^{'}_{\bar
a}$ of the gravitational field resulting from the Shanmugadhasan canonical
transformation applied to these 3-coordinates for $\Sigma_{\tau}$ (the $r^{'}
_{\bar a}$'s would be functions only of $\tau$ and of the 2-coordinates on
$S_{\tau \zeta}$, since there would be no dynamics in $\zeta$). This would
simultaneously implement the ideas of Ref.\cite{inverno} and give an explicit
realization of the statement of Christodoulou-Klainermann\cite{ckl} on the
independent degrees of freedom of the gravitational field.

Instead to make contact with the Newman-Penrose formalism for studying the
asymptotic behaviour of the gravitational field at null infinity, one should
look for a coordinate system on $\Sigma_{\tau}$ and for a gauge fixing to the
superhamiltonian constraint such that the resulting lapse and shift functions
$N=-\epsilon +n$, $N_r=n_r$ imply the existence of a foliation of $M^4$ with a
one-parameter family of null hypersurfaces ${\cal Z}_u$ labelled by a parameter
$u$ (a retarded time) such that $L_Ad\sigma^A$ is proportional to $du$ (see
Ref.\cite{stewart}). For instance, by asking \hfill\break
\hfill\break
$du =d\tau - {{{}^3e_{(a)r}
d\sigma^r}\over {N-\sqrt{\sum_{(c)}N^2_{(c)}} }}$, \hfill\break
\hfill\break
we get that $N$ and $N_{(a)}$ must satisfy \hfill\break
\hfill\break
$\partial_{\tau} {{{}^3e_{(a)r}}\over {N-\sqrt{\sum_{(c)}N^2_{(c)}}}}\quad
\quad\quad$, $\partial_r {{{}^3e_{(a)s}}\over {N-\sqrt{\sum_{(c)}N^2_{(c)}}}}=
\partial_s {{{}^3e_{(a)r}}\over {N-\sqrt{\sum_{(c)}N^2_{(c)}}}}$.

Let us finally remark that the definitions of ${\cal N}^A(\tau ,\vec \sigma )$,
of the null tetrad and of the previous construction become singular in
synchronous 4-coordinates [$N_{(a)}=0$].

\vfill\eject

\section{Connection with the Post-Newtonian Approximation.}

Since we are working in the 3-orthogonal gauges, we can easily make contact
with the ``post-Newtonian approximation" of general relativity in the recent
formulation of Refs.\cite{dam,dam1}. In this formulation one defines four
potentials $V$, $V_r$, starting from the 4-metric of metric gravity: in
$\Sigma_{\tau}$-adapted coordinates we have [we make the rescaling $\tau = c
\bar \tau$]

\begin{eqnarray}
{1\over {c^2}}\, {}^4g_{\tau\tau}&=& {}^4{\tilde g}_{\bar \tau \bar \tau}=
{{\epsilon}\over {c^2}} (N^2-{}^3g_{rs}N^rN^s)=
{{\epsilon}\over {c^2}} (N^2-{}^3g^{rs}N_rN_s)=\nonumber \\
&=&\epsilon e^{-{2\over {c^2}} V} =\epsilon [1-{2\over {c^2}} V +O(c^{-4})],
\nonumber \\
&&{}\nonumber \\
{1\over c}\, {}^4g_{\tau r}&=& {}^3{\tilde g}_{\bar \tau r}=-{{\epsilon}\over c}
\, {}^3g_{rs}N^s =- {{\epsilon}\over c} N_r =\epsilon {4\over {c^3}} V_r,
\nonumber \\
&&{}\nonumber\\
{}^4g_{rs} &=& -\epsilon \, {}^3g_{rs} =-\epsilon e^{{2\over {c^2}} V} \,
{}^3\gamma_{rs},\nonumber \\
&&{}\nonumber \\
&&\Downarrow \nonumber \\
&&{}\nonumber \\
V &=& -\Phi = - {{c^2}\over 2} ln\, {{N^2/c^2}\over {1+{1\over {c^2}}\,
{}^3\gamma^{rs}N_rN_s }}, (Newton\,\, potential)\nonumber \\
&&{}\nonumber \\
V_r &=& -{{c^2}\over 4} N_r, (gravitomagnetic\,\, potential).
\label{g1}
\end{eqnarray}

The study of Einstein's equations [with the time variables rescaled by $c$]
${}^4G^{\mu\nu}\, {\buildrel \circ \over =}\,
{{8\pi G}\over {c^4}} T^{\mu\nu}$ with a matter energy-momentum tensor
satisfying the post-Newtonian assumptions $T^{oo}=O(c^2)$, $T^{oi}=O(c)$,
$T^{ij}=O(c^o)$, implies that, independently from the choice of a coordinate
system for either $M^4$ or $\Sigma_{\tau}$, the 3-curvature of the auxiliary
3-metric $\gamma_{rs}$ is $O(c^{-4})$. Therefore, we can always choose
3-coordinates [algebraic spatial isotropy condition of Ref.\cite{dam1}:
$-{}^4g_{\bar \tau \bar \tau}\, {}^4{\tilde g}_{rs}=\delta_{rs}+O(c^{-4})$;
it contains both the harmonic and the standard post-Newtonian gauges] such that

\begin{eqnarray}
{}^3\gamma_{rs} &=& \delta_{rs} + O(c^{-4}),\nonumber \\
&&{}\nonumber \\
&&\Downarrow \nonumber \\
&&{}\nonumber \\
V &=& -{{c^2}\over 2} ln\, {{N^2/c^2}\over { 1+{1\over {c^2}} \delta^{rs}
N_rN_s+O(c^{-6})}};
\label{g2}
\end{eqnarray}

\noindent namely, such that ${}^3\gamma_{rs}$ is a flat 3-metric at the first
post-Newtonian approximation.

Since the 3-orthogonal gauges are a particular class of these 3-coordinate
systems, we get the following form of the potentials $V$, $V_r$ in our
rest-frame instant form of tetrad gravity without matter in the special
3-orthogonal gauge with $\rho (\tau ,\vec \sigma )\approx 0$ [see Eqs.
(\ref{co1})-(\ref{co6})  in the Conclusions]

\begin{eqnarray}
V_r(\tau ,\vec \sigma )&=&-{{c^3}\over 4} {\hat n}_r(\tau ,\vec \sigma
|r_{\bar a},\pi_{\bar a},\phi [r_{\bar a},\pi_{\bar a}]],\nonumber \\
&&{}\nonumber \\
V(\tau ,\vec \sigma )&=&-\Phi (\tau ,\vec \sigma )=-{{c^2}\over 2} ln\,
{{(-\epsilon +{\hat n})^2/c^2}\over
{1+{1\over {c^2}} \delta^{rs}{\hat n}_r{\hat n}_s
+O(c^{-6})}}(\tau ,\vec \sigma
|r_{\bar a},\pi_{\bar a},\phi [r_{\bar a},\pi_{\bar a}]],\nonumber \\
&&{}\nonumber \\
&&or\,\, (from\, {}^4g_{rs})\nonumber \\
&&{}\nonumber \\
V(\tau ,\vec \sigma )&=&-\Phi (\tau ,\vec \sigma )=c^2 \Big[ {1\over {\sqrt{3}}}
\sum_{\bar a}\gamma_{\bar ar} r_{\bar a}+2 ln\, \phi [r_{\bar a},\pi_{\bar a}]
\Big] (\tau ,\vec \sigma ) +O(c^{-2}).
\label{g3}
\end{eqnarray}

\noindent The two expressions of $V$ should agree at the first post-Newtonian
approximation, but it is not possible to make the check in absence of the
explicit knowledge of $\phi$, $\hat n$, ${\hat n}_r$ [moreover let us
remember that both $\hat n$ and ${\hat n}_r$ depend on both $G/c^3$ and $c^3/G$
simultaneously and this will complicate the check].

Instead, the relation with Sections 6 and 7 of Ref.\cite{paur} [with the field
$\Theta =1$ by rescaling the absolute time] is based on the equations of that
paper describing the Galileo generally covariant formulation of Newtonian
gravity as a limit $c\, \rightarrow \, \infty$ on the ADM action of metric
gravity. The starting point is the following parametrization of the 4-metric
(we show only the 26 terms which appear in the Newtonian action)

\begin{eqnarray}
{1\over {c^2}}\, {}^4g_{\tau\tau}&=& {}^4{\tilde g}_{\bar \tau \bar \tau}=
\epsilon [1-{{2A_o}\over {c^2}}+{{2\alpha_o}\over {c^4}}+O(c^{-6})]=
\epsilon e^{-{2\over {c^2}} V},\nonumber \\
&&{}\nonumber \\
{1\over c}\, {}^4g_{\tau r}&=& {}^4{\tilde g}_{\bar \tau r}=-\epsilon [-{{A_r}
\over c}-{{\alpha_r}\over {c^3}}+O(c^{-5})]=-\epsilon \Big( -{4\over {c^3}} V_r
\Big) ,\nonumber \\
&&{}\nonumber \\
{}^4g_{rs}&=&-\epsilon \, {}^3g_{rs} =-\epsilon [{}^3{\check g}_{rs}+{{{\check
\gamma}_{rs}}\over {c^2}}+{{{\check \beta}_{rs}}\over {c^4}}+O(c^{-6})]=
-\epsilon e^{{2\over {c^2}} V} \, {}^3\gamma_{rs},\nonumber \\
&&{}\nonumber \\
&&\Downarrow \nonumber \\
&&{}\nonumber \\
N^2&=&c^2 -2 A+{2\over {c^2}}(\alpha_o-{}^3{\check g}^{rs}\alpha_rA_s-{1\over
2}{\check \gamma}_{rs}\, {}^3{\check g}^{rm}\, {}^3{\check g}^{sn}A_mA_n)+
O(c^{-4}),\nonumber \\
&&A=A_o-{1\over 2}\, {}^3{\check g}^{rs}A_rA_s,\quad\quad {}^3{\check g}^{ru}\,
{}^3{\check g}_{us}=\delta^r_s,\nonumber \\
N_r&=&A_r+{1\over {c^2}} \alpha_r +O(c^{-4}).
\label{g4}
\end{eqnarray}

The final action with general Galileo covariance depends on the 26 fields
$A_o$, $\alpha_o$, $A_r$, $\alpha_r$, ${}^3{\check g}_{rs}$, ${\check
\gamma}_{rs}$, ${\check \beta}_{rs}$. There are 18 first class constraints and
8 pairs of second class ones. It turns out that $\alpha_o$, $A_r$, $\alpha_r$,
three components of ${}^3{\check g}_{rs}$, one component of the momentum
conjugate to ${}^3{\check g}_{rs}$, the trace ${\check \beta}^T$ and the
longitudinal ${\check \beta}^L_r$ parts of ${\check \beta}_{rs}$ in its TT
decomposition, and the longitudinal ${\check \gamma}^L_r$ part of ${\check
\gamma}_{rs}$ are Hamiltonian gauge variables, while $A_o$ (the Newton
potential) and the remaining components of ${}^3{\check g}_{rs}$, ${\check
\gamma}_{rs}$, ${\check \beta}_{rs}$ are determined, together with their
conjugate momenta, by the second class constraints. There are no propagating
dynamical degrees of freedom. The gauge variables describe the inertial forces
in arbitrary accelerated not-Galilean reference frames.

The post-Newtonian approximation of Refs.\cite{dam,dam1} implies the choices
\hfill\break
\hfill\break
$A_r=0,\quad\quad\quad$ ${}^3{\check g}_{rs}=\delta_{rs},\quad\quad\quad$
${\check \gamma}_{rs}=2 A_o \delta_{rs}$,\hfill\break
\hfill\break
which are consistent with the previous gauge freedom (it is a possible gauge
of the general Galileo covariant description of Newtonian gravity). Moreover,
we have\hfill\break
\hfill\break
$V=-\Phi =A_o+{1\over {c^2}} (\alpha_o-2A^2_o)+O(c^{-4})=A_o+O(c^{-2}),$
\hfill\break
\hfill\break
$V_r =-{1\over 4} \alpha_r +O(c^{-2}),$\hfill\break
\hfill\break
${}^3\gamma_{rs}=\delta_{rs}+{1\over {c^4}}[{\check \beta}_{rs}-2(\alpha_o+
2A^2_o)\delta_{rs}]+O(c^{-6})=\delta_{rs}+O(c^{-4}).$

\vfill\eject

\end{document}